\documentclass[a4paper,11pt]{article}

\usepackage[margin=1.5cm]{geometry}

\usepackage{mathptmx}

\usepackage[small,compact]{titlesec}
\usepackage{array}
\usepackage{amsmath}
\usepackage{amssymb}
\usepackage{tabularx}
\usepackage{graphicx}
\usepackage[comma,sort]{natbib}
\usepackage{setspace}
\usepackage[labelfont=bf]{caption}
\usepackage{upgreek}
\usepackage{listings}
\usepackage{longtable}
\usepackage{supertabular}
\usepackage{units}
\usepackage{booktabs}
\usepackage{multicol}
\usepackage{pifont}
\usepackage[dvipsnames]{xcolor}
\usepackage{subcaption}
\usepackage{booktabs}
\usepackage{wrapfig}
\usepackage{floatrow}
\PassOptionsToPackage{hyphens}{url}\usepackage[colorlinks=true, urlcolor=blue, linkcolor=blue, citecolor=blue]{hyperref}
\usepackage[export]{adjustbox}
\usepackage{soul}
\usepackage[symbol]{footmisc}
\usepackage{flafter}
\usepackage{outlines}
\usepackage[most]{tcolorbox}
\usepackage[all]{nowidow}
\usepackage{ragged2e}
\usepackage{enumitem}
\usepackage{pdfpages}

\appto{\bibfont}{\small}
\appto{\bibsetup}{\raggedright}

\titleformat*{\section}{\LARGE\bfseries}
\titleformat*{\subsection}{\Large\bfseries}
\titleformat*{\subsubsection}{\large\bfseries}

\titleformat{\paragraph}
{\normalfont\normalsize\bfseries}{\theparagraph}{1em}{}
\titlespacing*{\paragraph}
{0pt}{3.25ex plus 1ex minus .2ex}{1.5ex plus .2ex}

\titlespacing*{\section}
  {0pt}{1\baselineskip}{1\baselineskip}

\titlespacing*{\subsection}
  {0pt}{1\baselineskip}{1\baselineskip}

\titlespacing*{\subsubsection}
  {0pt}{1\baselineskip}{1\baselineskip}

\newtcolorbox[auto counter]{mybox}[2][]{float,title={Case Study: #2},#1}

\makeatletter
\newcommand*{\citelinktext}[2]{%
  \hyper@@link[cite]{}{cite.#1}{#2}}
\makeatother

\newcommand{\mnras}{MNRAS}
\newcommand{\aap}{A\&A}
\newcommand{\nat}{Nature}

\newcommand{\apj}{ApJ}
\newcommand{\apjs}{ApJS}
\newcommand{\apjl}{ApJL}
\newcommand{\aj}{AJ}
\newcommand{\pasp}{PASP}
\newcommand{\araa}{ARA\&A}
\newcommand{\icarus}{Icarus}
\newcommand{\jcap}{Journal of Cosmology and Astroparticle Physics}

\newcommand{\qjras}{QJRAS}
\newcommand{\solphys}{Solar Physics}
\newcommand{\grl}{Geophysics Research Letters}

\newcommand{\aapr}{Astronomy and Astrophysics Reviews}
\newcommand{\physrep}{Physics Reports}
\newcommand{\ore}{Open Research Europe}
\newcommand{\ssr}{Space Science Reviews}
\newcommand{\ao}{Applied Optics}

\begin{document}

\includepdf{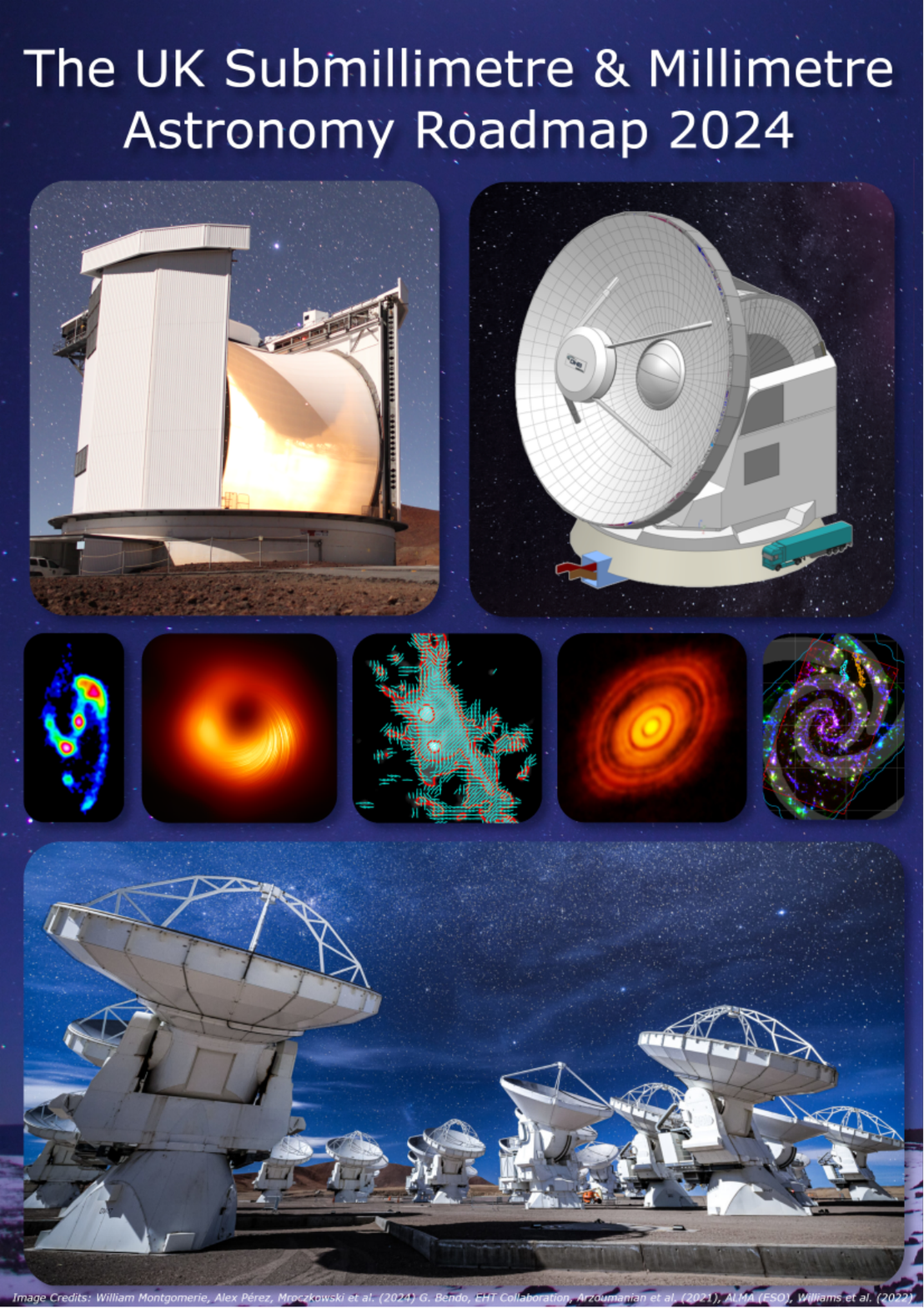}

%\frontmatter
%\maketitle
\pagenumbering{roman}

\begin{center}

\vspace*{5cm}

{\Huge The UK Submillimetre and Millimetre Astronomy Roadmap 2024}

\vspace{2cm}

{\Large August 2024}

\vspace{2cm}

\parbox{8cm}{\centering
\textit{\large  ``A whole field of astrophysics essentially founded by the UK, and where the UK leads in both science and instrumentation''}}  
\footnote[2]{Community quote from our UK submillimetre consultation, November 2023}

\vspace{3cm}

K.~Pattle$^{1}$, P.~S. Barry$^{2}$, A.~W. Blain$^{3}$, M.~Booth$^{4}$, R.~A.~Booth$^{5}$, D.~L.~Clements$^{6}$, M.~J.~Currie$^{7}$, S.~Doyle$^{2}$,
D.~Eden$^{8}$, G.~A.~Fuller$^{9}$, M.~Griffin$^{2}$, P.~G.~Huggard$^{7}$, J.~D.~Ilee$^{5}$, J.~Karoly$^{1}$, Z.~A.~Khan$^{1}$, N.~Klimovich$^{10}$, E.~Kontar$^{11}$, P.~Klaassen$^{4}$, A.~J.~Rigby$^{5}$, P.~Scicluna$^{12}$, S.~Serjeant$^{13}$, B.-K.~Tan$^{10}$, D.~Ward-Thompson$^{14}$, T.~G.~Williams$^{10}$, T.~A.~Davis$^{2}$, J.~Greaves$^{2}$, R.~Ivison$^{15,16,17}$, J.~Marin$^{9}$, M.~Matsuura$^{2}$, J.~M.~C.~Rawlings$^{1}$, A.~Saintonge$^{1}$, G.~Savini$^{1}$, M.~W.~L.~Smith$^{2}$, and D.~J.~Taylor$^{18}$

\vspace{0.1\baselineskip}

on behalf of the UK Submillimetre and Millimetre Astronomy Community

\end{center}

\clearpage

\vspace*{2cm}

\textbf{Affiliations}\\
$^{1}$ Department of Physics and Astronomy, University College London, London WC1E 6BT, UK\\
$^{2}$ Cardiff Hub for Astrophysics Research \& Technology, School of Physics \& Astronomy, Cardiff University, Queens Buildings, Cardiff, CF24 3AA, UK\\
$^{3}$ School of Physics and Astronomy, University of Leicester, Leicester LE1 7RH, UK\\
$^{4}$ UK Astronomy Technology Centre, Royal Observatory Edinburgh, Blackford Hill, Edinburgh EH9 3HJ, UK\\
$^{5}$ School of Physics and Astronomy, University of Leeds, Leeds LS2 9JT, UK\\
$^{6}$ Imperial College London, Blackett Laboratory, Prince Consort Road, London SW7 2AZ, UK\\
$^{7}$ RAL Space, STFC Rutherford Appleton Laboratory, Didcot OX11 0QX, UK\\
$^{8}$ Armagh Observatory and Planetarium, College Hill, Armagh BT61 9DB, UK\\
$^{9}$ Jodrell Bank Centre for Astrophysics, Department of Physics \& Astronomy, The University of Manchester, Manchester M13 9PL, UK\\
$^{10}$ Department of Physics (Astrophysics), University of Oxford, Denys Wilkinson Building, Keble Road, Oxford OX1 3RH, UK\\
$^{11}$ School of Physics \& Astronomy, University of Glasgow, Kelvin Building, Glasgow G12 8QQ, UK\\
$^{12}$ Centre for Astrophysics Research, University of Hertfordshire, Hatfield, UK\\
$^{13}$ School of Physical Sciences, The Open University, Walton Hall, Milton Keynes MK7 6AA, UK\\
$^{14}$ Jeremiah Horrocks Institute, University of Central Lancashire, Preston PR1 2HE, UK\\
$^{15}$ Institute for Astronomy, University of Edinburgh, Blackford Hill, Edinburgh EH9 3HJ, UK\\
$^{16}$ School of Cosmic Physics / Astronomy and Astrophysics Section, Dublin Institute for Advanced Studies, 31 Fitzwilliam Place, Dublin 2, D02 XF86, Republic of Ireland\\
$^{17}$ European Southern Observatory, Karl Schwarzschild Strasse 2, D-85748 Garching, Munich, Germany\\
$^{18}$ Centre for Extragalactic Astronomy, Department of Physics, Durham University, South Road, Durham DH1 3LE, UK\\

\vspace{2cm}

\textbf{Acknowledgements:} We thank Prof. Richard Ellis (UCL) and Prof. Serena Viti (U. Leiden) for constructive feedback on draft versions of this Roadmap.  We thank Ana Duarte Cabral (Cardiff), Rob Fender (Oxford), Hugh Hudson (Glasgow) and Olja Panic (Leeds) for helpful input.  K.P. is a Royal Society University Research Fellow, supported by grant number URF\textbackslash R1\textbackslash211322.

\vspace{2cm}

{\small \textit{\textbf{Cover image:} Top: the current and proposed future single-dish submillimetre facilities: the JCMT (left) and the AtLAST concept design (right).  Bottom: the ALMA interferometer. %The UK plays a leading role in JCMT and ALMA science and technology upgrades and aims to remain involved in the future as well as develop a leading role in the construction of AtLAST and/or its instruments. 
The middle row shows science results from these facilities. Far left: the dust emission of the M66 galaxy, observed by JCMT/SCUBA-2. Centre left: the magnetic field structure of the black hole M87, as imaged by the EHT. Centre: the magnetic field in the massive star forming region NGC 6334, observed by JCMT/SCUBA-2+POL-2.  Centre right: the HL Tau protoplanetary disc, observed by ALMA.  Far right: the cold molecular gas distribution in NGC 628 (in blue), observed by ALMA.  See front cover for image credits.}
}

\renewcommand{\thefootnote}{\arabic{footnote}}

\clearpage

\tableofcontents

\clearpage

\setcounter{page}{0}
\pagenumbering{arabic}

%\addcontentsline{toc}{section}{Executive Summary}
\phantomsection 
\addcontentsline{toc}{section}{Executive Summary}
\section*{Executive Summary}
%\addcontentsline{toc}{section}{\protect\numberline{}Executive Summary}
\label{sec:exec_summary}

In this Roadmap, we present a vision for the future of submillimetre and millimetre astronomy in the United Kingdom over the next decade and beyond.  This Roadmap has been developed in response to the recommendation of the Astronomy Advisory Panel (AAP) of the STFC in the AAP Astronomy Roadmap 2022. 

In order to develop our stragetic priorities and recommendations, we surveyed the UK submillimetre and millimetre community to determine their key priorities for both the near-term and long-term future of the field.  We further performed detailed reviews of UK leadership in submillimetre/millimetre science and instrumentation.

Our key strategic priorities are as follows:
\begin{enumerate}[leftmargin=!]
    \item The UK must be a key partner in the forthcoming AtLAST telescope, for which it is essential that the UK remains a key partner in the JCMT in the intermediate term.
    \item The UK must maintain, and if possible enhance, access to ALMA and aim to lead parts of instrument development for ALMA2040.
\end{enumerate}

Our strategic priorities complement one another: AtLAST (a 50\,m single-dish telescope) and an upgraded ALMA (a large configurable interferometric array) would be in synergy, not competition, with one another.  Both have identified and are working towards the same overarching science goals, and both are required in order to fully address these goals.

Our summary recommendations are as follows:

\vspace{0.25cm}

\textbf{Medium-term recommendations (2025--2030)}

\textbf{Recommendation M.1.} ALMA will continue to be vital to all areas of UK astronomical research. The UK must continue to play a significant role in both the instrumentation upgrades and the world-leading astronomy from ALMA.

\textbf{Recommendation M.2.} The JCMT will remain internationally excellent, in its unique position as the world's largest single-dish submillimetre telescope, at least until AtLAST is on sky. It is crucial that the UK maintains a key role in the JCMT, and widens access to all UK astronomers.

\textbf{Recommendation M.3.} An upgraded instrumentation suite for the JCMT will be required to maintain its world-leading position for the next 10 years in the run-up to AtLAST.  The UK should be at the heart of building a polarisation-sensitive MKID camera and a large-format heterodyne array for the JCMT, to avoid ceding our scientific and technological leadership in these fields.

\textbf{Recommendation M.4.} The Event Horizon Telescope EHT will continue to produce new insights into SMBHs and AGN as it moves to higher frequency and into time domain observations.  The UK should maintain and diversify its access to the EHT, which is currently contingent on access to the JCMT, potentially through involvement in the AMT.  The UK should be central to the EHT's move to submillimetre wavelengths.

\textbf{Recommendation M.5.} The UK should during the next 5 years be working towards AtLAST through the Horizon Europe AtLAST design consolidation study and beyond.

\vspace{0.5cm}

\textbf{Long-term recommendations (2030 and beyond)}

\textbf{Recommendation L.1.} The AtLAST Telescope should be seeing first light by the mid 2030s, and be beginning science operations shortly thereafter.  The UK should aim to be a key partner of an international consortium for the construction of AtLAST, to capitalise on and build on our world-leading expertise in submillimetre science and technology.
 
\textbf{Recommendation L.2.} UK instrumentation laboratories should take leading roles in the development and construction of first-light instruments for AtLAST.

\textbf{Recommendation L.3.} UK astronomers should play leading roles in planning and executing AtLAST science, building on our medium-term single-dish track record, and where possible using our technical contributions to leverage leading scientific positions.

\textbf{Recommendation L.4.} The UK must, as a minimum, participate in instrument development for ALMA2040, and ideally lead parts of it.

\vspace{0.5cm}

\textbf{Funding application recommendations}

\textbf{Recommendation F.1.} We recommend the UK community bid for UKRI support for JCMT operations to beyond 2031, supporting UK-wide access to JCMT up to the first light of AtLAST.

\textbf{Recommendation F.2.} We recommend the UK community bid for UKRI support for the UK share of a new MKID camera for the JCMT, which will be led from the UK.  
We also recommend the UK community bid for funding for technology development towards a new large-format heterodyne array.

\textbf{Recommendation F.3.} We recommend a coordinated UK community application for UKRI funding for AtLAST, to secure leading UK roles in this multi-national consortium.

\textbf{Recommendation F.4.} We recommend a community bid for upgrading the ALMA WSU 
to its full factor 4 enhanced bandwidth
to secure UK guaranteed observing in this very heavily oversubscribed, unique and internationally-leading facility.

\textbf{Recommendation F.5.} We recommend that the community bids for funding for the UK to play a leading role in ALMA developments as part of the ALMA 2040 plan.

\vspace{0.5cm}

\textbf{General recommendations}

\textbf{Recommendation G.1.} %Tensioning recommendation: although we can't with certainty 
It is not possible to predict with certainty the development timeline of new facilities, but there may be a point at which involvement in AtLAST becomes contingent on decommitting from JCMT.  When this point is reached, the UK must ensure that its critical single-dish capabilities are maintained throughout the transition from the JCMT to AtLAST.

\textbf{Recommendation G.2.} UK instrumentation laboratories must continue to be supported to allow them to be at the forefront of building first-light instruments for AtLAST and bidding for each round of ALMA instrumentation upgrades. 

\textbf{Recommendation G.3.} The UK should maintain and develop its high performance computing capabilities as the demands on compute increase from new submillimetre and millimetre-wave facilities and instruments. The facilities should continue to align their open data and open software practices with national, continental and international open science initiatives, and be supported by software engineering for the maintenance, development and support of software.

\textbf{Recommendation G.4.} We support the regional centre model for the provision of user support, such as the ALMA regional centre, for the development of expertise and provision of expert support for UK users of submillimetre/millimetre facilities.

\section{Introduction}
\label{sec:intro}

Submillimetre and millimetre astronomy is a field of research in which the UK has always led in both science and instrumentation.  This roadmap presents the current status and future aspirations of the submillimetre and millimetre astronomy community in the United Kingdom.  Our aim is to provide a clear and consistent voice for the UK submillimetre and millimetre community, and to chart a clear route from the current status of UK submillimetre/millimetre astronomy to a future in which the UK continues to play a leading role in the next generation of world-class submillimetre/millimetre instrumentation.

The submillimetre/millimetre wavelength regime is crucial to our understanding of the universe.  Approximately half of the energy output of star formation and black hole accretion in the history of the universe has been absorbed by dust, and re-radiated as thermal radiation in the far-infrared, submillimetre and millimetre regime. Submillimetre astronomy is therefore essential to answer many of the big questions in planet formation, star formation, galaxy evolution and large-scale structure formation, as part of our multi-wavelength observational capacity.

% Say some nice things about UK history of submm/mm

This roadmap arises from the Science and Technology Facilities Council (STFC) Astronomy Advisory Panel (AAP) Roadmap 2022, in response to their Recommendation 4.1:  \textit{``STFC (via AAP) should commission a review of UK submm/mm-wave science and technology, covering UK aspirations for current and future large single-dish facilities that feed the major international interferometers, and the underpinning aspirations for next-generation instrumentation, identifying areas of international excellence.''} \citep[][p. 21]{STFC_AAP_2022_Roadmap}.

The AAP also solicited community input for the UK Research and Innovation (UKRI) ``Preliminary Activity Wave 3'' infrastructure funding call in Summer 2022.  This resulted in five bids, of which two were from the submillimetre community, from a wide range of other astronomical topics.  The AAP were only permitted to recommend three of these bids upwards to STFC on grounds of demand management.  A further white paper on a bespoke instrument for line intensity mapping at millimetre/submillimetre wavelengths was also received as part of the wider consultation, the science of which was once again internationally excellent. Taken together, these three disparate submissions were felt to evidence that \textit{``the independent, internationally-excellent groups working on sub-mm/mm-wave science and technology do not yet have a single consistent voice''} (\citelinktext{STFC_AAP_2022_Roadmap}{ibid}., p. 21).  This roadmap provides this consistent voice.

In this roadmap, we define the submillimetre/millimetre wavelength regime as $\sim 3$\,mm--300\,$\mu$m ($\sim$100\,GHz--1\,THz).  This range spans from the crucial $^{12}$CO $J=1\to0$ transition at 115\,GHz to the shortest wavelengths observable from the ground.  Our report is mainly restricted to ground-based astronomy only, because UK involvement in space missions is largely the domain of the UK Space Agency (UKSA) and ESA rather than of the STFC.  Nevertheless, space missions are covered in the context of their commonalities in science themes, instrumentation and technology with ground-based astronomy, and in the ground-based follow-ups of space telescope surveys.  We also do not discuss dedicated Cosmic Microwave Background experiments as this topic has been covered by the recent CMB White Paper (M. Brown et al., priv. comm.). 

We first present the results of our submillimetre/millimetre community survey (Section~\ref{sec:consultation}), followed by detailed reviews of UK strengths in relevant science (Section~\ref{sec:science}) and instrumentation (Section~\ref{sec:uk_instrumentation}).  We then briefly summarise current submillimetre/millimetre instrumentation (Section~\ref{sec:current_instrumentation}), describing both the instruments to which UK astronomers have access, and the international context in which those instruments are situated.

Having presented the past and present of UK submillimetre astronomy, we then consider its future.  We first discuss single-dish astronomy in Section~\ref{sec:single-dish}, presenting the clear path to the UK playing a leading role in the forthcoming 50\,m Atacama Large Aperture Single-dish Telescope (AtLAST), through near-term support of the James Clerk Maxwell Telescope (JCMT).  We then consider interferometric instrumentation in Section~\ref{sec:interferometric}, discussing the UK's role in the future of the Atacama Large Millimeter/submillimeter Array (ALMA).  We next present the UK's role in the future of submillimetre/millimetre Very Long Baseline Interferometry (VLBI) in Section~\ref{sec:vlbi}, focussing on the future of the Event Horizon Telescope (EHT).  Section~\ref{sec:computing} discusses the significant improvements to current computing software and hardware required to maximise the scientific return on the advances described in the previous sections.  In Section~\ref{sec:synergies} we discuss how these submillimetre/millimetre advances will synergise with future space missions, and with future ground-based instrumentation at other wavelengths.

In Section~\ref{sec:swot}, we present a SWOT analysis for UK submillimetre/millimetre astronomy.  Finally, Section~\ref{sec:roadmap} states the strategic priorities for the field, our timeline for funding requests to the STFC and UKRI, and our summary recommendations.

\section{UK submillimetre/millimetre astronomy community consultation}
\label{sec:consultation}

In order to accurately represent the priorities of the UK submillimetre and millimetre astronomy community in this Roadmap, we conducted a community consultation that ran from 1$^{\rm st}$ to 30$^{\rm th}$ November 2023.  60 people responded to the consultation, which was circulated using the STFC's Astro Community mailing list.  In this section, we summarise the key results of the consultation.

\subsection{The UK submillimetre astronomy community}

\begin{figure}[b!!!]
    \centering
    \includegraphics[width=0.51\textwidth,valign=t]{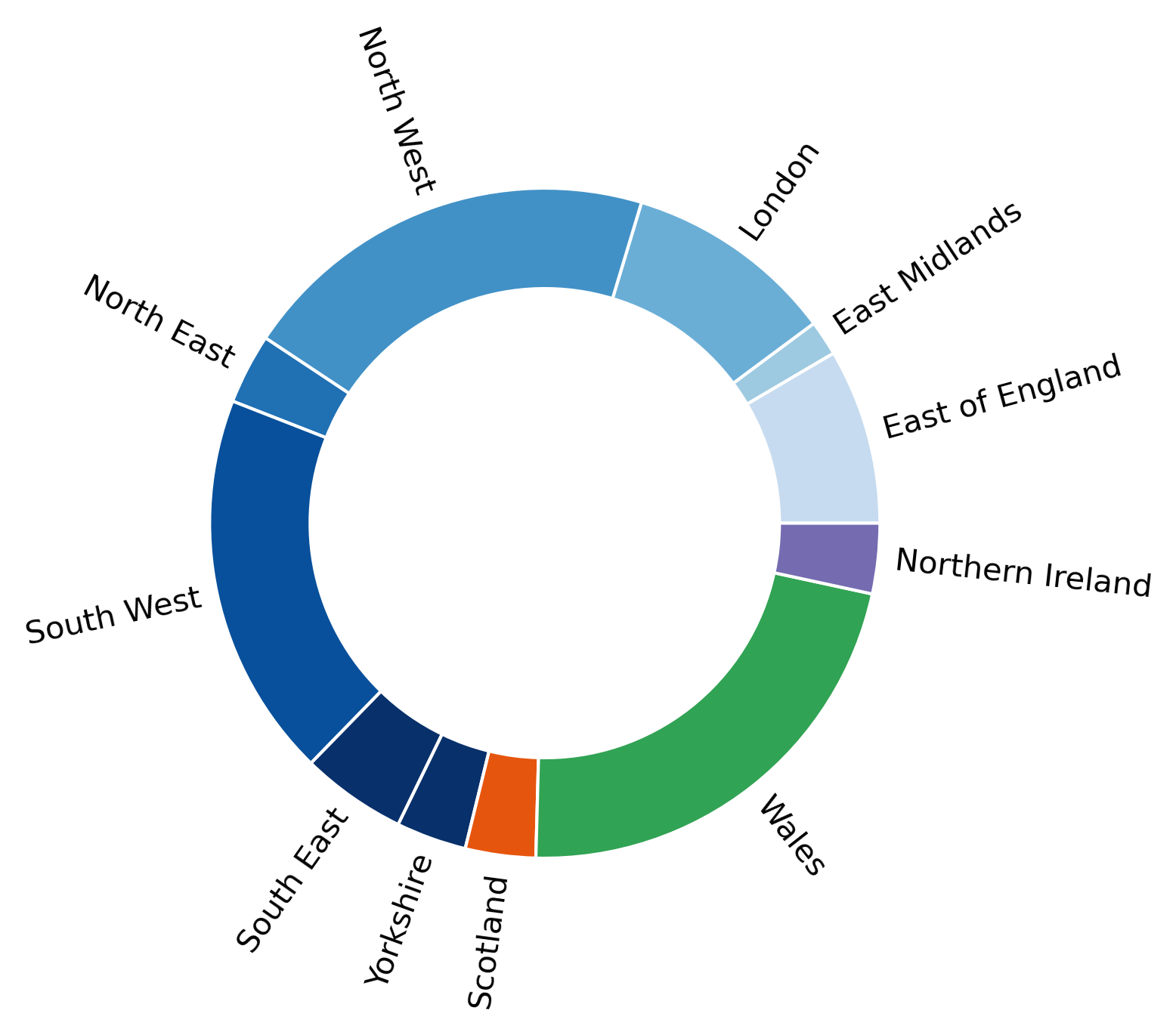}
    \includegraphics[width=0.47\textwidth,valign=t]{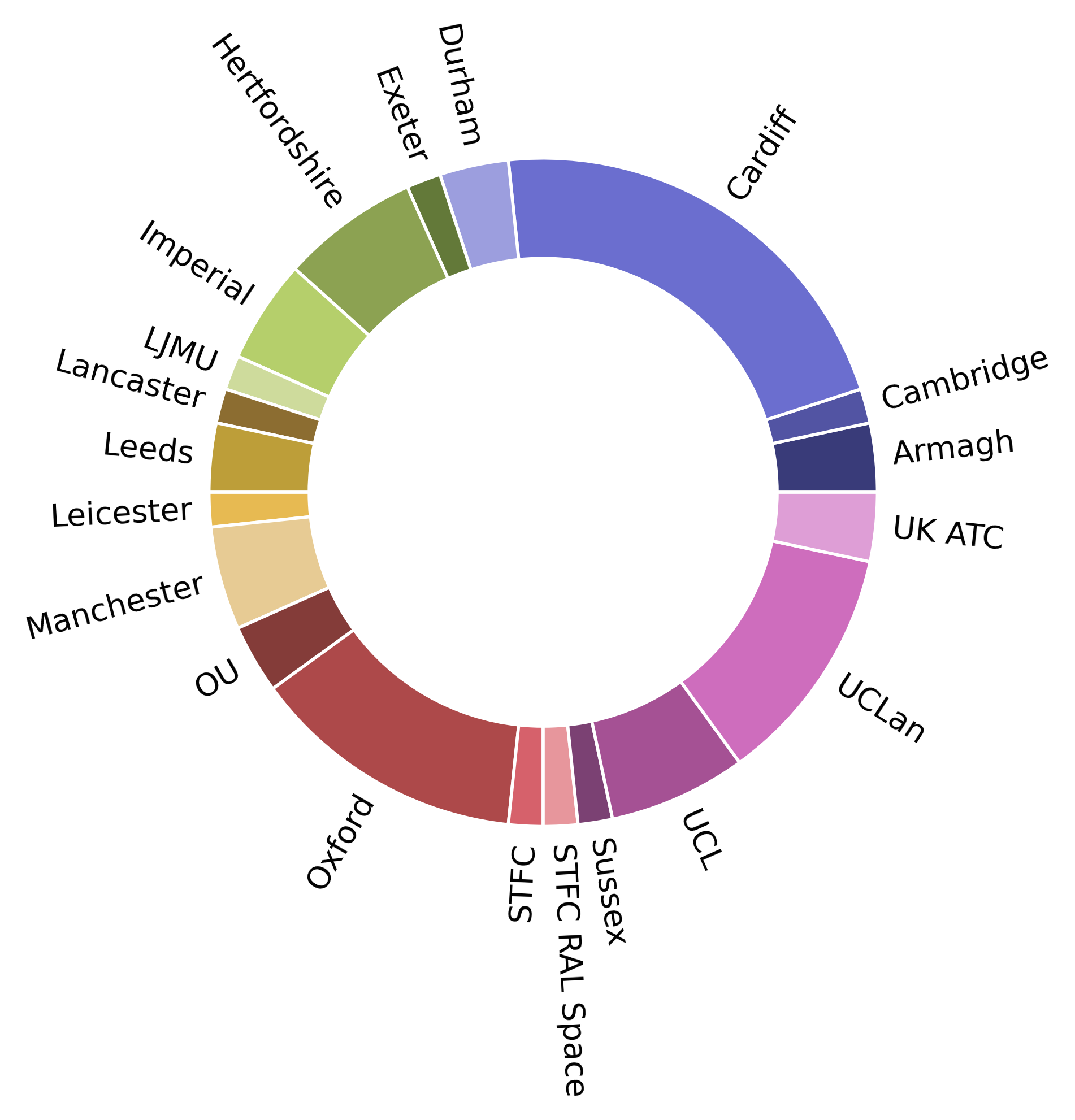}
    \caption{The distribution of survey respondents amongst UK nations (left), and broken down by institution (right).  Respondents from English institutions are further grouped by their region within England.}
    \label{fig:uk_pie}
\end{figure}

We found that submillimetre astronomy is a key part of astrophysics across the UK.  There are submillimetre astronomers at institutions in every UK nation  (Figure~\ref{fig:uk_pie}), with particular strength in Wales and the North West of England.  Submillimetre astronomy is also not restricted to any one group of universities, with Russell Group, University Alliance and MillionPlus universities all strongly represented.

\begin{figure}
    \centering
    \includegraphics[width=0.49\textwidth, valign=t]{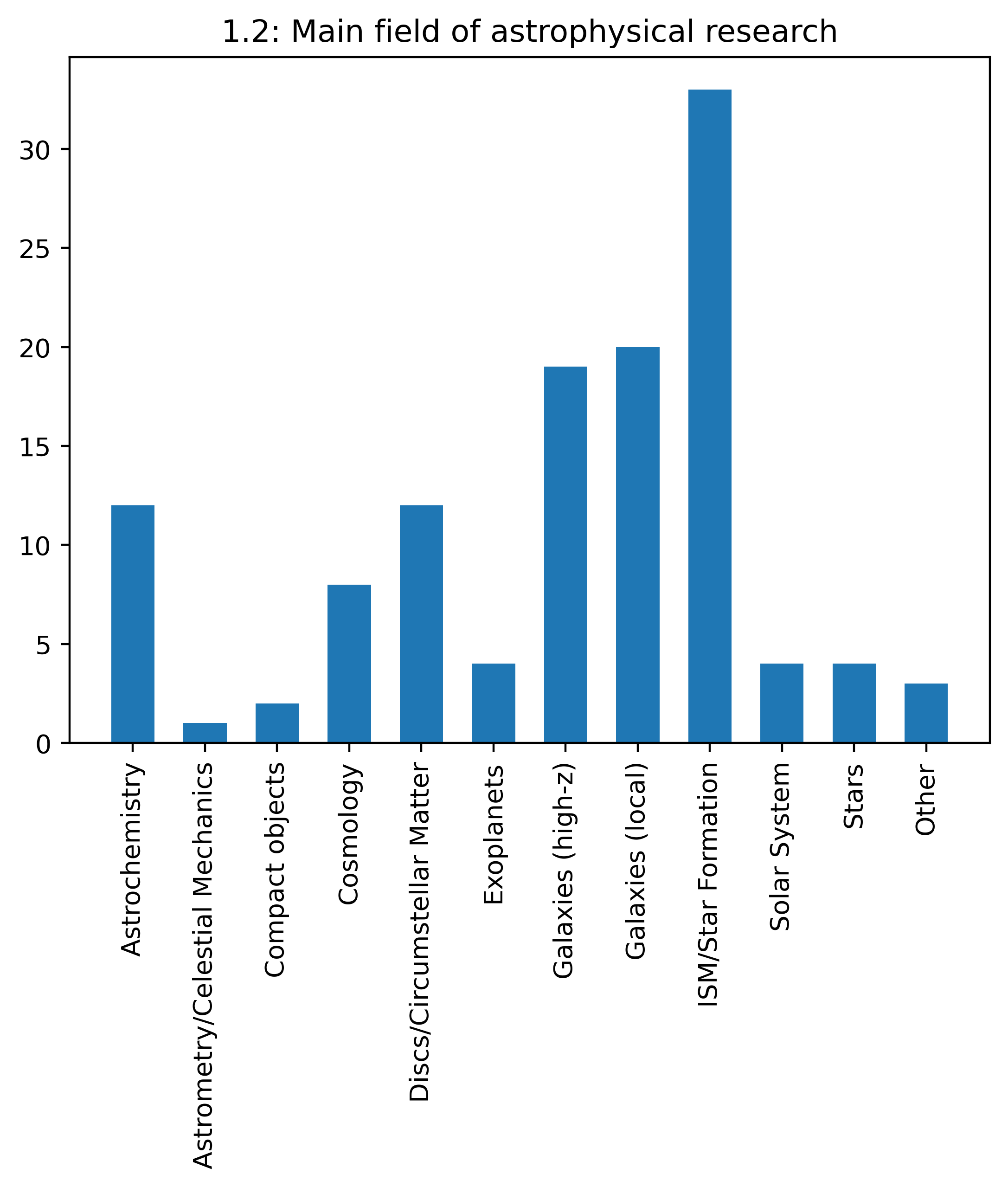}
    \includegraphics[width=0.49\textwidth, valign=t]{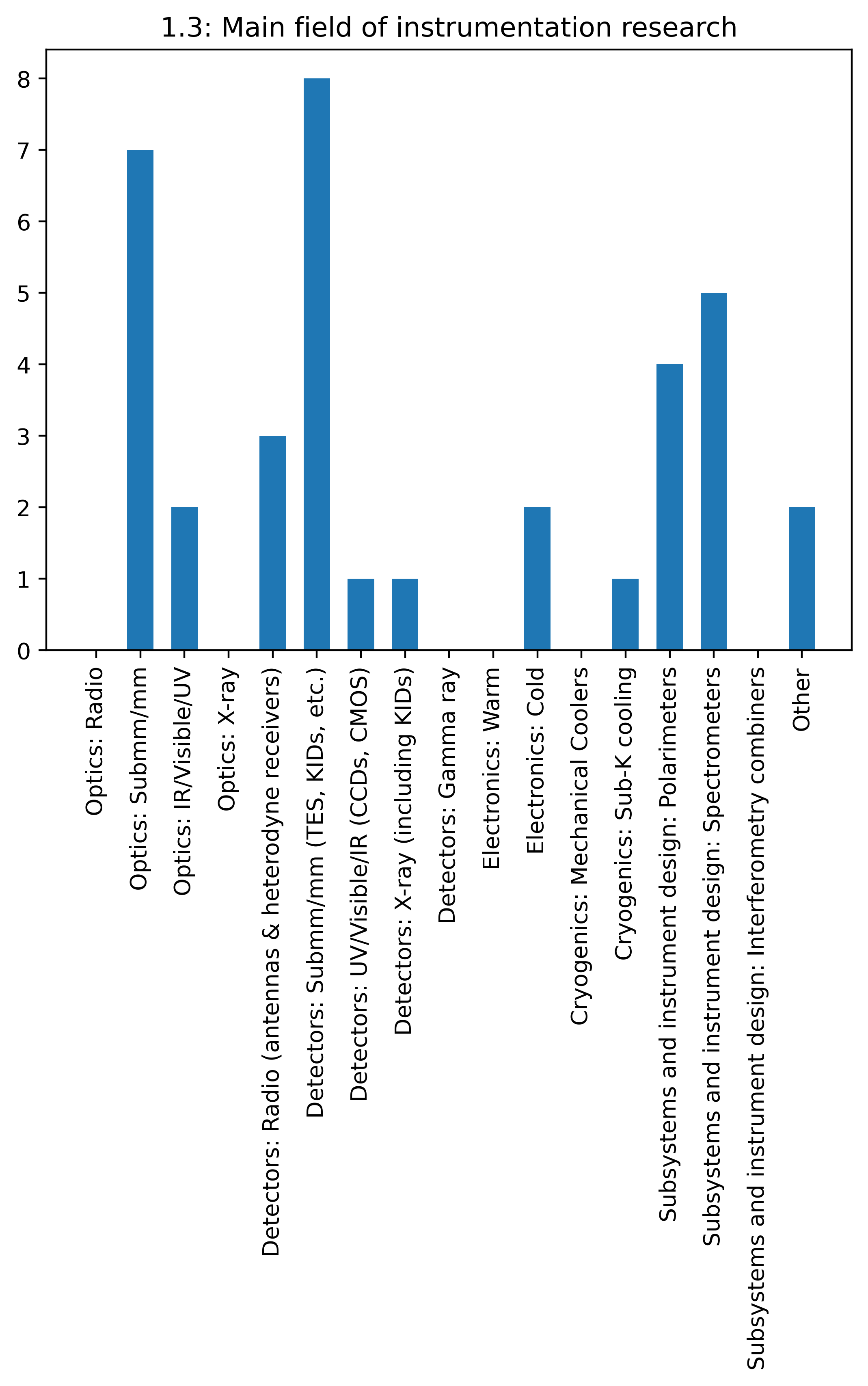}
    \caption{The fields of research of respondents to the submillimetre astronomy consultation.  Left: fields of astrophysical research.  Right: fields of instrumentation research and development.}
    \label{fig:fields_of_research}
\end{figure}

UK submillimetre astronomers work on topics spanning astrophysics and cosmology, as shown in Figure~\ref{fig:fields_of_research}a, with the most common fields of research being ISM and star formation studies, and the physics of galaxies, both local and at high redshifts.  The UK also has a strong submillimetre instrumentation community (Figure~\ref{fig:fields_of_research}b), with particular strength in the fields of submillimetre detectors and optics, as well as in spectrometer and polarimeter design.

\subsection{Current instrumentation}
\label{sec:consultation_current}

\begin{figure}
    \centering
    \includegraphics[width=0.5\textwidth]{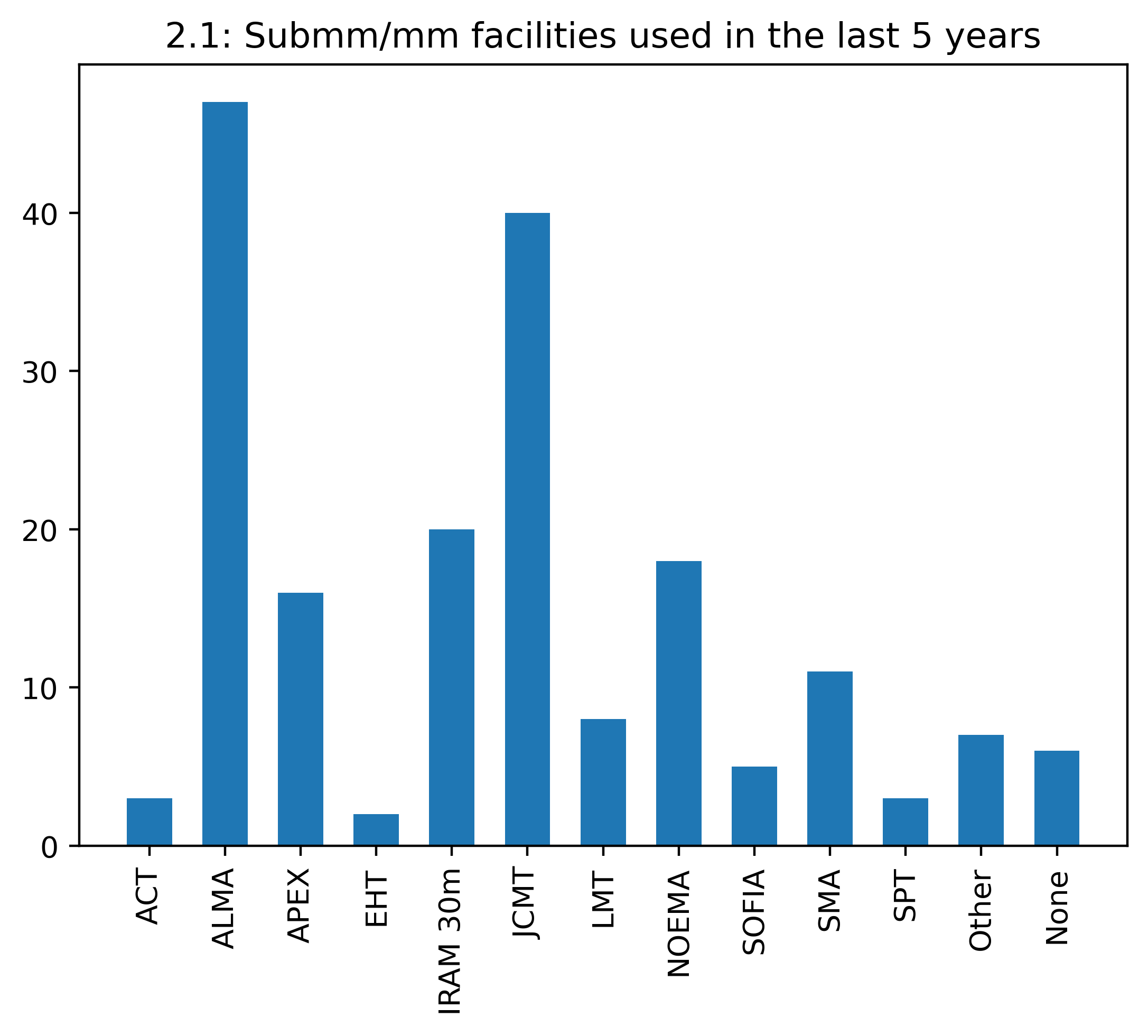}\includegraphics[width=0.5\textwidth]{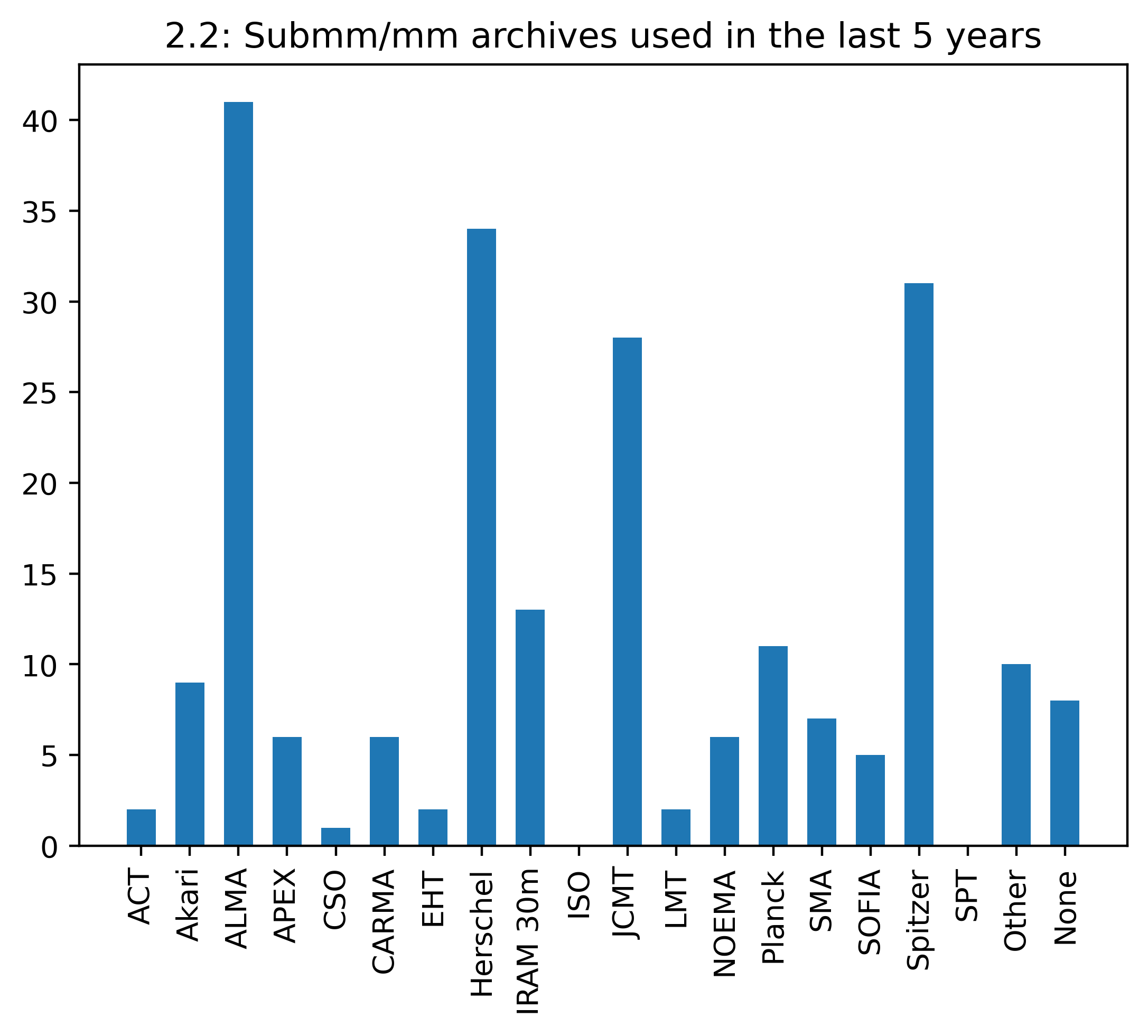}
    \caption{Responses to questions 2.1 and 2.2: use of facilities and their archives.}
    \label{fig:current_telescopes}
\end{figure}

The two facilities that are currently most used by the UK submillimetre astronomy community are ALMA (78\% of respondents) and the JCMT (67\% of respondents), as shown in Figure~\ref{fig:current_telescopes}a.  Maintaining access to these facilities is essential: 93\% of respondents consider ALMA with its current capabilities important to their research over the next 10 years, and 68\% of respondents consider the JCMT with its current capabilities important to their research over the next 10 years (Figure~\ref{fig:current_facilities}).  Improving the capabilities of both of these facilities is also considered essential for ongoing research over the next 10 years and beyond, as discussed below.

Other submillimetre telescopes are used less widely, but still have a significant user base: all of the extant submillimetre telescopes have users in the UK, as shown in Figure~\ref{fig:current_telescopes}a.  While the EHT user community in the UK is smaller than those of other instruments, the impact of its results is very significant, both in astrophysics and beyond.  When asked what the most high-impact result in submillimetre astronomy over the last 5-10 years has been, the most common response (32\% of respondents) was the imaging of the M87 and Sgr A* SMBHs with the EHT.

The consultation confirmed that submillimetre data sets have significant legacy value.  We found that the archives of submillimetre telescopes are well-used: the most-used archives are those of ALMA, \textit{Herschel}, \textit{Spitzer} and the JCMT (Figure~\ref{fig:current_telescopes}b).  Particularly, 57\% of the respondents have used the archive of the \textit{Herschel} Space Observatory in the last five years, demonstrating the legacy value of a mission that flew more than a decade ago.

\subsection{Future science goals}
\label{sec:consultation_future_science}

A typical description of submillimetre astronomy from a respondent to the consultation was ``A whole field of astrophysics essentially founded by the UK, and where the UK leads in both science and instrumentation''.  The UK was consistently described as `world-leading' in submillimetre astronomy, and the UK submillimetre astronomy community expects to be at the fore-front of significant breakthroughs from submillimetre observations in the coming years.  When asked what were the most important questions for submillimetre astronomy to answer, responses were broadly divided into three categories:

\textbf{Formation of exoplanets and the potential for life}  Submillimetre astronomy is ideally placed to investigate the formation of planetary systems: ALMA's imaging of protoplanetary discs has revolutionised this field over the last decade.  Future observations will allow investigation of both potential life and the potential for life.  Submillimetre observations will play a crucial role in exoplanet characterisation, particularly planetary atmospheres and the search for biomarkers.  Meanwhile, astrochemical studies will searches for prebiotic molecules in the ISM.  These searches extend into the Solar System, whether through observations of comets and asteroids, moons such as Titan and Enceladus (e.g., \citealt{1997Icar..126..170H}) or of planetary atmospheres, such as Venus \citep{greaves2021}, which is currently the subject of a UK-led JCMT long term project, JCMT-Venus. UK Schottky diode technology is also on its way to the Jupiter system as part of ESA's Juice mission. %Such searches are also proceeding in our own Solar System: the recent potential detection of phosphine on Venus was made using the JCMT and ALMA have in the last \citep{}

\textbf{Star formation and galaxy evolution through cosmic time}  Submillimetre astronomy is poised to answer questions on galaxy evolution at all redshifts, from galaxy assembly in the era of reionisation, through high-redshift star formation and dust production, to star formation in the modern universe.  Submillimetre observations are required to understand the baryon cycle through cosmic time, and to constrain the gas and dust budgets for galaxies.  Submillimetre observations are also crucial in order to understand the link between AGNs and their host galaxies, including understanding the physics of AGN jet launching, and measuring the masses of SMBHs at high redshift.

\textbf{Formation and evolution of cosmological structure}  Submillimetre observations are required in order to investigate fundamental questions of cosmology, such as resolving the Hubble tension and understanding the accelerating expansion of the universe, and searching the Cosmic Microwave Background for B modes, the signature of inflation, and for spectral distortions.

In order to accomplish all of these goals, improvements to current submillimetre facilities are required over the next few years, and in the longer-term, new submillimetre facilities must be developed.  If the UK does not invest in new submillimetre instrumentation, respondents predict ``A devastating loss of expertise, significantly reducing the UK's global standing in astrophysics.  Loss of the genuinely world-leading submillimetre astronomy and instrument development taking place in the UK.''.

The community consultation highlighted the great pride that the UK submillimetre community has in the UK's ground-breaking history and world-leading current work in this field, and their great hopes for the UK being an integral part of the future development of science and technology in this crucial wavelength range.

\subsection{Improvements to current instrumentation on timescales $<$10 years}\label{sec:improvements_to_current_instrumentation}

\begin{figure}[b!!]
    \centering\includegraphics[width=\textwidth]{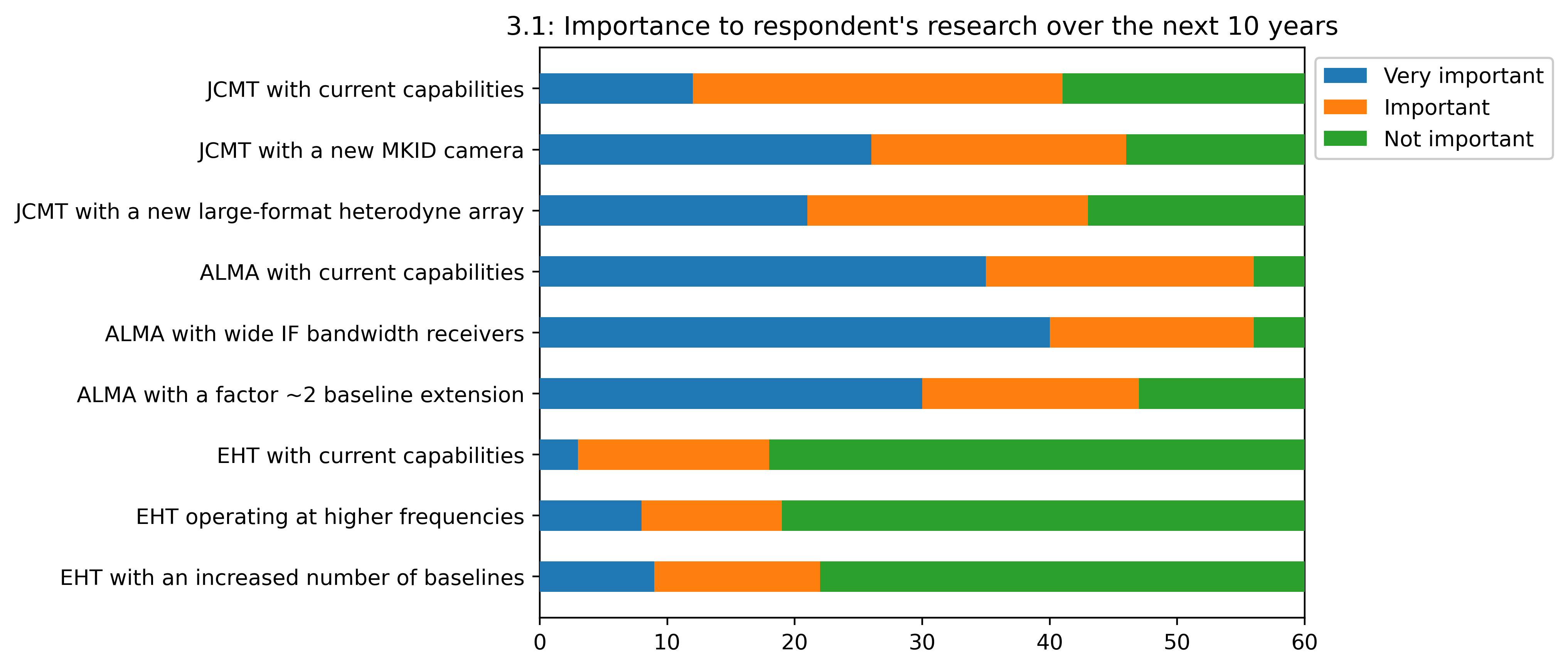}
    \caption{Facilities rated in terms of their importance to respondents' science goals over the next 10 years.}
    \label{fig:current_facilities}
\end{figure}

There is strong support from the submillimetre community for improvements to current telescopes, as shown in Figure~\ref{fig:current_facilities}.  The most strongly supported improvement to a current facility is wider IF bandwidth receivers for ALMA, currently in progress as part of the ALMA 2030 Development Roadmap \citep{2023pcsf.conf..304C}.  Other than this in-progress upgrade, the most strongly supported upgrades for existing facilities are the planned new camera for the JCMT with MKID detectors \citep{li2024}, and a factor $\sim2$ baseline extension for ALMA, a component of the ALMA Development Roadmap which is beyond the current Wideband Sensitivity Upgrade.   A new large-format heterodyne array for the JCMT is also strongly supported.

\subsection{New facilities on timescales $>$10 years}\label{sec:new_facilities_beyond_10_years}

\begin{figure}
    \centering
    \includegraphics[width=\textwidth]{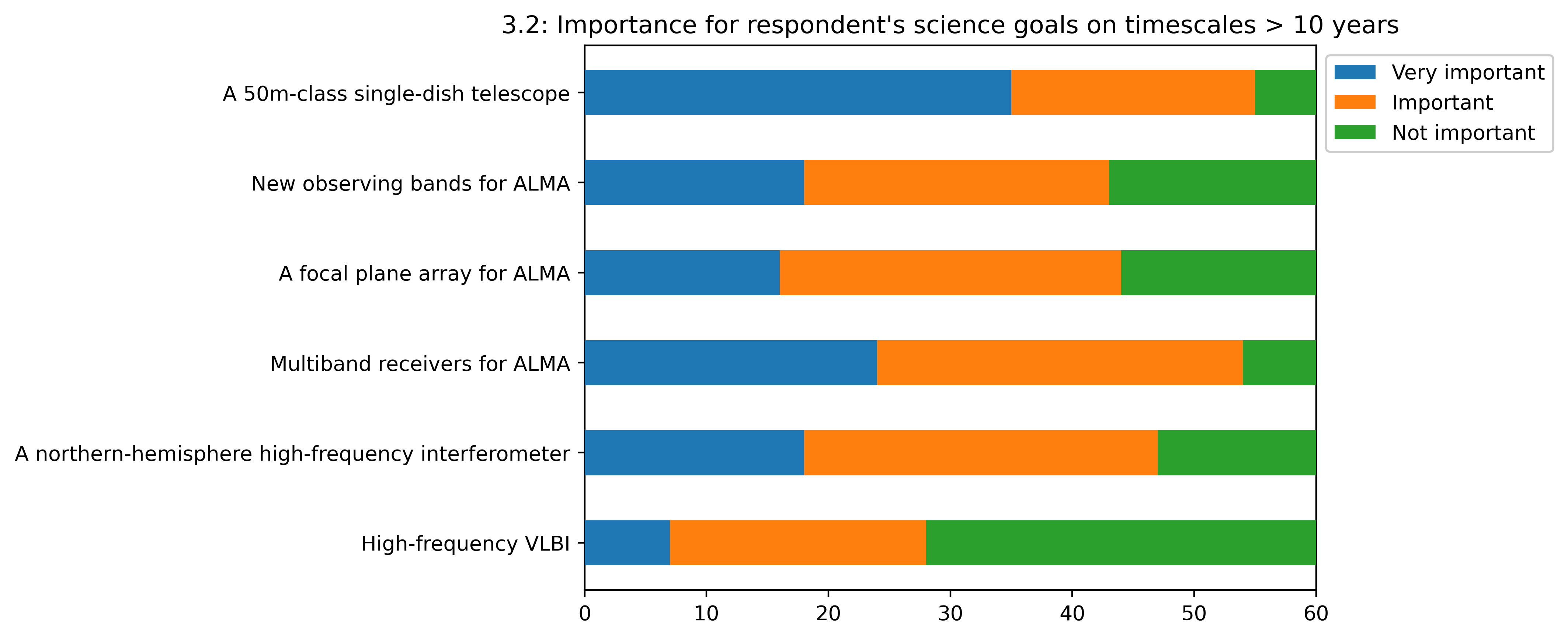}
    \caption{Hypothetical facilities rated in terms of their importance on timescales $>$ 10 years for the science goals of respondents to the consultation.}
    \label{fig:future_facilities}
\end{figure}

In the longer term, new facilities will be needed in order to answer the cutting-edge questions discussed above.  The new submillimetre facility considered most important for respondents' science goal on a timescale of 10+ years is a 50 m-class single-dish telescope such as AtLAST, as shown in Figure~\ref{fig:future_facilities}.  This was considered important by 92\% of respondents.  There is also significant community support for major upgrades to ALMA: of the three options presented in the consultation, multiband receivers were most strongly supported. (Note that the possibility of enhancing the sensitivity of ALMA through increasing the number of antennas, which is a proposed component of the ALMA 2030 Development Roadmap, was not included as an option in the survey.) However, a 50 m-class single-dish telescope was the only future facility that was considered very important by the majority (58\%) of respondents. 

\begin{figure}
    \centering
    \includegraphics[width=0.9\textwidth]{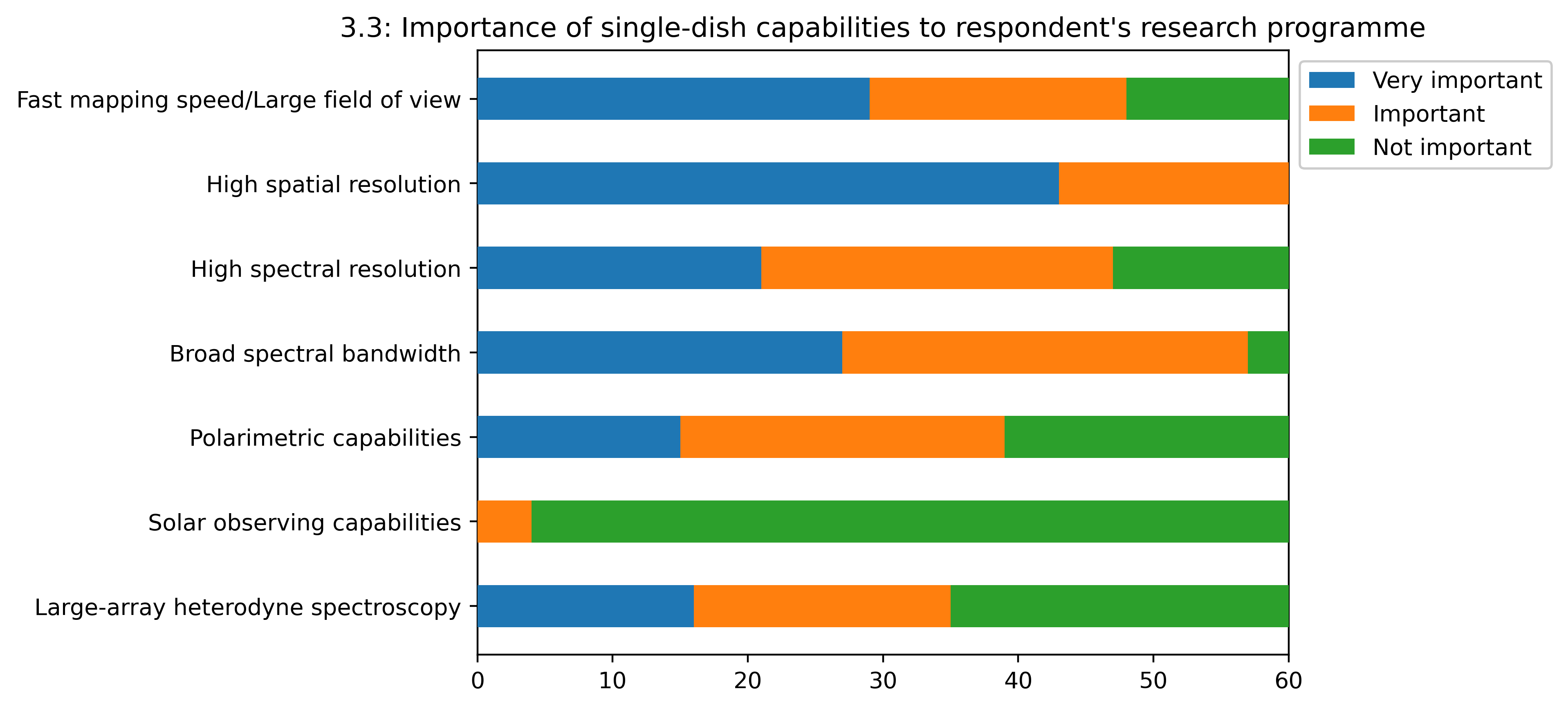}
    \caption{Importance of capabilities of a hypothetical new single-dish submillimetre telescope for the science goals of respondents to the consultation.}
    \label{fig:future_capabilities}
\end{figure}

What the submillimetre community would require from such a single-dish telescope is shown in Figure~\ref{fig:future_capabilities}.  The most important capability, agreed on by all respondents, is high spatial resolution, which emphasises the need for future submillimetre telescopes to be in the 50 m-class.  Beyond this requirement, broad spectral bandwidth, fast mapping speed and high spectral resolution are all considered key.  A majority of respondents also consider polarimetric capabilities and large-array heterodyne spectroscopy important to their scientific goals.

\subsection{Computing infrastructure for new facilities}\label{sec:computing_infrastructure_for_new_facilities}

\begin{figure}
    \centering
    \includegraphics[width=\textwidth]{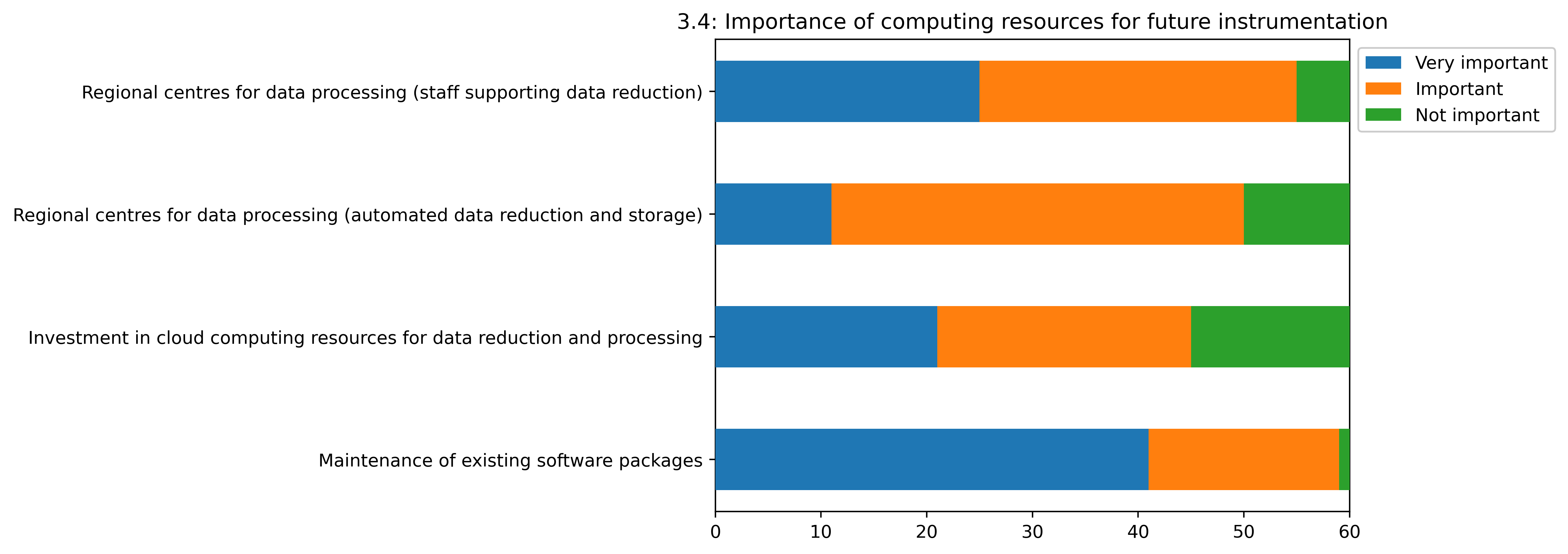}
    \caption{Importance of computing infrastructure and resources for future submillimetre instrumentation.}
    \label{fig:computing_resources}
\end{figure}

It is recognised by the community that support for computing infrastructure is essential if new facilities are to be properly exploited.  Figure~\ref{fig:computing_resources} shows that while all forms of such infrastructure are supported by the community, the `ALMA model' of regional centres for data processing is the most popular. (UK ALMA support is currently provided by the UK ALMA Regional Centre Node).  It is also almost universally agreed that it is important to provide support for the maintenance of existing software packages.

\begin{figure}
    \centering
    \includegraphics[width=0.8\textwidth]{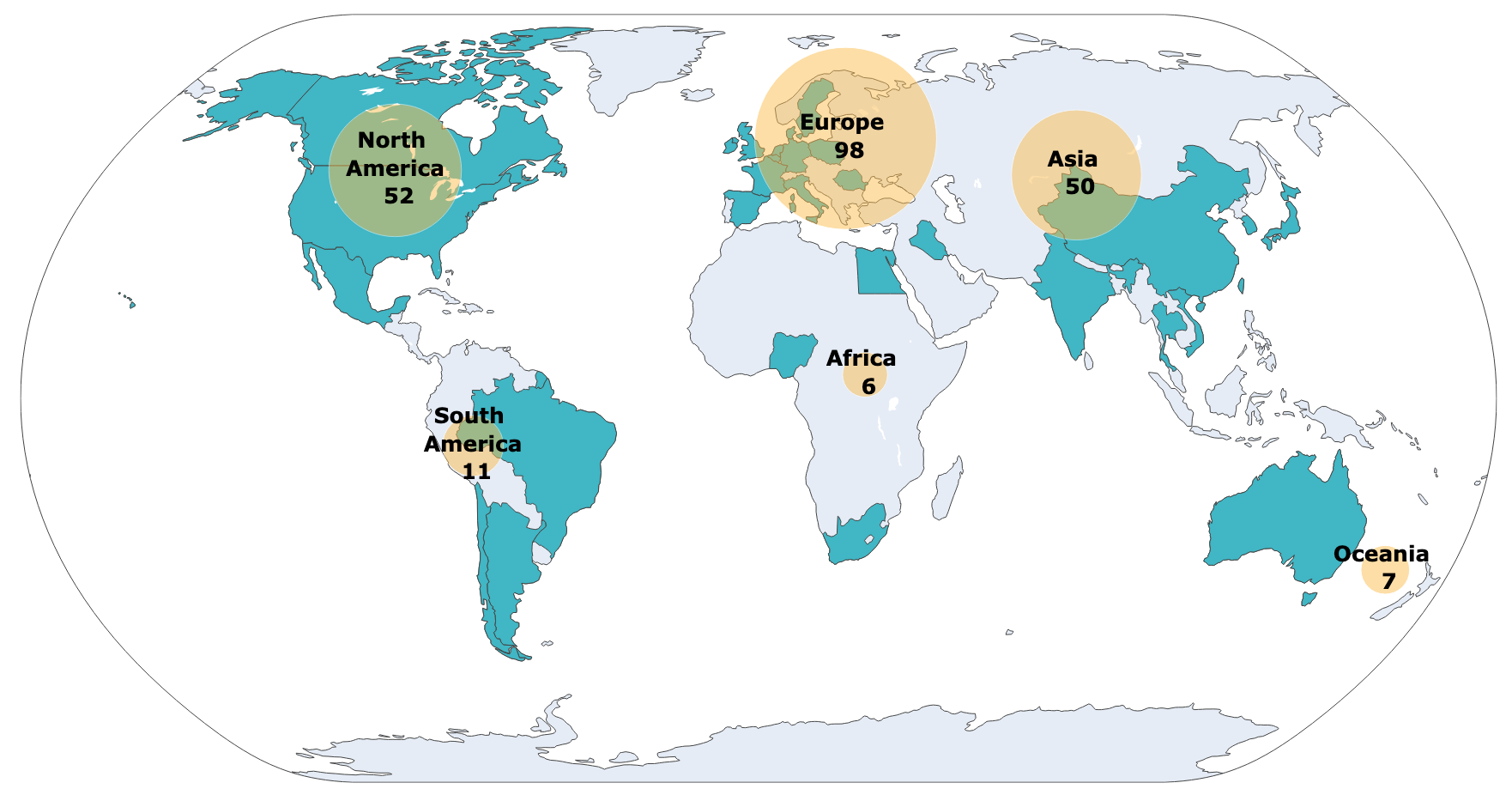}
    \caption{Collaborations.  N.B. totals include 5 collaborations listed as ``EU'' or ``Europe'' and 3 listed as ``East Asia'' or ``East Asian Observatory''.  One Russian collaboration was listed as currently suspended due to the war with Ukraine, and is not included in the totals listed.}
    \label{fig:collaborations}
\end{figure}

\subsection{Impact of submillimetre astronomy}

We further attempted to capture the impact of submillimetre astronomy through international collaboration, applications of submillimetre astronomy technology in other fields, and outreach and engagement programmes.

\subsubsection{Global reach}

Submillimetre astronomy is a global endeavour.  We found that UK submillimetre astronomers have collaborators in every populated continent, as shown in Figure~\ref{fig:collaborations}.  The most frequently listed countries for submillimetre collaborations were the United States and Germany.  However, significant numbers of collaborations were listed across Europe and East Asia, as well as a number of collaborations with astronomers in developing nations.  The East Asian Observatory (operating the JCMT) was particularly noted as driving collaborations with East Asia, India, and developing nations in South-East Asia.

\subsubsection{Non-astronomy applications}
\label{sec:consultation_nonastro}

A wide range of non-astronomical applications of the technology used in submillimetre instrumentation were highlighted by respondents to the survey.  These included:
\begin{itemize}
    \item \textbf{THz cameras} For next-generation security scanners, medical imaging, fusion plasma diagnostics,  non-destructive internal mapping of materials (e.g. for manufacturing quality control).
    \item \textbf{Earth observation} Including commercial and national weather forecasting and climate research.
    \item \textbf{Cryogenics} New submillimetre/FIR detectors will drive advances in cryogenic technologies.
    \item \textbf{Quantum computing} Amongst other potential developments, microwave travelling wave parametric amplifiers (TWPA) currently under development for
    millimetre-wave receivers are ideal for the ultra low-noise readout required for superconducting qubits. Qubit research is also anticipated to move upward into the millimetre/submillimetre range in the near future.
    \item \textbf{Telecommunications} Millimetre-wave and submillimetre-wave frequency ranges will be exploited in the next generation of technology for wireless communication (6G).
    \item \textbf{Big data} New facilities will require large data handling, and development of machine learning exploitation of those data sets. Supervised and semi-supervised machine-learning algorithms will need labelled training data, much of which will be provided by subject-specialist experts, and some of which may also be provided by crowd-sourcing the labelling or classification tasks with the help of a wider pool of volunteers \citep[e.g. citizen science,][]{Serjeant+24}
    \item \textbf{Dark matter searches} Current axion detection experiments, as ADMX and HAYSTAC, use resonant microwave cavities with a strong magnetic fields.  TWPA, currently under development for astronomical receivers, are ideal for use in these experiments. Submillimetre technologies such as superconductor-insulator-superconductor (SIS) mixers developed for astronomy are also beneficial in extending the search into the submillimetre regime. Other fundamental physics experiments such as neutrino mass determination experiments could benefit from the development of TWPA as well.
\end{itemize}

\subsubsection{Public outreach}

We asked respondents to suggest key priorities for public outreach work associated with future submillimetre facilities.  It was consistently suggested that any new submillimetre facilities should have an outreach programme planned from the very start.  At any site at which new facilities are planned, engagement with local communities at the earliest possible stage is crucial.  Communicating scientific results to a diverse range of audiences will be crucial.  Effort should be made to engage both children and young adults -- in the latter case, potentially by creating opportunities for secondary-school and undergraduate-level students to take part in research.  Citizen science projects, while being driven by primary objectives of particular science goals \citep[e.g.,][]{Serjeant+24}, may also have  secondary societal and educational benefits of engaging both amateur astronomers and the wider science-inclined public.  Observatory press offices should develop strong ongoing relationships with the media and public, so that the importance of the submillimetre regime, and the discoveries being made by new submillimetre facilities, can be clearly communicated to the general public.

\section{UK strengths in submillimetre and millimetre astronomy}
\label{sec:science}

Half the energy output of star formation and black hole accretion in the history of the universe has been absorbed by dust, and re-radiated as thermal radiation. In ultraviolet to near-infrared astronomy, this dust obscures many of the critical processes involved in the origins of stellar mass assembly and black hole growth. However, the regions in which these critical dust-obscured processes occur are also largely transparent at submillimetre and millimetre wavelengths. Submillimetre astronomy is therefore essential to answer many of the big questions in planet formation, star formation and galaxy evolution, as part of our multi-wavelength observational capacity. Submillimetre astronomy thus goes to the heart of the STFC Science Challenges in Frontier Physics\footnote{\url{https://www.ukri.org/publications/stfc-science-challenges/stfc-science-challenges-in-frontier-physics/}}:
\vspace{-6pt}
\begin{description}
\item[Challenge A:] How did the universe begin and how is it evolving?
\item[Challenge B:] How do stars and planetary systems develop and how do they support the existence of life?
\end{description}

\vspace{-6pt}
In this section, we summarise some of the key science questions that submillimetre/millimetre astronomy has answered in recent years, and highlight some of the key questions that it will address in the years ahead.  We organise these science questions under the three key categories identified by our community survey (Section~\ref{sec:consultation_future_science}).  The UK has an outstanding international reputation in submillimetre/millimetre science, and we here highlight examples of UK leadership in each of the topics that we discuss.

\subsection{Formation of planets and the potential for life}

\begin{flushright}
\textit{\textbf{STFC Science Challenge B3:} what processes govern how planetary systems form and evolve?}\\
\textit{\textbf{STFC Science Challenge B4:} what are the conditions for life and how widespread are they?}\\
\textit{\textbf{STFC Science Challenge B5:} how diverse are exoplanets and is our earth typical?}
\end{flushright}

The last decade has provided phenomenal advances in our understanding of how planets form around young stars, driven by advances in submillimetre/millimetre observations.  These advances, coupled with observations of prebiotic chemistry and potential biomarkers in our own Solar System, and of our Sun and its connection to space weather, are providing key insights into the origins of life.

\subsubsection{Protoplanetary Discs}
\label{sec:protoplanetary_discs}

\begin{figure}
    \centering
    \includegraphics[width=0.285\linewidth]{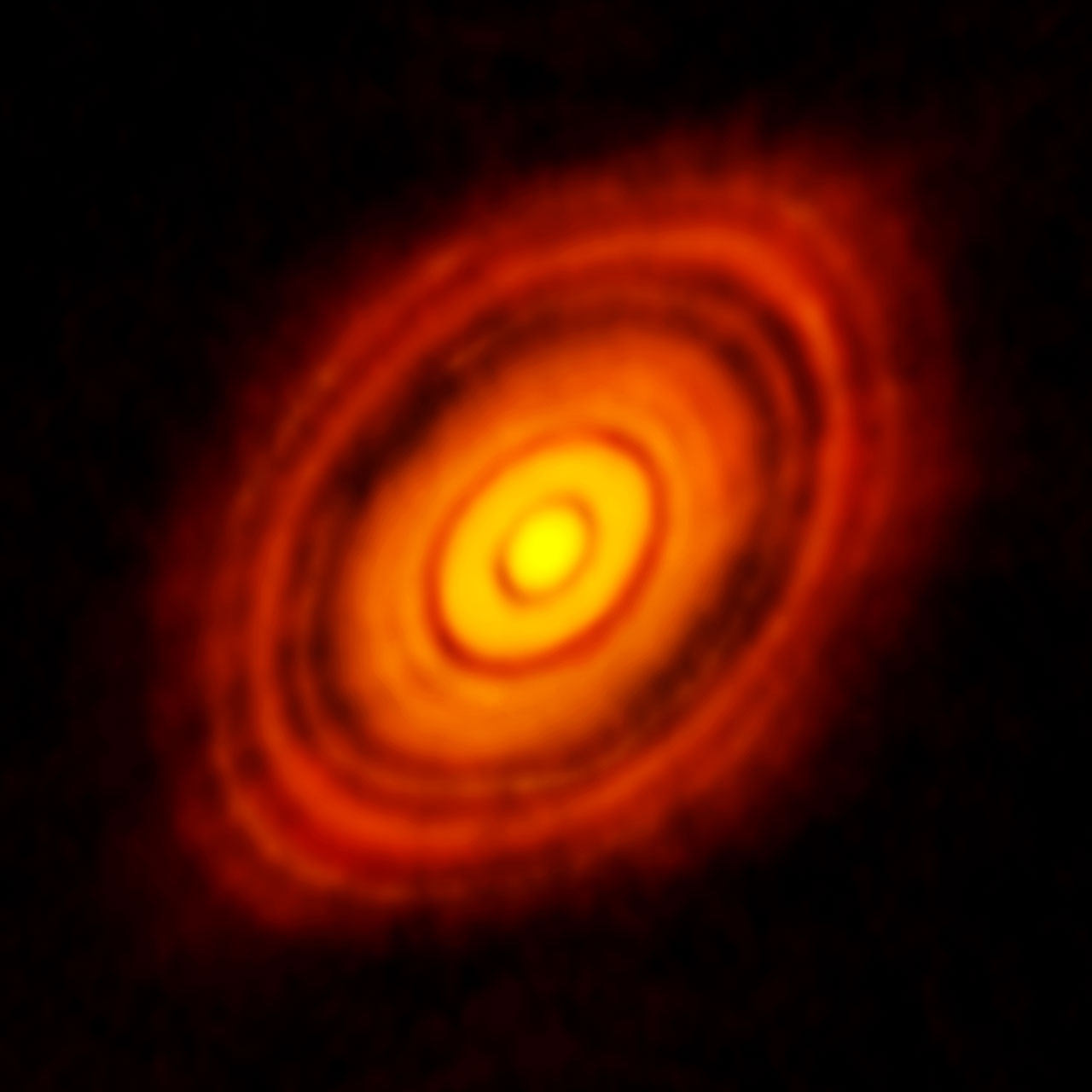}
    \hfill
    \includegraphics[width=0.695\linewidth]{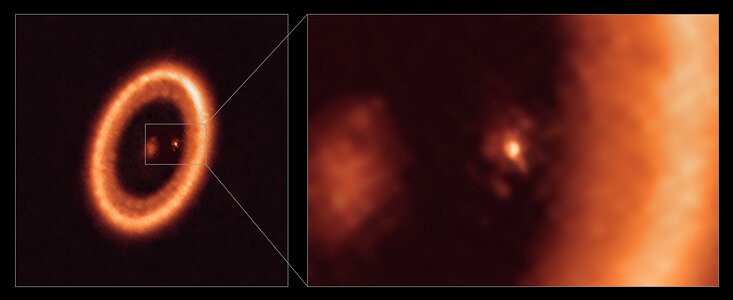}
    \caption{Left: the iconic image of the protoplanetary disc around HL Tau \citep{alma2015}.  Right: the circumplanetary disc around PDS 70c (\citealt{benisty2021}; ESO press release eso2111) .}
    \label{fig:circumstellar-disks}
\end{figure}

\textbf{Introduction}\\
Over the last decade, ALMA has revolutionised the study of protoplanetary discs, imaging nearby discs at 3--5 au resolution \citep[e.g.][]{alma2015, schwarz2016}.  These observations show characteristic gaps in the dust discs associated with planet formation, and in one remarkable case, the circumplanetary disc around a forming planet has been observed, showing planet formation in action \citep{benisty2021}, as shown in Figure~\ref{fig:circumstellar-disks}.  Observations of protoplanetary discs in nearby star-forming regions are increasingly being systematised through ALMA Large Programs such as the DSHARP \citep{andrews2018}, MAPS \citep{oberg2021} and eDisk \citep{ohashi2023} surveys, providing information with which to answer the many outstanding questions about how planets form in protoplanetary discs.

\textbf{UK Leadership}\\
The ALMA MAPS\footnote{\url{https://alma-maps.info}, co-PI: Catherine Walsh, Leeds} \citep{oberg2021} survey, co-led from the UK and with significant UK involvement, mapped multiple molecular lines to investigate the physical and chemical structure of the gas in discs at high spatial resolution.  Other ALMA large programmes with UK involvement include exoALMA\footnote{\url{https://www.exoalma.com}} and DECO (The ALMA Disk-Exoplanet C/Onnection).
The exoALMA programme is searching for for characteristic kinematic signatures of forming planets in 15 disks at the highest possible spectral resolution, and will be submitting its first publications in August 2024.  DECO is investigating the composition of a statistically significant number ($>80$) disks to explore the connection between disk and exoplanet atmosphere composition.  There is a strong synergy between DECO and the forthcoming UK-led ARIEL\footnote{\url{https://arielmission.space}; P.I. Giovanna Tinetti, UCL} space mission, which will survey the composition of at least 1000 exoplanetary atmospheres.

\textbf{Science questions and instrumentation drivers for the coming decade}\\
Accurately measuring the masses of protostellar discs, whether using molecular gas tracers or dust continuum emission, remains an unsolved problem \citep{miotello2023}.  Both gas and dust measurements show a ``missing mass problem'', in which protoplanetary disks appear not to be massive enough to generate the observed exoplanetary population \citep[e.g.][]{ansdell2016,manara2018,parker2022}.  Mapping the chemical composition of protoplanetary discs is essential for understanding their masses, evolution and how planets form \citep[e.g.][]{miotello2017,Zhang2021}. Similarly, a key goal of exoplanetary science is to use atmospheric compositions to probe how planets formed, for which a reliable picture of protoplanetary disc composition is required \citep[][]{madhu2019}.

Other important properties of protoplanetary discs include disc surface density, radius, temperature and scale height. Disc surface density distributions depend strongly on which disc evolution process is dominant \citep[e.g.][]{miotello2017,tazzari2017}, and measurements from a larger sample of discs are required to distinguish between current models.  The disc outer radius is a key property distinguishing between viscosity-driven and disc-wind-driven disc evolution, but gas radii have only been measured for a small fraction of protostellar discs \citep{ansdell2018,boyden2020}, with many discs not detected in CO.  Deeper CO observations of fainter protostellar discs are thus urgently needed.  The radial temperature profiles of discs also remain difficult to directly extract from observations \citep{miotello2023}.  New approaches include mapping optically thick CO lines \citep{dullemond2020} and thermochemical modelling \citep{calahan2021}.  However, deep and high angular resolution observations of many molecular lines in a large number of discs are required in order to understand this key disc property.  A more detailed understanding of the vertical structure of the gas and dust populations is also necessary in order to understand the physics and chemistry of the planet-forming disc midplane \citep[e.g.][]{kama2016}.  Large-scale surveys of disc chemistry and dust evolution tracers are required to understand the implications of disc structure for planet formation.

To understand these disc properties and their implications for planet formation, more systematic measurements with ALMA are required, supported by deep single-dish surveys of nearby star-forming regions to identify further low-mass disc candidates \citep[e.g.][]{herczeg2017}.  Better resolution is also crucial: longer baselines for interferometers will both resolve a larger sample of discs in more distant star-forming clouds, and also allow sub-au imaging of the nearest protoplanetary discs.  For such high-resolution imaging, submillimetre-wavelength observations are essential.  These science goals will be significantly aided by the ALMA Wideband Sensitivity Upgrade (WSU), which is currently underway, which will increase the number of lines it is possible to target simultaneously from about 20 to about 70, while also dramatically reducing the integration time required.

\subsubsection{Debris Discs}
\label{sec:debris_discs}

\begin{figure}
    \centering
    \includegraphics[width=\linewidth]{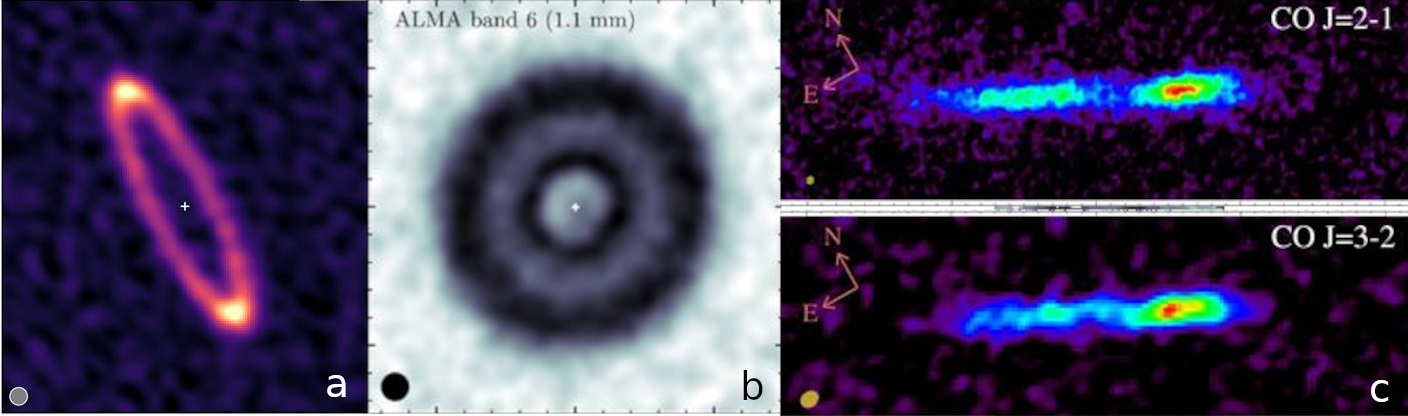}
    \caption{a: Narrow ring with a forced eccentricity likely due to a planet around the young star HR~4796A \citep{2018Kennedy}. b: Broad ring with varying density around the star HD~107146 \citep{2018Marino}. The gap is another potential indicator of the presence of a planet. c: CO gas in the edge-on disc around $\beta$ Pictoris with a clump on one side potentially indicating a giant collision \citep{2017Matra}.}
    \label{fig:debrisdisc}
\end{figure}

\textbf{Introduction}\\
Debris discs constitute the remnants of protoplanetary discs. Once the gas dissipates from the system, the planetesimals that formed during the protoplanetary phase but did not go into forming planets begin to collide, initiating a collisional cascade of debris that is then visible through its scattered light and thermal emission \citep[see][for recent reviews]{2021Wyatt,2022Marino,2024Pearce}.  Submillimetre and millimetre images of debris discs are particularly informative as the dust grains that dominate at those wavelengths are not impacted by the transport forces that affect smaller grains, which dominate at shorter wavelengths. Submillimetre/millimetre images therefore provide a trace of the population of parent planetesimals that are only impacted by gravitational effects, meaning that they also provide indications of where planets might be in the system \citep[e.g.][]{2006Wyatt}.

\textbf{UK leadership}\\
UK based researchers have played a leading role in both submillimetre and millimetre observations and the theoretical interpretation of them. First detected through their infrared excess, the installation of the UK-led SCUBA instrument at JCMT led to some of the first resolved images of debris discs \citep{1998Greaves, Holland1998}. 
SCUBA's follow-up, SCUBA-2 greatly increased the number of debris discs detected in the submillimetre \citep{2017Holland}, enabling a variety of analyses of population statistics to be undertaken and providing indications of the best candidates for detailed follow-up with ALMA. With the exquisite resolution of ALMA (see Figure \ref{fig:debrisdisc}), the details of debris disc structure can be unlocked \citep[e.g.][]{2018Kennedy,2018Marino,2021CroninColtsmann}, further increasing our understanding of planet-disc interactions \citep{2023ImazBlanco} and the evolution from protoplanetary to debris phases. The first ALMA large programme on debris discs (ARKS, led by Sebastian Marino of the University of Exeter) is currently capitalising on this by studying the detailed structure of 18 debris discs down to resolutions of around 5~au.

\textbf{Science questions and instrumentation drivers for the coming decade}\\
ALMA has also greatly increased our understanding of molecular gas in debris discs. Gas, long seen as a distinguishing factor between protoplanetary and debris discs, has now been discovered in many young systems \citep{2016Kral}. Questions remain over whether this gas is a remnant from the protoplanetary phase or produced from ices in the collisional cascade \citep{2017Kral}.
Continued ALMA observations will build our understanding of debris discs, particularly with regard to planet-disc interactions, the evolution of planetary systems and origin of gas in debris discs \citep{2018kral}. Yet, ALMA is in danger of becoming source-starved and it is not efficient for discovering new debris discs. The discs studied with ALMA so far represent the brightest members of the population. In order to understand the characteristics of discs as faint as our Solar System's Kuiper belt, a large aperture single-dish telescope is needed \citep{2019Holland, Klaassen2024}.

\subsubsection{Protostellar Variability}
\label{sec:protostellar_variability}

\begin{figure}
    \centering
    \includegraphics[width=0.8\linewidth]{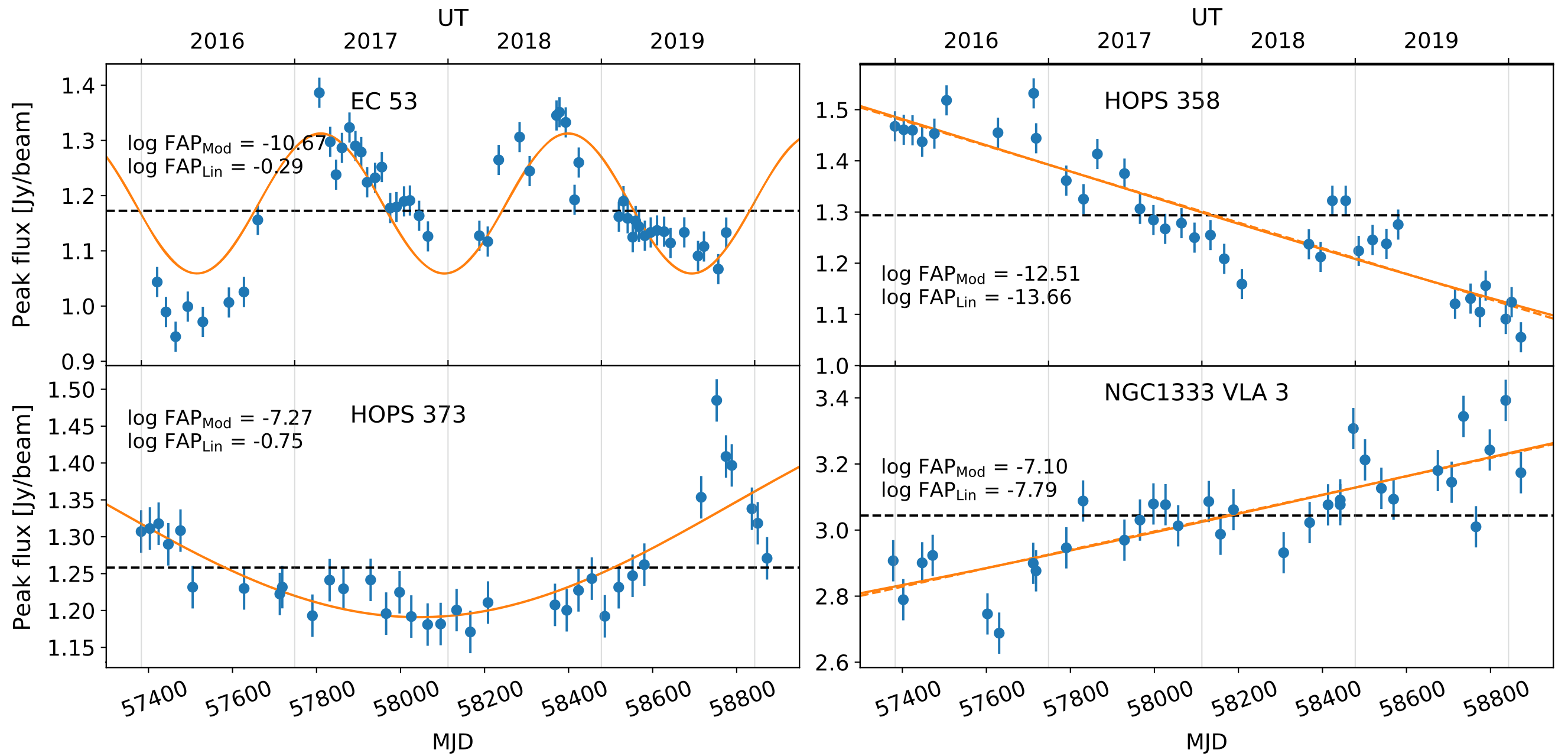}
    \caption{Examples of protostellar variability observed by the JCMT Transients Survey, adapted from \citet{lee2021}.}
    \label{fig:transients}
\end{figure}

\textbf{Introduction}\\
A classical problem of star formation is the ``protostellar luminosity problem'', in which the inferred accretion luminosities of low-mass protostars are significantly larger than the observed luminosities \citep{kenyon1990,enoch2009}.  This can be explained if protostars accrete much of their mass in short bursts; this accretion variability produces significant changes in their luminosity \citep[e.g.][]{fischer2023}.  While protostellar variability has been known about for a long time \citep[e.g.][]{hartmann1985}, it is only in the last decade that the extent to which young, deeply embedded (Class 0/I) protostars are variable in the submillimetre regime has been revealed. 

\textbf{UK Leadership}\\
The JCMT Transients Survey \citep{herczeg2017}, a project with significant UK involvement, 
has been monitoring eight 0.5-degree diameter fields in nearby star-forming regions with a monthly cadence since 2016.  The survey has identified years-long secular variation in more than 30\% of the protostars surveyed \citep{lee2021}, as well as identifying episodic accretion events in a Class I protostar \citep{yoo2017, lee2020}, a months-long accretion burst associated with a deeply embedded protostar \citep{yoon2022}, and an extraordinary stellar flare in a T Tauri binary system \citep{mairs2019}.  Examples of protostellar variability seen by the JCMT Transients Survey are shown in Figure~\ref{fig:transients}.  

\textbf{Science questions and instrumentation drivers for the coming decade}\\
In order to understand how protostars acquire their mass, it is necessary to quantify properly the frequency and amplitude of bursts as a function of protostellar mass and evolutionary stage.  Long-term mapping of a large number of protostellar sources, including those in more distant massive star-forming regions and the Magellanic Clouds, is required.  A single-dish instrument with high sensitivity and a large field of view would be ideally suited to achieve these goals. Meanwhile, the Simons Observatory Large Aperture Telescope (SO-LAT) will provide long-term monitoring at millimetre wavelengths of large areas of the sky, including star formation regions in our own galaxy. This will be able to provide a target list of millimetre-variable sources of all kinds, including protostars, for detailed followup with larger, more sensitive, telescopes. The UK is playing a leading role in SO through SO:UK\footnote{PI M. Brown, Manchester}. This includes work on sources and transients which is being led from Imperial College.

\subsubsection{Our Solar System}
\label{sec:solar_system}

\begin{flushright}
\textit{\textbf{STFC Science Challenge B2:} what effects do the Sun and other stars have on their local environment?}\\
\textit{\textbf{STFC Science Challenge B3:} what processes govern how planetary systems form and evolve?}\\
\textit{\textbf{STFC Science Challenge B4:} what are the conditions for life and how widespread are they?}
\end{flushright}

\begin{figure}
    \centering
    \includegraphics[width=0.8\linewidth]{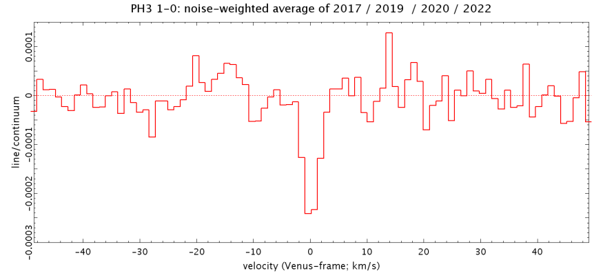}
    \caption{A multi-year coadd of the proposed PH$_{3}$ line in the Venusian atmosphere (\citealt{greaves2021}; J. Greaves, priv. comm.), combining JCMT and ALMA data.}
    \label{fig:phosphine}
\end{figure}

\textbf{Introduction}\\
Molecular species in the atmospheres of the planets of our Solar System can be detected and mapped using submillimetre and millimetre observations \citep{cordiner2024}.  For example, CO and HCN were first detected in the atmosphere of Neptune using the JCMT and the Caltech Submillimeter Observatory (CSO) \citep{marten1993}.  Such measurements can also be used to infer the properties of planetary interiors \citep[e.g.][]{lodders1994}.
In recent years, ALMA has significantly advanced  our understanding of the composition of planetary atmospheres.  CO, HCN, and HNC have been detected in the atmosphere of Pluto with recent ALMA observations \citep{lellouch2017}, while a variety of more complex molecules have been discovered in the atmosphere of Titan using ALMA \citep[e.g.][]{cordiner2015,cordiner2019,nixon2020}, some of which have astrobiological relevance \citep{palmer2017}.

Comets are thought to contain pristine material from the protosolar disk, and thus provide key information on the physical and chemical conditions in which the Solar System formed \citep[e.g.][]{mumma2011}.  Moreover, comets are rich in water and organic molecules, and so may have played an important role in delivering the ingredients for life to the Earth and other planets \citep{chyba1992}.  Submillimetre/millimetre spectroscopy of comets has resulted in the detection of a wide variety of molecules in cometary comas through their rotational transitions.  The JCMT has an illustrious history of measuring organic molecules such as CO, CS, HCN, CH$_3$OH, and H$_2$CO, amongst many others, in comets \citep[e.g.][]{coulson2020,yang2021}, including undertaking long-term monitoring projects \citep[e.g.][]{biver1999}.  ALMA is also undertaking comet observations with both its 12m array \citep[e.g.][]{cordiner2014, bogelund2017} and its compact array \citep[e.g.][]{roth2021}. These observations map the distribution of volatiles in the inner comas of comets at very high spatial and spectral resolution, thereby allowing parent and product species to be unambigiously distinguished between.

\textbf{UK Leadership}\\
One of the most high-profile results in recent years was the tentative detection of phosphine, a potential biomarker, in the upper atmosphere of Venus using the JCMT and ALMA \citep{greaves2021}, as shown in Figure~\ref{fig:phosphine}.  This result, led by UK astronomers, gained media attention around the world\footnote{\url{https://nature.altmetric.com/details/90068980}}.  While the detection and its interpretation remain under debate \citep[e.g.][]{lincowski2021,cordiner2022,mrazikova2024}, this work has energised discussion around interpretation of potential biomarkers \citep[e.g.][]{cockell2021}, and has informed planning for future Venus missions \citep[e.g.][]{gruchola2021,schulzemakuch2024}.  Meanwhile, the JCMT-Venus Large Program, led from the UK\footnote{Coordinators: D.L. Clements, Imperial; J. Greaves, Cardiff}, is performing long-term monitoring of the Venusian atmosphere and a UK SME spun-off from RAL, Teratech Components Ltd, has delivered submillimetre receiver hardware for ESA's mission to Jupiter and its moons, JUICE \citep{jupiter_moon}.

\textbf{Science questions and instrumentation drivers for the coming decade}\\
Key goals for submillimetre Solar System science include better characterising planetary wind fields and the thermal and chemical structures of planetary atmospheres, measuring the compositions of icy moon atmospheres and plumes, detecting new astrobiologically relevant gases in and performing isotopic surveys of comets, and synergising with future dedicated interplanetary space missions.

ALMA can place tens to hundreds of resolved pixels across Solar System planets, discovering atmospheric features such as jet streams and upwelling waves that connect to our understanding of Earth climate. ALMA also offers a huge span of frequencies to connect chemistry and dynamics across different layers of deep atmospheres,  such as Venus and the gas giants. JCMT can provide long-term monitoring of transient phenomena from known and newly discovered comets, through an agile target of opportunity programme. 

Interferometers and single dishes can work together to interpret the environment of plumes from icy moons, from searches for organics near vents to exotic chemistry in planetary plasma belts \citep{drabekmaunder2019}. A future larger single dish can also better detect the minor body population, where the submillimetre is vital to break the degeneracy of albedo and size as demonstrated by \textit{Herschel} for Trans-Neptunian Objects \citep{kovalenko2017}, and thus we will better understand solar system formation.

All of these observations offer synergy with space missions, including the JUICE flyby of Venus in 2025, and are capable of the rapid response needed for icy moon flybys and detections of venting episodes. For these goals, flexible telescope scheduling and instantaneous mapping over the entire area of Solar System bodies is critical, due to the rapidly rotating and evolving atmospheres of Solar System objects. Such observations could be provided by a new single-dish telescope with improved total power sensitivity and wide-field mapping capabilities.

\subsubsection{Solar physics}
\label{sec:solar_physics}

\begin{flushright}
\textit{\textbf{STFC Science Challenge B1:} how does the Sun and other stars work and what drives their variability?}\\
\textit{\textbf{STFC Science Challenge B6:} what are the processes that drive space weather?}
\end{flushright}

\textbf{Introduction}\\
The Sun at submillimetre and millimetre wavelengths is characterised by quiet-Sun continuum and bright sporadic emission during episodes of solar activity. 
The continuum radiation emitted by the Sun at millimetre
wavelengths arises from the chromosphere, the layer of the solar atmosphere
located between the photosphere and the corona \citep[e.g.][]{2016SSRv..200....1W},
while the sub-THz range of large solar flares is normally localised to small active 
regions. The origin of the solar flare sub-THz component remains a puzzle 
\citep{2010ApJ...709L.127F}.

The first observations of the Sun at submillimetre wavelength were made using
the JCMT, with UK involvement \citep{lindsey1995}, although the JCMT does not
have solar observing as a regular observing mode.  
The Solar Submillimeter Telescope (SST) in Argentina has been observing the Sun
since 2001 \citep{kaufmann2008}, and since 2016, ALMA has been observing the Sun
in both interferometric and total power modes \citep{shimojo2017}. 
There are many open questions about the Sun that require submillimetre observations,
including the thermal structure and heating of the solar chromosphere, 
the origin of flare submilimiter and milimiter and prominences, 
and the solar activity cycle.
Submillimetre astronomy may play a key role in our understanding of the origin 
of solar flare energetic particles and provide much needed diagnostics of the lower 
atmosphere and insights into the processes at the heart of space weather.
ALMA observations have revealed faint and localized sources,
likely thermal in origin, associated
with flaring emission in the extreme ultraviolet and soft X-ray regimes
\citep{shimojo2017a, skokic2023}.  
Submillimetre observations may thus be key to identifying the onset of, and
particle acceleration and transport in the lower atmosphere during 
solar flares and Coronal Mass Ejections \citep{fleishman2022}.

\textbf{UK Leadership}\\
The UK has a vibrant, internationally leading, solar physics community\footnote{\url{https://www.uksolphys.org/}}, 
conducting advanced MHD modelling \citep{2022NatCo..13..479S} to
develop the theory of millimetre emission in in solar 
flares \citep{2010ApJ...709L.127F},
as well as observations \citep{2019ApJ...875..163R}.  The first detection of the solar chromosphere at submillimetre wavelengths was made by UK astronomers \citep{ade1971}. 
The UK's solar physics research has a strong practical dimension contributing  
to the UK government's strategy to increase preparedness 
and resilience to severe space weather events\footnote{\url{https://www.gov.uk/government/publications/uk-severe-space-weather-preparedness-strategy}}.

\textbf{Science questions and instrumentation drivers for the coming decade}\\
Solving the longstanding coronal heating problem requires precise measurement of
the thermal, magnetic, and kinetic state of chromospheric plasma over time and
in three dimensions -- a problem that new submillimetre instrumentation would be
ideally suited to address \citep{2016SSRv..200....1W}.  
Solar prominence plasma is mostly optically thin at millimetre/submillimetre
wavelengths, which aids interpretation as there is a more direct relationship
between the observed flux and the plasma temperature of the emitting region than
at other wavelengths \citep{gunar2016,gunar2018}. In the future, simultaneous
observations in multiple millimetre/submillimetre bands may allow detailed
measurements of the kinetic temperature distribution of the prominence plasma to
be made. The SST has performed a sequence of observations of the Sun
demonstrating long-term radius variation indicative of changes in the solar
atmosphere that may be related to the solar cycle \citep{menezes2021}.
Observations of the full disc of the Sun in the submillimetre regime with a
daily cadence would allow investigation of the study of how the temperature (and
hence the energy content) evolves in Active Regions, Quiet Sun regions and
coronal holes, while extending such monitoring over many years would reveal how
both short- and long-term variability in submillimetre emission responds to the
solar cycle.
Solar flares and Coronal Mass Ejections, being manifestations of prompt energy releases 
and the drivers of space weather, produce electromagnetic radiation 
throughout the entire spectrum of electromagnetic emission \citep{2011SSRv..159..107H}.
However, the sub-THz range is probably the least understood due 
to the scarcity of the observations, complexity of the radiation mechanisms 
and bright-transient sources that are challenging to resolve 
at radio frequencies \citep{2019AdSpR..63.1404N}. 
For a dynamical sources like the Sun, temporal resolution is crucial,
with chromospheric submillimetre evolution sometimes happening 
on subsecond scales \citep{2018A&A...620A..95K}. 
The frequency-decreasing gyrosynchrotron spectrum produced by highest-energy 
electrons (and possibly positrons) is also emerging at millimetre wavelengths 
\citep{1985ARA&A..23..169D}.

Understanding the origin and processes behind acceleration of energetic particles 
is one of the main scientific objectives \citep{2020A&A...642A...3Z} 
of the recently launched ESA Solar Orbiter (SolO) missions, 
in which the UK has major instrument investment. 
These energetic particles constitute an important component 
of “Space Weather” understanding the nature of solar energetic particles at near the Earth 
is central to the prediction of such particle events and their associated
space weather effects, a key question being addressed in the UK, ESA and NASA.

\subsection{Star formation and galaxy evolution through cosmic time}
\label{sec:sf_gal_ev}

Star formation is the most important baryonic process that drives the evolution of galaxies.
Star formation converts the gas of the interstellar medium (ISM) into stars, depleting gas from the ISM, and feeds back energy, momentum and heavier elements into the ISM.
The physics of star formation is thus key to our understanding of how galaxies evolve.

\subsubsection{Star formation and the interstellar medium of the Milky Way}\label{sec:MW_star_formation}

\begin{figure}
    \centering
    \includegraphics[width=\textwidth]{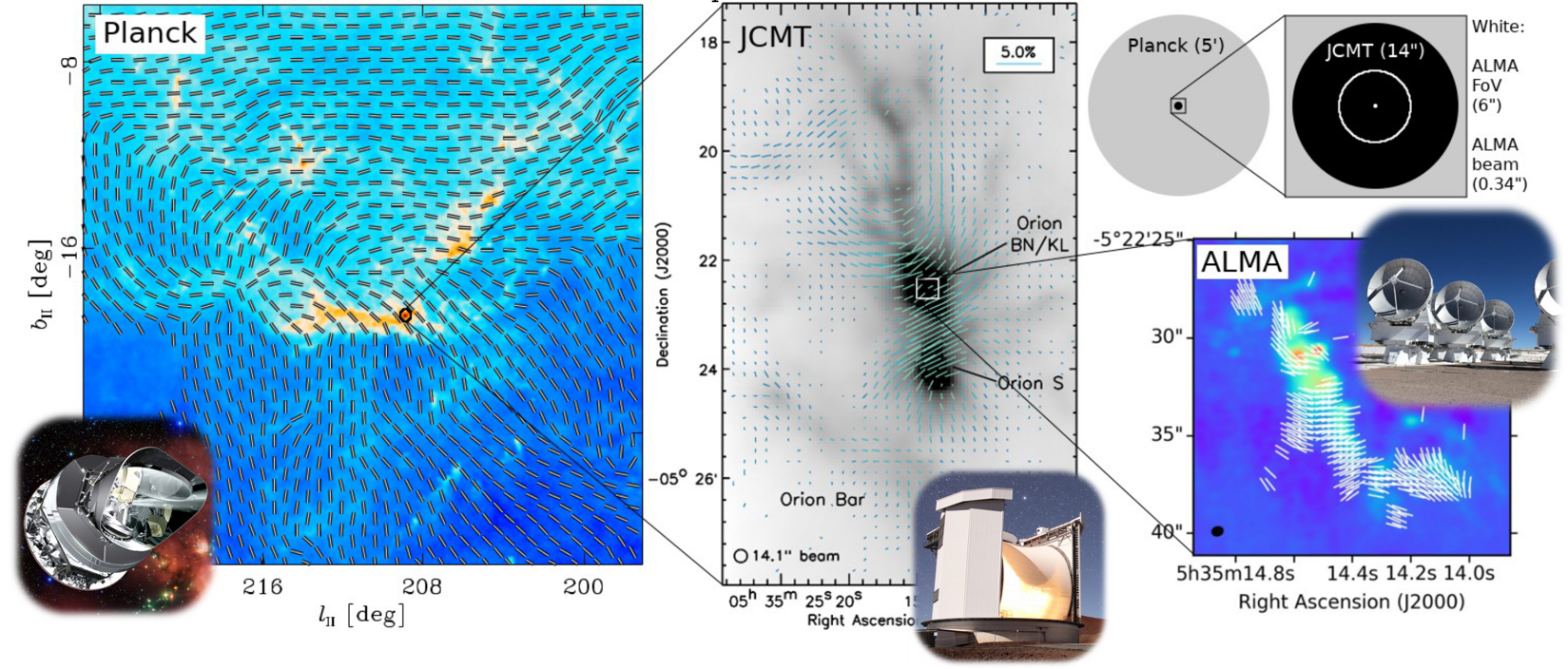}
    \caption{Dust polarization observations of the OMC-1 region of the Orion molecular cloud made at 350\,GHz/850$\mu$m over the last decade, demonstrating the need for both wide-area mapping and high-resolution observations in order to understand star formation and the physics of the ISM.  Left: \textit{Planck} observations at $\sim 5^{\prime}$ resolution \citep{planckintXIX}. Centre: JCMT observations at $14^{\prime\prime}$ resolution as part of the BISTRO Survey \citep{pattle2017}.  Right: ALMA observations made at $0.34^{\prime\prime}$ resolution \citep{pattle2021a}.  Figure adapted from \citet{furuya2020}.}
    \label{fig:bfield_scales}
\end{figure}

\begin{flushright}
\textit{\textbf{STFC Science Challenge A5:} How do stars and galaxies evolve?} \\
\textit{\textbf{STFC Science Challenge B3:} What processes govern how planetary systems form and evolve?}
\end{flushright}

\textbf{Introduction}\\
The submillimetre/millimetre regime is key to star formation and ISM studies.  Submillimetre continuum emission arises from cold dust grains, an excellent tracer of molecular hydrogen gas in regions of star formation \citep[e.g.,][]{Moore15}.  The polarization properties of this dust emission also allow mapping of interstellar magnetic fields \citep[e.g.,][]{andersson2015}.  Meanwhile, molecular line emission allows the dynamics and kinematics of these regions to be traced \citep[e.g.,][]{difrancesco2007}, and for the evolution of chemical complexity and the development of the molecules that are the building blocks of life to be mapped \citep[e.g.,][]{jorgensen2020}.  The last decade has seen the advent of several high-resolution Galactic Plane surveys by single-dish telescopes \citep[e.g.,][]{2016rigby,2016molinari,urquhart18}, significantly improving our understanding of the Milky Way, its star formation, and ISM cycle increase on all size scales.

Following the advent of \emph{Herschel}, a filamentary paradigm of star formation has received much attention \citep{Andre+14,Hacar+23}. In this scheme, interstellar filaments fragment into cores whose mass function (CMF) closely resembles the IMF \citep[e.g.][]{Konyves+15,Konyves+20,Ladjelate+20}. Meanwhile, hub-filament systems (HFSs) -- radial arrangements of filaments that converge on a central dense hub -- have emerged as likely sites of massive star formation \citep[e.g.][]{Peretto+13}, and efforts are underway to understand their evolution \citep{Anderson+21,Rigby+24}.
The star-formation efficiency (SFE), and rate (SFR), in star-forming clouds are measures of how much material is converted into stars in a given time, and is a key parameter in galaxy evolution.  SFE, as measured from submillimetre observations \citep{2016molinari,urquhart18} surveys, is found to not vary significantly as a function of Galactic environment \citep{Eden21}, 
and molecular clouds extracted from surveys such as CHIMPS, COHRS, and SEDIGISM, have found little contrast in properties between arms and inter-arms \citep[e.g.][]{Rigby19, Colombo19, Duarte-Cabral+21, Colombo22}, suggesting the MW’s spiral pattern might be more transient in nature.
Large variations are found from cloud to cloud \citep{Urquhart21}, implying that the internal physics and chemistry of molecular clouds is the most important scale setting the SFE.  This allows for detailed observations by targeted ALMA surveys, such as ALMAGAL\footnote{\url{http://www.almagal.org/}} \citep{2023atyp.confE..12J}, ALMA-IMF \citep{Motte22} and ATOMS \citep{Liu20}.

\begin{figure}[t!]
    \centering
    \includegraphics[width=\linewidth]{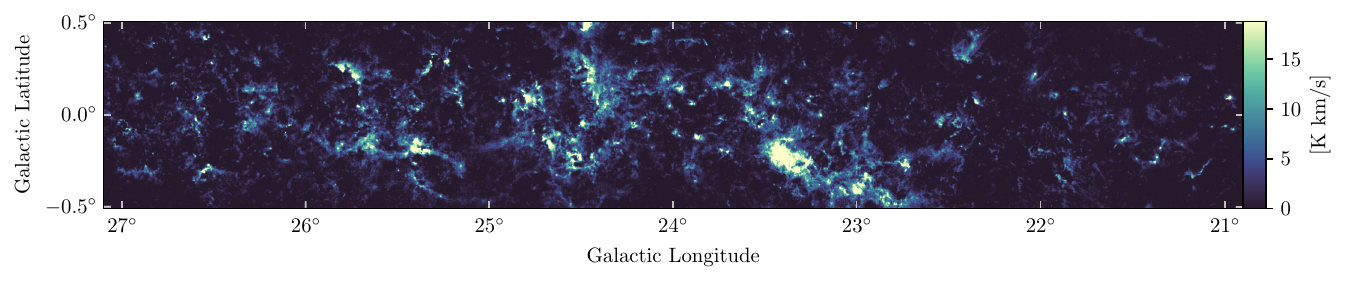}
    \caption{A six-square degree section of the UK-led JCMT Large Program CHIMPS2 in $^{13}$CO (3--2) integrated intensity. As part of CHIMPS2, the Inner Galaxy survey (Rigby et al. in preparation) extends the original CHIMPS survey \citep{2016rigby} to lower Galactic longitudes, probing molecular clouds and conditions within 4 kpc of the Galactic Centre (i.e. the bar-dominated region).}
    \label{fig:chimps2}
\end{figure}

The UK-led ACES ALMA Large Programme\footnote{\url{https://sites.google.com/view/aces-cmz/home}} 
will derive the properties of all potentially star-forming gas in the Galactic Centre, from global (100 pc) to proto-stellar core (0.05 pc) scales, down to sub-sonic ($<0.4$\, km/s) velocity resolution. The primary goal of ACES is to determine how global processes set the location, intensity and timescales for star formation and feedback in the Galactic Centre. 

Simultaneously, single-dish surveys such as the JCMT Large Program MAJORS\footnote{PI: D. Eden, Armagh} are investigating why empirical star-forming relationships survive all size scales from individual cores to entire galaxies. 

The ISM of the Milky Way and other galaxies is threaded by magnetic fields on all size scales \citep[e.g.][]{han2017}.  The rise of large-scale magnetic field surveys on single-dish telescopes and pointed high-resolution imaging with interferometers has led to a complex picture emerging, in which stars form from a magnetohydrodynamically (MHD) turbulent ISM \citep[e.g.][]{pattle2023,pineda2023}.  \textit{Planck} has provided all-sky dust polarization maps \citep{planckintXIX}, but cannot resolve the size scales within molecular clouds on which the transition to gravitational instability occurs.  The POL-2 polarimeter on the JCMT \citep{friberg2016} has been crucial for mapping magnetic fields on sub-parsec size scales in sites of star formation \citep[e.g.][]{pattle2017,Arzoumanian+21,karoly2023}. Meanwhile, ALMA's polarization capabilities \citep{nagai2016} have allowed magnetic fields to be mapped at extremely high resolution in both nearby \citep[e.g.][]{hull2017}, and distant massive \citep[e.g.][]{fernandezlopez2021} star-forming regions.

\textbf{UK Leadership}\\
The JCMT BISTRO Survey \citep{wardthompson2017}, a major international collaboration of approximately 200 astronomers led from the UK\footnote{PIs: 2016--2023 D. Ward-Thompson, UCLan; 2023--Present K. Pattle, UCL; co-Is from many other institutions.}, has performed the widest-area mapping of magnetic fields in molecular clouds to date, covering a total area of $\sim 1.5$ square degrees, at $\sim10^{\prime\prime}$ resolution, corresponding to 0.01 pc (1800 au) in the nearest star-forming clouds and 0.5 pc (110\,000 au) in the Galactic Centre. 

The UK-led ALMA Large Programme ACES\footnote{P.I.: Steve Longmore (LJMU)} is imaging the central molecular zone of the Milky Way. Imaging this region which extends for over 1.4$^\circ$ in galactic longitude has required the largest mosaic fields yet produced with ALMA.  The ALMAGAL ALMA Large Programme, which has imaged the population of cores in both spectral lines and continuum emission towards over 1000 star-forming massive molecular clumps in the Galactic Plane down to 1500 au linear resolution, has a UK representative on its steering group\footnote{G. Fuller, Manchester}. 

A number of UK-led projects have surveyed the Galactic Plane in spectroscopic CO rotational transitions. CHIMPS and CHIMPS2\footnote{PIs: 2017-2024 T. Moore, LJMU; 2024-Present D. Eden, Armagh Observatory} are observing 51 sq. deg of the central molecular zone and the inner Galactic disc and a portion of the outer disc in $^{12}$CO, $^{13}$CO and C$^{18}$O $J=3\to 2$ emission with the JCMT (Fig.~\ref{fig:chimps2}; \citealt{2016rigby,Eden20}), while CLOGS\footnote{PI: D. Eden, Armagh Observatory} is extending $^{12}$CO coverage in the outer disc. Complementary to this is the SEDIGISM programme\footnote{\url{https://sedigism.mpifr-bonn.mpg.de/index.html}; UK P.I.s: James Urquhart (Kent), Ana Duarte Cabral (Cardiff)} \citep{schuller2021}, a large-scale (84 sq. deg.) survey of the inner Galactic disc with the APEX telescope, primarily targeting the $^{13}$CO and C$^{18}$O $J=2\to 1$ rotational transitions that is  (largely) led from the UK.  Further UK leadership in spectroscopic surveys is found in the JCMT MAJORS project, which is investigating the role dense gas plays in the star-formation process by mapping star-forming regions in HCN and HCO$^{+}$ $J=3\to 2$ emission.

\textbf{Science questions and instrumentation drivers for the coming decade}\\ 
To truly understand the physics and chemistry of star formation, large samples of molecular clouds are required in many different tracers. Only by studying these are we in a position to determine how the conditions of the ISM vary across the Galaxy, and its impact on star formation (and vice-versa). 
Achieving such an understanding requires a move away from individual case studies, and towards statistical samples, identified through deep single-dish surveys with sufficient sensitivity and resolution to detect filamentary networks and their magnetic fields, and resolve any associated gas flows. High-angular resolution follow-up observations (e.g. with ALMA) of high-mass star-forming regions and their precursors will then allow us to determine how these gas flows and magnetic fields are organised at the smallest scales, and thus predict the probable sites of future high-mass star formation.

The ISM and its magnetic fields are continuous structures from Galactic scales ($\sim 10^{9}$\,au) to structure within protostellar discs ($<0.1$\,au).  While recent observations with the JCMT and ALMA have significantly enhanced our understanding, they are of necessity restricted in area and strongly biased towards regions of high molecular gas column density.  To fully understand where and how magnetic fields are important to the star formation process, unbiased surveys of polarized submillimetre dust emission covering a significant fraction of the Galactic Plane are required.  These surveys, which could be performed by a large single-dish telescope such as AtLAST, should have a resolution of a few arcseconds to allow feathering of these observations with sub-arcsecond-resolution ALMA images.

\subsubsection{Dust physics and evolved stars}
\label{sec:dust}

\begin{flushright}
\textit{\textbf{STFC Science Challenge A5:} How do stars and galaxies evolve?} \\
\textit{\textbf{STFC Science Challenge B1:} how does the Sun and other stars work and what drives their variability?}
\end{flushright}

\textbf{Introduction}\\
Thermal emission from cold ($\sim 3 - 30$\,K) dust is the major contributor to continuum emission in the millimetre and submillimetre ranges.
While this emission is often used to trace molecular hydrogen masks, it is also particularly important to studying the physics and properties of astrophysical dust itself \citep{Hensley2023}.
For example, the submillimetre spectral index constrains the size of dust grains -- critical to understanding how grains grow during planet formation \citep{Testi2014} -- and their composition.
Submillimetre imaging has been critical to estimating the total mass and composition of dust in galaxies both near \citep[e.g.][]{Smith2012, Cortese2012, Lamperti2019} and far \citep[e.g.][]{Beeston2018, Ward2024}, generally revealing much larger dust masses than can be explained by canonical models where dust is produced by evolved low- and intermediate-mass stars (e.g. asymptotic-giant-branch (AGB) stars).

This disconnect has motivated intensive study of dust producers in the submillimetre, in which ALMA and the JCMT have been instrumental.  
ALMA \& \textit{Herschel} observations of SN1987A demonstrated that core-collapse supernovae are capable of producing the $\sim 1$\,M$_\odot$ of dust required to explain dust production at very high redshift \citep{Matsuura2011, Indebetouw2014}, and JCMT data have been instrumental to estimating the dust yields of supernovae by exploring how effectively dust is destroyed in the forward and reverse shocks \citep[e.g.][]{Priestley2019}.
In parallel, ground-based submillimetre observations of AGB stars have shed light on the impact of binarity on the mass-loss process \citep[e.g.][]{Decin2020} and the chemistry involved in dust formation \citep[e.g.][]{Decin2018}, thanks to the high angular resolution of ALMA.
On the other hand, ongoing studies with single-dish telescopes are revealing peculiar dust properties from AGB stars \citep{Dharmawardena2018, Dharmawardena2019, Maercker2018, Maercker2022, Scicluna2022} suggesting that our understanding of submillimetre dust emission may have significant gaps; AGB dust appears to much brighter per unit mass than models calibrated on the mid-infrared have predicted, which may reflect differences in grain structure  or additional solid-state physics that must be incorporated in dust models. 

\textbf{UK Leadership}\\
The UK has a leading observational role in understanding dust from the submillimetre. The JCMT Large Programmes JINGLE \citep{2018saintonge}, HASHTAG \citep{2021smith} and DOWSING are revealing how dust properties change in the ISM of nearby galaxies, and are all led by UK PIs.
Meanwhile, the UK is a major contributor to the ALMA survey ATOMIUM \citep{Decin2020}, which is resolving a sample of AGB stars are very high resolution to measure the impact of companions down to planetary masses and the molecular content of their envelopes, including potential dust precursors. The Nearby Evolved Stars Survey \citep[NESS;][]{Scicluna2022}, on the other hand, is exploiting the JCMT and ALMA to measure the total dust and gas return by nearby AGB stars and the properties of the dust they produce; NESS has a significant UK contribution and a UK-based PI. 
The UK has also had a leading role in understanding dust production by supernovae: groups at both UCL and Cardiff have been particularly prolific \citep[e.g.][]{Matsuura2011,Priestley2019, Priestley2020, Kirchschlager2020, DeLooze2019, Chawner2019}.

\textbf{Science questions and instrumentation drivers for the coming decade}\\
Recent work has highlighted a number of key questions. These include whether stellar and sub-stellar companions actually alter the mass-loss properties of AGB stars or simply redistribute material and alter its chemical (and hence dust) properties, as well as the conditions required to actually initiate mass loss. While the first of these questions can be answered with a larger array of high-resolution ALMA observations and simulations, the latter requires fundamentally new capabilities. To date, mass-loss has only been effectively measured for samples of Galactic AGB stars; understanding the onset of mass loss will require observations of low-metallicity populations such as those of the Magellanic Clouds; only the combination of sensitivity and survey-speed of a large single-dish telescope like AtLAST can provide the required dataset. This will also have key implications for our understanding of chemical enrichment and the build-up of dust in the early universe.
Moreover, the emerging peculiarities of evolved-star dust raise further questions of our understanding of dust emission more generally. Establishing whether this is isolated to newly-formed dust or whether emissivities have been underestimated more generally will have significant implications for our interpretation of dust masses in galaxies both nearby and in the early universe, and will require both new observations measuring dust masses in large numbers nearby stars (and hence very high survey speeds in the submillimetre provided by future continuum instrumentation for JCMT or AtLAST) and close collaboration with laboratory groups to understand the physics underlying the dust emission itself.

\subsubsection{Nearby galaxies}
\label{sec:nearby_galaxies}

\begin{figure*} 
	\centering
   \includegraphics[scale=0.6,angle=0]{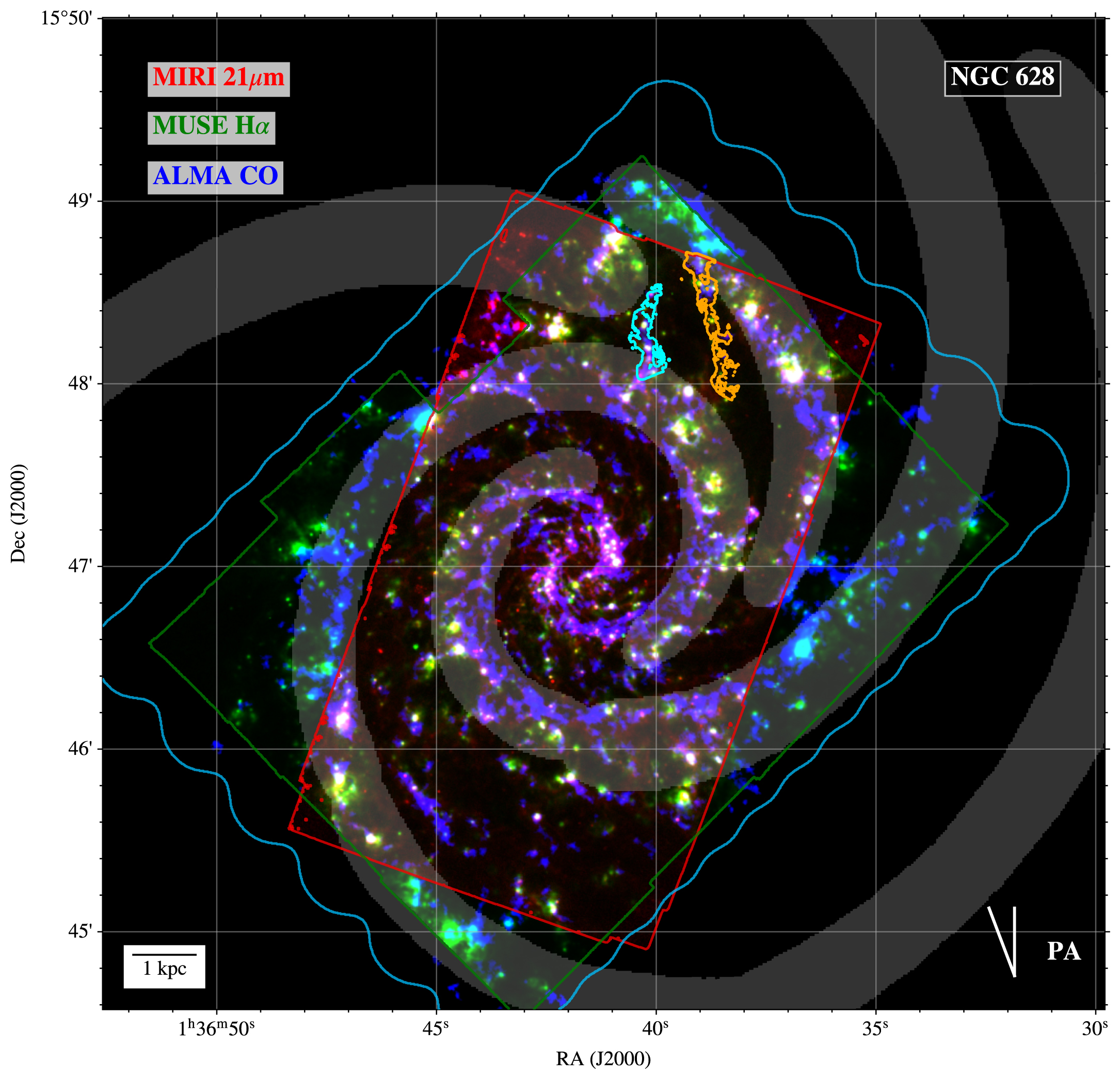}
	\caption{JWST/ALMA/VLT-MUSE three-colour image of NGC~628 (the Phantom Galaxy). JWST reveals filamentary structure between spiral arms (cyan and orange contours), and ALMA shows the molecular gas reservoir of this galaxy. From \citet{2022Williams}.}
    \label{fig:nearby_galaxies_alma_jwst}
 \end{figure*}

\begin{flushright}
\textit{\textbf{STFC Science Challenge A5:} how do stars and galaxies evolve?}
\end{flushright}

\textbf{Introduction}\\
Nearby galaxies (those with distances $\lesssim100$~Mpc) provide a unique opportunity to study galaxy evolution, offering a high-resolution external perspective.  Recent work has shown that the scale of individual molecular clouds ($\sim$100~pc) is a critical one -- at these scales, galaxy-integrated star formation laws start to break down \citep[e.g.][]{2010Schruba, 2014KruijssenLongmore, 2018Williams}, and the local processes driving galaxy evolution become apparent \citep[e.g.][]{2020Sun, 2021Leroy}. 
The last decade has been an extremely fruitful one for submillimetre observations of nearby galaxies, revealing a number of important results and providing tantalising questions to follow-up in the coming years.

\textbf{UK Leadership}\\
The UK has been at the forefront of much of the submillimetre efforts in observations of nearby galaxies. Projects such as JINGLE \citep{2018saintonge} mapping hundreds of nearby galaxies, and HASHTAG \citep{2021smith} covering M31 have both been led by UK PIs, taking advantage of our access to the JCMT. These projects are revealing real changes in the dust properties both within and between galaxies, and combined with other data will provide an important database for studies of the ISM.

There are also now members of the PHANGS collaboration in the UK. This is the largest multi-wavelength survey of nearby galaxies, and the work from this team highlights the power of combining data from various observatories (see Fig. \ref{fig:nearby_galaxies_alma_jwst}). The synergies with {\it JWST} are obvious, giving us our best look at the hot dust and gas content of galaxies to-date.

Work has also been led by UK PIs pushing to studies beyond the star-forming main sequence. Work from the WISDOM team (led primarily in the UK) has shown the variations in gas morphology \citep{2022Davis} and gas conditions \citep{2023Williams} in quiescent galaxies. It appears that despite the large reservoirs of molecular gas in these galaxies, this gas is unlikely to collapse and will disperse on timescales shorter than those required for star formation.

Finally, UK researchers have also been pushing to map the magnetic fields in nearby galaxies. The relatively new POL-2 instrument on the JCMT allows for mapping of magnetic fields, and recent results have shown that these typically align with the spiral arms \citep{2021Pattle}. These studies are complicated, however; magnetic fields are difficult to measure and multiple sources of the field may complicate the interpretation. Pushing forward here is critical, and the UK is ideally placed for this kind of science.

{\bf Science questions and instrumentation drivers for the coming decade}\\
Despite the breakthroughs made in the past decade, a number of open questions remain. The UK submillimetre community is well-placed to answer these given both current and upcoming observatories.

In particular, work on characterising the ISM has been focused almost exclusively on observations of $^{12}$C$^{16}$O. Isotopologues of CO (e.g. $^{13}$CO or C$^{18}$O) allow for complex modelling of the gas properties via radiative transfer. These lines are fainter and detecting them is significantly harder than CO, making ALMA nearly unique in this area. ALMA's upcoming wideband sensitivity upgrade will help with efficiently observing these isotopologues (and other more exotic molecules), as following this upgrade it will be possible to observe many of these transitions simultaneously.

Another topic that has received relatively little attention, despite its clear importance in shaping galaxies, is magnetism. We now have some observations of magnetic fields in galaxies that roughly resolve features such a spiral arms, as well as Milky Way observations that highly resolve individual clouds. How these are linked is currently completely unknown; is there order all the way down, or do the fields decouple at some point? The number of cloud-scale magnetic fields measurements we have in external galaxies is six \citep{2011LiHenning}, and misses the larger-scale magnetic field. The NASA FIR Probe mission candidate PRIMA, if selected, will be a game-changer here, and mapping the entire disc of M31 and M33 at cloud-scale will take on the order of 10 hours \citep[Williams et al., contribution in][]{moullet2023}. By tracing the magnetic fields at all scales, we will be able to see how they drive (or inhibit) star formation in a robust way.

Finally, the limiting factor in many extragalactic studies of the links between the ISM and star formation is becoming the resolution of dust maps. With {\it Herschel}'s relatively limited resolution in the submillimetre (36$^{\prime \prime}$ compared to the 1$^{\prime \prime}$ or better achievable with ALMA), the mismatch in these scales is rapidly becoming problematic for studying the links between dust and gas. An instrument such as AtLAST with its much larger mirror will help alleviate this, allowing for a much sharper look at the dust in nearby galaxies.

\subsubsection{Supermassive Black Holes and Active Galactic Nuclei}
\label{sec:smbh}

\begin{flushright}
\textit{\textbf{STFC Science Challenge A5:} how do stars and galaxies evolve?}\\
\textit{\textbf{STFC Science Challenge A7:} what is the true nature of gravity?}
\end{flushright}

\begin{figure*} 
	\centering
   \includegraphics[scale=0.6,angle=0]{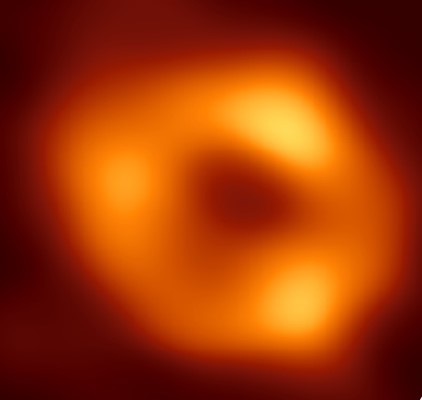}
	\qquad % Leaves a gap between the pictures. You could also use \hfill
	\includegraphics[scale=0.6, angle=0]{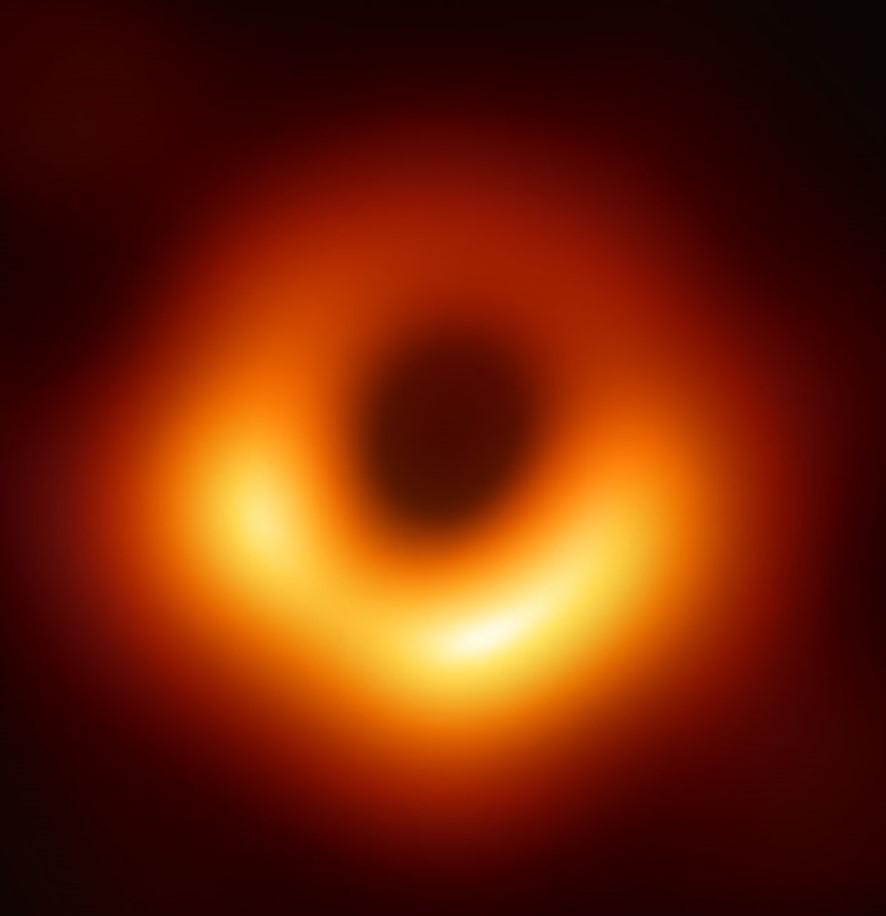}
	\caption{Left: An image made by the EHT of the region around the event horizon of the black hole at the centre of our Milky Way Galaxy, known as SgrA* \citep{eht2022}. 
 Right: An EHT image of the black hole event horizon in the galaxy M87, known as M87* \citep{eht2019}.
    Image credit: EHT collaboration.}
    \label{fig:SgrA*-M87*}
 \end{figure*}

\textbf{Introduction}\\
One of the most outstanding scientific results of millimetre astronomy over the last decade has been the imaging of the event horizons of the supermassive black holes (SMBHs) at the hearts of M87 \citep{eht2019} and our own Milky Way galaxy \citep{eht2022} by the Event Horizon Telescope (EHT), which has gained world-wide media attention\footnote{\url{https://www.altmetric.com/details/58823388}}.
The imaging of SMBHs was a stated goal of the EHT project \citep{eht2019}, and it has required a concerted global effort to advance the Very Long Baseline Interferometry (VLBI) technique to succeed.
The image of the silhouette of an SMBH in the centre of the relatively close galaxy Messier 87, which the EHT Collaboration published on April 10, 2019 \citep{eht2019} is the most striking outcome of this venture.  In agreement with theoretical predictions, it gives overwhelming evidence for the presence of a black hole.  This was followed by EHT imaging of the SMBH at the centre of our own galaxy, Sgr A$^{*}$ \citep{eht2022}.

The EHT Collaboration has investigated the morphology of sixteen AGN sources observed in 2017, focusing on the properties of the VLBI core, namely size, flux density and brightness temperature. Of these, seven have been published already and the remainder are being worked on.  EHT data of Centaurus A were published by \citet{janssen2021}, who found that the source structure of Centaurus A 
is consistent with a SMBH and jet. 
They identified the exact location of the Cen A SMBH with respect to its resolved jet core at a wavelength of 1.3\,mm, and concluded that the source's event horizon shadow should be visible at submillimetre wavelengths.  The most distant AGN observed by the EHT so far is NRAO530, which lies at z=0.902 \citep{jorstad2023}.

The EHT consists of a network of submillimetre and millimetre telescopes: ALMA, the JCMT, APEX, the Submillimetre Array (SMA), the Submillimetre Telescope (SMT), the Large Millimetre Telescope Alfonso Serrano (LMT), the IRAM 30-meter telescope, the South Pole Telescope (SPT), NOEMA, the 12m telescope on Kitt Peak, and the Greenland Telescope (GLT), with more EHT stations planned.

\textbf{UK Leadership}\\
The main UK groups involved in the EHT are UCL, MSSL, Oxford, Cardiff and UCLan.
The EHT is split into working groups and the UK has membership of several of these groups. The UK has the Chair of the Gravitational Physics Working Group (Z. Younsi, MSSL) and the Chair of the Publications Working Group (D. Ward-Thompson, UCLan), who is the communicating author for the major collaboration publications, and who has key editorial control over all outputs from the EHT. Younsi is also a member of the Science Board that determines the overall scientific direction taken by the EHT.

Younsi has also been working at	the forefront of radiative transfer in strong gravity systems,	developing theoretical formulations and numerical schemes	which now play a foundational role in predicting black hole images and	enabling tests of gravity near	event horizons.	His	numerical code,	BHOSS, is one of only two codes used by the EHT to perform these studies and is	now used in many research groups around the world.

\textbf{Science questions and instrumentation drivers for the coming decade}\\
Both M87* and SgrA* were observed in 2017 and 2018.  Both epochs were published for M87*, with so far only the 2017 image having appeared for SgrA*. From these data, coupled with a large amount of simulation work, the detailed properties of the black holes and their event horizons have been calculated. Going beyond these two to fainter sources requires expansion of the EHT network, particularly with more sensitive telescopes \citep{Akiyama2023}. With telescopes capable of observing the submillimetre, the resolution becomes high enough to separate out the multiple photon rings that are merged together in the current images \citep{Johnson2020, Tiede2022}.

Other major applications of VLBI include astrometric studies, for geodesy \citep{schuh2012}, and for the physics of the circumstellar and interstellar media, \citep[e.g.,][]{reid1980}.

This indicates the next step for the EHT: observing in the submillimetre regime.  The JCMT is crucial to the EHT successfully making images at such high frequencies, because it lies on the longest east-west baseline and is on one of only three sites that can achieve such high frequency observations.

\subsubsection{High redshift galaxies} 

\begin{flushright}
\textit{\textbf{STFC Science Challenge A2:} how did the initial structure in the universe form?}\\
\textit{\textbf{STFC Science Challenge A3:} how is the universe evolving and what roles do dark matter and dark energy play?}\\
\textit{\textbf{STFC Science Challenge A4:} when and how were the first stars, black holes and galaxies born?}\\
\textit{\textbf{STFC Science Challenge A5:} how do stars and galaxies evolve?}
\end{flushright}

\paragraph{To Cosmic Noon and beyond}\label{sec:cosmic_noon}

\textbf{Introduction}\newline 
All the processes at work in our Galaxy and in nearby galaxies discussed above are also found at cosmological redshifts, and thus to study detailed astrophysics in external galaxies, it is essential to be able to observe and compare the same key diagnostic lines and continuum emission that are studied in the context of our more local star-forming regions. 
These observations make a particularly useful comparison with semi-analytic cosmological models at ``cosmic noon'' \citep[e.g.,][]{Hodge_daCunha_2020}, when the comoving volume-averaged star formation density peaked, and when the black hole accretion comoving density also peaked.

\textbf{UK leadership}\newline 
The UK has an illustrious track record in the ground-based discovery and analysis of galaxies at cosmic noon at submillimetre wavelengths. 
There are many striking indications that submillimetre galaxies \citep[SMGs, originally known as ``SCUBA galaxies'' after the UK-built instrument that first detected them,][]{scuba-1,smail+97, Hughes+98} are the progenitor population of present-day giant elliptical galaxies. Firstly, their very high star formation rates can assemble a present-day giant elliptical in $\sim1$\,Gyr \citep[e.g.,][]{Ikarashi+17}. Secondly, ultraluminous
infrared galaxies
are on the ``main sequence''
galaxy scaling
relation by
redshift $z\sim2$ \citep[e.g.,][]{Elbaz+18} and provide a large fraction of the comoving volume-averaged star formation density \citep[e.g.,][]{casey+14}, while being very efficient at creating dense gas and converting it to stars \citep[e.g.,][]{gao+07}. Thirdly, the submillimetre galaxy number densities, bias parameter and clustering are all consistent with models of the formation of giant elliptical galaxies \citep[e.g.,][]{stach+21}. In summary, from the UK's inception of submillimetre-wave survey instrumentation with SCUBA and SCUBA-2 \citep{scuba-1,scuba-2}, to UK leadership of many large ground-based survey programmes \citep[e.g.][]{smail+97,Hughes+98,scott+02,geach+17}, the UK has driven many of the physical insights into dust-obscured star formation at cosmic noon over the past three decades. 

There is an analogous UK success story in space-borne FIR and submillimetre astronomy around cosmic noon and beyond, initiated by the UK-led SPIRE instrument \citep{Griffin_SPIRE} on the ESA \textit{Herschel} mission \citep{Pilbratt+10_Herschel} and by the \textit{Planck} high-redshift point sources \citep{Planck_PHz}, and driven by several UK-led legacy surveys (e.g., \citealp{Oliver+10}, \citealp{eales+10}; see also Section~\ref{sec:protoclusters}). These have led to a broad and diverse range of new insights into galaxy evolution \citep[e.g.,][]{wang+21_hermes_hlirgs,asboth+16_hermes_ultrared,amblard+11,burgarella+13_hermes,Ward+24,Eales+24,Eales+18_paradigm,Eales+18_green_mountain,Paspaliaris+23,Noboriguchi+22,Oteo+17,3000_ultra_reds}, strong gravitational lensing \citep[e.g.,][]{negrello+10,negrello+17,wardlow+13,Reuter+20,bakx+24_flash,Planck_GEMS_1,Planck_GEMS_2,Planck_GEMS_3,Planck_GEMS_4,Planck_GEMS_5,Planck_GEMS_6,Planck_GEMS_7,Planck_GEMS_8,swinbank+15,hezaveh+16_sdp81} and the interactions between the baryonic and dark matter sectors \citep[e.g.,][]{Vegetti+14,Despali+18,Li+16,Li+17}.

\begin{figure}
\includegraphics{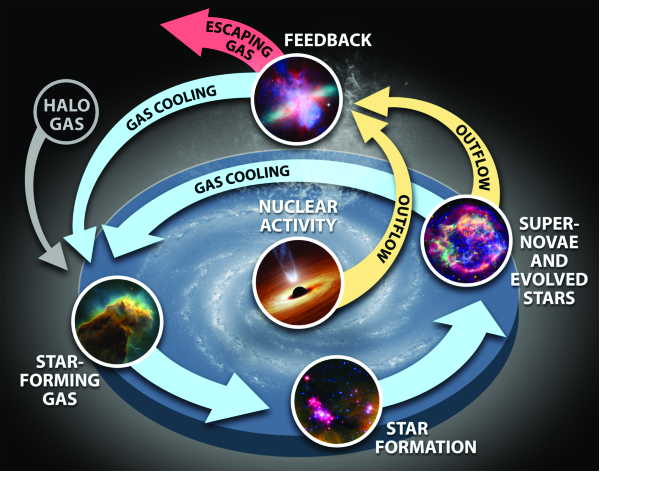}
\caption{Illustration of the baryon cycle in galaxy evolution, from 
\citet{origins_space_telescope_2019}.
Many of the energetic processes that shape this ecosystem are uniquely accessible to submillimetre and millimetre-wave observations.
\label{fig:baryon_cycle}
}
\end{figure}

\textbf{Science questions for the coming decade}\newline 
A consistent pattern in both the ground-based and space-based pioneering UK work in submillimetre astronomy has been that the wide-field continuum surveys have been needed to feed detailed follow-up programmes. 
Blank-field and cluster-lensed surveys of submillimetre galaxies have provided very large catalogues, but because of their dusty nature, spectroscopic redshifts tend to rely on submillimetre and millimetre-wave molecular and ionic lines, through large ALMA and NOEMA campaigns \citep[e.g.,][]{Neri+20,Reuter+20,Urquhart+22,Cox+23}; in the case of the \textit{Herschel} H-ATLAS project, target identification for the spectroscopic follow-up was via ground-based detection by the UK-built SCUBA-2 camera \citep{bakx+18}. These spectroscopic campaigns are in turn essential prerequisites for the majority of the science exploitation of these catalogues, especially in illuminating the physical processes of the baryon cycle in galaxies illustrated in Fig.\ \ref{fig:baryon_cycle} \citep[e.g.][]{origins_space_telescope_2019,Celine_Cristopher_2020_ARAA}.  Working backwards, the outflows driven by star formation and active nuclei have many molecular line observables in water, CO, OH, OH$^+$, etc, as well as in H$\alpha$ and X-ray; the feedback processes themselves are observable in principle through supernovae, supernova remnants and active nuclei; star formation has multiple observational signatures from the mid-infrared to submillimetre, as well as H$\alpha$, ultraviolet and radio; dense gas phases that fuel star formation are traceable with many molecular lines, including HCN, HNC, HCO$^+$ and more, while the bulk of the molecular gas is traceable in CO \citep[modulo the presence of CO-dark gas also traced by far-infrared lines, e.g.,][]{2020Madden,dunne+21_co_dark}; finally, the cold gas flows from the halo are difficult to observe directly, but their existence can be inferred from models of protoclusters (see below in Section~\ref{sec:protoclusters}). This has led to a very wide-ranging, diverse and deep characterisation of the dusty interstellar media at cosmic noon through submillimetre-wave and millimetre-wave observations \citep[e.g.,][]{zgal_2_dust,zgal_3_properties,reuter+23,bears_2_dust,bears_3_properties,dye+22}.
ALMA has also been key to the morphological and kinematic investigation of strongly gravitationally lensed submillimetre galaxies \citep[e.g.,][]{vlahakis+15}, and has been used to reveal the existence of extremely dusty, optically-dark submillimetre galaxies, often at $z\stackrel{>}{_\sim}4$ \citep[e.g.][]{simpson+14,chen+15,Ikarishi+15,ikarishi+17,cowie+18,smail+21,wang+19_dark_smgs}.  Much of this spatial and spectral characterisation is still at relatively early stages, so the key challenges in this area are therefore centred around the over-arching question, \textbf{how does the baryon cycle operate around cosmic noon?} 

\textbf{Instrumentation drivers for the coming decade}\newline 
There is a very clear and obvious scientific synergy between the wide-field submillimetre galaxy surveys with e.g. \textit{Herschel}, JCMT SCUBA-2, LMT TolTEC, SPT, ACT etc. (and in future CMB stage 4 experiments), and interferometric follow-ups with e.g. ALMA, NOEMA, JVLA, ASKAP and in future SKA, which in turn are also key to the detection of molecular and ionic species to characterise the interstellar media at cosmic noon. The ALMA wide-band sensitivity upgrade will also be enormously advantageous for the efficient characterisation of the interstellar media at cosmic noon.

\begin{figure}
\includegraphics[]{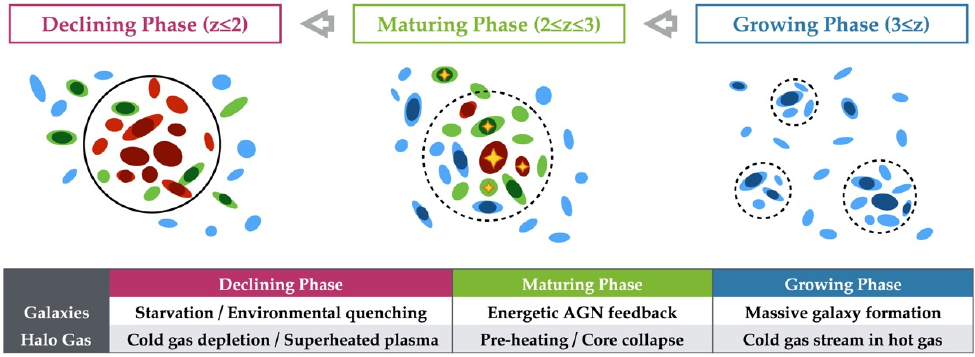}
\caption{Schematic description of the formation and evolution of galaxy protoclusters, as proposed by \citet{Shimakawa+18}. 
\label{fig:protocluster_schematic}}
\end{figure}

\paragraph{Protoclusters}\label{sec:protoclusters}
\textbf{Introduction}\newline 
Galaxy protoclusters provide some of the strongest challenges to semi-analytic
hierarchical models of structure formation \citep[see Figure\ \ref{fig:protocluster_schematic}, e.g.,][]{Overzier+16_protoclusters,Shimakawa+18,Chiang+17_protoclusters,Gouin+22_protocluster_theory}. In the approach to cosmic noon, models predict a transition in the core halos of protoclusters, from inflows of cold gas existing within hot massive halos, to a later regime in which all of the gas is accreted into the massive halos and shock-heated to the virial temperature \citep[e.g.,][]{Overzier+16_protoclusters}. These flows of low-metallicity cool gas are often difficult to detect directly in the intergalactic medium, but a filamentary stream of neutral carbon has been detected at rest-frame submillimetre wavelengths in the protocluster environment of the $z=3.8$ radiogalaxy 4C41.17 \citep{Emonts+21_CI_4c41.17}, and a filamentary Sunyaev-Zeldovich signal has been observed in the Spiderweb protocluster \citep{DiMascolo+23_spiderweb_SZ}. 

\textbf{UK leadership}\newline 
The UK again has an illustrious track record in the discovery and characterisation of these systems. \citet{Stevens+03_signposts} used SCUBA to demonstrate clearly that at least some high-redshift radiogalaxies exist in rich environments of star-forming galaxies. Physically, this was interpreted as a causal or inferential chain starting with systems ostensibly with the largest supermassive black holes, within what one would therefore expect to be the most massive spheroid or (proto-)elliptical host galaxies, and finally therefore within the richest protocluster environments. This ``AGN signpost'' approach to finding protoclusters has been enormously influential, leading to e.g. the JCMT Large Program ``RAdio Galaxy Environment Reference Survey'' \citep[RAGERS,][]{greve_ragers,RAGERS_Cornish,RAGERS_Zhou}, the SCUBA-2 High Redshift Bright Quasar Survey \citep{Li+23_SCUBA2_quasars_z6} that has already found evidence for richer-than-average submillimetre galaxy environments around $z\sim6$ quasars, and the discovery of a protocluster environment of star-forming galaxies around the Spiderweb radiogalaxy MRC1138-262
\citep{Dannerbauer+14_spiderweb}. 

\textbf{Science questions for the coming decade}\newline 
The key challenges in this area are centred around the over-arching questions, \textbf{how can protoclusters best be found, and are their properties consistent with semi-analytic cosmological models?} 
With the large numbers of high-redshift ultraluminous and hyperluminous galaxies now available from e.g. \textit{Herschel}, ACT, SPT and \textit{Planck}, there has been an increasing focus on using these instead as markers of potential protoclusters. The inferential chain in this case is more simple: protoclusters should be peaks in the cosmological star formation density field (Figure\ \ref{fig:protocluster_schematic}), confirmed by stacking analyses of $z\sim4$ optically-selected protocluster candidates in the far-infrared and submillimetre wavelengths \citep[][and curiously, the authors found no significant infrared flux excess around optically selected quasars at similar redshifts]{Kuko+19_planck_stacks_of_hsc_protoclusters}. Therefore, protoclusters should be more likely than the field to contain hyperluminous starbursts. Similarly, hyperluminous galaxies should be useful signposts of protocluster environments. This alternative approach to the ``AGN signpost'' has already had a number of successes \citep[e.g.][]{Lammers+22_Planck_SPIRE_protocluster_candidates,Polletta+21_Planck_protocluster_cosmos,Cheng+19_Planck_SCUBA2_images,Cheng+20_Planck_SCUBA2_overdensities,Polletta+22_protocluster_molecular_gas,Lewis+18_ultrareds_signpost_protoclusters,wang+21_spt_unlensed_smgs_signpost_protoclusters,Zhou+24_herschel_irac_overdensity_noema_protocluster,calvi+23_bright_smgs_signpost_protoclusters_goodsn,Arribas+23_z6.9_protocluster_SPT0311-58,bakx+24_herbs70_protocluster_scuba2_noema,RAGERS_Zhou}.

\textbf{Instrumentation drivers for the coming decade}\newline 
Unless there is abundant multi-wavelength data for photometric redshifts or spectroscopic follow-ups, long-wavelength mapping at wavelengths $\geq 850\,\mu$m  remains essential for demonstrating rich protocluster environments, because of the greatly suppressed foreground contaminant populations at longer wavelengths. There is therefore a strong scientific case for maintaining and developing UK capabilities in submillimetre continuum survey instrumentation, in the area of protocluster detection and characterisation. ALMA does not have the wide-field mapping efficiency to be able to achieve this; rather, ALMA's strength is in the subsequent pencil-beam spectroscopic analyses for redshift confirmations and for ionic and molecular diagnostics of the interstellar media, in targets supplied once again by the wide-field continuum mapping. As with galaxies at cosmic noon in general, the ALMA wide-band sensitivity upgrade will also be enormously advantageous for the efficient characterisation of the interstellar media in proto-cluster members.

\paragraph{Cosmic Dawn}\label{sec:cosmic_dawn}

\textbf{Introduction}\newline 
The window to a substantial number of galaxies at ultra-high redshifts ($z\stackrel{>}{_\sim}9$) has been opened by JWST, growing the multi-wavelength coverage in the cosmological survey fields established for cosmic noon. This ``cosmic dawn'' epoch covers the reionization of the universe by the first stars and the earliest black hole accretion. The first JWST images led to claims of photometric redshifts as high as 20, and while recalibrations of the photometry and JWST NIRSpec spectroscopy have moderated some of the more extreme early claims, confirmed spectroscopic redshifts are now available well into reionization \citep[e.g.,][and references therein]{Harikane+24}. There are many indications of different mechanisms for stellar mass assembly and/or SMBH growth at these early cosmic epochs. The surprising existence of supermassive black holes at these very early epochs indicates either phases of super-Eddington accretion, or very massive seed black holes \citep[e.g.,][]{maiolino+24_jwst_smbh_gnz11,dayal_2024_smbh_jwst_highz_models,Bogdan+24_z10_xray}. In the approach to cosmic dawn, populations of ``little red dots'' indicate either heavily dust-shrouded active nuclei or stellar mass assembly processes that have no obvious analogues in the later universe \citep[e.g.,][]{kocevski+24_lrd,Kokubo+24_lrd}. 

\textbf{UK leadership}\newline 
JWST's instrument complement includes the UK-led MIRI instrument, and the ESA-built NIRSpec instrument that also has UK instrument team members. The UK has led many pioneering results in the approach to cosmic dawn and into reionization, including and especially with results from ALMA working in the submillimetre and millimetre domain \citep[e.g.][]{Laporte+21,Laporte+19}. The JWST project PRIMER is UK-led \citealp{dunlop+21_primer,donnan+24_primer}, and there are many other major projects with UK involvement such as the JWST Advanced Deep Extragalactic Survey, JADES \citep{JADES_overview_Eisenstein+23}.

\textbf{Science questions for the coming decade}\newline 
Submillimetre and millimetre-wave observations are playing crucial roles in probing these early epochs, addressing key questions including: \textbf{When and how did reionization occur? What were the first galaxies, and what was their metallicity evolution during reionization?} ALMA is able to obtain detections of redshifted emission lines at even these very early epochs \citep[e.g.,][]{Harikane+24_ALMA_Keck,Fujimoto+23}. The redshifted [O{\sc iii}] $88\,\mu$m emission line is predicted to be the brightest emission line at cosmic dawn, partly due to the short $\sim50\,$Myr timescale for oxygen formation compared to $\sim500\,$Myr for carbon \citep[e.g.,][]{Bakx+23,Maiolino_Mannucci_2019,Inoue+16_88um_z=7.2120,Bouwens+22_REBELS}, and partly because it is harder for low-metallicity gas to cool through the emission lines, raising the excitation temperature \citep[e.g.,][]{Edmunds2024,Maiolino_Mannucci_2019}. ALMA unlocks a wide range of follow-up characterisation of the physical properties and ionization states in the high-redshift interstellar media. 
For example, there are already hints of a low carbon abundance and a top-heavy initial mass function approaching reionization \citep[e.g.,][]{Katz+22}. 

Another way in which submillimetre and millimetre-wave observations are proving crucial in the study of cosmic dawn is the identification of a photometric population of lower-redshift red galaxies, that is far more numerous on the sky than ultra-high-redshift galaxies \citep[e.g.,][]{Bouwens+23,SerjeantBakx23} and therefore a source of false-positives in the search for ultra-high redshift systems. 
\textbf{How many ultra-high-redshift galaxy candidates are in reality lower-redshift dusty star-forming galaxies?}
For example, \citet{Zavala+23} discovered SCUBA-2 emission in several $z>10$ galaxy candidates that had been selected photometrically on the basis of its very red JWST colours; the revised photometric redshifts are $z\sim5$. Since the dust in submillimetre galaxies typically leads them to be very red in the rest-frame optical and ultraviolet, they are obviously an important contaminant population. An analogous submillimetre galaxy contaminant problem for ultra-high-redshift galaxies discovered via strong gravitational lensing has also been noted by \citet[][]{Pearson+24}. 

\textbf{Instrumentation drivers for the coming decade}\newline 
Receiver upgrades for ALMA are a key priority for the study of the reionization epoch \citep[e.g.,][]{2023pcsf.conf..304C}, including the detection of e.g. [N{\sc II}] $121\,\mu$m out to redshift $z=11.3$, [O{\sc III}] $52\,\mu$m to $z=12.6$, and 
[O{\sc I}] $63\,\mu$m to  $z=10.2$. Searches for spectral lines in systems without redshift measurements will be more efficient with a new ALMA 8\,GHz IF receiver, while wide-field continuum surveys in ultra-deep galaxy survey fields still remain a driver for single-dish survey facilities in the submillimetre and millimetre range.

\subsection{Formation and evolution of cosmological structure}

%%%%%%%%%%%%%%%%% SIMONS OBSERVATORY BOX %%%%%%%%%%%%%%%%%%%%%%%
\begin{mybox}[floatplacement=t,label={box:so}]{The Simons Observatory}

The \textbf{Simons Observatory (SO)} \citep{2019JCAP...02..056A} will provide scientists with an unprecedented platform to study the nature of the fundamental physical processes that have governed the origin and evolution of the universe via high-precision measurements of the temperature and polarization of the Cosmic Microwave Background -- the oldest light in the universe.  Located at an 5190\,m altitude in the Atacama Desert, Chile, SO currently consists of three 0.4\,m-diameter Small Aperture Telescopes (SATs), and one 6\,m-diameter Large Aperture Telescope (LAT).

\vspace{0.5\baselineskip}

The Simons Observatory was very clearly identified by the UK CMB community as their top priority in the 2022 AAP Roadmap \citep{STFC_AAP_2022_Roadmap}, and consequently among the AAP Very High priorities. At the time of the most recent STFC Balance of Programmes Review there was no capacity in the core programme for such a major new commitment, but £17.9M of new funding was secured through the UKRI infrastructure call\footnote{\url{https://www.ukri.org/news/uk-joins-mission-to-search-for-the-origins-of-the-universe/}}, allowing the UK to provide a major enhancement of SO through the SO:UK project.

\vspace{0.5\baselineskip}

Funded through the UKRI Infrastructure Fund and STFC's PPRP, SO:UK is adding two further SATs, a UK-based data centre and major contributions to the data-processing pipeline. The institutions delivering the SO:UK project are Manchester, Cardiff, Oxford, Cambridge, Imperial College London and Sussex University. The project began in October 2022. All elements are now well underway. The data centre and pipeline groups are actively processing the data now arriving from the first SATs. Instrument development is ongoing in the UK with deployment of the two UK telescopes currently scheduled for the first half of 2026. 

\vspace{0.5\baselineskip}

The SO:UK project demonstrates how the UK's world-leading science and technology in this field secured UKRI investment beyond the core programme.  This was possible only because of a confluence of factors: the strong track record of the UK in CMB science and instrumentation, the strong CMB community consensus behind and commitment to the SO:UK project, and the appropriateness of the project for the UKRI call in question.

\end{mybox}
%%%%%%%%%%%%%%%%%%%%%%%%%%%%%%%%%%%%%%%%%%%%%%%%%%%%%%%%%%%%%%%%

\subsubsection{Line intensity mapping}

\textbf{Introduction}\newline 
Line intensity mapping uses low-angular-resolution spectroscopic surveys to probe the large-scale structure of the universe, from which fundamental cosmological inferences can be made. There are several line intensity mapping experiments existing or planned (e.g. SKA for 21cm, and TIME, CONCERTO, SPT-SLIM for millimetre wavelengths), though the first generation experiments probe low redshifts and lack sensitivity for fundamental cosmology. 

\textbf{UK leadership}\newline 
AAP received a white paper for its 2021 consultation exercise, proposing a UK-led millimetre-wave line intensity mapping experiment that would detect all large scale structure out to redshift $z\sim10$ using CO and CII lines\footnote{P.I.: P. Barry, Cardiff University}.  %The FIRSPEX\footnote{P.I.: D. Rigopoulou, University of Oxford} mission concept \citep{rigopoulou2015} was led from the UK.

\textbf{Science questions and instrumentation drivers for the coming decade}\newline 
A discipline can arguably only be regarded as mature when its fundamental parameters are over-constrained by multiple independent probes; line intensity mapping from 21\,cm and millimetre wavelengths complements optical galaxy surveys and CMB lensing. Fundamental science goals  
include primordial non-Gaussianity, dark energy using baryonic acoustic oscillations, beyond-GR theories via the growth of large scale structure, the sum of neutrino masses and the number of relativistic species. For such experiments, access to a large single-dish submillimetre telescope is obviously essential for wide-field submillimetre/millimetre imaging spectroscopy via a very large format focal plane array of spectrometers. 

\subsubsection{Cosmic Microwave Background experiments}

\textbf{Introduction}\newline 
The Simons Observatory \citep{2019JCAP...02..056A} was very clearly identified by the UK CMB community as their top priority, in the 2022 AAP roadmap \citep{STFC_AAP_2022_Roadmap}, and consequently among the AAP Very High priorities. At the time of the most recent STFC Balance of Programmes Review there was no capacity in the core programme for such a major new commitment, but £17.9M of new funding was secured through the UKRI infrastructure call. The science goals of the Simons Observatory include: ``the physics of primordial perturbations, identifying the correct model of an early inflationary epoch of the universe; effective number of relativistic species; the sum of neutrino masses; observable deviations from the cosmological constant paradigm; galaxy evolution; redshift and duration of the reionization epoch\footnote{\url{https://simonsobservatory.org/sotargets/}}''; as these fall mainly within the STFC Particle Astrophysics Advisory Panel remit rather than that of the Astronomy Advisory Panel, we do not discuss these CMB science goals further here. The Simons Observatory is nevertheless useful as an illustrative case study in a related field (see box, p.~\pageref{box:so}). 

\textbf{UK leadership}\newline
The UK Simons Observatory bid built on a very strong heritage of international leadership in CMB cosmology \citep[outside the scope of this review, but see e.g.,][]{STFC_PAAP_Roadmap_2022} and instrumentation development. UK astronomers are playing a leading role in preparing for SO point source and transient science through work at Imperial College led by D. Clements, alongside other UK contributions (see box,~\ref{box:so}).

\textbf{Science questions for the coming decade}\newline 
A key role for submillimetre observations is in measuring and testing the effects of foreground emission for studies of the microwave background: \textbf{what are the polarised foregrounds for CMB experiments?} \citep{2020A&A...642A.232L} Sensitive observations of polarization and information to complement baryon acoustic oscillations in terms of galaxy bulk flows require foreground contributions to be quantified, and wide field accurate surveys are required. Thus submillimetre observations offer the potential for checks on systematic errors on each of the key non-supernova techniques for measuring the shape and evolution of the universe in cosmology. 

Another non-cosmological science question opened by these new CMB facilities is: \textbf{what processes dominate the transient sky at millimetre and submillimetre wavelengths?}
Current and future generation CMB experiments will all repeatedly observe large areas of the sky to sensitive flux levels at millimetre and submillimetre wavelengths (see eg. \citet{2022ApJ...935...16E, 2023ApJ...956...36L}). This is necessary to reach the required sensitivities for, for example, the detection and characterisation of B-mode CMB polarization. The Simons Observatory LAT telescope, for example, will survey 40\% of the entire sky (16,000 sq. deg.) to 5$\sigma$ sensitivities from 7 to 25 mJy in six frequency bands from 95 to 280 GHz \citep{2019JCAP...02..056A, 2022ApJ...929..166H}. They will thus provide an unprecedentedly deep, wide area surveys of the submillimetre/millimetre sky. There has never been such a survey at submillimetre/millimetre wavelengths before. The nearest equivalents are the large area surveys with Herschel, which covered less than 1000 sq. deg. in total, and the IRAS all sky survey which only extended to 100\,$\mu$m in wavelength. The potential discovery space for these by-products of CMB observations is thus very large. This is especially the case when variable and transient sources are considered since the CMB surveys will repeatedly scan the sky with cadences ranging from days to years. Extragalactic transient and variable sources that will be found include AGN, gamma ray bursts, tidal-disruption events, and might potentially include gravitational wave sources. Galactic transients will include stellar flares and episodic accretion in young stars. Transient event rates of tens to hundreds per year are predicted for current and future CMB facilities \citep{2022ApJ...935...16E}.

\textbf{Instrumentation drivers for the coming decade}\newline
The primary drivers for UK involvement in the next generation of CMB experiments (Simons Observatory, CMB-S4) are cosmological, and out of scope for this review; however, there is a very wide range of legacy science that will follow from the detection of strongly gravitationally lensed galaxies (Section~\ref{sec:cosmic_noon}), protoclusters (Section~\ref{sec:protoclusters}) and polarization in multiple Galactic contexts (Section~\ref{sec:MW_star_formation}), with diverse follow-ups from both single-dish submillimetre surveys and the upgraded ALMA discussed in the sections above.

\section{UK strengths in submillimetre and millimetre instrument development and delivery}
\label{sec:uk_instrumentation}

For decades, the UK has led in developing submillimetre/millimetre continuum and spectroscopic instruments for international facilities like JCMT, ALMA, \textit{Planck}, and \textit{Herschel}. Building on this, the aim is to maintain the UK’s status as a major technological leader and pioneer the next generation of submillimetre/millimetre instruments for major ground-based and spaceborne facilities. As well as instrument hardware technology, software for operations and pipeline processing of data obtained at these wavelengths is also an historic strength of the UK community, and has been developed and demonstrated via the JCMT and ALMA on the ground and \textit{Herschel} and \textit{Planck} in space. With many new facilities and upgrades of existing observatories expected in the next 10-20 years, the UK can contribute significantly from initial telescope planning to the delivery and commissioning of UK-led instruments. 

\subsection{Past achievements and leadership}
\label{sec:instrumentation_past}

\subsubsection{Imaging photometry}
\label{sec:instrumentation_past_phot}

\begin{figure}[b!]
    \centering
    \includegraphics[width=0.28\textwidth]{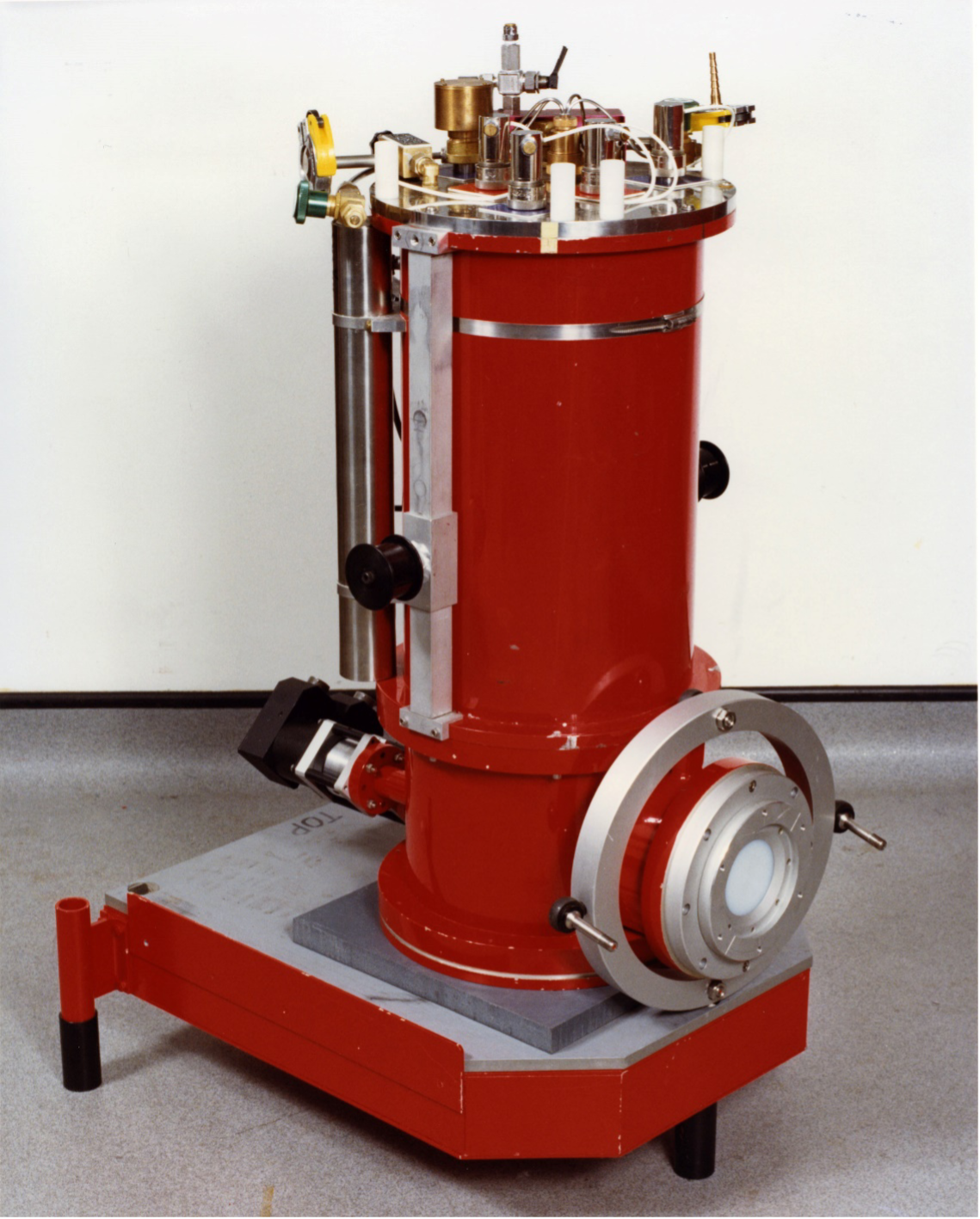}
    \includegraphics[width=0.38\textwidth]{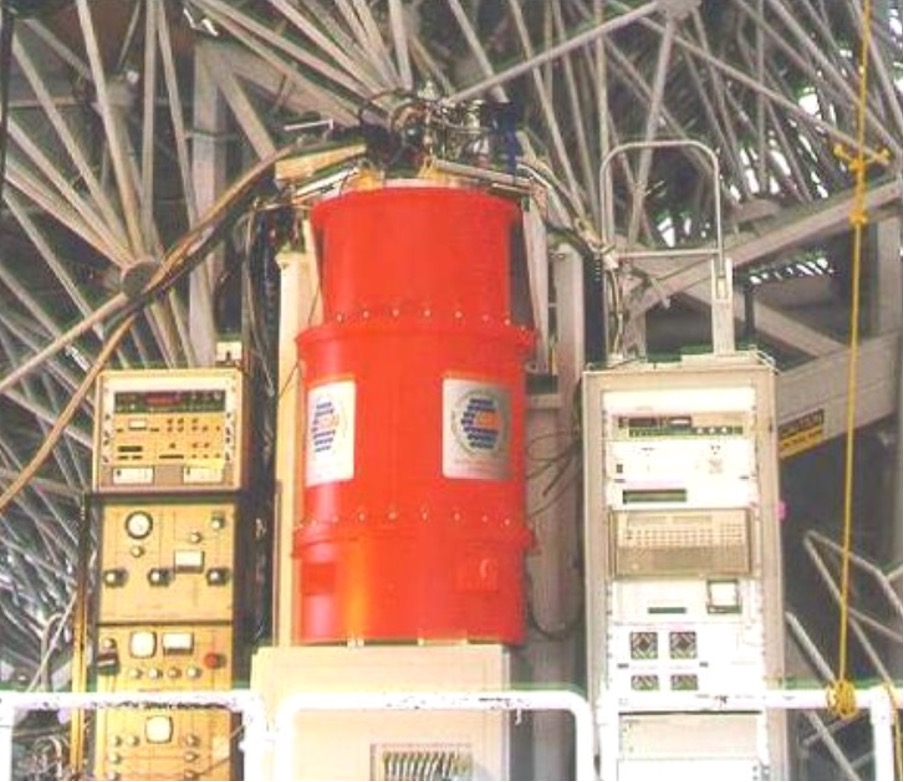}
    \includegraphics[width=0.25\textwidth]{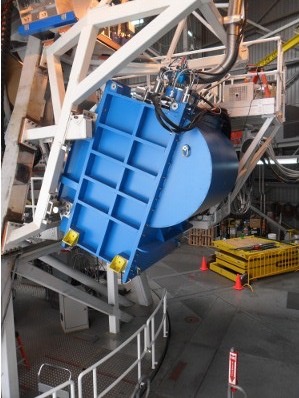}
    \caption{The three generations of UK-built submillimetre continuum instruments on the JCMT. \textit{Left: }UKT14, a single-pixel bolometer built in the 1980s. \textit{Centre: }SCUBA, the world’s first submillimetre camera. \textit{Right: }SCUBA-2, built in the 2000s.}
    \label{fig:submm_heritage}
\end{figure}

The UK has a long and distinguished heritage in building submillimetre continuum instruments and exploiting them scientifically, stretching back over 40 years. This history can best be seen through the three generations of UK-led continuum instruments on the JCMT -- UKT14, SCUBA, and SCUBA-2 (Figure~\ref{fig:submm_heritage}), representing the development from one pixel, through hundreds of pixels, to thousands of pixels.

Early efforts were led by the Queen Mary College (QMC) group with balloon flights and pioneering single-pixel photometers on telescopes including UKIRT. The first common user bolometer instrument on any telescope was UKT14 \citep{duncan1990}, a single-channel bolometer built by ROE and QMC, and operated first on UKIRT and subsequently on JCMT. UKT14 was world-leading at the time in terms of sensitivity and importantly in availability to the community as a sell-supported common user instrument – establishing a practice that was continued with subsequent JCMT submillimetre instrumentation, enabling maximum scientific productivity. UKT14 was also equipped with a polarimeter \citep{flett1991}.

As a single-pixel instrument, UKT14 had very limited mapping capabilities. It was superseded by SCUBA \citep{cunningham1994,Holland1998}, the world’s first genuine submillimetre camera, which was also built by ROE in collaboration with QMC, and which became the world-leading instrument of its kind. 

\begin{figure}[b!]
     \centering
     \includegraphics[width=0.28\textwidth]{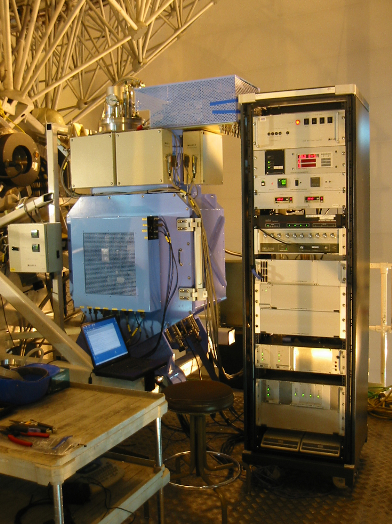}
     \includegraphics[width=0.3\textwidth]{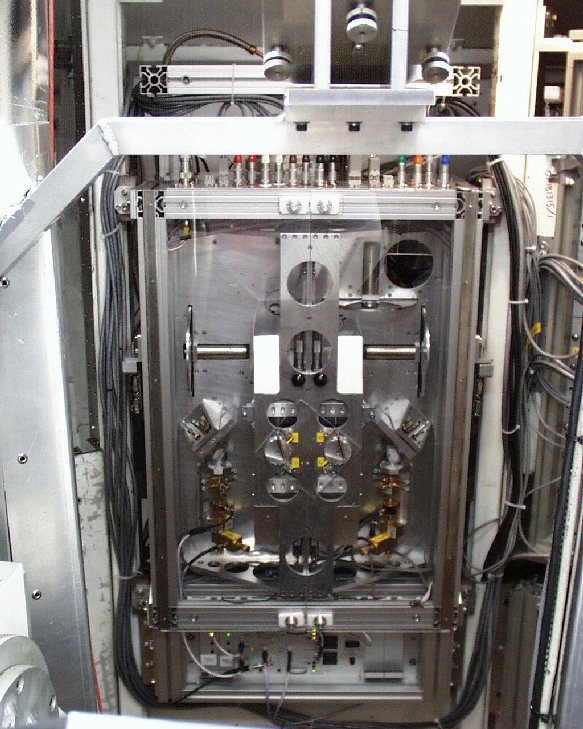}
     \includegraphics[width=0.32\textwidth]{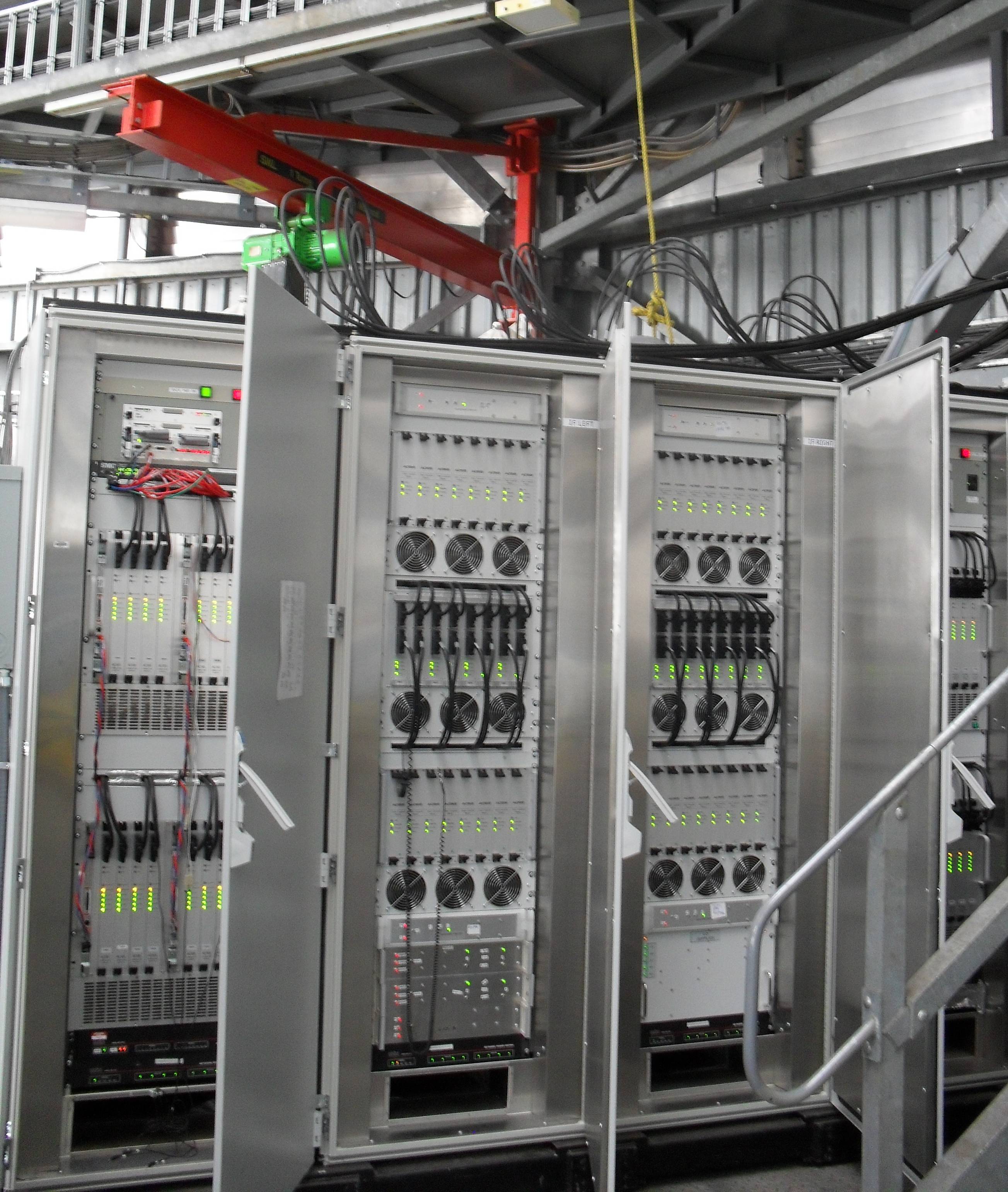}
     \caption{UK-built submillimetre spectroscopy instruments on the JCMT. \textit{Left: }HARP built by the University of Cambridge (325--375 GHz). \textit{Centre: }RxW receiver built by MRAO in Cambridge (currently operational at 620--710 GHz). \textit{Right: }ACSIS, the digital autocorrelation spectrometer that provides the backend for all of the spectral instruments at the JCMT, which was built in collaboration with the UK ATC.}
     \label{fig:submm_spec}
\end{figure}

SCUBA transformed submillimetre astronomy on the global stage, making many important discoveries, including a new population of very high redshift galaxies that are  known to this day as ‘SCUBA galaxies’  \citep[e.g.][]{Hughes+98} and submillimetre bright debris around nearby stars \citep{Holland1998}. SCUBA's polarimeter, SCUPOL \citep{murray1997,greaves2003}, also revolutionised studies of magnetic fields in the ISM and in star-forming regions through ground-breaking imaging polarimetry.

SCUBA-2 \citep{scuba-2} replaced SCUBA on the JCMT in the early 2000s, maintaining the UK’s pre-eminent position in submillimetre imaging. It was built by UKATC and RAL, again in collaboration with Cardiff, and, with its large-format arrays of ultrasensitive 100\,mK transition edge sensor (TES) bolometers, once again revolutionised the field. It remains today (despite many competing instruments), the world’s largest and most scientifically productive submillimetre camera. Its polarimeter, POL-2, together with the upgraded sensitivity obtained with SCUBA-2, has enabled mapping of the magnetic fields in our own galaxy down to the faintest dense molecular clouds where stars have yet to form \citep{wardthompson2017}.

Together UKT14, SCUBA and SCUBA-2 have produced nearly 2,000 publications, with almost 50,000 citations (NASA ADS), highlighting the strength of UK heritage in submillimetre continuum astronomy. The JCMT remains the largest purely submillimetre single dish telescope in the world, and SCUBA-2 is still intensively used. But it is now an old instrument and work has been started on designing a successor, which will provide an order of magnitude increase in mapping speed.

The UK has led the submillimetre polarimetry field for the last four decades with the design and manufacture of achromatic half-wave plates.  These include SCUPOL for SCUBA \citep{greaves2003}, ROVER \citep{leech2005}, POL-2 for SCUBA-2 \citep{savini2009,friberg2016},  NIKA \citep{ritacco2017} and NIKA2-Pol \citep{ritacco2020} from the ground and BLASTPol \citep{moncelsi2014} and BLAST-TNG from balloon, as well as half-wave plates for CMB experiments (e.g. \citealt{savini2006}; \citealt{pisano2006}).  All of these half-wave planes were designed and build at Cardiff.

The pre-eminent success of the UK in submillimetre continuum imaging has been based on several important factors: (i) the combination of detector, filter, optics, and cryogenics expertise available in University groups and the National Labs, supported and nurtured through technology development funding, primarily from STFC and its predecessors; (ii) well-optimised collaboration between the National Labs and University groups from project conception to execution; and (iii) the building and deployment of well-engineered and professionally supported common-user facility instruments enabling easy UK community access and maximising scientific productivity of the instruments and facilities.

The technical and scientific strengths that enabled the UK’s leading position in ground-based submillimetre instrumentation and astronomy have also ensured major roles in FIR-submillimetre space missions, especially \textit{Herschel} and \textit{Planck}, which revolutionised FIR astrophysics and CMB cosmology in the 2010s. The UK led the \textit{Herschel}-SPIRE instrument, with participation by Cardiff (PI institute), RAL-Space, UKATC, MSSL, and Imperial College, and Planck-HFI also involved major contributions from Cardiff and Imperial College, complemented by involvement by Manchester, Cambridge, and Imperial in the LFI instrument. These missions were hugely successful: to date \textit{Herschel} has produced over 3,800 papers (with around 60\% using SPIRE data), and \textit{Planck} over 3,000.

\subsubsection{Spectroscopy}
\label{sec:instrumentation_past_spec}

The UK also has a rich history in the development of heterodyne instrumentation for astronomy. In the early days of ground-based submillimetre/millimetre observations, visiting receivers based on Schottky diode and indium antimonide (In:Sb) HEB mixers, from Cambridge and QMC, were operated on UKIRT. With the advent of the JCMT, facility receivers using first Schottky or In:Sb, and subsequently superconductor-insulator-superconductor (SIS) mixers were built in the UK for A-band (210--280 GHz), B-band (320--380 GHz), C-band (450--490 GHz) and W-band (430--500; 630--700 GHz) by various groups including Cambridge, RAL, QMC and Kent. The JCMT’s 16-element SIS-based receiver array for B-band, HARP \citep{buckle2009}, was led by Cambridge, and is still operational today as the workhorse heterodyne instrument on the telescope. All of these instruments, which have produced important science, were based on the comprehensive expertise of UK laboratories in all aspects of astronomical heterodyne spectroscopy including mixers, local oscillators, optics, cryogenics, and back-end spectrometer electronics and software. 

RAL played a major role during the development and construction of ALMA. It hosted the ALMA UK Project office, which had oversight of UK-wide activities for ALMA, including pipeline software, digital IF data transfer to the correlator, and telescope control code. RAL’s Technology Department designed and built all the ALMA cryostats, and RAL Space supplied hundreds of phase locking photomixers and hosted one of three global ALMA Front End integration centres. %The RAL Millimetre Wave Technology (MMT) group subsequently supplied photomixers and prototyped local oscillator technology for ALMA Band 5. 

\begin{figure}[b!]
     \centering
     \includegraphics[width=\textwidth]{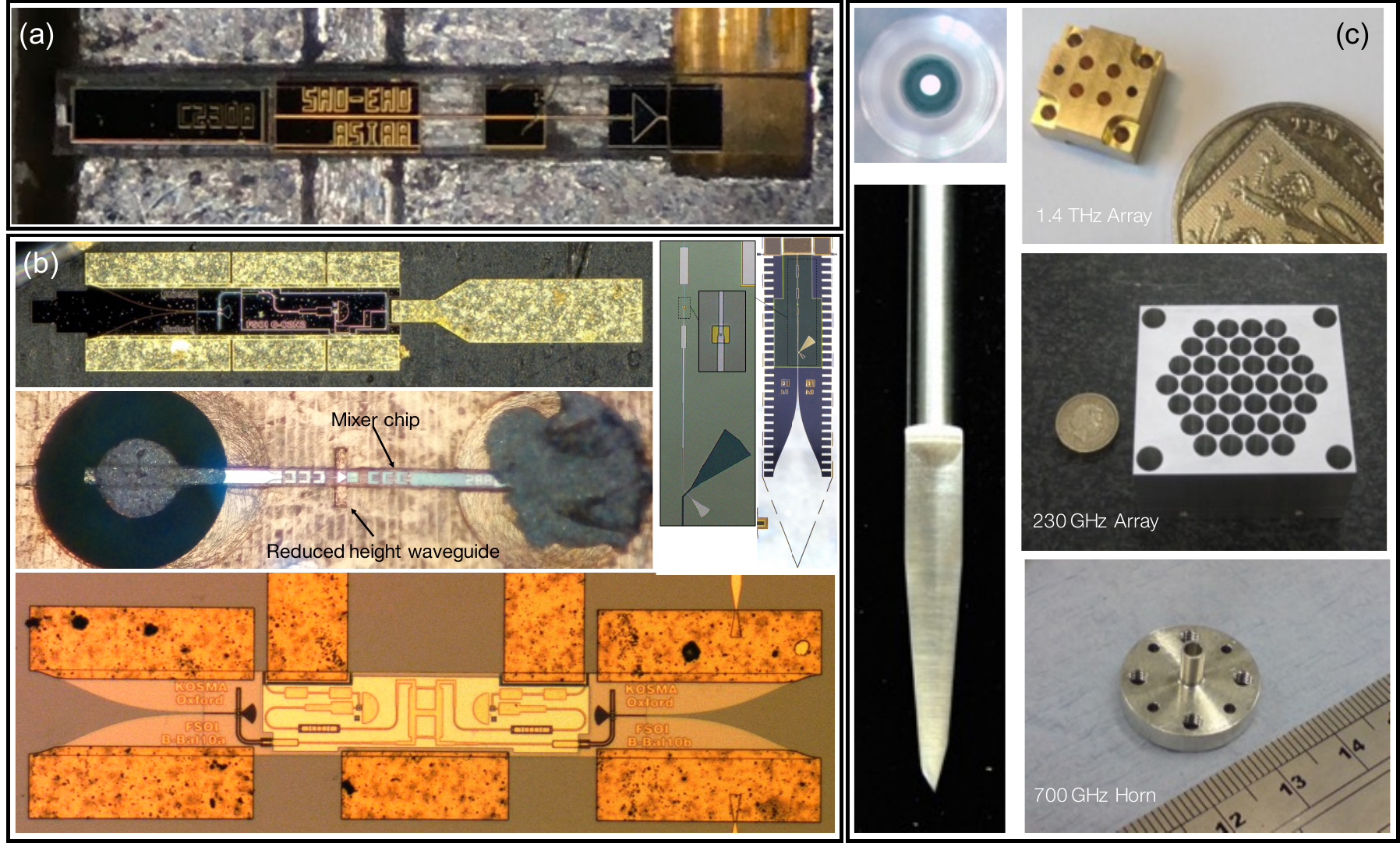}
     \caption{(a) HARP-b mixer. (b) Modern SIS mixers from 200--950\,GHz, with on-chip circuit integration for wider RF/IF bandwidth and advanced capability such as LO-noise rejection balanced architecture. (c) Oxford-pioneered smooth-walled horn technology from millimetre to THz range.}
     \label{fig:mixer_horn}
\end{figure}

The development of superconducting mixers, now used in most submillimetre heterodyne instruments, has been strong in the UK for many years. Fig.\,\ref{fig:mixer_horn}(a) shows the SIS tunnel junction mixer developed for HARP \citep{buckle2009}, pioneering the use of an on-chip probe antenna for the first time (Withington \& Yassin), subsequently used as the de-facto on-chip antenna for all of ALMA's SIS receivers. Compared to the HARP-B SIS mixer produced many years ago for JCMT, the modern mixers developed by Oxford integrate much more on-chip functionality to enable wider RF and IF performance, as well as permitting advanced capabilities, such as sideband separating schemes. As well as a series of single-ended SIS mixers developed to cover 200–950 GHz (Fig.\,\ref{fig:mixer_horn}(b)), the state of the art balanced SIS mixer with substantial on-chip circuit component integration is a significant achievement. Oxford has also developed easily manufacturable smooth-walled horn technology for the FIR-millimetre range; see Fig.\,\ref{fig:mixer_horn}(c). This technology has been adopted by many astronomical experiments, including international CMB projects.

The UK also has important expertise in air-bridged GaAs and InGaAs Schottky diode technology for local oscillator systems, an essential part of all heterodyne receivers up to at least 1 THz, with a world-leading production facility at RAL. Schottky devices support high-power, high-stability applications, functioning as multi-stage frequency multipliers, and mixers for sub-harmonic, fundamental, and sideband-separating processes, naturally partnering low noise amplifier based front ends. RAL has developed a local oscillator system for ALMA Band 5, supplied space-qualified Schottky devices for meteorology satellites and provided waveguide photomixers for ALMA's laser synthesizers and all of its submillimetre-wave front-end receivers. 

\subsection{Current developments and capabilities}
\label{sec:instrumentation_current}

\begin{figure}
    \centering
    \includegraphics[width=\textwidth]{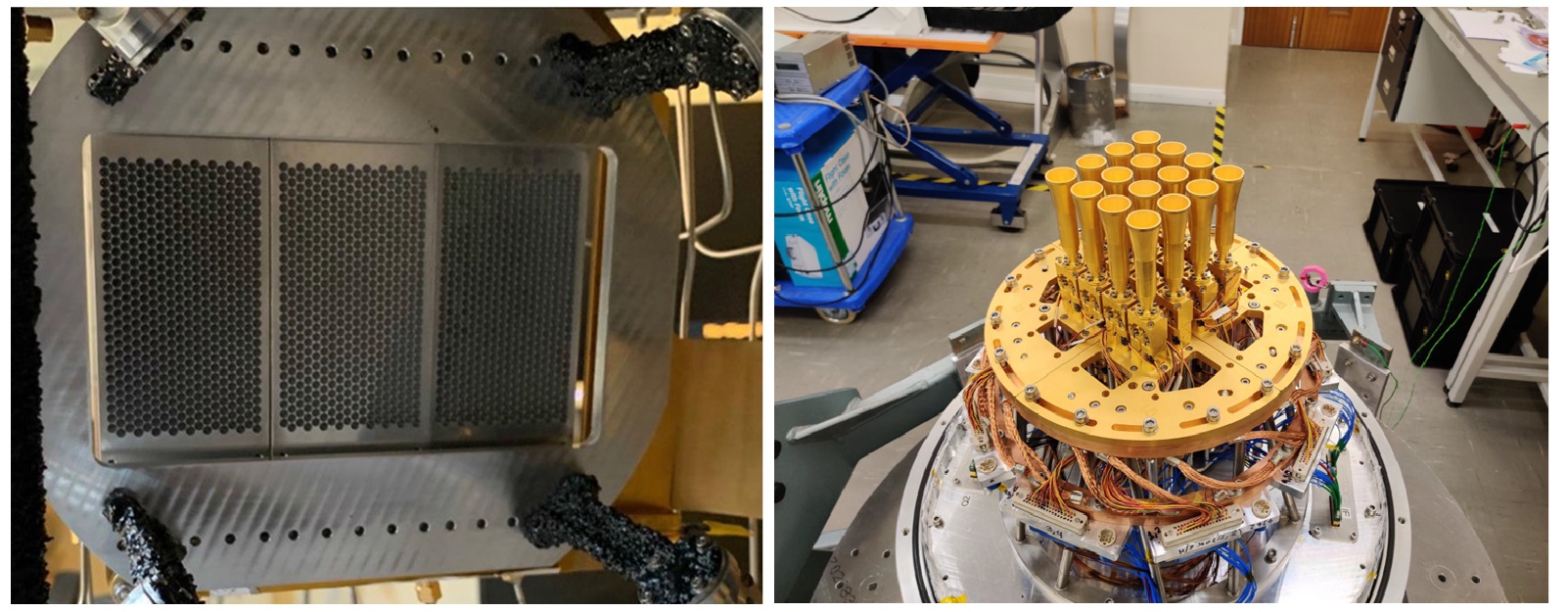}
    \caption{Left: The MUSCAT focal plane arrays comprising a total of 1500 LEKID detectors. Right: The 16 beam RAL-Manchester developed CARUSO receiver for the Sardinia Radio Telescope.}
    \label{fig:MUSCAT-CARUSO}
\end{figure}

\subsubsection{Direct detection imaging and spectroscopy technology}
\label{sec:instrumentation_current_direct}

\textbf{Kinetic Inductance Detector arrays:} Superconducting kinetic inductance detectors (KIDs) are an alterative to TES arrays for large-format focal planes.  Their main attractions are ease of manufacture and the ability to read out many detectors on a single feedline with frequency-division multiplexing. KID-based systems are not yet as mature as operational TES-based instruments, but the situation is rapidly evolving and it is possible that over the next decade KIDs will be come the technology of choice on the ground and in space.

Currently operational KID cameras include the Cardiff-built MUSCAT \citep{tapia2020muscat} on the Large Millimeter Telescope (LMT) (see also Figure~\ref{fig:MUSCAT-CARUSO}) and NIKA-2 \citep{Calvo2016} on the IRAM 30-m telescope (which also had substantial Cardiff involvement). The Dutch (SRON) led AMKID camera has also been operated on the APEX telescope. MUSCAT will be superseded by the US-led TolTEC camera with a larger KID-based focal plane. MUSCAT’s design provides a route to future simplified focal planes with scalable pixel counts operating at the photon noise limit. Such devices have also demonstrated an elegant polarisation sensitive architecture in line with the requirements of the next generation of submillimetre/millimetre instruments.

Current KID-based focal planes are typically read out with multiplexing ratios of 250--1000 detectors per channel (for example the 1500 pixels of MUSCAT are read out over 6 channels, and the BLAST-TNG balloon experiment demonstrates multiplexing ratios nearing 1000 \citep{Blast_TNG}. But with innovations in fabrication and post-processing as well as readout hardware, led by UK institutes like Cardiff and Oxford, multiplexing ratios as high as 5000 pixels per channel should be achievable given dedicated technology development. This would enable the development of focal planes into the hundreds of thousands to million pixel arrays modular systems.

The Cardiff-invented Lumped Element KID (LEKID; \citealt{LEKID}) used in MUSCAT has been adopted by most deployed KID based instruments worldwide, and has also been implemented by Cardiff in the world’s first passive millimetre-wave security imaging camera. High multiplexing ration demand large readout bandwidth. By furthering the development of Radio Frequency System on Chip (RF-SoC) boards, like those being developed by Cardiff and Oxford for the Simons Observatory, currently achieved bandwidth can be increased by factors of $\sim$4, enabling a corresponding increase in multiplexing ratio.

Even with the high multiplexing ratios envisaged, power dissipation from cryogenic low noise amplifiers in future large-format imaging systems will be a limiting factor. To address this the Cardiff group is developing LEKID architectures optimised for use with amplifiers that can be operated at higher temperatures, reducing the required cooling capacity. The potential for this has been demonstrated using the device architecture deployed by MUSCAT in terrestrial imaging systems developed by Cardiff.  Related to this development, Cardiff has built a large-area magnetron sputtering system capable of growing uniform superconducting films on 8-inch wafers, and of growing aluminium or titanium nitride (TiN) films, for large arrays.

\textbf{Direct detection spectroscopy:} In addition to cameras for wide-area surveys, imaging direct-detection spectrometers will be needed for redshift determination, galaxy ISM characterisation, and cosmology via line intensity mapping. A number of medium-resolution on-chip spectrometers (Super-Spec, DESHIMA, and $\mu$Spec; \citep{Jovanovic2023}) have been developed based on filter bank technology. Cambridge and Cardiff have technology development and demonstration programmes underway including CAMELS, the CAMbridge Emission Line Surveyor \citep{CAMELS}; SPT-SLIM, a spectroscopic imager for line intensity mapping on the SPT \citep{Barry2022}; and the Superconducting On-Chip Fourier Transform Spectrometer (SOFTS) \citep{SOFTS} technology.  SOFTS is a novel technique of achieving direct detection spectroscopy using TWPA technology without filter-banks that is currently being pursued at Oxford, which could achieve ultra-broad RF bandwidth with high spectral resolution similar to the PIXIE experiment \cite{kogut2014primordial}.

\textbf{Filters and quasi-optical components:} Quasi-optical components (spectral filters, polarisers, beam dividers, waveplates, etc.) are essential for all astronomical instruments. In the FIR-millimetre range Cardiff University’s Astronomy Instrumentation Group (AIG) remains as the main supplier of such components for the worldwide astrophysics and CMB communities.  The demands of bigger component size for large-format arrays, increased lithographic accuracy and lower dielectric losses for shorter-wavelength performance, and tooling and quality control for large-scale production, require expansion and enhancement of production of production facilities beyond what can be achieved in a University environment.  This has motivated the formation of an industrial spin-out company, Celtic Terahertz Technology (CTT) Ltd., which will provide the normal route for future component provision. CTT will maintain a strong design and R\&D relationship with the Cardiff AIG.

\subsubsection{Heterodyne spectroscopy technology}
\label{sec:instrumentation_current_heterodyne}

In heterodyne technology worldwide, there is a strong push towards the development and operation of arrays with hundreds, or even thousands, of pixels. This will require on-chip integration and miniaturisation as well as advances in low-noise amplifiers with reduced power dissipation, and back-end data handling capability. The UK’s strong capabilities in all of these areas make it well-placed to take the lead in these developments. 

\begin{figure}[h!]
     \centering
     \includegraphics[width=\textwidth]{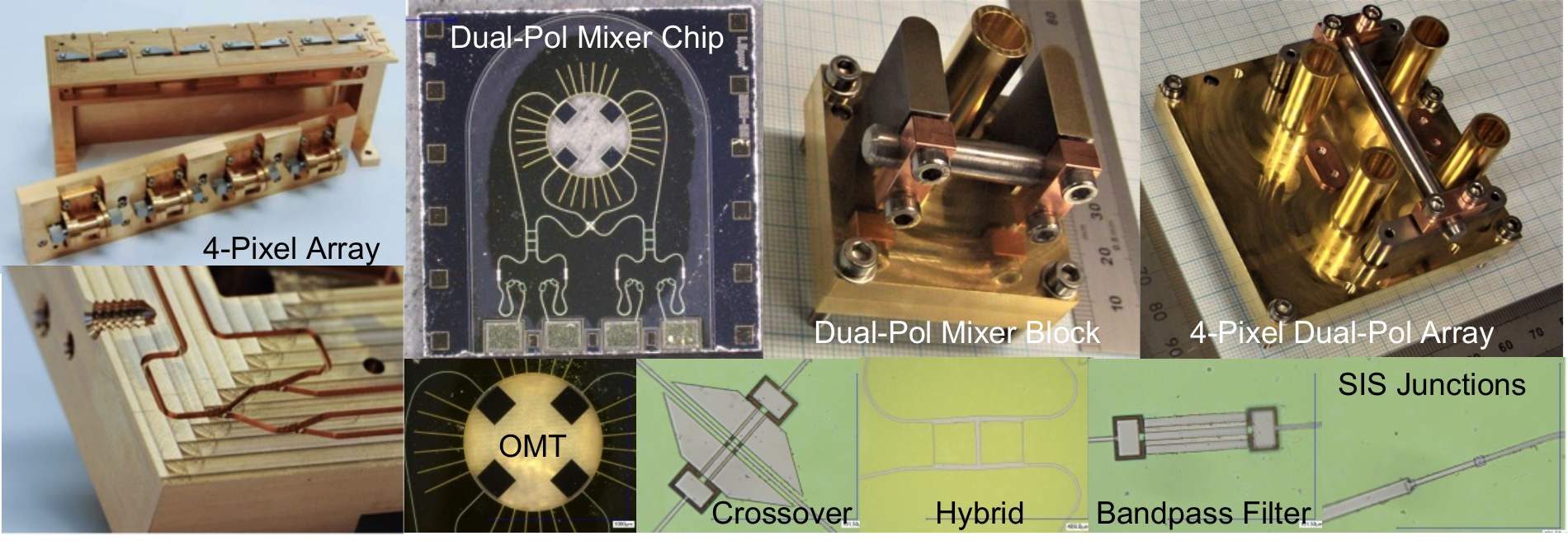}
     \caption{Heterodyne array receiver demonstrators including the dual-polarisation on-chip integrated mixer.}
     \label{fig:mixer_array}
\end{figure}

Along with the easy-to-machine high-performance feedhorn array, Oxford's ongoing exploration of planar-on-chip circuit technology in pursuit of an architecture that can realise compact array receivers has led to the adoption of orthomode transducer (OMT) technology, integrating two polarised (2-pol) receivers into one unit, as shown in Figure~\ref{fig:mixer_array}. This approach is much more efficient and compact compared to traditional methods using an optical diplexer or waveguide OMT with a separate SIS receiver for each polarisation. A successful demonstration with both balanced and 2-pol mixers demonstrates the strong potential to integrate numerous RF components on-chip, paving the way for a compact large-format heterodyne array receiver with hundreds to thousands of pixels. 

The current research aim is to make these systems modular and array-able for future deployment of kilo-pixel receivers, i.e. improving the TRL (Technology Readiness Level) in preparation for near-future instrument deployment. Oxford and RAL have proposed to explore pioneering cutting-edge array technologies by developing a 25-pixel demonstrator to bring the TRL to 4. Advantages of the demonstrator will include  1) capturing both polarisations to retain full signal strength and polarisation information, crucial for tasks like mapping magnetic fields near black holes, 2) separating the sidebands to avoid spectral contamination and 3) widening each sideband's IF bandwidth to 16\,GHz, catering to the most demanding science requirements such as those for research into high-redshift galaxies. This ambition means that each pixel covers a 32\,GHz RF bandwidth in a single tuning, allowing multiple spectral lines to be mapped simultaneously. Each pixel requires 4 IF outputs, resulting in a total of 100 IF chains. Essentially, this is equivalent to a 100-pixel single-polarisation double sideband (DSB) receiver, making it one of the most advanced arrays in the field. A key design principle of the demonstrator will be scalability to much larger pixel counts.

Other on-going effort in this regime including widening the RF and IF bandwidths permitting combination of adjacent ALMA bands in a single receiver, and potentially going further than just a two-band combination. This effort aligns with the ALMA Wideband Sensitivity Upgrade (WSU; see Section~\ref{sec:wsu}), and the need for new heterodyne receivers for upcoming EHT stations such as the Greenland Telescope (GLT) and the Africa Millimeter Telescope (AMT).

Precision machining while maintaining tight tolerances over large areas is also crucial for manufacturing large-scale heterodyne arrays. RAL’s Precision Development Facility can machine features as small as a few µm over scales of up to 40 cm, enabling the fabrication of submillimetre/millimetre waveguides, horns, and cavities, complementing the planar circuit on-chip integration technologies. Another significant challenge in building heterodyne arrays is developing a suitable backend spectrometer architecture. Duplicating existing electronics is impractical. The UK has heavily invested in scalable digital signal processing (DSP) systems for the SKA project, which involves digitising over 250,000 RF streams. This expertise would be extremely useful in advancing heterodyne array technology.

Traditional semiconducting HEMT readout amplifiers are unsuitable for large array applications due to their significant heat dissipation and sub-optimal noise performance. Superconducting travelling wave parametric amplifiers (TWPAs) offer a solution, achieving quantum-limited added noise with broad band high gain while dissipating negligible heat (three orders of magnitude less than HEMT amplifiers), critical for array applications. Oxford and Cambridge have developed microwave (1--20 GHz) superconducting parametric amplifiers for astronomical receivers and other experiments. The UK can fabricate a wide range of superconducting devices using various films (Al, Nb, Ti, NbN) and composites for different applications. The Cambridge/Oxford team also excels in the theoretical understanding of superconducting quantum materials, electromagnetic simulations, and experimental setups to achieve quantum-limited performance of ultra-sensitive devices across microwave to submillimetre range.

The cross-school (Engineering and Physics \& Astronomy) Advanced Radio Instrumentation Group (ARIG) at the University of Manchester also has world-leading expertise in low noise amplifiers, especially at frequencies between 20 GHz and 373 GHz. With support from ESO and STFC, it has developed amplifiers for ALMA Band 2 \citep{2020A&A...634A..46Y} and is working on component integration towards a LNA-based single receiver system to cover the current ALMA Bands 4 plus 5, and also Bands 6 plus 7.

Current RAL MMT technology development for ALMA is a continuation of the  collaboration with ARIG to supply the CARUSO receiver (Figure~\ref{fig:MUSCAT-CARUSO}) to the Sardinia Radio Telescope \citep{2017ITMTT..65.1589C}. CARUSO is a 16--pixel, low noise amplifier based, dual polarisation 70--116 GHz heterodyne camera, RAL also has facilities and expertise covering frequencies from radio to approximately 5\,THz. These include design, fabrication and test facilities ranging from device level designs to sub-systems and full systems for ground-based telescopes as well as Earth observation satellites.

\subsection{Future opportunities}
\label{sec:instrumentation_future}

A new generation of FIR-submillimetre observational facilities will be crucial for realizing strategic plans outlined in initiatives like the US Decadal Review (Astro2020), and EU-ASTRONET, and they will complement forthcoming facilities at longer and shorter wavelengths such as SKA, ELT, SPHEREx, Rubin, Roman and JWST. The key science driver for future submillimetre instrumentation is increased observing speed for both continuum and spectroscopic observations, and ultra-sensitive, broad-band, large-pixel-count ($>10^5$ for continuum and $>10^3$ for heterodyne) capabilities will be needed. New observatories like the Africa Millimetre Telescope (AMT) and AtLAST (see Section 6) and new instruments such as successors to SCUBA-2 and HARP on the JCMT and facility instrumentation for AtLAST and ALMA upgrades will provide opportunities for the UK to lead in the development and exploitation of these instruments and facilities. \textit{Herschel} and \textit{Planck} have also firmly established the UK as an essential partner for any future FIR space missions. Although the ESA-JAXA SPICA mission was cancelled, attention is now focused on the opportunity for a NASA Probe-class FIR mission in the early 2030s (Section 10). 

For direct detection instrumentation, the proven capability of KID-based instrument development and deployment along with the vision for addressing the limiting factors of future large-scale instruments, demonstrate that the UK is leading the way in developing these technologies for the next generation of continuum cameras and on-chip spectrometers. New and unique fabrication facilities at Cardiff coupled with readout development at Oxford place the UK in a strong position for major roles in future instruments for the JCMT and AtLAST.  The unique capabilities of the Cardiff AIG also provide the UK with guaranteed roles in these important international projects, and also for NASA’s FIR Probe mission, for which all three contenders (FIRSST, PRIMA, and SALTUS) are baselining Cardiff contributions.   

For heterodyne instrumentation, crucial for all the observatories highlighted in this roadmap, Oxford’s pioneering work in superconducting mixer technologies represents a significant advance. This development is complemented by their innovative, cost-effective, high-quality feedhorns, design of integrated on-chip high-frequency components, and the development of quantum-limited broadband superconducting amplifiers. Development of band-combination capabilities for ALMA and the ability to exploit integration of on-chip superconducting quantum circuits to construct large heterodyne arrays for both single-dish and interferometer applications further solidifies the UK’s position at the forefront of heterodyne technology development. In parallel and with support from ESO, ARIG at the University of Manchester is pursuing LNA-based solutions for both very wide RF band receivers and large format heterodyne arrays. RAL is the unique UK institute with the ability and appetite to develop the ultra-broadband 
LO (local oscillator) systems needed as two or more single pixel ALMA bands are combined into one cartridge, and to provide high power LO sources and distribution systems needed for large pixel count 
arrays.

Oxford’s development of superconducting quantum microwave amplifiers has a strong potential for many applications in submillimetre/millimetre instrumentation. They can be used to read out heterodyne detectors as well as bolometric MKID cameras, replacing HEMT amplifiers to improve sensitivity. They also address the heat dissipation issue, a crucial factor when deploying large format heterodyne and bolometric arrays. These superconducting amplifiers can also be optimised as pre-amplifiers for first-stage detectors for frequencies in the submillimetre/millimetre range, further minimising noise. Exploiting their highly nonlinear properties also opens up new technological avenues e.g., constructing an on-chip Fourier Transform spectrometer that can achieve ultra-wide RF bandwidth with moderately high spectral resolution. 

UK developments are firmly focused on the technical challenges for future facility instruments.  With well-thought-through and properly-funded technology development the future will be bright, with the capabilities of institutions like Cardiff, Imperial, Manchester, MSSL, Cambridge, Oxford, RAL and UCL. This expertise and leadership in the specialised technologies needed for direct detection and heterodyne instruments is complemented by the advanced project engineering capabilities of UKATC, RAL-Space, and various University groups, in opto-mechanical and electronic design and manufacturing, instrument-level assembly, integration and testing, and in data processing systems.

\section{Current submillimetre and millimetre facilities}
\label{sec:current_instrumentation}

\begin{figure} [t]
    \centering
    \includegraphics[width=\textwidth]{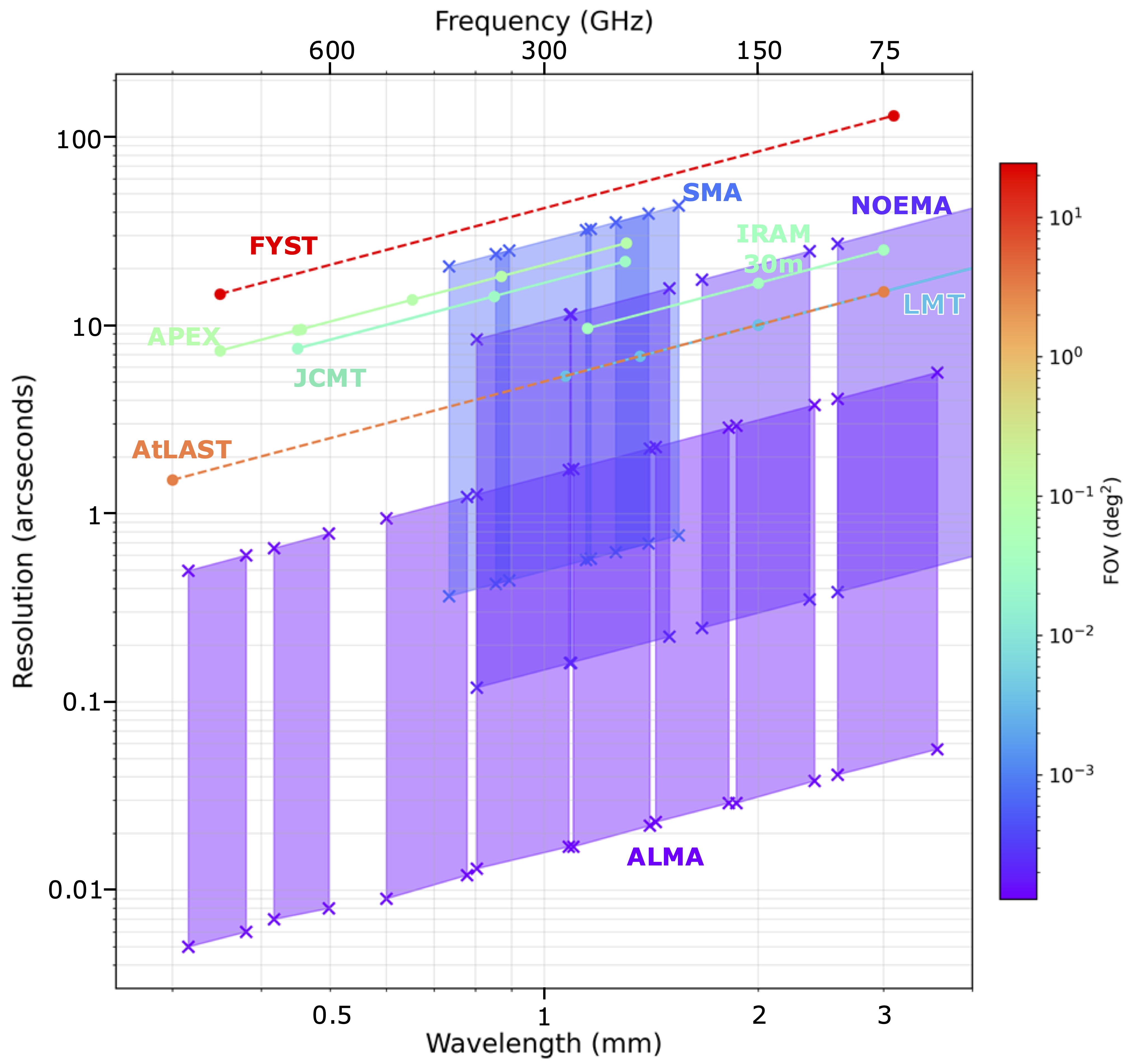}
    \caption{A summary of current and near-future instruments with submillimetre/millimetre capabilities.  Solid lines show existing facilities; dashed lines show future facilities. Circles show single dish facilities; crosses show interferometric facilities. AtLAST points show the extremes of the proposed wavelength range. 
    Configurable interferometers are plotted as a range within each band for the the compact 
    and extended 
    configurations.  The ACA is excluded for clarity. 
    The colour bar represents the field of view (FoV) for each telescope. For the interferometers the FoV varies with band; we use the value at 250\,GHz. FYST FoV is given at 1.3\,mm.}
    \label{fig:wave_res}
\end{figure}

In this section, we briefly summarise the extant submillimetre and millimetre facilities, emphasising those facilities to which UK astronomers have access.  We first discuss single-dish telescopes, followed by interferometers and VLBI (Very Long Baseline Interferometry) networks.  The resolutions and operating frequencies of the telescopes described are summarised in Figure~\ref{fig:wave_res}.  In subsequent sections, we describe future plans for those instruments that the UK has access to, as well as planned new facilities.

\subsection{Single-dish instrumentation}

\begin{figure}
  \centering
  \includegraphics[width=0.8\textwidth]{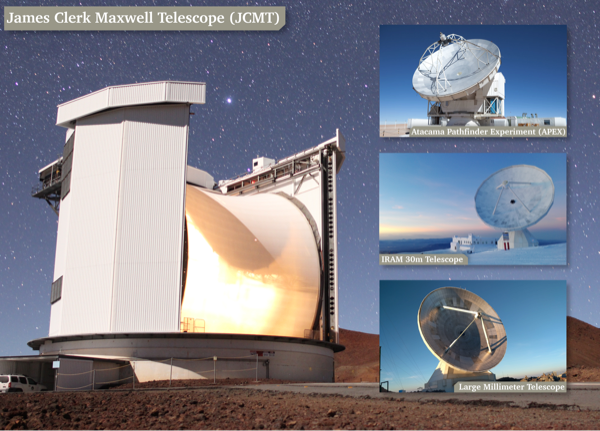}
  \caption{A gallery of current single-dish instrumentation. Image Credits: JCMT -- William Montgomerie (EAO), APEX -- ESO, IRAM 30m Telescope --  K. Zacher (IRAM) LMT -- Gopal Narayanan.) }
  \label{fig:single-dish}
\end{figure}

Single-dish telescopes have large fields of view and moderate angular resolution, allowing large areas of the sky to be mapped.  Key submillimetre and millimetre single-dish telescopes are shown in Figure~\ref{fig:single-dish}.

\subsubsection{Telescopes with UK access}
\label{sec:current_instrumentation_jcmt}

Currently, there is only one general-purpose single-dish submillimetre-wavelength telescope to which UK astronomers have significant access. The 15\,m \textbf{James Clerk Maxwell Telescope (JCMT)}\footnote{\url{https://www.eaobservatory.org/jcmt/}} is the largest extant single-dish submillimetre telescope, and is located near the summit of Maunakea in Hawaii. It is currently operated by the East Asian Observatory (EAO) and funded by its members and partners, having until 2014 been owned by the Joint Astronomy Centre, which was a collaboration between the UK, Canada and the Netherlands.  The JCMT is currently accessible to astronomers at UK universities that are members of the UK JCMT Consortium, match-funded by the STFC through PPRP.  A bid to restore access to all UK universities will be submitted to PPRP in September 2026.  Please see Section~\ref{sec:single-dish_funding} for details.

The JCMT has both continuum and heterodyne capabilities.  The available instruments are:
\begin{itemize}
\item \textbf{SCUBA-2:} a 10,000-pixel bolometer array \citep{scuba-2} which simultaneously observes at 850\,$\mu$m (353\,GHz) and 450\,$\mu$m (667\,GHz).
\item \textbf{POL-2:} a half-wave plate and grid analyser that can be inserted in front of the SCUBA-2 cryostat window \citep{friberg2016}, allowing detection of linearly polarised continuum emission at 850\,$\mu$m and 450\,$\mu$m. 
\item \textbf{HARP:} a $4\times4$-pixel heterodyne array \citep{buckle2009}, operating at 325--375 GHz (925--800\,$\mu$m) with an instantaneous bandwith of $\sim 2$\,GHz.
\item \textbf{N\=amakanui}: a single-pixel instrument \citep{mizuno2020}, loaned from the Greenland Telescope (GLT), with three heterodyne receiver inserts:
    \begin{itemize}
        \item `\={U}`\={u} (`Soldierfish'): 230 GHz (1.3\,mm); fully commissioned
        \item `\={A}weoweo (`Big Eye'): 345 GHz (850\,$\mu$m); shared-risk observing
        \item `Ala`ihi (`Squirrelfish'): 86 GHz (3.5\,mm); to be commissioned
    \end{itemize}
\end{itemize}

\begin{figure}
    \centering
    \includegraphics[width=0.48\textwidth]{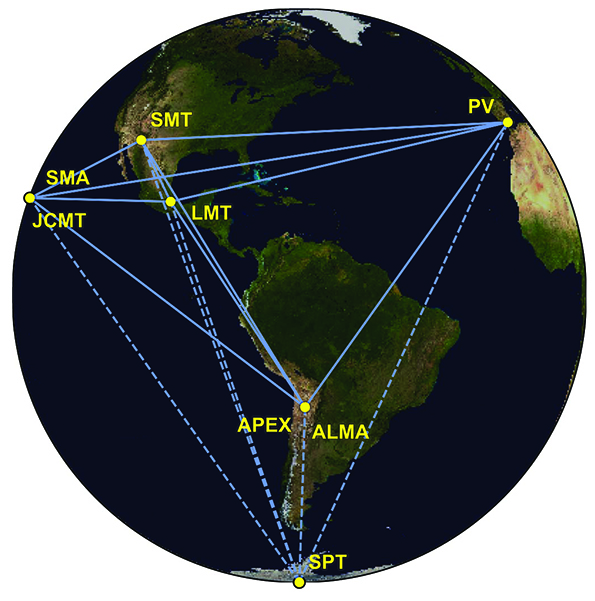}\hfill
    \includegraphics[width=0.48\textwidth]{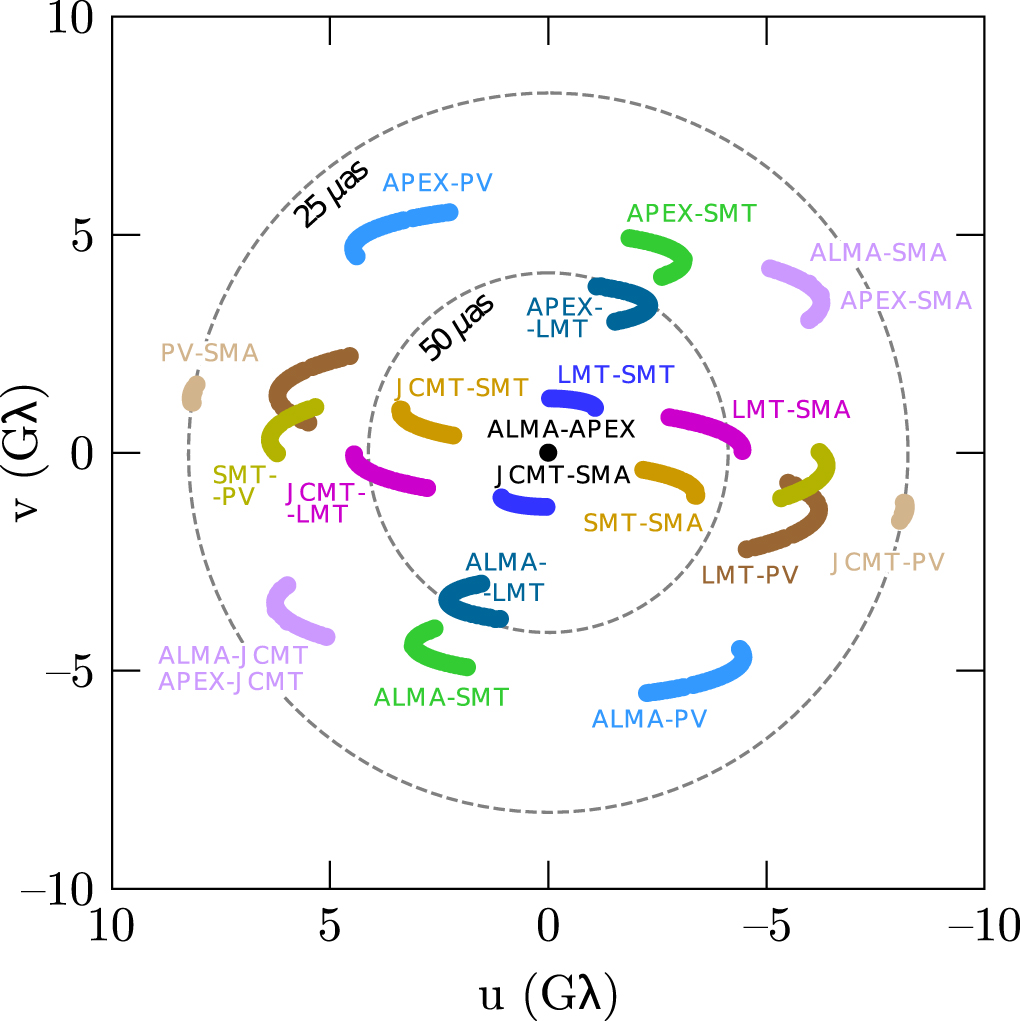}
    \caption{The configuration of the Event Horizon Telescope (EHT) for the imaging of M87* \citep{eht2019}.  \textit{Left:} The EHT stations and their baselines (baselines to the SPT are shown as dotted lines as this station was used for calibration observations only).  \textit{Right:} The \emph{uv} plane for the imaging of M87*.}
    \label{fig:eht_stations}
\end{figure}

The ACSIS (Auto Correlation Spectral Imaging System; \citealt{buckle2009}) digital autocorrelation spectrometer is used as the backend for HARP and N\={a}makanui.  ACSIS can be configured with up to two (HARP) or four (N\={a}makanui) spectral windows, and supports a variety of bandwidth modes ranging from 250 to 3,200\,MHz. The spectral resolution of ACSIS varies from 30\,kHz to $\sim$1\,MHz, depending on the configuration used. The JCMT is also a part of the Event Horizon Telescope (EHT), forming the shortest baseline with the Submillimeter Array (SMA) on the summit of Maunakea (see Figure~\ref{fig:eht_stations}).

\subsubsection{Telescopes with open-skies time}

\textbf{APEX (The Atacama Pathfinder Experiment)} is a 12m telescope located in the Atacama desert in Chile at 5,100\,m on the Chajnantor Plateau.  APEX is principally operated by the Max Planck Institute for Radio Astronomy (MPIfR), Germany, and offers some informal open-skies time. The suite of instruments on APEX is extensive\footnote{\url{https://www.apex-telescope.org/instruments/}} and includes openly available instruments and instruments which are PI-led and can be used in collaboration with the instrument teams.  The APEX spectroscopic instruments include:
    \begin{itemize}
        \item SEPIA, a single-pixel heterodyne receiver operating at 597--725, 272--376 and 159--211\,GHz.
        \item nFLASH, a dual-polarisation, dual-sideband, receiver operating at 196--281\,GHz.
        \item LASMA, a 7-pixel heterodyne array operating at 268--375\,GHz.
    \end{itemize}
As a replacement for the productive LABOCA camera, MPIfR is currently commissioning the 25,000-pixel APEX A-MKID camera, which will operate simultaneously at 347 and 850 GHz with a field of view of $15\times15$~arcmin$^2$. %The effort to install high-frequency ALMA Band 9 and 10 side-band separating SIS mixers for technological demonstration is also currently underway.

The \textbf{IRAM 30~m Telescope} is located on the Pico Veleta in the Spanish Sierra Nevada, and operated by the Institut de Radioastronomie Millim\'etrique (IRAM).  15\% of its time is available to non-IRAM partner countries. The telescope has three primary instruments:
\begin{itemize}
    \item EMIR, a spectrometer operating at 73--117, 125--184, 202--274 and 277--350\,GHz.
    \item HERA, a 3x3 pixel spectrometer operating at 215--272\,GHz.
    \item NIKA-2, a continuum camera operating at 1.15 and 2.0\,mm (260--150\,GHz).  Commissioning of 1.15\,mm polarimetry is ongoing.
\end{itemize}

\subsubsection{Telescopes with Guaranteed Time for UK institutions}

The \textbf{Large Millimeter Telescope (LMT)}\footnote{\url{http://lmtgtm.org}} is a 50\,m telescope in Mexico operating at 1.3\,mm -- 1.3\,cm.
In general, LMT projects require a P.I. based in Mexico, the USA or Canada.  However, Cardiff University built the MUSCAT camera for the LMT (\citealt{Muscat2018}; \citealt{tapia2020muscat}; see Section~\ref{sec:uk_instrumentation}, above), and so has some Guaranteed Time on this instrument.

\begin{figure}
  \centering
  \includegraphics[width=0.8\textwidth]{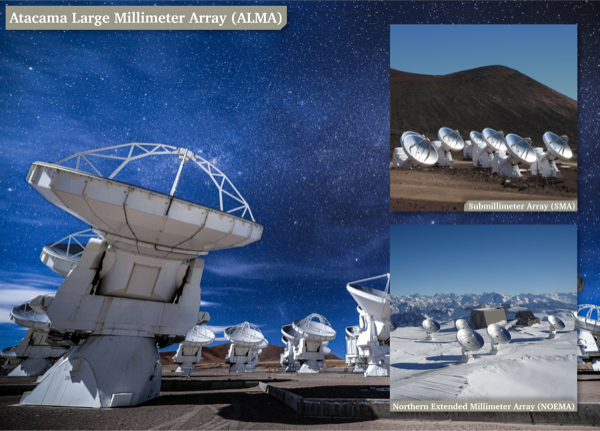}
  \caption{A gallery of current interferometric instrumentation. Image Credits: ALMA -- Alex Perez, SMA -- Glen Petitpas, NOEMA -- Andre Rambaud (IRAM).}
  \label{fig:interferometric}
\end{figure}

\subsubsection{Dedicated cosmology survey facilities}

There are also dedicated cosmology facilities which operate in the submillimetre/millimetre wavelength regime.  While these facilities are largely out of the scope of this report, we briefly summarise them here for completeness.

Cardiff University and the University of Oxford were partner institutions of the recently-closed Atacama Cosmology Telescope (ACT)\footnote{\url{https://act.princeton.edu/collaboration/our-partners}}, which operated at 27--220\,GHz until 2022.  All ACT data is made available through public data releases.  ACT is being incorporated into the Simons Observatory (see case study, p.~\pageref{box:so}), a major new cosmology facility with significant UK involvement.

The Cosmology Large Angular Scale Surveyor (CLASS; \citealt{essinger2014}; \citealt{harrington2016})\footnote{\url{https://sites.krieger.jhu.edu/class/}} is a US-funded telescope array in the Atacama Desert observing the CMB at 40--220 GHz (7.5--1.4\,mm).  The South Pole Telescope (SPT; \citealt{carlstrom2011})\footnote{\url{https://pole.uchicago.edu/public/Home.html}} is a dedicated cosmology facility operated by a number of US institutions, operating at 95--220 GHz (3.1--1.4\,mm).  The SPT is also used as an EHT station.

\subsubsection{Other single-dish submillimetre/millimetre facilities}

A number of other millimetre-wavelength observatories exist around the world, including Purple Mountain Observatory (China), the Nobeyama 45m Telescope (Japan), the TRAO (South Korea) and Mopra (Australia).  The Greenland Telescope (GLT), currently operating at frequencies up to 230\,GHz (1.1\,mm), is a dedicated EHT station.

Other submillimetre facilities include the Submillimeter Telescope (10m, Mt Graham), and the Kitt Peak 12m Telescope, both operated by the University of Arizona, and the Solar Submillimeter Telescope, a dedicated 1.5m facility for solar observations, located in Argentina \citep{kaufmann2008}.

\subsection{Interferometric instrumentation}

Interferometers have high angular resolution and small fields of view, allowing detailed imaging of targets of interest.  Key submillimetre and millimetre interferometers are shown in Figure~\ref{fig:interferometric}.

\subsubsection{Interferometers with UK access}

The currently extant submillimetre/millimetre interferometer to which UK astronomers have unrestricted access is the \textbf{Atacama Large Millimeter/submillimeter Array (ALMA)}\footnote{\url{https://www.almaobservatory.org/en/home/}}, located on the Chajnantor Plateau in Chile.  ALMA operates at frequencies from from 35\,GHz (8.6\,mm; Band 1) to 950\,GHz (330\,$\mu$m; Band 9), although Band 2 is under development.  ALMA has polarization capabilities in Bands 1 and 3--7 (84--373\,GHz; 3.6--0.8\,mm).  ALMA consists of:
\begin{itemize}
    \item The \textbf{12m Array}, a large array of 12-m antennas which cycles through a range of baseline configurations.  In the most compact configuration, with baselines up to 160\,m, resolutions range from 0.5$^{\prime\prime}$ at 950\,GHz to 11.9$^{\prime\prime}$ at 40 GHz. In the most extended configuration, with baselines up to 16\,km, resolutions range from 0.0048$^{\prime\prime}$ at 950 GHz to 0.11$^{\prime\prime}$ at 40 GHz.
    \item The \textbf{Atacama Compact Array (ACA)}, also known as the Morita Array, is an array of twelve 7\,m antennas (the `7-m array') and four 12\,m antennas (the `Total Power array'), with a fixed configuration with a maximum baseline of 45\,m, and resolutions ranging from 1.44$^{\prime\prime}$ at 950\,GHz to 31.5$^{\prime\prime}$ at 40\,GHz.
\end{itemize}
ALMA can deliver data cubes with up to 7,680 frequency channels. The width of these channels can range between 3.8\,kHz and 15.6\,MHz, but the total bandwidth cannot exceed 8\,GHz.  UK astronomers have access to ALMA through UK membership of ESO.

UK astronomers also have limited access to the \textbf{Submillimeter Array (SMA)}\footnote{\url{https://lweb.cfa.harvard.edu/sma/}}, which consists of eight 6\,m dishes located near the summit of Maunakea in Hawaii.  The SMA operates at frequencies from 180 GHz to 420 GHz, with configurations with baselines of up to 509m, and up to 48 GHz total bandwidth.  The SMA is operated by ASIAA (Taiwan) and the Center for Astrophysics (USA). The SMA is also currently undergoing an upgrade into the wide-band (w-)SMA \citep{2024arXiv240617192G}, discussed in Section~\ref{sec:interferometric_context}. UK astronomers at universities that are members of the JCMT consortium have some access to the SMA via the East Asian Observatory.
Access is also available to all UK astronomers through the CfA queue via an `open skies' policy.

\subsubsection{Interferometers with open-skies time}

The \textbf{Northern Extended Millimetre Array (NOEMA)}\footnote{\url{https://iram-institute.org/science-portal/noema/}} is operated by IRAM and located on the Plateau de Bure in France.  The array is composed of twelve 15\,m antennas and operates at frequencies of 70--120\,GHz, 127--183\,GHz and 196--276\,GHz, with an effective bandwidth of 15.5\,GHz per polarization.  The array has four configurations, the most compact of which has a resolution of 3.7$^{\prime\prime}$ at 100 \,GHz, while the most extended has a resolution of 0.5$^{\prime\prime}$ at 100\,GHz.  Planned upgrades to NOEMA are discussed in Section~\ref{sec:interferometric_context}.
15\% of NOEMA time is available to non-IRAM partner countries.

\subsection{VLBI}

Very Long Baseline Interferometry (VLBI) links telescopes around the world to undertake ultra-high-resolution (tens of $\mu$as) imaging of astrophysical targets of interest.

\subsubsection{Submillimetre and millimetre VLBI networks with UK access}

The \textbf{Event Horizon Telescope (EHT)}\footnote{\url{https://eventhorizontelescope.org/}} is the only millimetre VLBI network to which the UK has access.  For the images of M87* and Sgr A*, the EHT consisted of ALMA, APEX, IRAM 30m Telescope, JCMT, LMT, SMA, SMT, and the SPT.  Three more instruments have since been added to the network: GLT, NOEMA, and the Kitt Peak 12m Telescope.  The current operating frequency of the EHT is 230\,GHz (1.3\,mm).  For the initial observations of M87* and Sgr A*, baselines ranged from 160\,m to 10,700\,km, resulting in a resolution of $\simeq 25$\,$\mu$as \citep{eht2019}. 

The UK can nominate members of the EHT consortium through UK membership of the JCMT consortium.  There are currently 6 EHT consortium members in the UK, but this number could be substantially increased; for example, Taiwan has 20--30 EHT consortium members.

\subsubsection{Other submillimetre and millimetre VLBI networks}

The Very Long Baseline Array (VLBA)\footnote{\url{https://public.nrao.edu/telescopes/vlba/}} is a network of ten observing stations located across the USA, with a highest frequency of 96\,GHz (3.1\,mm).  The Korean VLBI Network (KVN; \citealt{lee2014})\footnote{\url{https://radio.kasi.re.kr/kvn/main.php}} is a network of three telescopes in South Korea, with a highest frequency of 129 GHz (2.3\,mm).  There are also a number of radio VLBI networks around the world operating at centimetre wavelengths or longer, including the UK's own e-MERLIN array\footnote{\url{https://www.e-merlin.ac.uk/index.html}}.

\section{A roadmap for single-dish submillimetre instrumentation: from the JCMT to AtLAST}
\label{sec:single-dish}

\textit{\textbf{STFC AAP Roadmap 2022, Recommendation 5.3\footnote{\citet{STFC_AAP_2022_Roadmap}, p.28}:} ... STFC should maintain and develop a broad portfolio of high/very high priority science facilities, including ... JCMT ..., as well as fostering roles in currently non-STFC-supported projects such as ... AtLAST ... }
\vspace{0.5\baselineskip}

\begin{figure} [b!]
    \centering
    \includegraphics[width=0.48\linewidth]{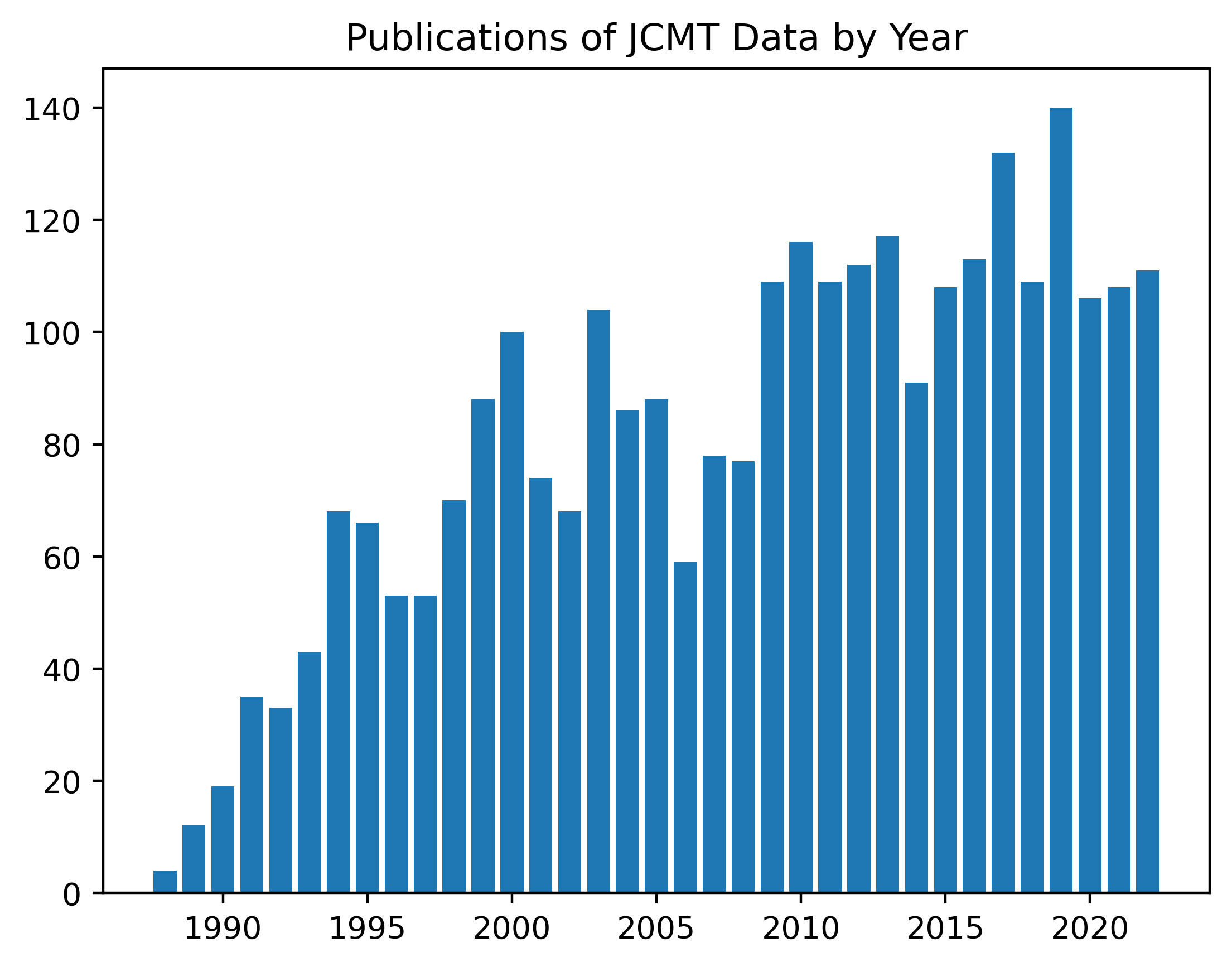}\hfill\includegraphics[width=0.48\linewidth]{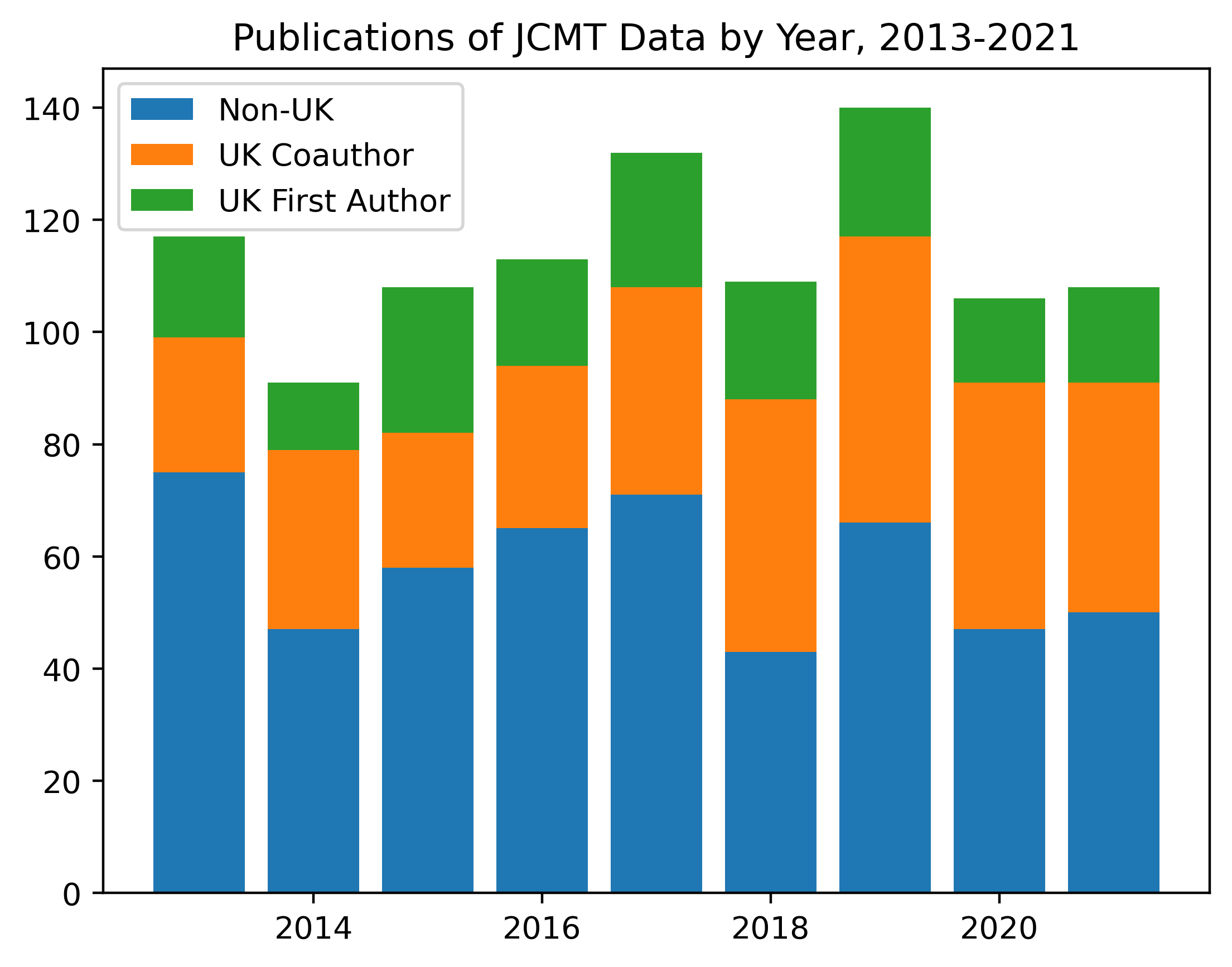}
    \caption{A summary of JCMT publication statistics.  \textit{Left:} All JCMT publications since the telescope opened in 1987 (source: \url{https://www.eaobservatory.org/jcmt/science/publications/}).  \textit{Right:} Papers published between 2013--2021, showing the fractions that have UK-affiliated lead authors and co-authors (source: \url{https://ui.adsabs.harvard.edu/public-libraries/FFUnBRxWROK-NauemkAjlQ}).}
    \label{fig:jcmt_pubs}
\end{figure}

\begin{figure}[t]
    \centering
    \includegraphics[width=0.7\linewidth]{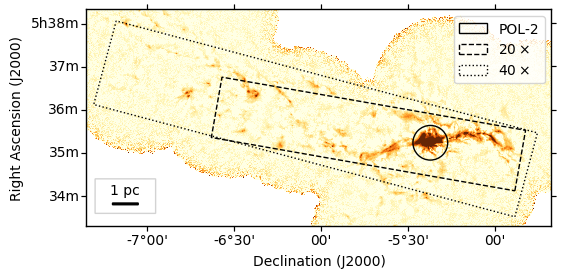}
    \caption{The expected improvement of mapping speed in polarized light with the new JCMT MKID camera, from \citet{furuya2020}.  With the new MKID camera it will be possible to map the entire Orion A molecular cloud in the time it currently takes to map the central OMC-1 region with POL-2.}
    \label{fig:scuba3_area}
\end{figure}

It is clear from our community survey that a new, 50m-class, single-dish submillimetre telescope is crucial to submillimetre science in the 2030s and beyond.  In this section we discuss how the UK can be at the forefront of the development of such a facility.

The UK has a strong heritage in both continuum and high-spectral resolution instrumentation for submillimetre wavelengths, with a reputation for world leading technology development.  There is a clear path from maintaining the UK's current involvement in the JCMT for the UK astronomers to play a leadership role in both science and technology with AtLAST.  Building new instruments for the JCMT is the next step in maintaining that legacy while looking towards the next big challenge in submillimetre single dish facilities: the expected sensitivity, resolution and field of view of AtLAST. Building these instruments now puts the UK in a strong position to fill the 1+ square degree field of view of AtLAST, which will require orders of magnitude increases in the number of pixels in both continuum and spectroscopic cameras.

In this section, we focus first on upgrades which will keep the JCMT as a world-leading facility over the next 10 years, before discussing AtLAST and the step-change in submillimetre astronomy capabilities that it will bring.  We then discuss the international context for single-dish submillimetre astronomy, before finally discussing the funding landscape and pathways for both instruments.

\begin{figure} [t]
    \centering
    \includegraphics[width=0.45\textwidth]{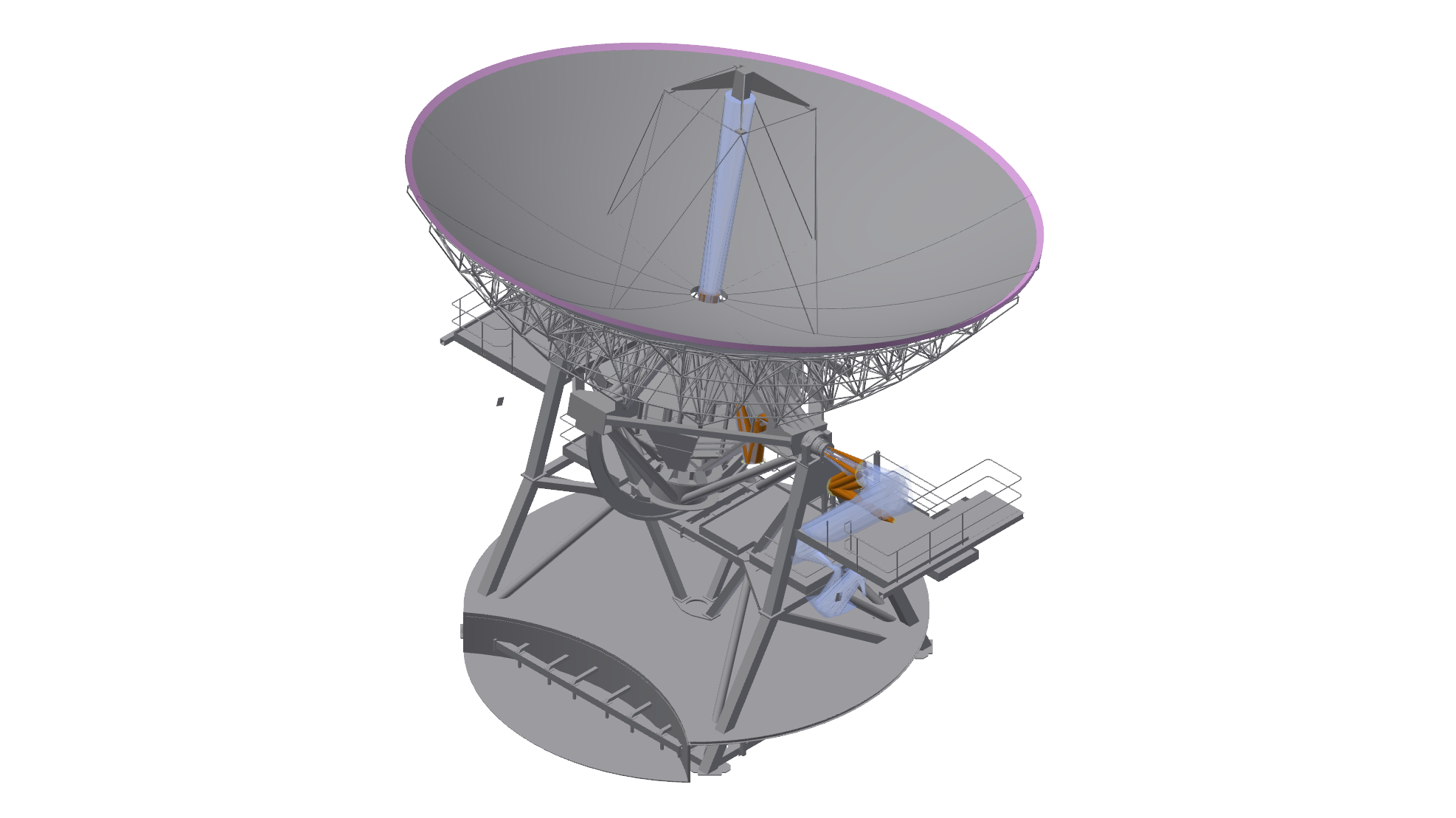}
    \includegraphics[width=0.52\textwidth]{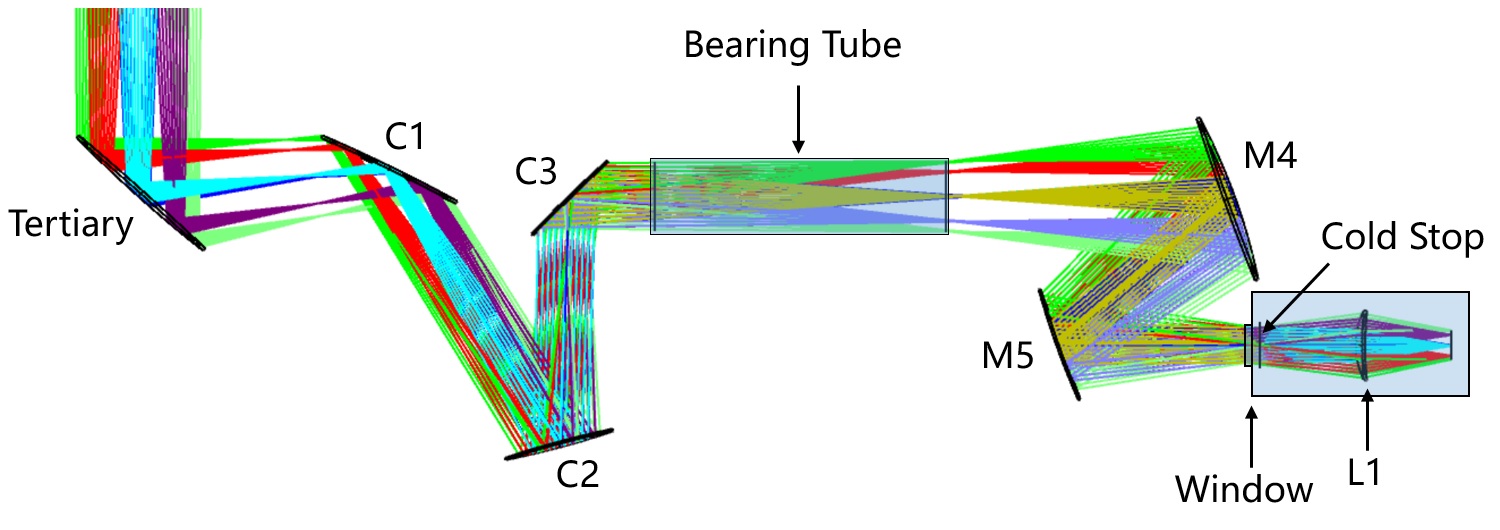}
    \includegraphics[width=0.47\textwidth]{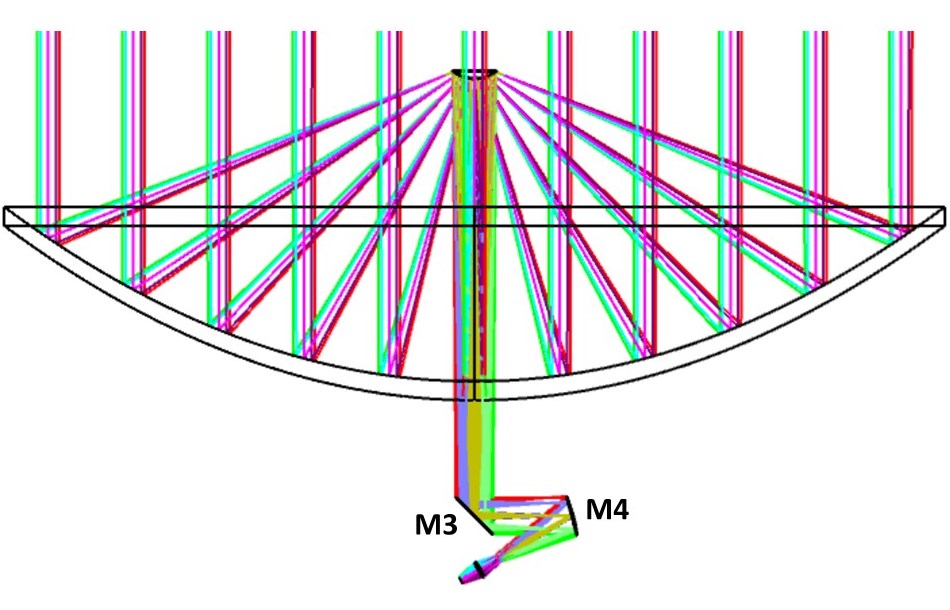}
    \includegraphics[width=0.47\textwidth]{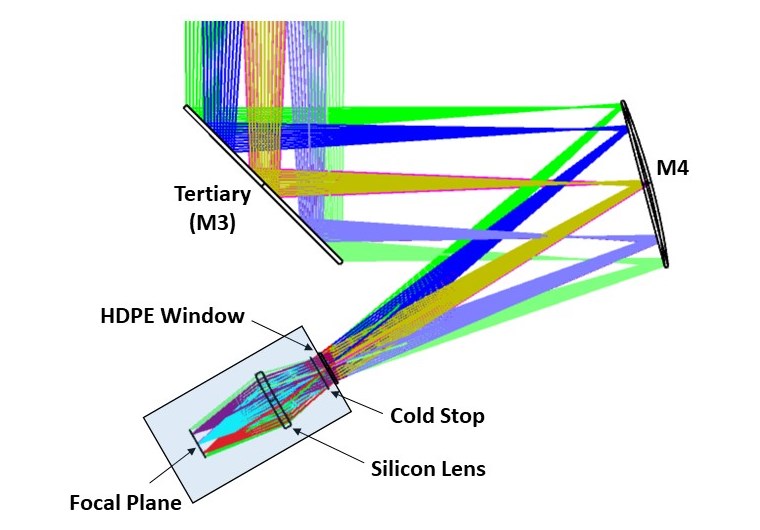}
    \caption{Potential light paths for the new JCMT MKID camera, from \citet{li2024}.  Top row: the out-of-cabin light path.  \textit{Top left:} The JCMT, showing the existing SCUBA-2 light path in light blue and the proposed MKID camera light path in orange.  \textit{Top right:} the light path, showing the existing tertiary mirror and the three ambient temperature mirrors C1-3. Two new mirrors would be added (M4 and M5) which would direct the light into the MKID camera.  Bottom row: the in-cabin light path.  \textit{Bottom left:} The light path from the the primary to the tertiary mirror (labelled as M3), and from there to the camera.  \textit{Bottom right:} An enlarged schematic of the in-cabin design light path.}
    \label{fig:scuba3in}
\end{figure}

\subsection{The JCMT: leading submillimetre astronomy into the 2030s}
\label{sec:JCMT_future}

The 15m JCMT is currently the largest single-dish submillimetre telescope in the world, and is likely to remain so until the 2030s.  The JCMT is also the only extant single-dish submillimetre telescope to which the UK has a significant amount of access, and over the future of which the UK is in a position to exert a significant amount of control.  Our community survey found that the JCMT is strongly supported by UK submillimetre astronomers, with 68\% of respondents considering it important for their research over the next 10 years.  However, UK access to the JCMT is currently only guaranteed until April 2027.  If UK access to the JCMT were not extended, there would be no general-purpose single-dish facility operating at submillimetre wavelengths for UK astronomers, as well as significant loss of UK expertise in submillimetre instrument design.

Until 2015, the JCMT was operated by the Joint Astronomy Centre, a collaboration between the UK, Canada and the Netherlands.  In 2015, operations were taken over by the East Asian Observatory (EAO), a consortium of universities in China, Japan, South Korea and Taiwan, with the UK as a partner in the JCMT consortium.  UK engagement with the JCMT remains strong: the UK has a representative on the JCMT Board (S. Eales, Cardiff); the current chair of the JCMT Time Allocation Committee (TAC) is at a UK institute (K. Pattle, UCL); and UK institutes, particularly Cardiff, Oxford, UCLan and UCL, are in discussion with the JCMT about potential instrument upgrades.  Figure~\ref{fig:jcmt_pubs} shows the high publication rate of the JCMT, and that a very significant fraction of JCMT papers have UK authors: over the 2013-2021 period, for which complete data is available\footnote{\url{https://ui.adsabs.harvard.edu/public-libraries/FFUnBRxWROK-NauemkAjlQ}}, 49\% of JCMT papers had one or more UK authors, and 17\% of JCMT papers were led from the UK.

Despite its world-leading position, the JCMT's current flagship instruments, SCUBA-2 and HARP, are more than a decade old and require replacement for the JCMT to remain world-leading over next 10--15 years.  We here describe the work in progress for their replacement, in which the UK is playing a leading role.

\subsubsection{Continuum instrumentation: a new MKID camera for the JCMT}
\label{sec:jcmt_mkid}

The proposed MKID camera on the JCMT to replace SCUBA-2 \citep{li2024} will be an entirely new instrument which will maintain the JCMT's and the UK's leading role at the forefront of submillimetre astronomy. The new camera will build on the success of the previous SCUBA-2/POL-2 camera, but will be an entirely new design and built with new technology.  It will make use of recent advances in Kinetic Inductance Detectors (KIDs) technology to build a camera with a working frequency of 353\,GHz (850\,$\mu$m), with increased sensitivity over SCUBA-2 (a target NEP of $<4\times10^{-17}$ W\,Hz$^{-1/2}$), and a 12-arcminute field of view, double that of SCUBA-2.  These advances will lead to significantly increased mapping speeds for large surveys of both our galaxy and of extragalactic sources. 

The array, which will contain 7272 MKID detectors, will be one of the largest of its kind.  The capability of designing, developing and fabricating such an array has already been demonstrated at Cardiff \citep{Muscat2018,tapia2020muscat}. The array will be feedhorn-coupled, which limits the number of pixels but better controls stray light and electromagnetic interference.  Feedhorn-coupled MKID arrays have also been more thoroughly tested than alternatives.
The new camera is targeting a 10--20$\times$ increase in mapping speed for continuum observations and a 20--40$\times$ increase for polarization observations over SCUBA-2.  As shown in Figure~\ref{fig:scuba3_area}, this increase in mapping speed will allow entire molecular clouds to be mapped in polarized light in the time it currently takes to map a single POL-2 field.

There are two proposed designs for this new system, an out-of-cabin design and an in-cabin design, which are shown in Figure~\ref{fig:scuba3in}.  The in-cabin design has fewer optical components, and therefore points for losses along the optical path. In addition, the in-cabin design is better suited to achieve the 12 arcminute FOV which helps to increase the mapping speed of the new instrument. Furthermore, an in-cabin design would allow SCUBA-2 to operational as the new instrument is developed, thereby minimising downtime.  However, fitting the camera into the instrument cabin would be difficult, and the Cassegrain design of the JCMT would require the cryostat of the new camera to be tipped as the telescope changes elevation.  Thus, an out-of-cabin design is also being developed, in which the new camera would be placed on the the Nasmyth platform currently occupied by SCUBA-2.  

The UK will be a major partner in the construction of the new camera, with the most significant UK contribution being the MKID focal plane array, as well as warm electronics and pipeline software contributions.  Other components will be built by the East Asian Observatory (EAO) partners, including China, Japan, South Korea, Taiwan and Thailand.  The funding being sought is described in Section~\ref{sec:single-dish_funding}, below.

\subsubsection{Spectroscopic instrumentation: upgrading and replacing HARP}
\label{sec:jcmt_superharp}

\textbf{Near-term HARP upgrade:} HARP (cf.~Section~\ref{sec:current_instrumentation_jcmt}) is a 16-pixel 345\,GHz heterodyne array with single-sideband (SSB) mixers, operating from 325--375\,GHz with an instantaneous bandwidth of $\sim$2 GHz and an Intermediate Frequency (IF) of 4--6\,GHz. 
HARP is currently operating with only 12 of its original 16 pixels. It is almost 20 years old, and there are both short- and long-term plans to upgrade it. A UK team from Oxford is leading these efforts, in collaboration with the EAO and partners in Thailand and Taiwan.

A staged upgrade is planned to avoid disrupting HARP operations.  Within approximately one year, an intermittent upgrade of HARP would keep the interferometric optics and the backend, but replace all 16 SIS mixers with modern front-end detectors with much wider RF and IF capabilities, covering the entire ALMA Band 7 window from 275--375\,GHz and beyond, as well as achieving a better noise temperature. Work is currently in progress (K.-Y. Liu, EAO) to develop a 220\,GHz broadband finline mixer based on Oxford's design. Devices have been fabricated (M. J. Wong, ASIAA), and tests are currently underway at NARIT, Thailand (P. Kittara; Mahidol University). The team in Thailand have recently begun developing millimetre/submillimetre superconducting mixers (with previous support from the UK STFC-NARIT Newton Grant, led from Oxford), and have achieved impressive results in a short time with successful measurement of high-quality current-voltage (IV) curves, and good mixer noise temperatures.  The next step is to scale the design to 345\,GHz as a potential replacement for the existing HARP mixers.

Alongside this development, the Oxford team is also currently developing a probe-based SIS mixer focusing on band-combining (Bands 5 and 6) and Band 7 as a potential contribution to the JCMT and other submillimetre observatories such as the AMT, an upgraded ALMA, the GLT or the SMA.  The team expects to finalise the design and start fabrication and testing in early 2025.  If successful, this device could be deployed to replace the current HARP mixers. The second stage of the HARP upgrade will involve upgrading the DSB SIS mixers with sideband separating mixers. This effort will both free up receiver cabin space and immediately double the observable bandwidth.  Upgrading the backend IF spectrometer to meet the capability of the front-end receiver is expected to be undertaken by EAO member states such as Taiwan, South Korea, Thailand, Vietnam and Malaysia.

\textbf{Longer-term HARP replacement} The JCMT aims to replace HARP with a 25-pixel dual-polarisation sideband separating SIS array, effectively a 100-pixel singly polarised DSB heterodyne array. This would be the most advanced heterodyne array in the world operating in the submillimetre regime. However, the TRL level of constructing a truly large and expandable heterodyne array is low compared to the TRL of the corresponding bolometric MKID array.  With current technologies, it is challenging to bring large heterodyne array technology to a level suitable for deployment. Such a large array presents a number of technical challenges, including ensuring efficient LO distribution, machining feed horns at high frequencies, cooling to the required operating temperatures, extending from a single-polarisation receiver to a dual-polarisation array, and the need to provide all accesses via two interfaces \cite{wenninger2023design}. The Oxford and RAL team have therefore formulated a strategy to achieve this ambition and is seeking support to bring the approach to fruition. Upon successfully validating the underlying technologies with a demonstrator prototype at TRL 4 and establishing the feasibility of all technologies required to form hundreds or thousands of heterodyne-pixel arrays, including superconducting readout amplifiers, the next stage would be to extend and populate the array, hence constructing the new 100-pixel HARP instrument.

This replacement for HARP would allow truly blind, wide-area spectral line surveys, such as Galactic Plane surveys and blind cosmology heterodyne surveys which are currently too time-intensive to be achievable.  It would also act as a pathfinder instrument to pave the way for UK-led development of the kilo-pixel heterodyne arrays at multiple frequency bands which are needed for AtLAST.

\begin{figure}[t]
    \centering
    \includegraphics[width=1.0\linewidth]{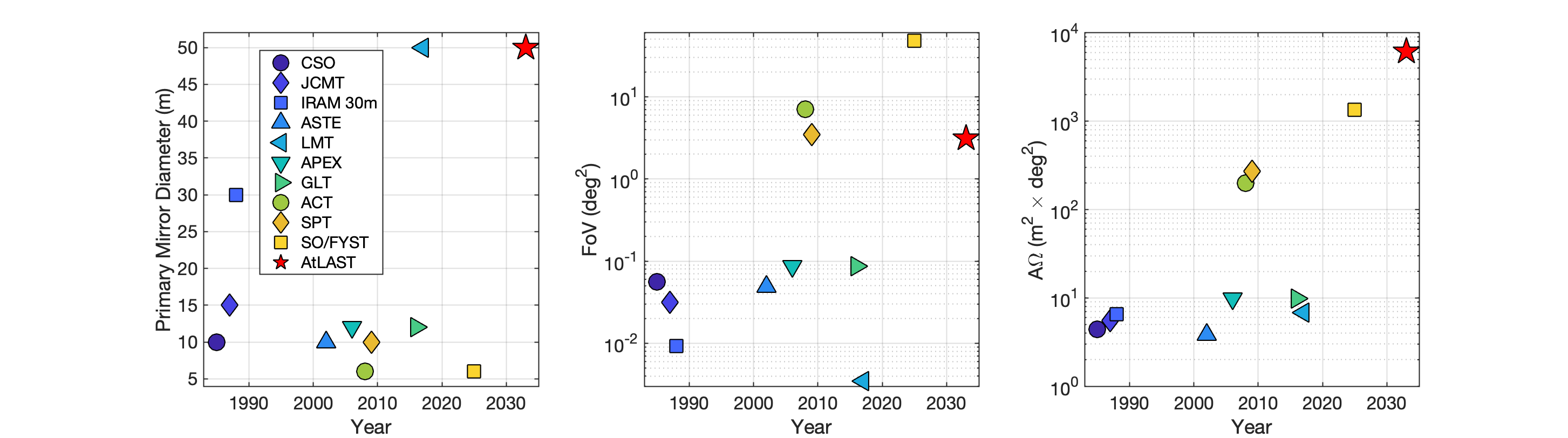}
    \caption{Throughput of existing and upcoming submillimetre single-dish telescopes operating at wavelengths shorter than 1\,mm.  \textbf{Left:} The diameter of each observatory. \textbf{Centre:} Field of View (FoV) of each telescope; note that many of the large FoV facilities at the top of the plot are dedicated survey facilities generally focused on specific CMB experiments. \textbf{Right:} Throughput of each observatory which takes into account both field of view and spatial resolution. Figure modified and updated from \citet{ramasawmy2020}.}
    \label{fig:SD_throughput}
\end{figure}

\subsection{AtLAST: The future of single-dish submillimetre astronomy}
\label{sec:single-dish_atlast}

The Atacama Large Aperture Submillimetre Telescope (AtLAST) concept is for a 50\,m class single dish facility to be built on the Atacama Plateau (within either the wider Atacama Astronomy Park or the ALMA concession) which will have a large focal plane, be able to observe in the same atmospheric windows as ALMA and host up to 6 large format (highly-multiplexed) instruments. The AtLAST consortium is a European-led international team of astronomers, telescope engineers and sustainability experts who are bringing about a design for an observatory powered through sustainable resources.

\begin{figure}[t]
    \centering
    \includegraphics[width=0.45\textwidth]{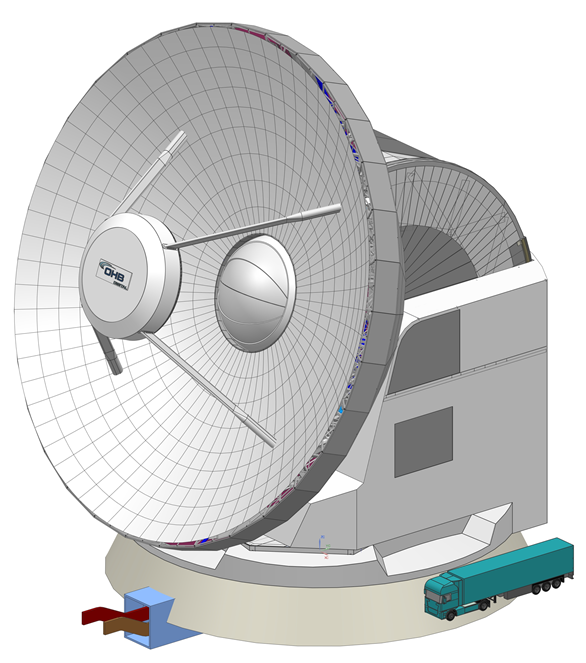}\hfill
    \includegraphics[width=0.52\textwidth]{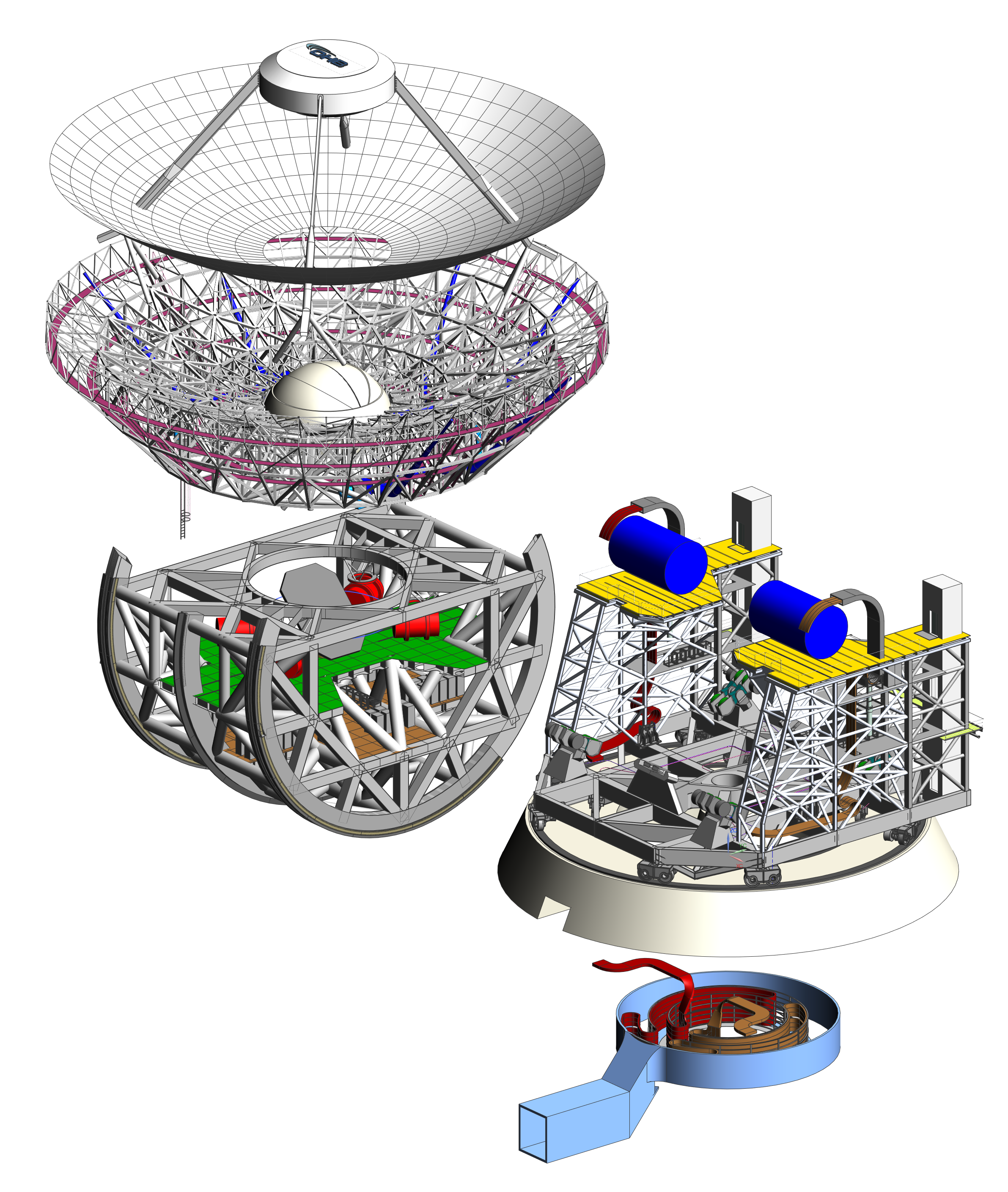}
    \caption{Models of AtLAST adopted from Figures 7 and 17 of \citet{mroczkowski2024}. \textit{Upper left:} The CAD design of AtLAST with a truck shown as scale. \textit{Upper right:} An exploded diagram of the structure of the dish and the internal components of AtLAST.}
    \label{fig:atlast_cad}
\end{figure}

\begin{figure}[t]
    \centering
    \includegraphics[width=0.95\textwidth]{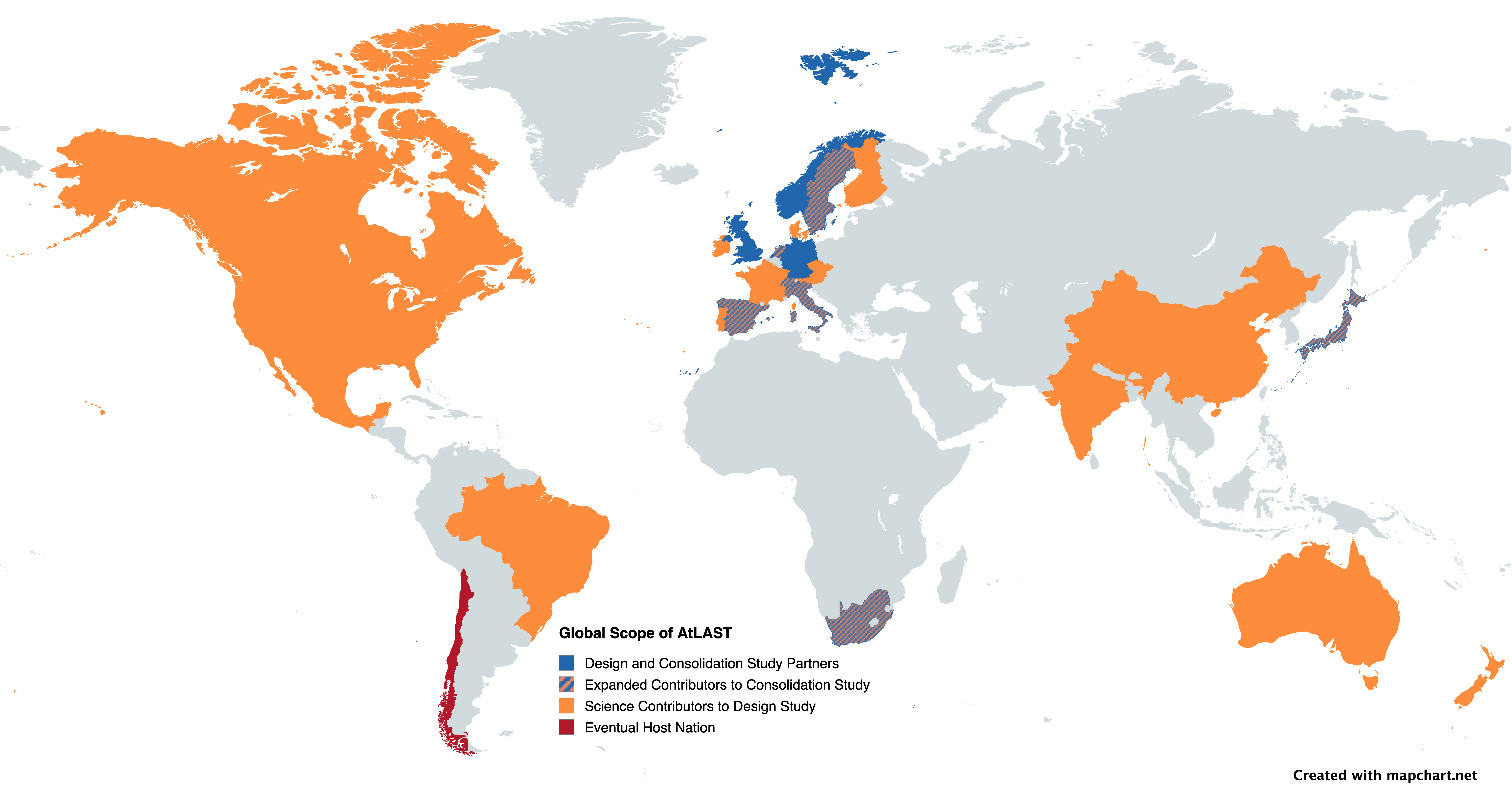}
   
    \caption{Global reach of the AtLAST consortium, both through the Horizon 2020 Design Study and anticipated Horizon Europe Design Consolidation as the contributing and coordinating partner countries expand.}
    \label{fig:atlast_reach}
\end{figure}

The initial design study\footnote{Funded through Horizon 2020 grant agreement 951815}, has contributions from University, government, IGO and industrial partners, is led from the University of Oslo, with ESO and STFC/UKATC representation on the coordination committee.  The 3.5 year design study, which is concluding as this report is being written, is looking at long term funding/operating frameworks, telescope design, site selection, operations, sustainability and science community engagement.  The initial goals of this design study are summarised in \cite{klaassen2020}. 

Following on from this study there will be a Horizon Europe funded design consolidation (starting in Q1 2025) study which expands the international collaboration, critically reviews the expected designs and plans, incorporates instrumentation, prototypes sustainability initiatives and operations plans, further investigates site selection and derives a science reference plan. The Horizon Europe project is continuing to be led from the University of Oslo, with coordination committee representation expanded to not only ESO and STFC/UKATC, but also MPIfR (DE), IAC (ES) and OHB Mechatronic (DE). The full collaboration has been expanded to 20 participants spanning Europe, Asia and Africa (see Figure~\ref{fig:atlast_reach}).

The telescope design \cite[outlined in][see also Figure~\ref{fig:atlast_cad}]{mroczkowski2024} has been distilled to a rocking chair design with a 50\,m primary, 12\,m secondary, a 4.7\,m focal surface, and a field of view of at least 1 sq. deg at all expected operational wavelengths (0.3 -- 10\,mm). When comparing these specifications to current submillimetre facilities, the diameter gives an increase of an order of magnitude in both resolution and sensitivity, while the field of view, is more than an order of magnitude greater than general purpose observatories, and is only surpassed by 6\,m class survey telescopes such as SO, ACT, SPT and CCAT/FYST (see Figure~\ref{fig:SD_throughput}).

In the next phase of development, the telescope design will be critically reviewed by external experts. Instrument teams, including the UKATC, Cardiff University and institutes across Europe, Japan and South Africa will contribute their design experience to that review and refinement.

The sustainability portion of the current design study has focused on understanding the power needs of the observatory and its instrumentation to derive a series of `levelised cost of electricity' metrics across various power generation /storage systems \cite[][including everything from diesel generators to full photovoltaic production with various types of battery storage]{viole2023}. These predictions take into account the full lifetime of the observatory and the supply chains required to get hardware and consumables in place, and have shown that the most economical way of running the observatory is via photovoltaics with battery storage. In parallel to this, has been a community engagement program aimed at understanding the power needs of the local communities including the town of San Pedro de Atacama and other observatories in the region \citep{valenzuela2023}.

The next stage of development will see refinement of these processes and a series of prototyping exercises to be carried out on existing facilities including power generation and regenerative breaking\footnote{\url{https://spie.org/astronomical-telescopes-instrumentation/presentation/Energy-recovery-system-for-large-telescopes/13094-13}}.

The science engagement, led from the UK \cite[see, e.g.,][]{ramasawmy2020, booth2024} in the current design study has the remit of engaging the worldwide astronomical community (with more than 100 participants across 19 countries) in building the science case for AtLAST. The rationale here is twofold; to increase the global visibility of the project and grassroots support for the telescope and to derive a set of requirements for the observatory that are science driven. These requirements  for the telescope, instrumentation suite and operations will then be used to further refine the telescope and instrument designs in future studies. The key science drivers for the telescope, as defined in the current study are presented in Table ~\ref{tab:atlast_key_science}, which is very much aligned with the science goals presented in Section~\ref{sec:science} and outcomes of the consultation presented in Section~\ref{sec:consultation}.  In the next phase, a more formal science reference plan will be derived.

\begin{table}[hbt]
    \centering
    \begin{tabular}{|p{1.6cm}p{4.5cm}p{4.5cm}p{4.5cm}|}
\hline
& \textit{\textbf{Where are all the baryons?}} 
& \textit{\textbf{How do structures interact with their environments?}}
& \textit{\textbf{What does the time-varying (sub-)mm sky look like?}}\\ 
  \hline
 
{\bf Detailed science goal} 

&  Measuring the total gas and dust content of the Milky Way and other galaxies, in the interstellar, circumgalactic, and intergalactic media, reaching down to the sensitivities required to probe the typical populations of sub-mm sources.

&  Understanding the lifecycle of gas and dust near and far; mapping the baryon cycle on multiple-scales; observing the interplay between gravity, radiation, turbulence, magnetic fields, and chemistry and their mutual feedback.  

& Identifying the mechanisms responsible for time variability across astrophysical sources: from the Solar corona and other objects in our solar system to luminosity bursts in everything from protostars to active galactic nuclei.\\
\hline

{\bf Detailed technical specification}

&  \textbf{\textit{High sensitivity to the faint signals}} (at sub-mK levels)  \textit{\textbf{on large scales}} ($\geq$1 deg$^2$) from even the most diffuse and cold gas through sub-mm line tracers. Wide field ($>$500 deg$^2$)  continuum surveys capturing the plane of our galaxy and resolving 80\% of the cosmic infrared background, probing typical populations and looking back over 90\% of the age of the universe.   

& \textit{\textbf{High spectral resolution}} and \textit{\textbf{polarisation }}measurements on the relevant size scales for cores (0.1 pc, our galaxy), clumps (10 pc, nearby galaxies) and cloud complexes ($\sim$ few kpc, distant universe) to quantify the chemistry, disentangle the dynamics, and measure the magnetic fields working together to shape the evolution of structures within their larger-scale environments. 

& An operations model that allows for \textit{\textbf{highly cadenced and rapid response observations}} and data reduction pipelines with in-built \textit{\textbf{transient detection algorithms}}\textbf{; }high time-resolution (few seconds) observations of our Sun and other stars.\\
\hline
    \end{tabular}
    \caption{Key Science Drivers for AtLAST, reproduced from \cite{booth2024}.}
    \label{tab:atlast_key_science}
\end{table}

In parallel to the current efforts there have been designs drawn up by a consortium of East Asian communities (led by Japan) to create a similar telescope: The Large Submillimeter Telescope (LST).  At the end of the Horizon 2020 design study, these two consortia will merge into a single project under the Horizon Europe design consolidation project. 

As noted above, the UK is heavily involved in this project, and the instrumentation projects listed in Section~\ref{sec:JCMT_future} will enhance UK lead technical competencies in these areas, as well as more generally advance these technologies towards the type of multiplexing capabilities required to fill the large field of view expected for AtLAST.

\subsection{International context}

While there are a number of single-dish telescopes planned or operating in the millimetre regime, only the JCMT and APEX are currently operating at $< 1$\,mm.  While the IRAM 30m telescope can in principle operate at submillimetre wavelengths, it is in practice restricted to $>$1\,mm by the atmospheric conditions at Pico Veleta.  The LMT is restricted to $>$1\,mm by design.  There are a number of dedicated cosmology facilities with millimetre or submillimetre capabilities, but these are not available for the wide range of science goals discussed in Section~\ref{sec:science}.

The Fred Young Submillimeter Telescope (FYST; formerly known as CCAT-prime), which is currently under construction, will be a 6\,m millimetre/submillimetre telescope located at 5,600\,m altitude on Cerro Chajnantor, overlooking the ALMA array.  FYST is designed as a survey instrument, with stated aims of measuring the kinematic Sunyaev-Zel’dovich effect of galaxy clusters, mapping of [CII] emission in the epoch of reionization, and spectral line mapping of the ISM and nearby galaxies, all at $\sim$0.5--1 arcminute resolution, considerably lower than the $\sim$1--10 arcsecond resolutions of the other telescopes considered here.

In the intermediate term, the Africa Millimetre Telescope (AMT) and the Greenland Telescope (GLT) are both expected to have submillimetre capabilities. As both facilities would contribute their observing time to the EHT project, and are largely funded in this capacity, we will discuss them in Section~\ref{sec:vlbi}, below.

AtLAST is the only planned general-purpose submillimetre telescope to be built in the foreseeable future.  As discussed above, the Japan-led LST will merge with AtLAST in early 2025.

\subsection{Funding landscape for single-dish astronomy}
\label{sec:single-dish_funding}

\subsubsection{JCMT access funding}
\label{sec:single-dish_funding_jcmtops}

UK access to the JCMT is currently funded by contributions from a consortium of UK universities\footnote{The UK JCMT Consortium: Cardiff University (lead institute), University of Central Lancashire, University College London, University of Edinburgh, University of Hertfordshire, Imperial College London, Liverpool John Moores University, Armagh Observatory and Planetarium, The Open University, University of Manchester, University of Oxford, University of St. Andrews, Durham University}, match-funded by STFC funding through the PPRP.  The current grant\footnote{\url{https://gtr.ukri.org/projects?ref=ST\%2FV000268\%2F1}} runs until April 2027.  The end date of this grant was originally April 2024; funding was generously extended by STFC to allow time to prepare the bid described below.  At the present time, only UK astronomers affiliated with one of the universities in the UK consortium have the right to submit proposals to the JCMT, or to be a member of a JCMT Large Program.

The UK JCMT Consortium has recently proposed a new paradigm would widen access to the JCMT to all astronomers affiliated with UK universities.  A Letter of Intent for this proposal was accepted, with the full proposal due to be submitted in September 2024.  However, in August 2024, STFC extended the current funding model for two years, to March 2027, awaiting a new system for assessing operational grants.  The UK JCMT Consortium will thus apply for further JCMT operations funding in 2026.

The current level of UK JCMT funding gives the UK 25\% of the Principal Investigator time on a world-class facility, which is leveraged to access more than 50\% of JCMT time through UK access to JCMT Large Programs.  It must be emphasised that if UK access to the JCMT is not continued, UK astronomers will have no direct access to a general-purpose single-dish submillimetre telescope from 2027 onwards.

Under the proposed new funding model, the STFC would fund JCMT operations, and UK university contributions will allow PhD students to travel to Mauna Kea and EAO partner institutions, to assist with telescope operations and instrument development.  The new funding paradigm would thereby give UK postgraduate students an opportunity to gain key skills in astronomical observation and instrumentation that would otherwise be inaccessible.

Membership of the JCMT consortium is also currently the only route by which UK astronomers have access to the EHT.  Access to a single-dish submillimetre telescope is also key to identifying targets for interferometric follow-up.  This was noted by the AAP 2022 Roadmap, which notes that \textit{``In many cases it is not enough for there to be only access to the largest international facilities; national access to world-leading smaller specialist facilities can provide feeder programmes to the larger facilities and provide the UK with tactical and strategic advantages (e.g. JCMT feeding ALMA and e-MERLIN feeding SKA).''} (\citealt{STFC_AAP_2022_Roadmap}, p. 5).
  
Secure access to the JCMT over the next 5--10 years will make it possible for the UK community to develop the new MKID submillimetre camera and the new large-format heterodyne array for the JCMT described above.  These developments are strongly supported by our community survey: 76\% of respondents consider the JCMT with a new MKID camera important to their future research, and 71\% consider the JCMT with a new large-format heterodyne array important to their future research, as discussed in Section~\ref{sec:consultation}.  Developing these instruments would put the UK in a strong position to take a leading role in AtLAST. 

\subsubsection{JCMT MKID camera funding}
\label{sec:single-dish_funding_mkid}
  
A bid is planned for within the next 1--3 years for PPRP funding for the UK to play a significant role in the construction of the new MKID camera.  The UK consortium will be led by the University of Central Lancashire (P.I. D. Ward-Thompson), and will include scientists at Cardiff University, University College London, and the UKATC.  We expect to apply for $\sim$£1.5M over 3 FY, to provide a UK in-kind contribution to the new MKID camera, a similar level of commitment as other JCMT partners are giving. We emphasise that if the UK is not a key partner in this project, the camera will be built more slowly, and the UK will risk losing its competitive edge in submillimetre instrumentation, and particularly in MKID array fabrication, to East Asia. 

\subsubsection{Heterodyne instrument development for the JCMT and AtLAST}
\label{sec:single-dish_funding_heterodyne}

Line intensity mapping at submillimetre wavelengths was identified by the AAP Roadmap 2022 as an emerging priority, where it was noted that \textit{``A bespoke instrument for the JCMT (where UK access is currently funded by PPRP and university contributions) would provide UK leadership in this area.''} (\citealt{STFC_AAP_2022_Roadmap}; Section 5.2.2).

Support is currently being sought to develop the key fundamental technologies required to build an extremely large heterodyne array, envisioned not only to fulfill the need of the JCMT but also for AtLAST, ALMA near-future upgrades\footnote{A study on the feasibility of developing large heterodyne array capability on ALMA stations is ongoing.}, and beyond. This technological development is crucial to pave the way to bid to build truly large-format heterodyne arrays for the JCMT, ALMA and AtLAST.  This project will develop the advanced fundamental units and technologies required to construct a large focal plane array: both the front end superconducting and back end warm readout electronics (based on SKA technologies), and construct a prototype demonstrator comprising the entire heterodyne array receiver chain, to demonstrate the feasibility of integrating these components to form an advanced array receiver. 
 
Once these underlying technologies are developed, with the leadership of the UK, construction of complementary components as well as mass production of fundamental building units will be coordinated with East Asian Observatory (EAO) partners, particularly groups at ASIAA in Taiwan and NARIT in Thailand.  This will allow deployment of a new large-format heterodyne array at the JCMT, which will also serve as a technology testbed for AtLAST. The funding for the UK participation in actually building the instruments is likely to be sought from PPRP, UKRI infrastructure funding or other suitable funding schemes so that the UK continue to play a leading role in this development.  This would likely be led by researchers at Oxford and RAL, and is expected to comprise a 25-pixel, extendable to 100-pixel, dual-polarisation sideband separating heterodyne array; the latter is an equivalent to 400-pixel traditional array, a truly revolutionizing instrument.  This would represent the most advanced heterodyne array in the world operating in the submillimetre regime.

\subsubsection{Summary of JCMT funding timeline}
\label{sec:jcmt_funding_summary}

The community is seeking support, and is expected to continue to apply to various suitable funding schemes to support many of these activities such as the STFC Large Award, PPRP, Infrastructure Fund and others:

\textbf{JCMT Operations:}

\textbf{September 2024:} A Letter of Intent for a proposal to extend JCMT Operations and extend JCMT access to all UK universities was accepted by the STFC for submission 09/2024.  However, in August 2024, STFC extended the current funding model for two years, to March 2027, awaiting a new system for assessing operational grants.

\textbf{September 2026:} Renewed JCMT Operations grant, providing $\sim$25\% of PI time on the JCMT and access to JCMT Large Programs for all UK astronomers for the 2027--2032 period, with a likely value of $\sim 1.5$M GBP over 5 FY (subject to inflation).

\textbf{September 2031:} Renewed JCMT Operations grant, providing access for all UK astronomers for the period until AtLAST begins operations.  Value currently unknown, but the request per annum will likely be similar to the 2026 request (subject to inflation).

Thereafter, we would expect AtLAST to be on-sky, and we do not currently envisage making significant further funding requests for JCMT operations or instrumentation beyond this point.

\textbf{JCMT Instrument Development:}

\textbf{Next 1-3 years:} Funding for UK share of the new MKID camera for the JCMT, which will be led from the UK.  This will have a likely value of approximately 1.5M GBP over 3 FY.  See Section~\ref{sec:single-dish_funding_mkid} for details.

\textbf{Next 3-5 years:} Funding for the UK share of a new large-format heterodyne array to replace HARP.  Value currently uncertain.

These instruments will also serve as a demonstration of new technologies for AtLAST.

Prior to seeking funding for a new large-format heterodyne array, funding for basic technology development associated with scalable large-format heterodyne arrays will be sought. This could be exploited both for the JCMT and for other instruments, including AtLAST and ALMA, but is aimed at maintaining national positioning for provision of the technologies involved.

\subsubsection{AtLAST}
\label{sec:single-dish_funding_atlast}

The UK submillimetre/millimetre community very strongly supports the development of AtLAST.  As discussed in Section~\ref{sec:consultation}, the new submillimetre facility considered most important for survey respondents' science goals on a timescale of 10+ years is a 50\,m-class single-dish telescope such as AtLAST, which was considered important by 92\% of survey respondents.  Concerted effort needs to be made over the next five years to ensure that AtLAST is funded and built. 
  
The AtLAST Horizon 2020 Design study\footnote{\url{https://cordis.europa.eu/project/id/951815}}, with a value of €3.5M, is currently concluding, with an end date of August 2024.  A further Horizon Europe Design Consolidation study has just been approved\footnote{\url{https://www.ukatc.stfc.ac.uk/Pages/Further-funding-for-AtLAST-announced.aspx}}.  This project is funded for 3.5 yrs, with a value of €4M, of which $\sim$10\% of direct funding will go to the UK.  The project has 20 international partners\footnote{Europe: Norway, UK, Denmark, Spain, Italy, the Netherlands, Sweden, Switzerland; Asia: Japan -- NAOJ, U Tokyo, Nagoya U, and Kitami Institute of Technology; Africa: South Africa -- University of Pretoria.}, and a timeframe of Q1 2025 – Q3 2028.  
  
At the end of the Horizon 2020 design study, the AtLAST consortium will merge with a consortium of East Asian communities, led by Japan, that has been developing a similar telescope, the Large Submillimeter Telescope (LST).  The two consortia will merge into a single project under the forthcoming Horizon Europe design consolidation project.  Involvement in AtLAST will therefore maintain and develop the collaborative links with East Asian astronomers forged by the UK’s membership of the JCMT Consortium.  The UK has a seat at the table in the AtLAST consortium in the form of participation in the coordination committee of both the Horizon 2020 and Horizon Europe projects. 

The European Southern Observatory (ESO) has announced that it will begin discussion of its next flagship project, as the Extremely Large Telescope (ELT) nears completion.  It is thus vital that submillimetre science and technology is well-represented in this discussion, to ensure that this crucial wavelength range is at the heart of ESO’s future plans, and to drive ESO engagement with the AtLAST project.  The AtLAST project is already beginning to engage with ESO and the UK representative on the ESO User Committee on this topic. 
 
The estimated total cost of AtLAST is $\sim$\$300M for the telescope itself and $\sim$\$520M for its instrumentation (\citealt{klaassen2020}; note that these numbers are indicative only), and so AtLAST must be funded by a multinational consortium.  The UK has the opportunity to be at the heart of this consortium, and so to maintain its place at the forefront of submillimetre astronomy.

\section{A roadmap for interferometric instrumentation: The UK's role in the future of ALMA}
\label{sec:interferometric}

\textit{\textbf{STFC AAP Roadmap 2022, Recommendation 3.6\footnote{\citet{STFC_AAP_2022_Roadmap}, p.19}:} The UK must remain a member of the European Southern Observatory and play leading roles in its development of its world-class instrumentation, including ... the development of ALMA instrumentation. }

\begin{figure}
    \centering\includegraphics[width=0.9\linewidth]{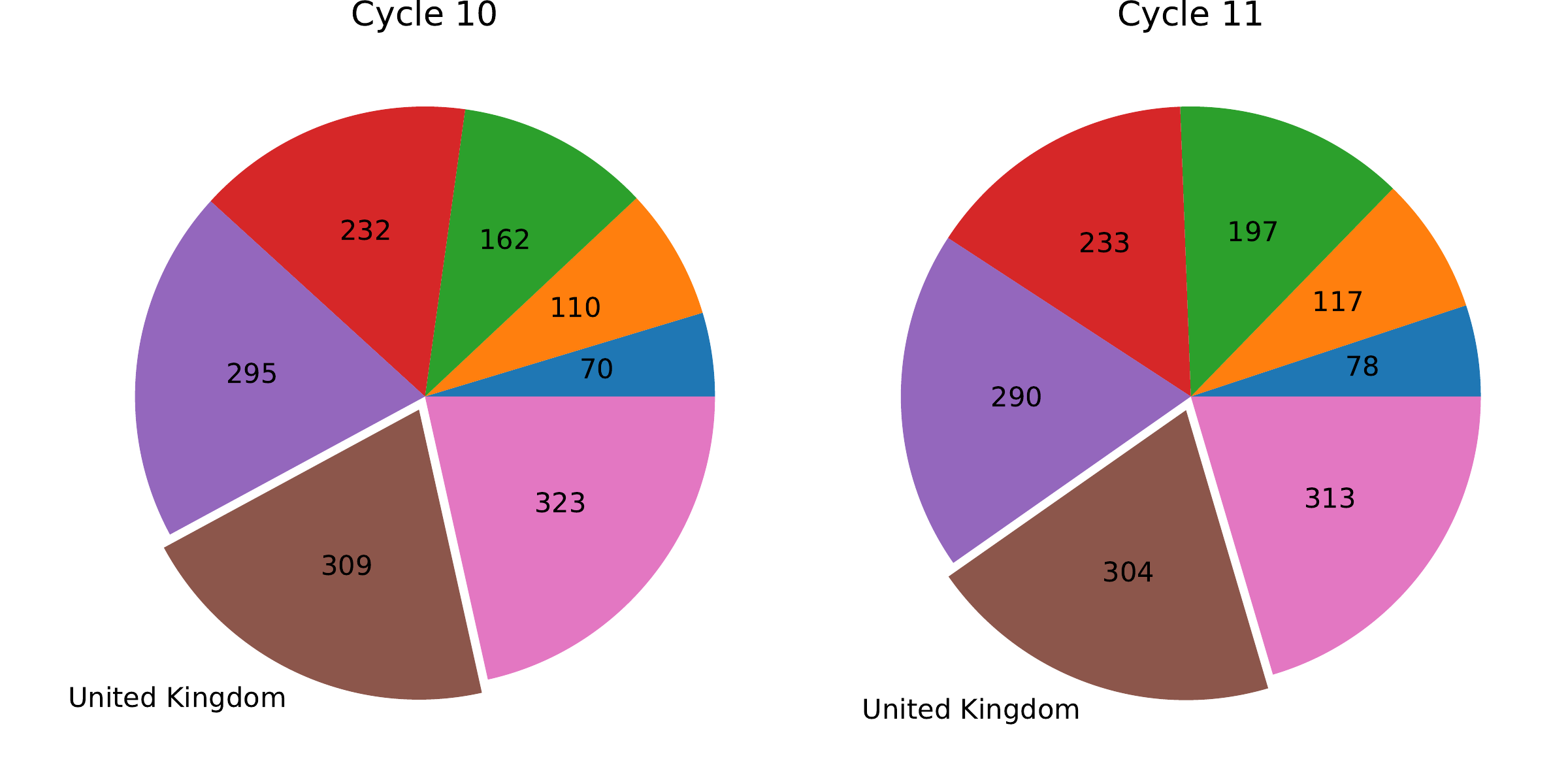}
    \caption{The number of unique PIs and Co-Is per country on proposals submitted to  ALMA  for the most recent two ALMA cycles, Cycle 10 (left) and Cycle 11 (right). Only countries/regions supported by the EU ALMA Regional Centre with more than 50 PIs plus Co-Is are shown.}
    \label{fig:alma-proposals}
\end{figure}

Our community survey shows that ALMA is the submillimetre instrument that is most widely used by UK astronomers, and is expected to be a crucial facility for the foreseeable future. Figure~\ref{fig:alma-proposals} shows the number of investigators in Europe on European proposals submitted in the last two proposal submission cycles (Cycles 10 and 11). In each of these cycles over 300 different UK astronomers were either principal investigator or a co-investigator on ALMA proposals. For these two cycles the over-subscription for European time (which is 33.7\% of the total available time) on the main array of 12m telescopes was over 8 (8.4 in Cycle 10 and 8.2 in Cycle 11). 

\begin{wraptable}{r}{8.5cm}
    \centering
    \begin{tabular}{lccc}
    \toprule
          Year & Total & UK lead author & UK PI Projects\\
         \midrule
         2022 & 168 & 32 & 41\\
         2023$^*$ & 133 & 21 & 28\\
         \bottomrule
    \end{tabular}
    \caption{Number of publications  using ALMA data with UK authors for the last two years.  The columns show the total number of papers, the number with UK lead authors and the number based on UK PI ALMA project. $^*$ The data for 2023 only covers publications up until 17 October 2023 (when the last census of publications was completed). }
    \label{tab:alma_papers}
\end{wraptable}

The survey results also showed strong support for upgrading the facilities of ALMA, to further enhance its scientific capabilities. In this section, we summarise the likely upgrades to ALMA over the next 10+ years.
ALMA's Development Roadmap\footnote{\url{https://www.almaobservatory.org/en/publications/the-alma-development-roadmap/}}, published in 2018, identified three science drivers for ALMA in the coming decade. 
From these, the technical developments necessary for ALMA to achieve these goals were derived. Subsequently the highest priority of these developments have been collected together in to the ALMA Wideband Sensitivity Upgrade (WSU; \citealt{2020arXiv200111076C}). The WSU is now underway and scheduled to be complete in Q4 2029.

\subsection{10-year plan: The ALMA Wideband Sensitivity Upgrade (WSU)}
\label{sec:wsu}

\begin{figure}
    \centering
    \includegraphics{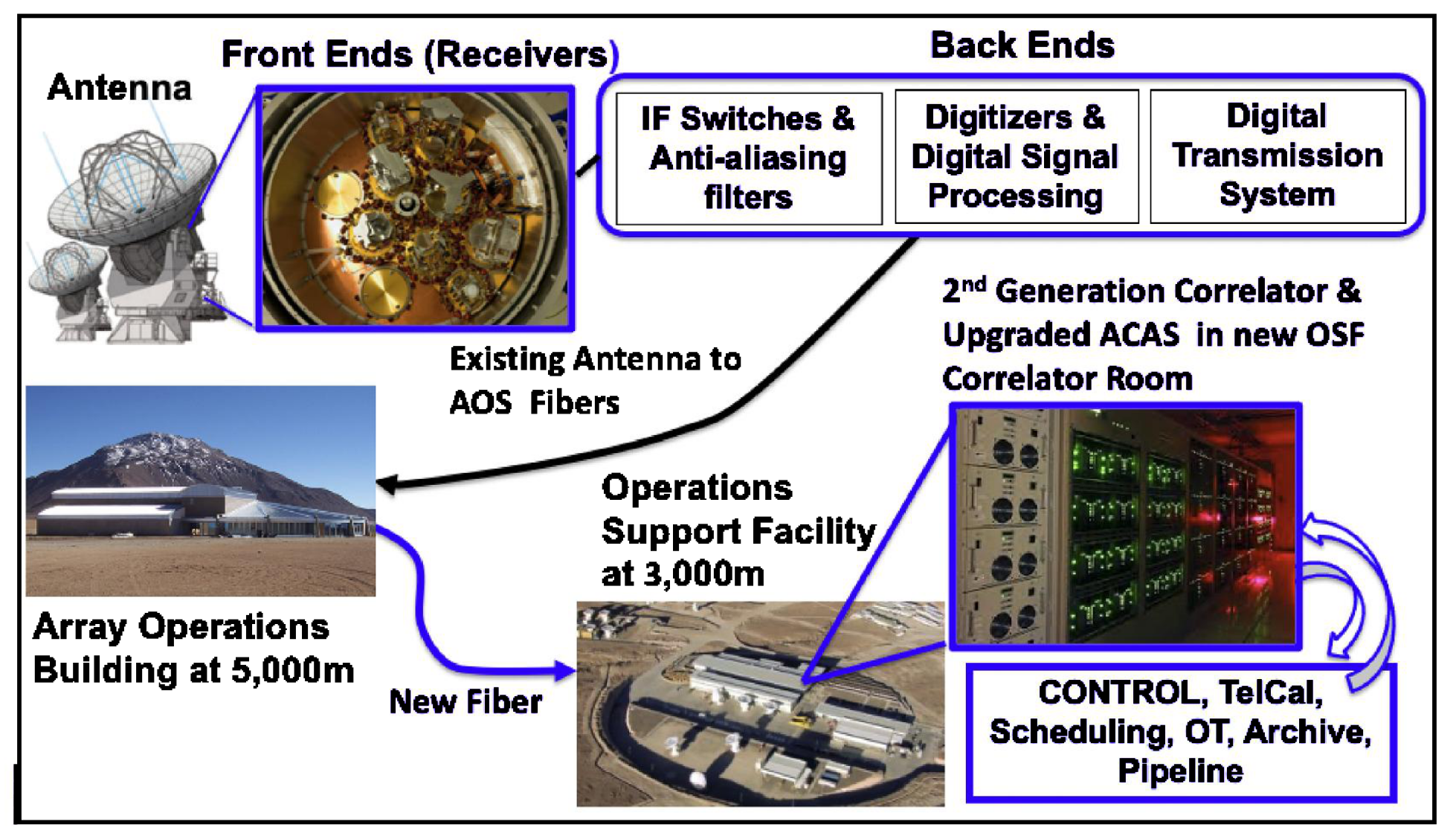}
    \caption{The ALMA Wideband Sensitivity Upgrade (WSU), from \citet{2023pcsf.conf..304C}.  All components marked in blue are either new or being upgraded.}
    \label{fig:alma_wsu}
\end{figure}

The goals of the WSU (Figure~\ref{fig:alma_wsu}) include an increase in instantaneous bandwidth of ALMA by a factor of at least 2, increased digital sensitivity, and improved sensitivity of key receivers.  

The initial stage of the WSU is an upgrade of the signal chain for increased bandwidth and improved digital efficiency.  The large majority of the hardware components in the existing signal chain are being replaced, including the IF Switch, Digitizer, Data Transmission System, and the fibre optic cable linking the Array Operations Building at 5,000m to the Operations Support Facility at 3,000m.  ALMA will also gain a new digital correlator, the Advanced Technology ALMA Correlator (ATAC), and a new Total Power GPU Spectrometer (TPGS). 

The ATAC which will provide at least double the instantaneous bandwidth of ALMA (from the current 8\,GHz per sideband to $\ge16$\,GHz per sideband). The new correlator will not only provide  an increased bandwidth, but will also be able to provide a spectral resolution of 0.1 km/s across the entire spectrum and provide full polarisation products with no loss of bandwidth, neither of which are possible with the current correlator.  The WSU is being implemented so that it could deliver a factor of 4 increased bandwidth, but financial constraints in the project currently limit the correlator to a factor of 2 increased bandwidth.

The ATAC will be located at the Operations Support Facility (OSF), at 3,000 m elevation, rather than the Array Operations Site at 5,000 m where the current Baseline Correlator is located.  This will provide improved power efficiency and ease of support, but requires a new correlator room to be created in the OSF. 

The WSU also includes a major upgrade of both online and offline software systems, including a new calibration and imaging pipeline and data processing system. The signal chain upgrade will include at least one new, high-bandwidth, receiver, in Band 2+3 (67--115\,GHz) with upgrades to the Band 6 (211--275\,GHz) and Band 8 (385--500\,GHz) receivers. Future stages of the WSU will implement receivers with broader bandwidth and higher sensitivity for other bands.

Together, the increased bandwidth with uniform, high spectral resolution and increased sensitivity which the WSU will provide will significantly reduce the observing time required for blind redshift surveys, chemical spectral scans, and deep continuum surveys.

\subsection{10-year+ plan: ALMA2040}
\label{sec:alma2040}

The WSU is only the first component of the ALMA Development Roadmap and with the WSU underway, plans are now starting to be explored for ALMA 2040, the route to maintaining ALMA at the forefront of submillimetre/millimetre astrophysics in the coming decades.  In the ALMA Development Roadmap, three further enhancements (which are not part of the WSU) were identified to enable ALMA to continue to expand the frontiers of submillimetre/millimetre astrophysics:

\textbf{Extended Baselines} A key aim is to increase angular resolution by a factor 2--3, by extending the maximum baselines of ALMA to 30--50\,km.  The key scientific aim is to reach 1\,au linear resolution in the nearest star-forming regions at $\sim 130\,$pc (Ophiuchus, Taurus, Lupus, Corona Australis).  This would allow the terrestrial planet zone of a significant number of circumstellar discs to be resolved, providing important insights into the formation of Earth-like planets.

Extending the ALMA baselines by a factor of 2--3 would require the construction of new antenna pads and associated infrastructure, as well as new antennas, as transporting antennas from the current array to the remote pads of the extended baseline array would not be feasible.

\textbf{Focal Plane Arrays}  Large format heterodyne focal plane arrays (FPAs) could significantly increase ALMA’s wide-field
mapping speed.  This would allow ALMA to survey large regions of molecular clouds, image nearby galaxies, and conduct deep-field cosmological surveys. However, this would be technically complex to achieve: such receivers would likely to occupy a significant fraction of the available focal plane space in the ALMA antennas, and would require the ALMA correlator to have enhanced bandwidth capacity compared to that required for a single pixel receiver.

Despite these challenges, the scientific advantages of FPAs are significant, and a study to build the science case for and assess the technical specifications and of ALMA FPAs  is ongoing (PI: Giorgios Magdis, DTU), with UK participation.

\textbf{Additional 12m Antennas} Adding 12\,m antennas to the current array would increase the sensitivity and/or decrease integration time for a given observation, while improving the image fidelity and quality. The longest-baseline configurations would have significantly improved $uv$ coverage with which to image regions of complex extended emission at high resolution.  ALMA was originally envisaged as having 64 12\,m diameter antennas, which was later de-scoped to 50 antennas for the 12\,m array. However, the correlator requirements for a 64-antenna array were retained, and so 14 antennas could be added to the array relatively straightforwardly.  Additional antennas would bring a number of operational benefits, including allowing the antenna configurations to cover a wider range of resolution, thereby saving time and making better use of the best weather conditions, making more calibrators accessible, and allowing for better phase correction and self-calibration of observations.

The ALMA Development Roadmap also notes the need for a new  \textbf{Large Single-dish Submillimetre Telescope} in order to survey the sky in the submillimeter continuum thousands of times faster than ALMA is capable of, thereby identifying large samples of Galactic and extragalactic sources.  While the ALMA Development Roadmap considers this outside the remit of current ALMA operations, it is important to note that development of new single-dish submillimetre instrumentation such as AtLAST would be of significant benefit to ALMA and other interferometers.

\subsection{International context}
\label{sec:interferometric_context}

% Planned NOEMA upgrades

Currently NOEMA is implementing two upgrades. The first is to provide full polarisation capabilities, allowing its use for the first time for the study of polarised dust emission tracing the structure of the magnetic fields. The second upgrade will allow simultaneous observations in the 3\,mm and 1\,mm bands. This will allow the simultaneous observation of two rotational transitions of simple linear molecules such as CO. 

Plans for upgrading the SMA have recently been published by \citet{2024arXiv240617192G}. These include increasing the instantaneous bandwidth of the receivers and correlator to up to 60\,GHz and improved capabilities for simultaneous observations at two frequency bands. The receivers for the first stage of this upgrade are expected to be in operation in 2026/27 with a 24\,GHz IF bandwidth.  

See Future Synergies section for a discussion of the capabilities of the proposed ngVLA.

\section{A roadmap for submillimetre VLBI: The UK's role in the future of the EHT}
\label{sec:vlbi}

General relativity predicts that black hole images ought to display a bright, thin ring. This ``photon ring'' is produced by photons that explore the strong gravity of the black hole before escaping along geodesic trajectories that experience extreme bending within a few Schwarzschild radii of the event horizon. The shape of the photon ring is largely insensitive to the precise details of the emission from the astronomical source surrounding the black hole and therefore provides a direct probe of the Kerr geometry and its parameters.
 
Measuring these parameters is the first goal of submillimetre VLBI with the EHT. Images of the photon ring have been published for M87* \citep{eht2019} and SgrA* \citep{eht2022}, and gained global media attention. However, even though the hole in the centre of the ring has been resolved, the ring itself has not yet been resolved. Current and future plans for VLBI involve improving the resolution of the existing VLBI experiment to resolve the ring. There are two ways of improving resolution: going to higher frequencies or to longer baselines.
Other improvements include adding in more dishes to fill in the \emph{uv} plane and improve image fidelity.

The UK has access to the Event Horizon Telescope only through its subscription to the JCMT. For this reason alone, it is vital that the UK remains an integral part of the JCMT for the foreseeable future. Presumably in the long term the AtLAST telescope will take over from the JCMT as the foremost single-dish submillimetre telescope in the world, and at that point the UK will need to be a part of AtLAST. Nevertheless, the strategic location of JCMT in the Northern hemisphere means that it will likely still play an important role in EHT observations.

So far, the EHT has only scratched the surface of the science it can do. It has taken images of the two black holes it can see best, at a single wavelength 1.3\,mm (230\,GHz). At the time of writing, the EHT is in the process of publishing maps at a second wavelength of 0.8\,mm (345\,GHz) and there are future plans to go to 0.4\,mm (690\,GHz), with the consequent improvements in angular resolution and potential to see multiple photon rings, as well as the link between the disk and the jet. In addition, there are other AGN jet systems that should become observable at higher resolutions, specifically the base of the jet in each case and its relation to the AGN itself, with the ultimate goal of understanding the launching mechanisms of the jets. Other longer-term goals involve testing GR at ever finer scales.

\subsection{Immediate improvements: $<$ 5 years}
\label{sec:vlbi_immediate}
 
\subsubsection{The move to submillimetre wavelengths }
 
All of the images that have been published so far by the EHTC have been at 230 GHz (1.3\,mm). This provides a resolution of about 25 $\mu$as, sufficient to produce the images of the black holes in M87 and the Milky Way, as well as imaging a number of quasars.  However, higher-frequency (submillimetre) observations will allow more detailed modelling of the structure of the photon ring and the accretion disk in each case, including measuring the ellipticity of the ring to obtain the black hole spin.  The EHT observations at 0.8\,mm (345\,GHz) will have a resolution of about 15 $\mu$as.
%first images should start to appear in the next 1--2 years. These images will have a resolution of about 15 $\mu$as.
 
Work has already begun on a further upgrade to 690 GHz (0.4\,mm) to double the angular resolution again to 5-10 $\mu$as.  The JCMT will play a pivotal role in these observations, because it is one of only a few sites in the EHT array from which such high-frequency observations are possible: Hawai'i, the Atacama Plateau, the South Pole and Greenland (see below).  The UK has a unique opportunity to lead this worldwide effort, through our access to the JCMT and our expertise in high-frequency heterodyne receiver technology.
 
\subsubsection{The Greenland Telescope (GLT) and the South Pole Telescope (SPT)}
 
The Greenland Telescope (GLT) has recently been commissioned and integrated into the EHT. It is being upgraded with new instruments. It is a 12m ALMA antenna, currently located at the Thule Air Base in Greenland, close to sea level.  Hence it is currently restricted to wavelengths $>\,1\,$mm. The GLT is planned to be moved to the Greenland Summit Station at 3,210\,m on a timescale of 1--2 years.  At that point it will be capable of achieving submillimetre wavelengths down to 0.4\,mm (690~GHz).

The South Pole Telescope (SPT) has also been used in the EHT array. Together, the SPT and GLT represent the longest possible north-south baseline available from the ground.
Having these two telescopes in the array also optimises the EHT beam circularity.  Observations have already been made with both the SPT and GLT in the EHT array simultaneously. This is planned to happen on a routine basis in the near future.

All of the above is already in hand and included in the short-term forward plans of the EHT.

\subsection{The medium term: 5-10 years}
\label{sec:vlbi_medium}

Looking further forward there are a number of planned upgrades that are as yet unfunded, and would require significant further investment. The UK Community is already playing a leading role in aspects of these activities, as described below.
%For example, Rob Fender in Oxford is leading an ERC Synergy grant contributing to the development of the Africa Millimetre Telescope (see below). 

Most importantly, the UK needs to remain a key part of the JCMT in order to retain a role in the EHT. Further developments that would be critical include heterodyne instrument upgrades to JCMT and upgrades to ALMA.

\subsubsection{New EHT sites}
 
As well as going to higher frequencies, the EHT plans to bring in more dishes to fill in the \emph{uv} plane and improve image fidelity \citep{raymond2021}. There are a number of initiatives currently underway. One of these is the EU/Africa consortium, which plans to add more dishes in Europe and Africa into the array, an example of which is discussed below. Various other dishes around the globe have also expressed interest in joining in. The planning for these are well underway and should be starting to come on-line in around 5--10 years.

\subsubsection{The Africa Millimetre Telescope: UK contributions to the ngEHT}
\label{sec:amt}

The Africa Millimetre Telescope (AMT; \citealt{backes2016}) will have a 15\,m dish, the same size as the JCMT, and will be located in the Southern hemisphere in Namibia, at a similar latitude to ALMA.  
One of the key objectives of the AMT is to fill in the baseline gap for the EHT and produce much sharper and `movie-like' imaging of black hole event horizons.  Other science goals include a transients rapid response and monitoring programme, which will be initially led by Oxford.

Construction of the AMT is currently partially funded by the `BlackHolistic' ERC Synergy Grant, one of the P.I.s of which is based in the UK (R. Fender, Oxford).  Oxford has also participated in potentially providing some of the quantum heterodyne detectors for the AMT.  NOVA from the Netherlands has also just approved funding for the cartridge building for the AMT.  The expected instruments and timeline for deployment are as follows:
\begin{itemize}
    \item First generation receivers: 230\,GHz (1.3\,mm, EHT), 86\,GHz (3\,mm, GMVA and ngVLA)
    \item Second generation receivers: 345\,GHz (0.8\,mm, EHT), 43\,GHz (7\,mm, EVN)
    \item Third generation receivers: 22\,GHz (14\,mm, EVN),  12.5\,GHz (24\,mm, SKA/Geodesy)
\end{itemize}
Site and dish design decisions for the AMT will be made in the next 6 months.  First light for the AMT is expected in 2028.

\subsection{The long term: 10+ years}
\label{sec:vlbi_long}

It is difficult to predict exactly how the EHT will evolve on a timescale of 10--20
years. It will depend on many variables, including technological advances, 
the funding landscape for science globally, and the feasibility of carrying out 
some of the ideas that astronomers might have. However, there are two sets of plans currently being worked on. Both are in the very earliest stages of development.

\subsubsection{The Next Generation Event Horizon Telescope (ngEHT)}

The ngEHT will use state-of-the-art technology to modernize the existing instrumentation and develop new capabilities while expanding the geographical footprint of the array with roughly 10 new dishes. With this transformative enhancement, the ngEHT will use the technique of very long baseline interferometry (VLBI) to unite an array of dishes spread across numerous continents into a single virtual  telescope. Taking advantage of an additional observing frequency and modern high-speed data transfer protocols, data from this array will be used to form images through advanced data processing algorithms.
With high-resolution black hole images, the ngEHT will detail the size, shape, and variability of the accretion disk.

\subsubsection{The Black Hole Explorer Telescope (BHEX)}
 
The long-term goal for the EHT is space VLBI. Plans are already started for extending the EHT into space in the form of the Black Hole Explorer Telescope (BHEX)\footnote{\url{https://www.blackholeexplorer.org}}. This is an orbiting, multi-band, millimetre radio-telescope, in hybrid combination with millimetre terrestrial radio-telescopes. It is designed to discover and measure the thin photon ring around the supermassive black holes M87* and Sgr A*.  The proposed science instrument for BHEX is a dual-band coherent heterodyne receiver system for 80--320\,GHz, coupled to a 3.5\,m antenna. The BHEX receiver will observe the 80--106\,GHz and 240--320\,GHz bands simultaneously in dual polarization.  In preparation for the BHEX, work has already begun to explore various aspects of the photon ring, and tracking, through visual simulations, photons as they course along geodesics. Ultimately, the aim of these visualizations is to advance the foundational aims of the BHEX instrument, and through this experiment to articulate spacetime geometry via the photon ring.
 
This will ultimately test General Relativity as a whole, and will specifically test the Kerr metric for a rotating black hole.
The BHEX will be proposed as a NASA Small Explorers Mission in 2025.  If selected, it is expected to fly in the early 2030s.  An alternative mission concept, the Event Horizon Imager \citep{kudriashov2021}, is also under development.

\section{Computing resources for submillimetre and millimetre astronomy}
\label{sec:computing}
% Big picture: 

At submillimetre wavelengths, atmospheric and instrumental noise prevents us from simply ``taking pictures'' of the sky. 
Extensive computational processing is required to convert the timeseries and interferometric visibilities measured by the telescopes' detectors into images, spectra or data cubes.
As a result, submillimetre astronomy relies heavily on access to appropriate software and hardware to transform the data.
The UK has for a long time led the development of software for submillimetre astronomy, such as the {\sc starlink} package \citep{starlink} which has long facilitated exploitation of JCMT data.
However, continued maintenance of these packages and further development is essential, 
Not only do new facilities require new computing facilities to use them at all, but enhanced software or hardware can enable new insights from existing facilities, by improving data quality with more expressive or expensive processing, or by enabling previously impossible analyses.
Compared to the large cost of telescopes and instrumentation, computing is a small investment but essential to maximising the scientific return on facilities.
Without appropriate computing resources, data taken by instruments would languish unprocessed and uninterpreted; conversely, relatively small investments in further computing resources can significantly improve the scientific return of a facility.

\subsection{Software requirements}\label{sec:computing_software}

% Two themes: 1) continued maintenance of existing software and 2) required improvements in software.

The UK has led development of a range of software integral to submillimetre astronomy,  whose continued maintenance has been identified as of critical importance by the community (see Section~\ref{sec:consultation}).
These existing software packages include data processing and analysis tools for single-dish (e.g. {\sc starlink} contains {\sc smurf, orac-dr, splat} and more) and interferometer (e.g. CASA) data, proposal preparation tools (like the original JCMT {\sc observing tool} and more recently the ALMA {\sc observing tool}), as well as more general tools (e.g. {\sc topcat}, {\sc gaia}, {\sc kappa}) and associated low-level libraries, e.g. {\sc AST} and {\sc NDF} in the {\sc starlink} context. 
These tools form the backbone of proposing for and processing data for submillimetre astronomy.

Just as with physical infrastructure, digital infrastructure and software requires ongoing maintenance, otherwise problems remain unfixed and as hardware advances, incompatibilities creep in, eventually rendering the software useless and the original investment in its creation meaningless.
A major component of UK leadership in submillimetre software has been the existence of dedicated support through the JCMT, rather than relying entirely on community efforts. 
Continued support in terms of personnel and funding for software will be essential going forward to ensuring the existing software stack continues to function effectively.

While the existing software stack is excellent, further improvements will be needed in future to maximise scientific output.
This is driven from five directions: 1) the need to minimise the carbon footprint of astronomical computing; 2) new software and algorithmic developments (e.g. applications of machine learning, optimised software stacks, improved data-processing algorithms); 3) new computing hardware developments (e.g. GPUs); 4) the increasing complexity of observatory specifications, astronomical datasets, models and analyses; and 5) the needs of new facilities and instruments resulting in more complex datasets.
Indeed, many of these directions are related: one way to minimise carbon footprints is the widespread adoption of GPU computing, since it is much more energy efficient for a subset of problems, and GPUs also alleviate issues of data complexity as instruments grow.

Submillimetre astronomy is well-placed to lead the way in exploiting these advances for the wider astronomical community. 
Many algorithms in use have significant potential for parallel and GPU computing, for example interferometric image reconstruction \citep[e.g.][]{Baron2010}.
Since nearly all submillimetre data relies on reconstruction algorithms, new algorithms can easily be retrofitted and compared \citep[e.g.][]{Taniguchi2021, Terris2022}.
Thanks to the growth of machine learning, a large stack of software tools have developed for accelerated computing in high-level languages (e.g. JAX and Pytorch for python).
Exploiting these tools can facilitate a wider exploration of accelerated algorithms for submillimetre astronomical data, as well as easy comparison of different algorithms.
This has the potential to dramatically cut processing time and cost for existing approaches (e.g. constructing images from SCUBA-2 observations often takes hours; this could be reduced to minutes) as well as allowing the use of more expensive algorithms; for example, regularised maximum-likelihood approaches were used to reconstruct the EHT images \citep{Chael2019}, but these have yet to gain traction in the wider interferometry community due to the computing cost -- wider adoption of GPU computing would make this feasible for interferometry, improving the sensitivity and resolution of reconstructed images.
These approaches could also be applied to single-dish bolometer camera observations, potentially resulting in more stable image reconstructions with fewer artefacts, and the scope to exploit the multi-wavelength capabilities of cameras like SCUBA-2 and prior information such as large-scales from Planck.

As well as processing the observational data itself, the results must be interpreted through the lens of numerical models to extract scientific insights.
Models of continuum emission are well-developed, with a wide variety of codes for e.g. dust, although there are fewer tools which support the intrinsically more complex interpretation and modelling of the observations of the emission from molecular lines.

New developments are required to enable the interpretation of line emission, especially from spectral surveys, and  to streamline the process of comparing models to data both in the continuum and spectral lines.
The required improvements are primarily in two regimes.
Firstly, there is a constant stream of algorithm development in Data Science to optimise for different factors - e.g. best credible-interval coverage, fastest convergence, fewest model evaluations.
New software tools that make these advances easy for astronomers to use will dramatically improve and accelerate many analyses.
This must also be accompanied by high-level documentation and learning resources.
This combination would mirror the way that {\it emcee} \citep{emcee} so successfully brought MCMC and Bayesian approaches into the mainstream.
However, many of the new algorithms work best when the gradients of the model are available, while many astrophysical models are ``black boxes'' in that they only take input and provide an output. 
Machine learning has driven huge advances in differentiable computing, and the development of differentiable models is the second development needed to drive forward modelling software.

\subsection{Hardware requirements}\label{sec:computing_hardware}

Computing hardware needs are growing across astronomy, and submillimetre astronomy is no exception. 
As instruments become larger and more powerful, greater hardware capacity is required to process the data, and this is often redoubled by increasing complexity of analyses.
The need for increased computing power was highlighted in the community survey, suggesting several different approaches to achieving the community's needs.

AtLAST instrumentation will produce data orders of magnitude larger than current data (e.g. SCUBA-2), thanks to the vastly larger number of detectors in the arrays. 
Realistically, this places compute needs beyond the scope of individual grants or institutions, as has already been recognised by future facilities at other wavelengths including the SKA and ELT.
These observatories are moving to use {\it Science Platforms} where the user never downloads data, but processes it in the cloud by moving their code to the data, e.g., CANFAR\footnote{\url{https://www.canfar.net/en/}}.
Our community survey (Section~\ref{sec:consultation}) highlighted both the investment in cloud computing, potentially on a similar Science Platform model, and the creation of support centres following the Alma Regional Centre model as important for future data processing.
Regional centres can combine both expertise and compute to make it easy for downstream users to process data remotely, even if they are unfamiliar with the telescope or instrument.

However, given the increasing size of datasets across all wavelengths, and the increasing diversity of datasets in analyses, siloing data and resources would re-introduce the need to download data.
Federated resources are required, not just across sites but across projects and observatories; modern analyses feature data from, e.g., ALMA, JWST and VLA at the same time, and hence if in future if one needs to analyse SKA, AtLAST and Roman data at once, we need a platform that can access all of them.
Hence, while significant additional compute resources will be required to process data from future submillimetre facilities, these resources much be coordinated across fields, for example building on efforts to provide the compute power needed for future SKA data (e.g. IRIS\footnote{\url{https://www.iris.ac.uk}}).
Such resources will be essential both to the successful processing of data from future observatories and maximising their scientific return by enabling comprehensive analyses of the data.

\section{Synergies with future ground- and space-based telescopes}
\label{sec:synergies}

We here highlight some key synergies between current and future submillimetre/millimetre instrumentation, and forthcoming ground- and space-based telescopes operating at other wavelengths.  Many other possible synergies exist, including with the UK-led exoplanetary atmospheres mission \textit{Ariel}\footnote{\url{https://arielmission.space}} (see Section~\ref{sec:protoplanetary_discs}), future X-ray missions such as as ATHENA\footnote{\url{https://www.the-athena-x-ray-observatory.eu/en}}, the Rubin Observatory's transient searches\footnote{\url{https://rubinobservatory.org}}, and the forthcoming near-infrared Roman Observatory\footnote{\url{https://science.nasa.gov/mission/roman-space-telescope/}}.

\subsection{The Square Kilometre Array (SKA)}

The Square Kilometre Array (SKA)\footnote{\url{https://www.skao.int/}} will be two complementary radio telescope arrays, operating in Australia and South Africa.  The low-frequency array, SKA-Low, will consist of 131,072 log-periodic dipole antennas, operating in a frequency range 50--350 MHz.  It will be located in the Murchison Radio-astronomy Observatory, Western Australia. 
%Around 50\% of the stations will be located within a 1 km diameter core, with the remaining stations organised in clusters of 6 stations on three modified spiral arms. The maximum baseline length will be around 70 km.
The higher-frequency array, SKA-Mid will consist of 133 15m SKA dishes and 64 13.5m Meerkat dishes at the Karoo site in South Africa, operating in the frequency range 350 MHz -- 24 GHz.
%The core will be composed of around 50\% of the dishes, randomly distributed within 2 km. There are 3 logarithmic spiral arms with a maximum baseline extending out to 150 km.
Construction of the SKA is well underway, with Science Verification expected to begin in 2026, and shared-risk observations expected to begin the following year.  Full operation will begin in the late 2020s.

Multiple synergies exist for observations with the SKA (centimetre and metre wavelengths) and AtLAST and ALMA (millimetre and submillimetre wavelengths).   Many objects emit a broad spectrum due to different physical processes, and to fully understand their properties requires observations at multiple wavelengths.  We give a few examples here.

Understanding the star-formation history of the universe, where continuum observations of both thermal (primarily AtLAST/ALMA) and non-thermal (primarily SKA) processes are needed to directly measure star-formation rates out to redshifts of $z\sim10$.   Understanding the evolution of the gas content of galaxies across cosmic time requires information on the content and kinematics of all phases of the ISM (ionised, atomic, molecular), so resolved observations of galaxies of the (redshifted) lines of e.g., HI (SKA), CO/HCN/HCO+ (ALMA), and [CII]/[OI] (AtLAST) are needed.

In our own galaxy understanding both star formation and the cycle of material within the multi-phase ISM requires observations of e.g., HI, OH, Radio Recombination lines (RRLs), H$_{2}$CO and NH$_{3}$ (SKA); CO, HCO$^{+}$, N$_{2}$H$^{+}$ and H$_{2}$CO (ALMA), and [CI], N[II] and H$_{2}$D$^{+}$ (AtLAST), as well as continuum observations of thermal dust emission (ALMA/AtLAST), and free-free and non-thermal emission (SKA).

Studies of planet-forming disks around young stars requires high-resolution observations of thermal and non-thermal emission to track the growth of dust grains and the evolution of disk structure.

% \begin{figure} [h!]
%     \centering
%     \includegraphics[width=1\textwidth]{SKA_night.jpg} \\
%     \includegraphics[width=1\textwidth]{KSG1_V2_antennas_galevo.jpg}
%     \caption{\textit{Upper:} A schematic of the SKA array. On the left is the MeerKAT array which is in South Africa. On the right is an artist's impression of the SKA-Low array in Australia. \textit{Lower:} An artist's impression of the  ngVLA superimposed an image illustrating galaxy evolution, one of the key science drivers of the ngVLA, along with expolanets, black holes and gravity.}
%     \label{fig:ska_ngvla}
% \end{figure}

\begin{figure} [h!]
    \centering
    \includegraphics[width=1\textwidth]{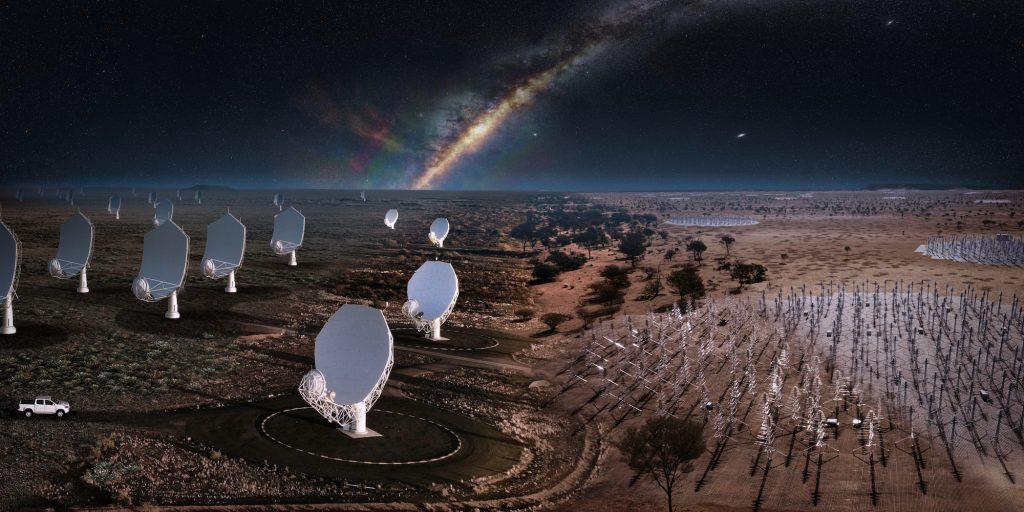} \\
    \includegraphics[width=0.42\textwidth]{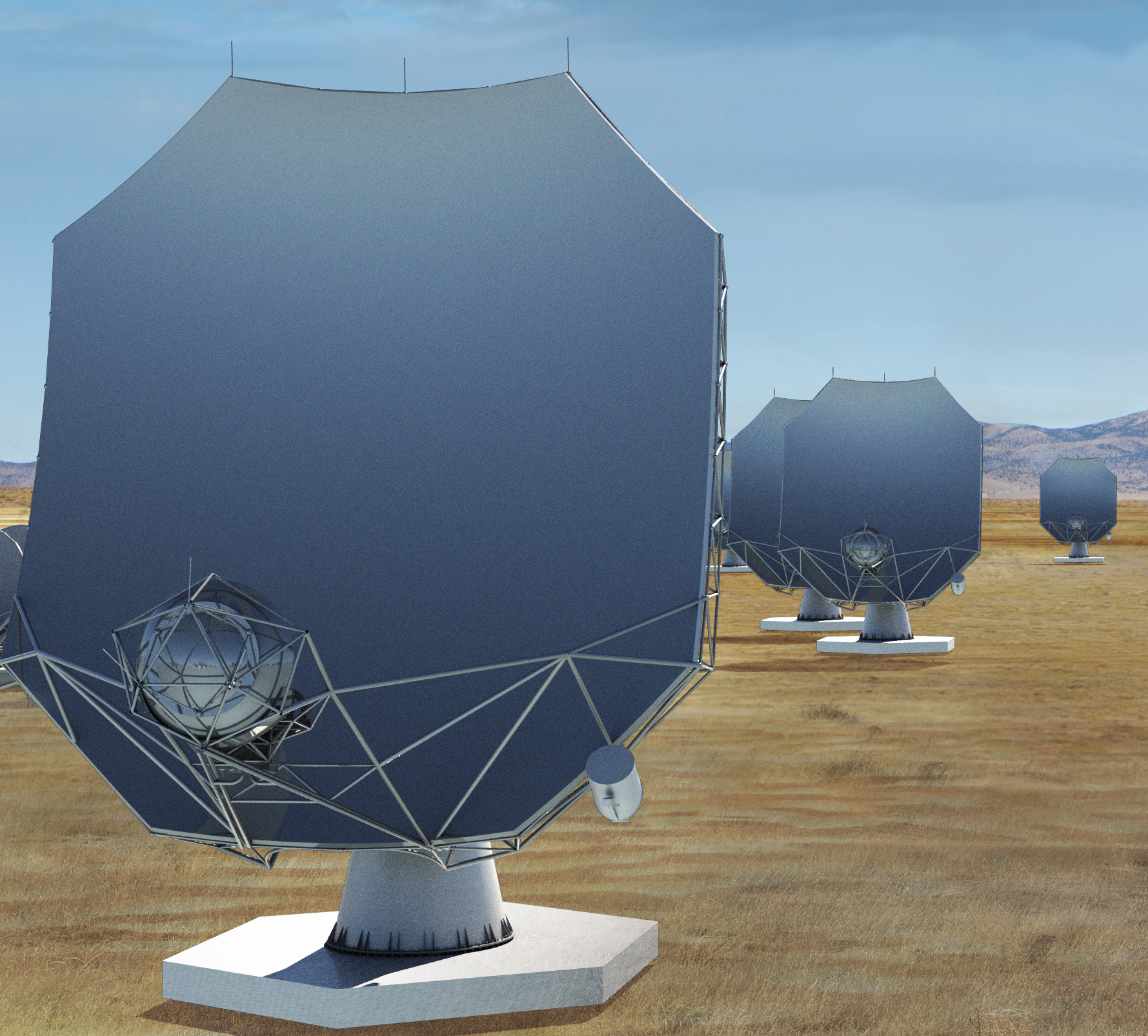}
    \includegraphics[width=0.52\textwidth]{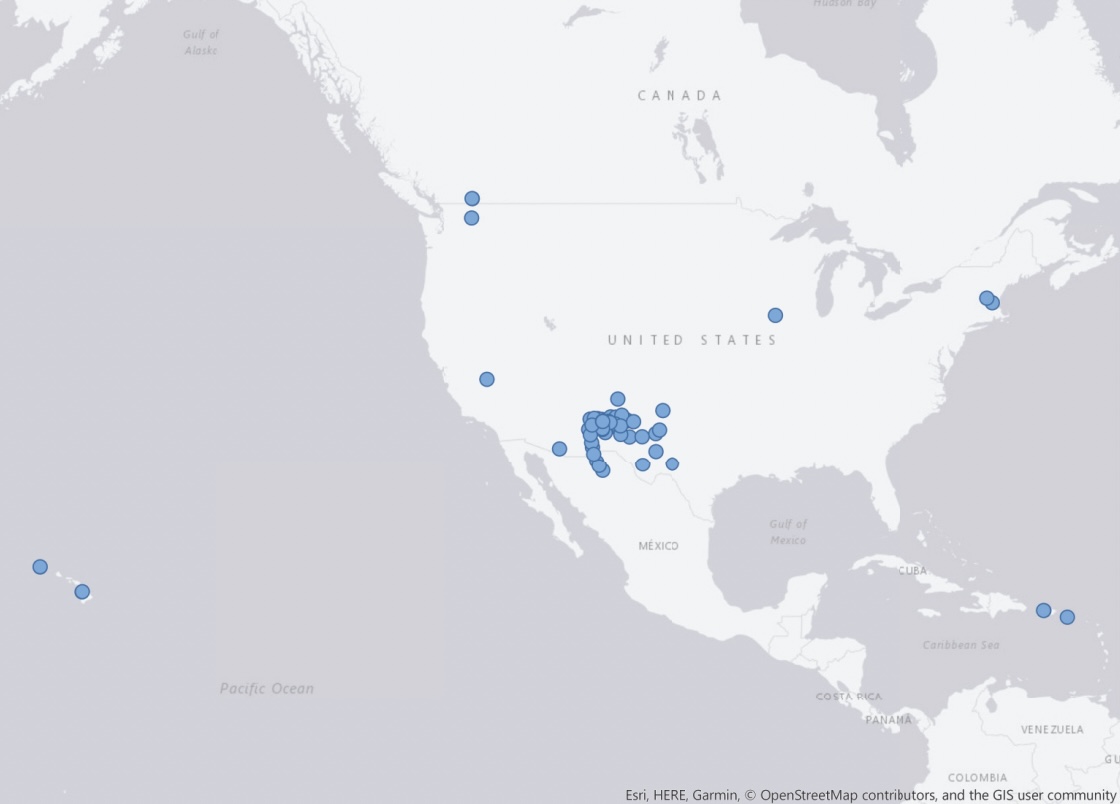}
    \caption{\textit{Upper:} A schematic of the SKA array. On the left is the MeerKAT array which is in South Africa. On the right is an artist's impression of the SKA-Low array in Australia. \textit{Lower:} Left is a design of the 18\,m dish that will be one of 244 antennas of this size in the ngVLA. On the right is the configuration of the antennas. The short baseline array is the cluster shown in New Mexico while the long baseline array is made up of the locations scattered around the edges of the map. Each dot will contain multiple antennas.}
    \label{fig:ska_ngvla}
\end{figure}

\subsection{The ngVLA}

The next-generation Very Large Array (ngVLA)\footnote{\url{https://ngvla.nrao.edu}} is a planned interferometric array operating at frequencies of 1.2\,GHz (21\,cm) to 116\,GHz (2.6\,mm). This new radio interferometer aims to provide ultra-sensitive imaging of thermal line and continuum emission down to milliarcsecond resolution, broadband continuum imaging, and polarimetry of non-thermal emission. 

The ngVLA will comprise an array of 244 antennas each 18m in diameter, supplemented with a short baseline array of 19 antennas of 6m in diameter.  The ngVLA will operate in frequency ranges 1.2--50.5\,GHz and 70--116\,GHz, and will image on angular scales down to a milliarcsecond.  The ngVLA core will be located in New Mexico, USA, with additional mid-baseline stations spread over the USA, Mexico and Canada.  The longest baselines, reaching across North America and Hawaii, will deliver 0.1 milliarcsecond resolution and enable microarcsecond precision astrometry.

The ngVLA is currently undergoing design study, with construction planned to begin in the late 2020s and operations to begin in the late 2030s.  If built, it will be highly complementary to both ALMA and AtLAST.  The ngVLA would be significantly more sensitive than ALMA at wavelengths $\geq 2.6$\,mm.  However, unlike the ngVLA, ALMA is able to operate in the submillimetre regime.  This will continue to distinguish ALMA from the ngVLA, and will make ALMA a unique facility into the late 2030s and beyond. 

\subsection{The Deep Synoptic Array}

The Deep Synoptic Array-110 (DSA-110)\footnote{\url{https://www.deepsynoptic.org/overview}} is a radio interferometer purpose-built for fast radio burst (FRB) detection and direct localization. The array is currently under construction at the Owens Valley Radio Observatory (OVRO), and a 63-antenna deployment is being commissioned. When construction is completed, 110 4.65-m dishes will continuously survey for FRBs at frequencies between 1.28--1.53 GHz.

The DSA-2000 is a proposed radio survey telescope intended to be a multi-messenger discovery engine. The array will consist of 2,000 5m dishes instantaneously covering the 0.7--2\,GHz frequency range.

A sufficiently flexible submillimetre telescope, such as AtLAST, would be well-placed to follow up and characterise transient events detected by DSA-110 and DSA-2000.

\subsection{A Future NASA FIR Probe Mission}
\label{sec:nasa_fir}

Following the recommendation of the 2020 US Decadal Survey\footnote{\url{https://nap.nationalacademies.org/catalog/26141/pathways-to-discovery-in-astronomy-and-astrophysics-for-the-2020s}}, NASA is expected to launch a far infrared (FIR) astrophysics Probe mission, with \$1B budget, in either the early 2030s or 2040s (with an X-Ray mission in the other slot)\footnote{\url{https://explorers.larc.nasa.gov/2023APPROBE/}}.
Three FIR mission concepts are being developed by US-led teams for the 2030s launch:
\begin{itemize}
    \item \textbf{PRIMA (PRobe far-Infrared Mission for Astrophysics)\footnote{\url{https://prima.ipac.caltech.edu}}} is a 1.8-m telescope, cryogenically cooled to 4.5\,K.  Key instruments: PRIMAger and FIRESS.  PRIMAger is an imager with KID arrays operating at 100\,mK, performing hyperspectral imaging  between 24--84\,$\mu$m with $R=10$, and polarimetric imaging in four bands from 80--261\,$\mu$m.  FIRESS is a multimode survey spectrometer operating at 1\,K in the 24--235\,$\mu$m spectral range at $R>85$, with a more than $10\times$ point source sensitivity improvement over previous missions, and a 1\,000--100\,000$\times$ improvement in spatial-spectral mapping speed.
    \item \textbf{FIRSST (Far-InfraRed Spectroscopy Space Telescope)} is a 2-m class telescope with 30--600\,$\mu$m wavelength coverage, cryogenically cooled to $<8$\,K.  Key instruments: DDSI, a multi-mode, direct detection FIR spectrometer ($\sim$30--300\,$\mu$m), using 100-mK KIDs, and HSI, three dual-polarisation long-wavelength heterodyne array receivers ($\sim$200--600\,$\mu$m) which would be the first heterodyne arrays in space. FIRSST has a large instantaneous field of view of more than 2$\pi$, which would give it particular strength in time-domain astronomy.
    \item \textbf{SALTUS (Single Aperture Large Telescope for Universe Studies)} is a planned telescope with a deployable 14-m primary mirror with an inflatable parabolic membrane, with a sunshield that will radiatively cool the mirror to 45\,K. Key instruments: SAFARI-Lite, providing  $R\sim$300-resolution spectroscopy over 34--230\,$\mu$m, also using 100-mK KIDs, and the High-Resolution Receiver (HiRX), performing high resolving power ($R\sim 10^{5} - 10^{7}$) heterodyne spectroscopy in four frequency bands ranging from 455\,GHz (660\,$\mu$m) to 4.7\,THz (56\,$\mu$m).
\end{itemize}

\begin{figure} [h!]
    \centering
    \includegraphics[width=0.29\textwidth]{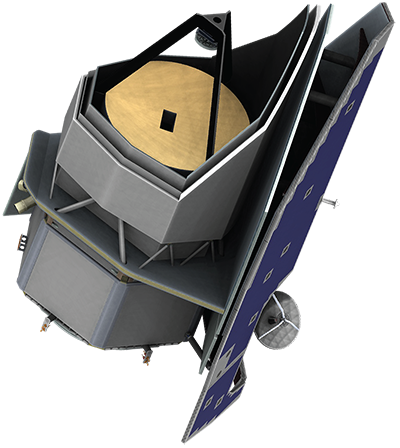}
    \includegraphics[width=0.32\textwidth]{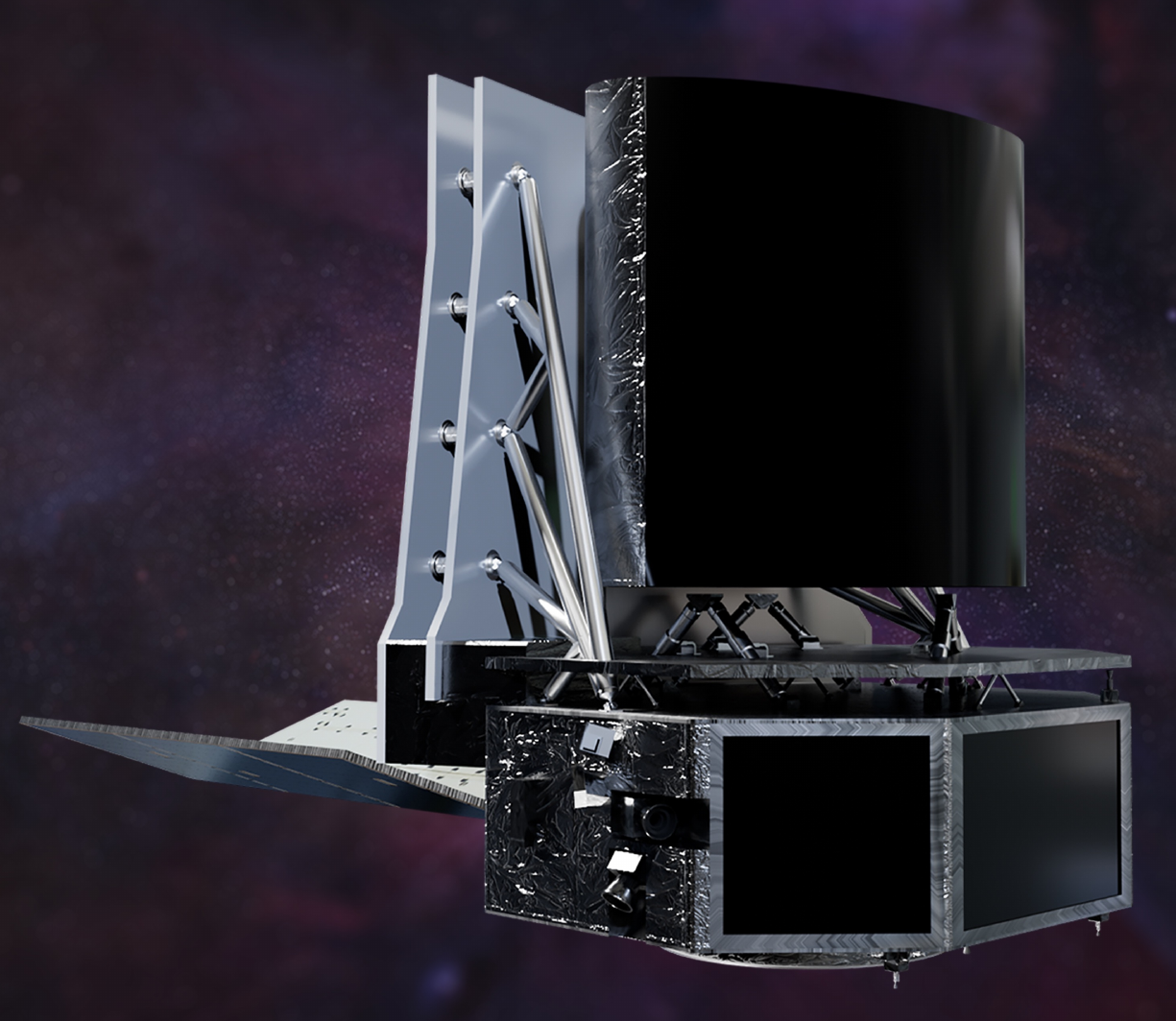}
    \includegraphics[width=0.31\textwidth]{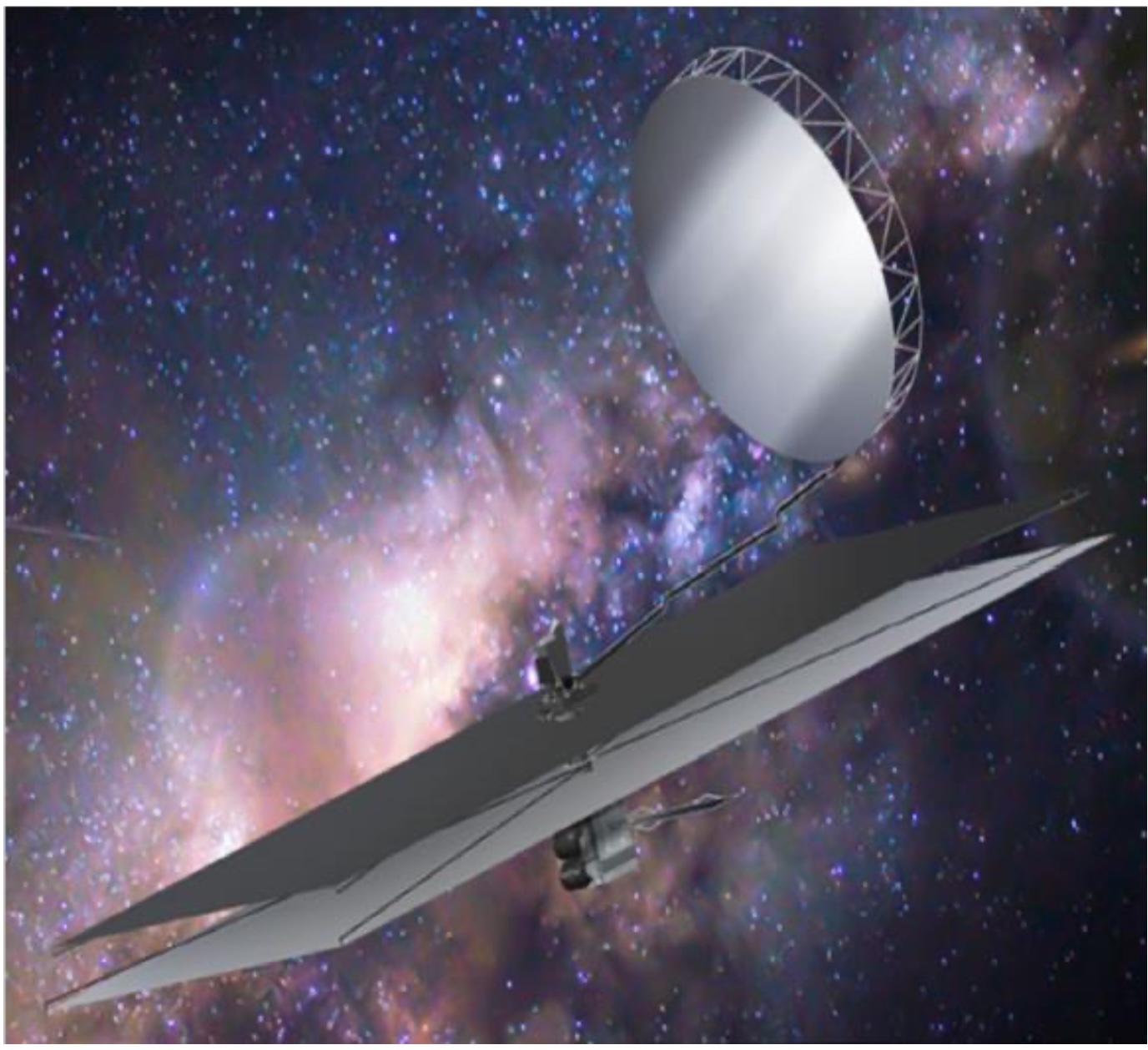}
    \caption{\textit{Left:} Image of PRIMA from CalTech (see \url{https://prima.ipac.caltech.edu/}) \textit{Centre:} Image of FIRSST from a presentation given by the PI Asantha Cooray at the NASA Infrared Science and Technology Integration Group (IRSTIG)
%    \footnote{
    (see \url{https://cor.gsfc.nasa.gov/stigs/irstig/events/webinars/04-Dec-2023/04-Dec-2023.php}).
    %} 
    \textit{Right:} Image of SALTUS from Figure 1 of \citet{chen2024}.}
    \label{fig:nasa_missions}
\end{figure}

All of these telescopes would be in a Sun-Earth L2 orbit, and would have a lifetime of at least 5 years.  Proposals for the three missions were submitted to NASA in November 2023. Announcement of the selection for competitive Phase A studies is expected in Autumn 2024, and it is envisaged that at least one FIR mission will go forward to Phase~A. Final selection of the mission to fly will be in late 2025. UK astronomers have involvement in all three proposals, including filter provision and Ground Segment participation for any selected mission and other potential elements of hardware provision depending on the selected candidate. UK involvement, including filter technology development, instrument design participation, and Ground Segment planning, is supported by the UK Space Agency (UKSA; PI: S. Oliver, Sussex).  %Particularly,...

The main science drivers for these missions are the evolution of galactic ecosystems through cosmic time, the build-up of dust and metals in the universe, and planet formation and the evolution of planetary atmospheres, as described by the PRIMA Science Book \citep{moullet2023} and the SALTUS Science Overview \citep{chen2024}.  These science goals are thus very congruent with the big questions that submillimetre/millimetre astronomers wish to address over the coming decades.  A FIR mission will provide access to tracers inaccessible from the ground, such as HD, which can provide direct model-independent measurements of molecular hydrogen mass in planet-forming discs, solid-state FIR features of dust and ice which can probe the solid material in PPDs, and redshifted atomic and ionic fine structure lines from galaxies at redshifts out to $\sim$3, enabling characterisation of the ISM conditions in which most of the stars in the current
universe were formed.  

Whichever of these missions is selected for flight in either the 2030s or the 2040s, it will be highly complementary to ground-based submillimetre/millimetre astronomy. The FIR regime does not trace the coldest material in the interstellar medium, and space-based telescopes cannot match the resolution of either AtLAST or ALMA.  A synergy between AtLAST, an upgraded ALMA and a NASA FIR mission flying in the 2030s would thus be an invaluable tool for understanding the cold universe through cosmic time.

\subsection{Balloon-borne FIR-submillimetre astronomy}

Earth-based observations from aircraft (reaching $\sim$20 km altitude) and stratospheric balloons ($\sim$40 km altitude) enable the effects of the atmosphere to be mitigated to a significant extent without going into space. Although this can never match the large apertures achievable on the ground or the cold apertures and zero atmospheric contamination achievable in space, it can nevertheless enable many science investigations. With the termination of operations of SOFIA, which was the world’s only airborne observatory, balloon experiments are now the only available intermediate between ground and space. 

For many years, NASA has operated a substantial and systematic balloon programme, with the objectives of carrying out novel observational science and promoting technology development and prototyping, with balloon projects often acting as scientific and technical pathfinders for space missions. Another important objective is facilitating the training and personal development of early career researchers. Balloons are launched from Antarctica and the continental US, with Antarctic flights capable of long duration (up to $\sim$60 days). NASA’s balloon programme has included several successful and high-profile FIR-millimetre projects including the BOOMERanG \citep{McTavish2006}, and EBEX \citep{Reichborn-Kjennerud2010} CMB experiments, and astrophysics experiments such as BLAST \citep{Truch2009} and GUSTO \citep{Walker2022}.  These are mainly NASA-led projects with some UK involvement (although not substantially funded by STFC), usually through provision of filters and optical components by the Cardiff Astronomy Instrumentation Group.

\begin{itemize}
    \item BLAST-TNG (Balloon-borne Large Aperture Submillimeter Telescope – The Next Generation; \citealt{Coppi2020}): a submillimetre mapping polarimetry experiment for galactic magnetic fields and CMB foreground characterisation. Its flight in 2020, although curtailed by a technical problem with the platform, provided proof-of-concept data. A follow-up experiment, the BLAST Observatory, is proposed. 
    
    \item GUSTO (Gal/Xgal U/LDB Spectroscopic/Stratospheric THz Observatory) a high-frequency heterodyne spectroscopy experiment to study the Galactic Plane and the Magellanic Clouds via mapping of FIR fine structure lines from ionised nitrogen and carbon, and atomic oxygen.  It had a successful long-duration flight in 2023. Instrument technology developed for GUSTO is relevant for potential future space missions including the SALTUS FIR Probe candidate mission. There is no UK participation in GUSTO.

    \item ASTHROS (Astrophysics Stratospheric Telescope for High Spectral Resolution Observations at Observations at Submillimeter-wavelengths; \citealt{Pineda2022}): another high-frequency heterodyne experiment, to be launched in 2025. It will carry out large-scale mapping of the fine structure line of ionised nitrogen, and use HD rotational lines for model-independent molecular hydrogen mass measurements in protoplanetary discs.  Its science and technology are also very relevant to a future FIR Probe. There is no UK involvement.
    
    \item PIPER (Primordial Inflation Polarisation Explorer; \citealt{Pawlyk2018}): a CMB polarisation experiment to search for B-mode signatures of cosmic inflation. The UK is involved via the Cardiff AIG.

    \item TIM (Terahertz Intensity Mapper; \citealt{Marrone2022}): a line intensity mapping instrument to track star-formation activity over cosmic time using the ionised carbon 158--$\mu$m line. There is no UK involvement. 
\end{itemize}

There have been some European (usually French) led balloon programmes. In the past these have included Archeops \citep{Benoit2004}, a CMP polarisation experiment which, like BOOMERanG, was a pathfinder for \textit{Planck}, and PILOT \citep{Bernard2019}, a FIR polarimetric mapper for study of Galactic molecular clouds. The UK was involved in Archeops and PILOT via the Cardiff AIG.  BISOU \citep{Maffei2022} is a proposed balloon project to measure CMB spectral distortions, with UK involvement via the University of Manchester. Similar techniques are also being adopted for Earth observation -- for example, OSAS-B is a German balloon experiment which will make 4.7--THz observations of the neutral oxygen fine structure line in the Earth’s mesosphere and thermosphere.

UK participation in US and European balloon experiments has strategic benefits: it offers opportunities to collaborate on front-rank science and technology development, and it facilitates working engagement with the FIR-millimetre communities on both sides of the Atlantic – important in paving the way for future collaborations on major ground-based and space-borne facilities.

\section{SWOT analysis for submillimetre and millimetre astronomy}
\label{sec:swot}

We next discuss the strengths and weaknesses of, opportunities for and threats to UK submillimetre/millimetre astronomy.  This discussion is summarised in the boxes on p.~\pageref{box:swot}.

\subsection{Strengths}

The UK has an outstanding track record in submillimetre and millimetre astronomy, and is at the forefront of global partnerships in the field, as evidenced by leadership of JCMT and ALMA Large Programs (cf. Section~\ref{sec:science}).  The UK also has extensive experience of, a track record of leadership in, and current capability in all aspects of direct detection and heterodyne instrumentation for space-borne and Earth-based facilities, including hardware elements (detectors, filters, optics, cryogenics, readout electronics, control software etc.), integration and test facilities, operations and data processing (cf. Section~\ref{sec:uk_instrumentation}).  The UK is thus a credible international partner for major future submillimetre and millimetre instrumentation projects.

The UK is home to several internationally leading institutes capable of and with appetite for new technology development and deployment.  These include national institutes with permanent specialised engineering and project management staff to undertake build of new telescope and space instrumentation.

\subsection{Weaknesses}

The current funding landscape for both astronomy and astronomical instrumentation is challenging, with many calls on finite resources.

There is currently a lack of UK investment in, and consequently access to, submillimetre/millimetre single-dish facilities.  Access to the JCMT is currently restricted to astronomers at universities that are members of the UK JCMT Consortium.  Hence, the forthcoming JCMT access funding bid (see Section~\ref{sec:single-dish_funding}) proposes moving from a university-funded model to an STFC-funded model, thereby widening JCMT access to all UK-based astronomers.

In submillimetre/millimetre instrumentation, the UK has a low level of investment in astronomical superconducting detector technologies compared to the US and European comparator nations (e.g., Dutch investment in SRON).  There is also a lack of connection with the extensive superconducting/semiconductor materials and device fabrication facilities within the UK, most of which are EPSRC-funded. This missing link in the design-to-product (receiver) pipeline results in reliance on foundries in the Asia, the US or Europe with consequent loss of control of development directions, time scales and costs.  A more general concern is the risk that lack of continuity of funding might result in loss of technical expertise in the field.

It is also increasingly uncommon for PhD students to have the opportunity to travel to telescopes to perform observations.  This is creating a significant skills gap for early-career researchers.  Under the proposed new paradigm for UK JCMT funding, UK university contributions will allow PhD students to travel to the JCMT, to assist with telescope operations and instrument development.  The new funding paradigm, if approved, will thereby give UK postgraduate students an opportunity to gain key skills in astronomical observation and instrumentation that would otherwise be inaccessible.

\subsection{Opportunities}

The European Southern Observatory (ESO) has announced that it will begin discussion of its next flagship project, as the Extremely Large Telescope (ELT) nears completion.  This discussion, which will take place from July 2024 -- July 2026, is a vital opportunity to put submillimetre science and technology at the heart of ESO’s future plans.  The AtLAST project (Section~\ref{sec:single-dish_atlast}) is already beginning to engage with ESO and the UK representative on the ESO User Committee on this topic. 
 
The UK has the opportunity to become a more major partner in the JCMT (Section~\ref{sec:JCMT_future}) and to shape the future of the only general-purpose single-dish submillimetre telescope to which UK astronomers have access.  The UK has the opportunity to lead the development of a new instrumentation suite for the JCMT, particularly a new polarization-sensitive MKID camera (Section~\ref{sec:jcmt_mkid}), and a large-format heterodyne array (Section~\ref{sec:jcmt_superharp}).  UK astronomers also have the opportunity to lead JCMT Large Programs and to forge new collaborative links with new JCMT partners in Southeast Asia (Thailand, Vietnam, Malaysia and Indonesia) and South America (Argentina and Brazil), both in terms of science exploration and technological developments. 
  
ALMA (Section~\ref{sec:interferometric}) continues to present a vital opportunity for science exploitation by the UK astronomers.  The ALMA Wideband Sensitivity Upgrade (Section~\ref{sec:wsu}) presents the opportunity for science \& heterodyne technology leadership roles in upgraded ALMA projects, which will significantly reduce the time required for blind redshift surveys, chemical spectral scans, and deep continuum surveys.  The UK will also have the opportunity to play a key role in future ALMA developments, as part of the ALMA 2040 plan (Section~\ref{sec:alma2040}), such as baseline extensions, additional antennas, or deployment of focal plane arrays. 
  
AtLAST (Section~\ref{sec:single-dish_atlast}) presents an opportunity for the UK to be a key stakeholder in a world-leading new facility.  It also presents the opportunity to leverage UK strengths in submillimetre instrumentation to build AtLAST instruments in the UK, and the opportunity for UK scientists to play leadership roles in AtLAST science programmes.  By being a key stakeholder in AtLAST, the UK has the opportunity to maintain and advance its world-leading position in submillimetre science and instrumentation. 
  
The UK’s ongoing involvement in the EHT (Section~\ref{sec:vlbi}) via the JCMT Consortium gives us the opportunity to play a leading role in moving the EHT to submillimetre wavelengths.  This also gives UK astronomers the opportunity to forge new collaborative links with researchers in Africa through involvement in the Africa Millimetre Telescope (Section~\ref{sec:amt}), through the BlackHolistic ERC grant. 
  
The UK has the opportunity to leverage our submillimetre science and instrumentation leadership for future NASA or ESA far-infrared space missions, particularly the NASA Probe Mission currently undergoing selection for Phase A study (Section~\ref{sec:nasa_fir}).  The scientific objectives of ground-based submillimetre/millimetre astronomy and FIR space astronomy are complementary, and indeed mutually reliant -- access to ground and space facilities is essential to address comprehensively the key research questions in star and planet formation and galaxy evolution.  The results of NASA Probe Mission selection for Phase A study are expected in Autumn 2024.  The UK community and agencies should work together to promote strong UK participation in any future NASA FIR Probe mission that is selected.

There are a range of opportunities associated with submillimetre instrumentation in the UK (Section~\ref{sec:uk_instrumentation}). This includes the opportunity to work within UKRI with EPSRC to set up a multi-disciplinary manufacturing facility for high frequency transistor circuits for low noise amplifiers and superconducting thin film foundries, and the opportunity to exploit synergies with UK-wide quantum technology development e.g., the development of superconducting parametric amplifiers crucial for quantum-computation platforms and the utilisation of submillimetre heterodyne mixer technology for high frequency qubit applications.  There is also the opportunity to develop a dedicated structure to enable coordination of existing facilities and activities to optimise the design to receiver delivery cycle.  Finally, there are very significant opportunities to enable the exploitation of submillimetre and millimetre technology beyond astrophysics, in fields such as security, telecommunications, environmental sensing, quantum computing and many other fundamental physics experiments such as dark matter searches and neutrino mass determination experiments (Section~\ref{sec:consultation_nonastro}).

\subsection{Threats}

The key threats to the UK submillimetre astronomy are the lack of confirmed UK access to a single-dish submillimetre telescope beyond Q1 2027 (Section~\ref{sec:single-dish_funding_jcmtops}), and the lack of an ESO roadmap for submillimetre astronomy.

If the UK were to lose access to the JCMT, or the JCMT were to cease operations, without AtLAST being built, there would be no general-purpose single-dish facility operating at submillimetre wavelengths for UK astronomers (Sections~\ref{sec:current_instrumentation_jcmt}, \ref{sec:single-dish}).  It is therefore essential to proactively support both the JCMT in the near and intermediate term, and AtLAST in the intermediate and long term.
  
The JCMT's current flagship instruments, SCUBA-2 and HARP (Section~\ref{sec:current_instrumentation_jcmt}), are more than a decade old and require upgrading if the JCMT is to remain world-leading over the next 10 years (Sections~\ref{sec:jcmt_mkid}, \ref{sec:jcmt_superharp}).  Moreover, the JCMT requires stable funding from its partners, both the UK and partners in East and Southeast Asia, both for operations and to maintain its current instrument suite before upgrades are available.  The JCMT's funding is dependent on a number of international partners in a complex political landscape.  A letter from Prof. Paul Ho (JCMT Director) on the current status of the JCMT and its funding partners is provided to the AAP along with this document.

The JCMT operations grant to be submitted to the STFC in September 2026 will stabilise the UK funding route by moving from a university-funded model to an STFC-funded model (Section~\ref{sec:single-dish_funding_jcmtops}).  This will provide a stable income stream for the telescope and guarantee UK access to the JCMT until 2031, as well as widening the UK userbase of the JCMT. Building new instruments for the JCMT will keep it as a world-leading facility for the next decade or more (Section~\ref{sec:JCMT_future}), and will encourage more engagement and investment from new JCMT partners (e.g. Thailand, Vietnam, Malaysia and Indonesia).

Loss of access to the JCMT would also result in a loss of UK access to the EHT (Section~\ref{sec:vlbi}).  It is important both to support the JCMT, and to explore alternative routes to EHT access, potentially through UK involvement in the AMT (Section~\ref{sec:amt}).

The long-standing dispute around governance and land rights on Maunakea, which has in the past disrupted telescope operations, is a minor risk.  However, in 2022 a new Maunakea governance body was formed\footnote{\url{https://www.capitol.hawaii.gov/sessions/session2022/Bills/GM1358_.PDF}}\textsuperscript{,}\footnote{\url{https://aas.org/posts/news/2022/08/new-stewardship-paradigm-maunakea}}, founded on a paradigm of mutual stewardship in which ``ecology, the environment, natural resources, cultural practices, education, and science are in balance and synergy'', and astronomy has been declared a policy priority of the state.  The renewal of the Master Lease for Mauna Kea Observatories will take place in 2033, by which time AtLAST should have achieved or be approaching first light.

It is vital that concerted efforts are made to over the next five years both within the UK and internationally to ensure that AtLAST is funded and built (Sections~\ref{sec:single-dish_atlast}, \ref{sec:single-dish_funding_atlast}).  This would be even more imperative if the UK were to lose access to the JCMT.  The estimated total cost of AtLAST is $\sim$300M USD for the telescope itself and $\sim$520M USD for its instrumentation (\citealt{klaassen2020}; note that these numbers are indicative only), and so must be funded by a multinational consortium.  The UK has the opportunity to be at the heart of this consortium, and so to maintain its place at the forefront of submillimetre astronomy. 

Without a single-dish facility to find new objects for ALMA to follow up on, its scientific return would also suffer (Section~\ref{sec:interferometric}).  We do not associate any other immediate risks with ALMA.  The ALMA WSU (Section~\ref{sec:wsu}) should in principle be funded from ESO budgets, although costs could overrun.
  
ESO has announced that it will begin discussion of its next flagship project, as the Extremely Large Telescope (ELT) nears completion, but does not currently have a Submillimetre Roadmap.  It is thus vital that submillimetre science and technology is well-represented in this discussion, to ensure that this crucial wavelength range is at the heart of ESO’s future plans.  The AtLAST project (Section~\ref{sec:single-dish_atlast}) is already beginning to engage with ESO and the UK representative on the ESO User Committee on this topic. 
  
More broadly, the UK submillimetre/millimetre community requires ongoing support from the STFC and other funding bodies in order to sustain the community itself.  This support is required in order to retain the human resources and expertise required to build on the strong UK heritage in both astronomy and instrumentation, and to further explore opportunities to develop new techniques and novel instruments. Ongoing support for smaller-scale lab-based activities and experiments, and for science exploitation, including through STFC Astronomy Small and Large Awards, is required in order to train the next generation of UK leaders in this field.

\clearpage

\begin{tcbraster}[raster columns=2, raster equal height,nobeforeafter,raster column skip=0.2cm]
  \begin{tcolorbox}[title=Strengths, colback=CornflowerBlue!25!white, colbacktitle=CornflowerBlue, coltitle=black, fonttitle=\bfseries]
  \label{box:swot}

    \raggedright 
    \begin{itemize}[leftmargin=*]
    
    \item Outstanding track record in submillimetre and millimetre astronomy%\par
    %\vspace{0.3\baselineskip}
    \item Outstanding track record in ground- and space-based submillimetre and millimetre instrumentation%\par
    \item Current involvement in active and proposed major international projects
    \item UK astronomers leading global partnerships and research collaborations in the field
    \item Several world-leading institutes driving new technology development and deployment
    \item Availability of national institutes with permanent specialised engineering and project management staff
    
    \end{itemize}

  \end{tcolorbox}
  \begin{tcolorbox}[title=Weaknesses\phantom{g}, colback=Goldenrod!25!white, colbacktitle=Goldenrod, coltitle=black, fonttitle=\bfseries]
    \raggedright 
    \begin{itemize}[leftmargin=*]
    
    \item Challenging funding landscape for both astronomical science and instrumentation
    \item A lack of investment in and access to single-dish submillimetre and millimetre facilities
    %\item Concentration of specialised expertise in individuals risks single-point failures
    \item Low level of investment in superconducting detector technology relative to comparator nations
    \item Lack of opportunity for PhD students to perform observations is creating a significant skills gap for early-career researchers 
    \item Lack of continuity of funding risks loss of technical expertise in the field
    
    \end{itemize}
  \end{tcolorbox}
\end{tcbraster}
\begin{tcbraster}[raster columns=2, raster equal height,nobeforeafter,raster column skip=0.2cm]  
  \begin{tcolorbox}[title=Opportunities\phantom{g}, colback=LimeGreen!25!white, colbacktitle=LimeGreen, coltitle=black, fonttitle=\bfseries]
    \raggedright 
    \begin{itemize}[leftmargin=*]
    \item Being a key stakeholder in AtLAST, a world-leading new submillimetre/millimetre facility, thereby maintaining global leading role.  
    \item Leading the development of a new MKID camera and large-format heterodyne array for the JCMT.
    \item Opportunity to invest in the ALMA WSU, gaining Guaranteed Time for UK users.
    \item Playing a leading role in future ALMA developments as part of the ALMA 2040 plan.
    \item Leading the EHT move to submillimetre wavelengths through the JCMT and the AMT.
    \item Representing submillimetre astronomy in ESO forward planning, and developing an ESO Submillimetre Roadmap.
    \item Building and developing global research collaborations, including with developing economies through the JCMT and the AMT.
    \item Submillimetre instrumentation opportunities include developing multi-disciplinary manufacturing facilities, exploiting synergies with quantum technology development.
    \item Leveraging submillimetre science and instrumentation leadership for future NASA or ESA FIR space missions.
    \item Enabling the exploitation of submillimetre and millimetre technology beyond astrophysics.
    
    \end{itemize}
  \end{tcolorbox}
  \begin{tcolorbox}[title=Threats\phantom{g}, colback=OrangeRed!25!white, colbacktitle=OrangeRed, coltitle=black, fonttitle=\bfseries]
    \raggedright 
    \begin{itemize}[leftmargin=*]
    \item There is no guaranteed UK access to a single-dish telescope beyond the current end date of UK JCMT funding in Q1 2027.
    \item ESO does not currently have a Submillimetre Roadmap.
    \item There is no confirmed general-purpose single-dish submillimetre telescope in the 2030s.
    \item The JCMT's funding arrangements are dependent on a number of international partners in a complex political landscape.
    \item The aging instrumentation suite on the JCMT will require upgrading over the next few years.
    \item Without a single-dish facility to find new objects for interferometric follow-up, ALMA's scientific return would suffer.
    %\item The ALMA WSU should in principle be funded from ESO budgets, but costs could overrun.
    \item UK membership of the EHT Consortium is contingent on its membership of the JCMT Consortium: loss of JCMT access would also remove access to the EHT .
    \item A concerted international effort is required if AtLAST is to be funded and built.
    \item Continuing support from UK funding bodies is required to sustain the submillimetre/millimetre community itself.
    \end{itemize}
  \end{tcolorbox}
\end{tcbraster}

\begin{figure}[ht!]
    \centering
    \includegraphics[width=\linewidth]{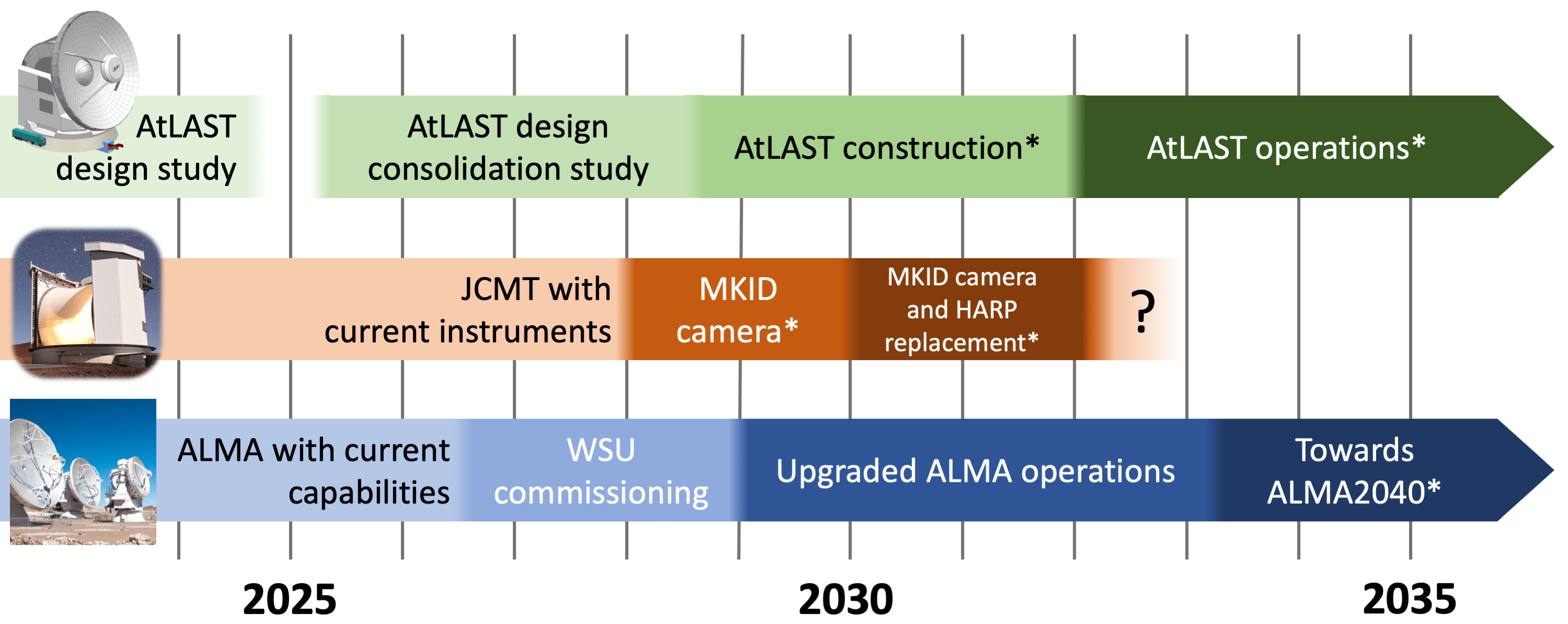}
    \caption{Indicative timeline. A * indicates an item that is not yet funded. The anticipated sequence of facility and instrument developments is as follows: JCMT instrument upgrades, followed by AtLAST, then ALMA 2040 (in other words: an essential upgrade to single-dish capabilities is earlier in the timeline, followed by the anticipated major upgrade of ALMA).} %Note all of these telescopes are/will be part of the EHT
    \label{fig:timeline}
\end{figure}

\section{Strategic Priorities and Summary Recommendations}
\label{sec:roadmap}

We here summarise this report to present a clear set of priorities and recommendations for UK submillimetre and millimetre astronomy over the next decade and beyond.  Throughout this Roadmap, we have demonstrated the UK's excellence in submillimetre and millimetre astronomy and instrumentation.  We have further demonstrated how the UK's world-leading position in the field can be leveraged to play a leading role in the next generation of submillimetre and millimetre facilities.

We present our key strategic priorities for UK submillimetre and millimetre astronomy in Section~\ref{sec:strategy}, and present our summary recommendations in Section~\ref{sec:recommendations}.  An indicative timeline for the key facilities discussed is shown in Figure~\ref{fig:timeline}.

\subsection{Strategic Priorities}
\label{sec:strategy}

As shown in our Community Survey (Section~\ref{sec:consultation}), the UK submillimetre/millimetre community's highest priority for a \textit{new} facility is a 50\,m-class single-dish telescope (Figure~\ref{fig:future_facilities}), while the highest-priority \textit{current} facility is ALMA (Figure~\ref{fig:current_facilities}).  This synergy of single-dish and interferometric instruments is key to advancing the field, as shown by the wide range of science cases in Section~\ref{sec:science}.

Our key strategic priorities are as follows:
\begin{enumerate}[leftmargin=!]
    \item The UK must be a key partner in the forthcoming AtLAST telescope, for which it is essential that the UK remains a key partner in the JCMT in the intermediate term.
    \item The UK must maintain, and if possible enhance, access to ALMA and aim to lead parts of instrument development for ALMA2040.
\end{enumerate}

The UK is already a key member of the AtLAST Horizon 2020 design and Horizon Europe design consolidation studies (see Section~\ref{sec:single-dish_atlast}).  The science engagement for the Horizon 2020 design study was led from the UK, and UK astronomers participate in the coordination committee of both the Horizon 2020 and Horizon Europe projects.

Working towards a future in which the UK is playing a leading role in an AtLAST international consortium requires ongoing investment into the JCMT.  Maintaining and expanding UK links to the JCMT while AtLAST is being designed and built is crucial both so that the JCMT can be used as a technology testbed for AtLAST instruments designed in the UK, and also to maintain UK science and technology expertise in the field.  The unprecedentedly large bolometer cameras and heterodyne arrays planned for AtLAST (Section~\ref{sec:single-dish_atlast}) will require significant technology development, especially the latter.  The UK can be in prime position to build these instruments by building an upgraded instrumentation suite for the JCMT, as described in Sections~\ref{sec:jcmt_mkid} and~\ref{sec:jcmt_superharp}.

It is important to note that investment into smaller facilities leads to the development of larger ones.  A prime example of this is the ongoing ESO investment into adaptive optics technology: the New Technology Telescope (NTT) led to the Very Large Telescope (VLT), which in turn has led to the forthcoming Extremely Large Telescope (ELT).  Conversely, in the USA, there was little investment in VLBI facilities while the Very Large Baseline Array (VLBA) was being built, leading to a lower science return from the VLBA than could otherwise have been expected.  A gap in facilities operating in a given wavelength range leads to a corresponding loss of expertise in both science and technology.  The UK can thus maintain its world-leading position in submillimetre/millimetre technology by continuing to invest in the JCMT until AtLAST is on-sky.

Maintaining UK access to ALMA is essential to submillimetre/millimetre astronomy in both the short and the long term.  The ALMA Wideband Sensitivity Upgrade (WSU) will be tranformative for ALMA science.  This upgrade is funded from ALMA Development Fund, but the possibility of the UK investing in the WSU in exchange for Guaranteed Observing Time on ALMA should be explored. 
The UK should also aim to lead instrument development for the ambitious ALMA2040 development plan.

By investing in AtLAST, ALMA and the JCMT, the UK will also be investing in the future of the EHT.  Currently, UK access to the EHT is only possible through the UK's membership of the JCMT consortium.  It is thus important both to maintain access to the JCMT, and to  explore alternative access routes to the EHT, possibly through UK involvement in the AMT.

It must be emphasised that our strategic priorities for single-dish and interferometric instrumentation complement one another: AtLAST and an upgraded ALMA would be in synergy, not competition, with one another.  Both have identified and are working towards the same science goals (cf. Sections~\ref{sec:single-dish_atlast} and \ref{sec:interferometric}), and both are required in order to fully address these goals, as shown by the range of scientific questions in Section~\ref{sec:science}.  This complementarity is noted by the ALMA Development Roadmap\footnote{\url{https://www.almaobservatory.org/en/publications/the-alma-development-roadmap/}}, which notes the need for a new large-aperture single-dish submillimetre telescope, while declaring it beyond the scope of ALMA operations.  The ALMA Development Roadmap states: \textit{``A large single dish submillimeter telescope of a diameter of at least 25\,m would enable deep, multi-wavelength images of the sky and provide many scientific synergies with ALMA.''}.

\subsection{Summary recommendations}
\label{sec:recommendations}

\subsubsection{Medium-term recommendations (2025--2030)}

ALMA has revolutionised submillimetre/millimetre science over the last decade.  Its impact has been felt across the field, as discussed throughout Section~\ref{sec:science}.  It has been particularly vital to the field of planet formation, for the first time allowing resolved imaging of protoplanetary discs (Section~\ref{sec:protoplanetary_discs}) and debris discs (Section~\ref{sec:debris_discs}), as well as providing high-resolution imaging of regions of both high- and low-mass star formation (Section~\ref{sec:MW_star_formation}), insights into dust production through imaging of supernova remnants (Section~\ref{sec:dust}, and key information on prebiotic chemistry and biomarkers through observations of Solar System bodies (Section~\ref{sec:solar_system}).  Large ALMA campaigns have also allowed multiple diagnostics of feedback processes, morphologies, kinematics and the physical and chemical compositions of the interstellar medium in galaxies at Cosmic Noon 
%\hl{something to do with cosmic noon} 
(Section~\ref{sec:cosmic_noon}),  
as well as the detection of redshifted emission lines from galaxies at cosmic dawn (Section~\ref{sec:cosmic_dawn}).  ALMA is the most widely-used submillimetre/millimetre instrument in the UK (Figure~\ref{fig:current_telescopes}), and the one that UK astronomers think will be most important to their research over the next ten years (Section~\ref{sec:improvements_to_current_instrumentation}; Figure~\ref{fig:current_facilities}).  The ALMA Wideband Sensitivity Upgrade (WSU; Section~\ref{sec:wsu}), which is currently in progress, funded by the ALMA Development Fund, will upgrade ALMA's bandwidth by a factor of at least 2. 

\begin{tcolorbox}[colback=white]
\textbf{Recommendation M.1.} ALMA will continue to be vital to all areas of UK astronomical research. The UK must continue to play a significant role in both the instrumentation upgrades and the world-leading astronomy from ALMA.
\end{tcolorbox}

The JCMT is the largest single-dish submillimetre telescope in the world, and remains crucial to submillimetre astronomy in the UK, being the second-most used facility after ALMA (Figure~\ref{fig:current_telescopes}).
The JCMT can perform wide-area mapping of dust and molecular gas, and is the only telescope that can map polarized dust emission at submillimetre wavelengths over large areas (Section~\ref{sec:current_instrumentation_jcmt}), as well as being at the forefront of protostellar variability monitoring (Section~\ref{sec:protostellar_variability}) and exploring sites of dust production (Section~\ref{sec:dust}).  JCMT Large Program leadership has led to the UK being at the forefront of star formation research over recent years (Section~\ref{sec:MW_star_formation}).  The JCMT has also been key to recent work mapping potential biomarkers in the Solar System (Section~\ref{sec:solar_system}).
The JCMT is also crucial for finding new sources for interferometric follow-up, including protostellar discs and debris discs (Sections~\ref{sec:protoplanetary_discs}, \ref{sec:debris_discs}) and high-redshift galaxies and protoclusters (Sections~\ref{sec:cosmic_noon}, \ref{sec:protoclusters}).  The JCMT is also able to map nearby galaxies that are largely inaccessible by interferometers due to their large angular size (Section~\ref{sec:nearby_galaxies}).

\begin{tcolorbox}[colback=white]
\textbf{Recommendation M.2.} The JCMT will remain internationally excellent, in its unique position as the world's largest single-dish submillimetre telescope, at least until AtLAST is on sky. It is crucial that the UK maintains a key role in the JCMT, and widens access to all UK astronomers.
\end{tcolorbox}

The JCMT's flagship instrumentation suite, the SCUBA-2 camera and the HARP heterodyne array (Section~\ref{sec:current_instrumentation_jcmt}), are now more than 10 years old, and require upgrading (Section~\ref{sec:JCMT_future}).  
A MKID camera for the JCMT was identified by our community as a key upgrade for the next 10 years (Figure~\ref{fig:current_facilities}), with a new large-format heterodyne array being deemed only marginally less urgent.  The UK has a strong heritage in building submillimetre cameras (Section~\ref{sec:instrumentation_past_phot}) and spectrometers (Section~\ref{sec:instrumentation_past_spec}).  Cardiff's globally-recognised leadership in KID technology, used in the MUSCAT camera built for the LMT, and in quasi-optical components for the FIR to millimetre regime, of which Cardiff is the leading supplier (Section~\ref{sec:instrumentation_current_direct}), makes the UK the obvious choice to lead the construction of a new MKID camera for the JCMT (Section~\ref{sec:jcmt_mkid}), for which a pathfinder instrument already exists (Section~\ref{sec:instrumentation_current_direct}).  Meanwhile, Oxford and RAL's leadership in heterodyne spectroscopy technology, along with Manchester's expertise in low-noise amplifiers (Section~\ref{sec:instrumentation_current_heterodyne}), makes the UK well-placed to lead the construction of a large-format heterodyne array for the JCMT (Section~\ref{sec:jcmt_superharp}).  Building these instruments will maintain the UK's global leadership in submillimetre/millimetre instrumentation, and will allow the JCMT to serve as a technology testbed for AtLAST.

\begin{tcolorbox}[colback=white]
\textbf{Recommendation M.3.} An upgraded instrumentation suite for the JCMT will be required to maintain its world-leading position for the next 10 years in the run-up to AtLAST.  The UK should be at the heart of building a polarisation-sensitive MKID camera and a large-format heterodyne array for the JCMT, to avoid ceding our scientific and technological leadership in these fields.
\end{tcolorbox}

The Event Horizon Telescope (EHT) has produced some of the most high-profile results in astrophysics in recent years (Section~\ref{sec:smbh}), with its imaging of the supermassive black holes at the centres of M87 and the Milky Way.  Future science goals of the EHT include resolved imaging of multiple photon rings around SMBHs, and monitoring of AGN variability (Sections~\ref{sec:smbh}, \ref{sec:vlbi}).  To achieve these, an expanded EHT network and submillimetre observations are required (Section~\ref{sec:vlbi_immediate}).  The JCMT is key to the move to submillimetre wavelengths, as it is one of few sites that can observe at these high frequencies (Sections~\ref{sec:current_instrumentation_jcmt}, \ref{sec:vlbi_immediate}).  The UK is involved in building receivers for the Africa Millimetre Telescope (AMT), which will provide a key new baseline for the EHT (Section~\ref{sec:amt}).  UK access to the EHT is currently dependent on UK membership of the JCMT consortium (Sections~\ref{sec:current_instrumentation_jcmt}, \ref{sec:vlbi}).

\begin{tcolorbox}[colback=white]
\textbf{Recommendation M.4.} The Event Horizon Telescope (EHT) will continue to produce new insights into SMBHs and AGN as it moves to higher frequency and into time domain observations.  The UK should maintain and diversify its access to the EHT, which is currently contingent on access to the JCMT, potentially through involvement in the AMT.  The UK should be central to the EHT's move to submillimetre wavelengths.
\end{tcolorbox}

The Atacama Large Aperture Submillimeter Telescope (AtLAST; Section~\ref{sec:single-dish_atlast}) will revolutionise single-dish submillimetre/millimetre astronomy in the same way that ALMA has revolutionised submillimetre/millimetre interferometry.  It is important to note that the AtLAST design consolidation study begins in Q1 2025 (Section~\ref{sec:single-dish_funding_atlast}), and has significant UK involvement (Section~\ref{sec:single-dish_atlast}).  If AtLAST is to be built, it is imperative that work that has begun towards it continues, and that the UK remains a key contributor.

\begin{tcolorbox}[colback=white]
\textbf{Recommendation M.5.} The UK should during the next 5 years be working towards AtLAST through the Horizon Europe AtLAST design consolidation study and beyond.
\end{tcolorbox}

%The JCMT will continue to play a key role in the Event Horizon Telescope, as will ALMA, as they are located at 2 of the only 3 sites from which high-frequency (sub-millimetre) observations can be taken with the EHT. These observations have already started, with an upgrade to 690~GHz being planned. It is vital that the UK remains at the forefront of this research.}

\subsubsection{Long-term recommendations (2030 and beyond)}

The results of our Community Survey (Section~\ref{sec:new_facilities_beyond_10_years}; Figure~\ref{fig:future_facilities}) show that the highest priority of the UK submillimetre/millimetre community is a new, large (50\,m-class) single-dish telescope.  The only candidate for such a telescope is AtLAST (Section~\ref{sec:single-dish_atlast}).  The science questions and instrumentation drivers discussed in Section~\ref{sec:science} make the case for AtLAST abundantly clear.  AtLAST will have key applications across astrophysics: for studies of protostellar discs (Section~\ref{sec:protoplanetary_discs}) and debris discs (Section~\ref{sec:debris_discs}), surveys of nearby star-forming regions with AtLAST would identify further low-mass disc candidates.  These surveys would also be crucial for understanding protostellar variability and how stars acquire mass (Section~\ref{sec:protostellar_variability}).  AtLAST will also provide key insights into Solar System bodies (Section~\ref{sec:solar_system}), and will provide key insights into the link between the Sun's chromosphere and space weather (Section~\ref{sec:solar_physics}).
Unbiased mapping of the Galactic Plane and nearby galaxies in polarized light with AtLAST will also provide key insights into star formation and its link to galactic evolution (Sections~\ref{sec:MW_star_formation}, \ref{sec:nearby_galaxies}), as well as key insights into dust production mechanisms (Section~\ref{sec:dust}).  At higher redshifts, AtLAST will also perform wide-field submillimetre galaxy surveys (Section~\ref{sec:cosmic_noon}), and protocluster detection and characterisation (Section~\ref{sec:protoclusters}).  AtLAST will also be a part of the EHT, performing time-resolved VLBI observations of SMBHs and AGN (Section~\ref{sec:smbh}).

\begin{tcolorbox}[colback=white]
\textbf{Recommendation L.1.} The AtLAST Telescope should be seeing first light by the mid 2030s, and be beginning science operations shortly thereafter.  The UK should aim to be a key partner of an international consortium for the construction of AtLAST, to capitalise on and build on our world-leading expertise in submillimetre science and technology.
\end{tcolorbox}

The UK leads the world in submillimetre/millimetre instrument development (Section~\ref{sec:uk_instrumentation}) and therefore is very well placed to make leading contributions to the development and construction of first-light instruments for AtLAST (Section~\ref{sec:single-dish_atlast}). Not only will this maintain and consolidate the UK leadership in technological developments in this domain, but there is a long track record in this wavelength domain and beyond of the UK using its technical contributions to secure scientific leadership roles \citep{STFC_AAP_2022_Roadmap}.

\begin{tcolorbox}[colback=white]
\textbf{Recommendation L.2.} UK instrumentation laboratories should take leading roles in the development and construction of first-light instruments for AtLAST.
\end{tcolorbox}

\begin{tcolorbox}[colback=white]
\textbf{Recommendation L.3.} UK astronomers should play leading roles in planning and executing AtLAST science, building on our medium-term single-dish track record, and where possible using our technical contributions to leverage leading scientific positions.
\end{tcolorbox}

Our community survey shows strong support for major upgrades to ALMA (Section~\ref{sec:new_facilities_beyond_10_years}; Figure~\ref{fig:future_facilities}).  The ALMA2040 component of the ALMA Development Roadmap (Section~\ref{sec:alma2040}) is the route to maintaining ALMA at the forefront of submillimetre/millimetre astronomy in the coming decades.  The UK is already involved in preparatory work for ALMA2040, through UK involvement in the ALMA-FPA design study.

\begin{tcolorbox}[colback=white]
\textbf{Recommendation L.4.} The UK must, as a minimum, participate in instrument development for ALMA2040, and ideally lead parts of it.
\end{tcolorbox}

These four recommendations are not in competition, but are based on the complementarity between AtLAST and ALMA.

\subsubsection{Funding application recommendations}
\label{sec:funding_timeline}

We here outline our recommended funding requests for submillimetre/millimetre instrumentation over the next 10+ years.  We note that further opportunities may arise, and that the funding timelines described here are indicative only.

The JCMT is the only general-purpose single-dish submillimetre telescope to which UK astronomers have access, and so will be essential to UK submillimetre/millimetre astronomy until AtLAST is on-sky.  It is important to emphasise that there is no guaranteed UK access to a single-dish telescope beyond the current end date of UK JCMT funding in Q1 2027.  As well as securing telescope access through JCMT operations funding, it is important that UKRI support is sought to allow UK astronomers to lead the development of a new MKID camera and a new large-format heterodyne array for the JCMT.  These instruments will serve as technology demonstrators for AtLAST.   We detail the timeline for JCMT funding requests in Section~\ref{sec:jcmt_funding_summary}.

\begin{tcolorbox}[colback=white]
\textbf{Recommendation F.1.} We recommend the UK community bid for UKRI support for JCMT operations to beyond 2031, supporting UK-wide access to JCMT up to the first light of AtLAST.
\end{tcolorbox}

\begin{tcolorbox}[colback=white]
\textbf{Recommendation F.2.} We recommend the UK community bid for UKRI support for the UK share of a new MKID camera for the JCMT, which will be led from the UK.  
We also recommend the UK community bid for funding for technology development towards a new large-format heterodyne array.
\end{tcolorbox}

The estimated total cost of AtLAST is $\sim$\$300M for the telescope itself and $\sim$\$520M for its instrumentation (\citealt{klaassen2020}; note that these numbers are indicative only), and so AtLAST must be funded by a multinational consortium (Section~\ref{sec:single-dish_funding_atlast}).  We envisage applying for UKRI Research Infrastructure funding for AtLAST, with the agreement and support of the STFC. 

\begin{tcolorbox}[colback=white]
\textbf{Recommendation F.3.} We recommend a coordinated UK community application for UKRI funding for AtLAST, to secure leading UK roles in this multi-national consortium. 
\end{tcolorbox}

Access to ALMA is paid for via the UK contribution to the European Southern Observatory (ESO).  The UK currently contributes approximately 16\% of ESO’s revenue (2021 contribution), with an average contribution of GBP 22.7M per year (2020 prices) since the UK joined ESO in 2002\footnote{\url{https://www.ukri.org/wp-content/uploads/2022/02/STFC-240222-SocioEconomicImpactEvaluationStudyUKSubscriptionESO-FinalReportSummary.pdf}}.  The ALMA Wideband Sensitivity Upgrade (WSU, Section~\ref{sec:wsu}), which is funded through the ALMA Development fund, will upgrade ALMA’s bandwidth by a factor 2--4, amongst other improvements, and so will significantly enhance ALMA’s scientific capabilities.

Currently, the funding for the WSU is only sufficient to enable the implementation of a factor 2 increase in processed bandwidth. Contributing funds to enhance the correlator to process the full factor 4 enhanced bandwidth in exchange for guaranteed observing time would provide an opportunity to address the factor 8 oversubscription faced by potential UK ALMA users (Section~\ref{sec:interferometric}).

\begin{tcolorbox}[colback=white]
\textbf{Recommendation F.4.} We recommend a community bid for upgrading the ALMA WSU 
to its full factor 4 enhanced bandwidth
to secure UK guaranteed observing in this very heavily oversubscribed, unique and internationally-leading facility.
\end{tcolorbox}

In the longer term, the UK has the opportunity to play a key role in potential future ALMA developments as part of the ALMA 2040 plan (Section~\ref{sec:alma2040}), such as baseline extensions, additional antennas, receiver upgrades, software development and focal plane arrays. The first steps in this include the ALMA-FPA study funded by an EU ALMA Development Grant which has  UK involvement and initial funding from  ESO and an EU ALMA Development Grant to The University of Manchester to develop technologies for the next generation of ALMA receivers. 

\begin{tcolorbox}[colback=white]
\textbf{Recommendation F.5.} We recommend that the community bids for funding for the UK to play a leading role in ALMA developments as part of the ALMA 2040 plan.
\end{tcolorbox}

\subsubsection{General Recommendations}

The anticipated timeline in Figure~\ref{fig:timeline} is driven partly by the results of our community consultation. In particular, ALMA and its WSU have the most community support for near-term and $<10$ year timescales (Figure~\ref{fig:current_facilities} and Section~\ref{sec:improvements_to_current_instrumentation}), followed by support for the JCMT and its potential improved instrumentation. In the longer term, the community preference switches (Figure~\ref{fig:future_facilities}), with a $50\,$m-class single dish telescope being given slightly more preference than long-term ALMA upgrades (Section~\ref{sec:new_facilities_beyond_10_years}). Therefore, the sequencing reflects this community preference as well as external constraints. Nevertheless, none of the scientific objectives in Section~\ref{sec:science} anticipates a decline in appetite for a single-dish facility to feed ALMA. 

\begin{tcolorbox}[colback=white]
\textbf{Recommendation G.1.} %Tensioning recommendation: although we can't with certainty 
It is not possible to predict with certainty the development timeline of new facilities, but there may be a point at which involvement in AtLAST becomes contingent on decommitting from JCMT.  When this point is reached, the UK must ensure that its critical single-dish capabilities are maintained throughout the transition from the JCMT to AtLAST.
\end{tcolorbox}

The 2022 STFC AAP roadmap \citep{STFC_AAP_2022_Roadmap} implicitly advised against ``picking winners'' in instrumentation technologies. Instead, it emphasised the strategic benefits of maintaining diversity: \textit{``We do not believe anyone can unambiguously identify the technologies needed to address
STFC’s key science challenges in the next decade. So to remain competitive there must be
strong UK investment in a broad program of advanced instrumentation and its supporting
technologies''}. 
Partly, this reflected the benefits of \textit{``trading unique UK technical capabilities, often mission-critical,
for a seat at the table setting the scientific agenda of a mission''}, a sentiment that applies specifically to the submillimetre and millimetre-wave domains as it does to UK astronomy instrumentation in general. This led AAP to a recommendation that \textit{``there must be strong UK investment in a broad programme of advanced
instrumentation and its supporting technologies''}. Nothing in our community consultation (Section~\ref{sec:consultation}) or our reviews of scientific and technological strengths (Sections~\ref{sec:science} and \ref{sec:uk_instrumentation}) indicated that this wavelength domain would be an exception to this overarching recommendation.

\begin{tcolorbox}[colback=white]
\textbf{Recommendation G.2.} UK instrumentation laboratories must continue to be supported to allow them to be at the forefront of building first-light instruments for AtLAST and bidding for each round of ALMA instrumentation upgrades. 
\end{tcolorbox}

There is abundant evidence for the growing demands for compute in this wavelength domain driven by the increasing volume and complexity of data (Section~\ref{sec:computing_hardware}), as well as for software maintenance and sofware engineering (Section~\ref{sec:computing_software}). This is broadly in concordance with the wider situation in astronomy, reflected in the 2022 AAP recommendation 3.7: \textit{``UK HPC capabilities such as DiRAC \& IRIS underpin a wide range of world-leading UK theoretical astrophysics and data science that must be continually supported and upgraded to remain competitive''} \citep{STFC_AAP_2022_Roadmap}. Furthermore, the community clearly both values and benefits enormously from open science and open data practices (e.g. Figure~\ref{fig:current_telescopes}) as well as there being a breadth of support for the user community using the regional centre model (e.g. Section~\ref{sec:computing_infrastructure_for_new_facilities}, Figure~\ref{fig:computing_resources}).

\begin{tcolorbox}[colback=white]
\textbf{Recommendation G.3.} The UK should maintain and develop its high performance computing capabilities as the demands on compute increase from new submillimetre and millimetre-wave facilities and instruments. The facilities should continue to align their open data and open software practices with national, continental and international open science initiatives, and be supported by software engineering for the maintenance, development and support of software.
\end{tcolorbox}

\begin{tcolorbox}[colback=white]
\textbf{Recommendation G.4.} We support the regional centre model for the provision of user support, such as the ALMA regional centre, for the development of expertise and provision of expert support for UK users of submillimetre/millimetre facilities.
\end{tcolorbox}

%\newpage
%\setcounter{section}{0}
%\renewcommand{\thesection}{\Alph{section}}
\phantomsection 

\addcontentsline{toc}{section}{Appendix A: STFC PPAN Science Board queries and answers}
\section*{Appendix A: STFC PPAN Science Board queries and answers}

\textbf{Recommendation M.2 proposes widening access to JCMT data to the whole community (a noble aim, certainly), but is there evidence for demand to JCMT data from researchers beyond the Consortium institutes listed on p49? The majority of respondents to the questionnaire (which I appreciate is not the same as interest in JCMT) seem to be from Consortium member institutes.}

The match-funded access model for the JCMT is almost unique amongst major telescopes, and as a result, access for UK astronomers is currently inherently inequitable.  JCMT access is currently effectively contingent on holding an affiliation with a university that has a strong history of submillimetre astronomy, making it difficult for new users to enter the field.

Moreover, membership of the UK JCMT consortium requires a financial subscription, which not all UK universities are able to resource.  At the last renewal round of the current JCMT funding arrangements, every university in the UK was invited to join the consortium. There were 13 universities that were able to find the money to join or remain part of the consortium, but there were 8 universities that either dropped out or expressed an initial interest but did not ultimately have the money to join the consortium, suggesting that more than one third of the UK submillimetre community are currently being denied access to the JCMT, who would otherwise be active users.

As further evidence of interest in the JCMT from universities not currently affiliated with the JCMT consortium, we have examined JCMT papers published in 2021 and 2022 (the most recent years for which information is available in the JCMT ADS library).  We found authors affiliated with 16 UK institutes that are not part of the consortium: Cambridge, Exeter, Keele, Kent, Lancaster, Leeds, Leicester, the National Physical Laboratory, Nottingham, Portsmouth, Royal Observatory Greenwich, RAL Space, South Wales, SKAO, Sussex, and Warwick.  These authors were either co-authors on papers led by consortium members (whether UK-based or in another EAO partner region), or were making use of the JCMT’s public data archive.  This broad range of affiliations again indicates strong interest in the JCMT across the UK community.

As discussed in the Roadmap, an SOI for a revised, more equitable, funding model was accepted earlier this year, and the JCMT Consortium was preparing to submit the full proposal this month.  However, the STFC has recently imposed a two-year delay before the new funding model can be considered, requiring the UK JCMT Consortium to again provide funding from their own budgets.  In the current climate, it will be difficult for existing consortium members to raise the sums required, and more so for new members to join.  The Office for Students reports that 40 percent of UK universities are operating at a loss this year (\href{https://www.officeforstudents.org.uk/publications/financial-sustainability-of-higher-education-providers-in-england-2024/} {officeforstudents.org.uk}).  Faced with such financial pressures, there is a significant risk that some members of the current consortium will be forced to withdraw, further restricting UK access to this unique and internationally important facility. 

We thus consider that there is an evidence base for a more equitable resourcing model for the JCMT, which as demonstrated in the Roadmap is a facility of international importance. 

\textbf{On p52 there is an indicative cost estimate for the AtLAST telescope, but this does not seem to include an estimate for the proposed 1.7 MW photo-voltaic + battery + back-up power system. Is this a significant cost? Would this be paid up-front or be part of the operations costs? (Very possibly this has still to be studied, but it looks to be roughly the same power budget as ALMA).}

The numbers quoted from 
\cite{klaassen2020}
%Klaassen et al. (2020) 
consider only hardware needed for construction, not power or running costs.  Studies are ongoing for the costs of not only building the observatory and its instrument suite, but also operating and powering it. 

Those indicative costs were in 2019 US dollars, and more than half of that budget was focused on instrumentation using estimates of scaling up the technology of the time. While hardware costs have increased significantly since COVID, the state of the art for instrumentation has evolved since that review, and newer technologies are expected to enable smaller instruments delivering higher quality (and quantities of) data. This will decrease their construction costs, their sizes and their energy consumption, which will mitigate some of the risk in the cost estimates derived in 2019.  Raising the TRL on these technologies (i.e. building instruments for JCMT and maintaining our technological advantages as recommended in this Roadmap) will further mitigate cost risks on the instruments and telescope costs for AtLAST.

\addcontentsline{toc}{section}{Appendix B: List of Acronyms}
\section*{Appendix B: List of Acronyms}

AAP -- Astronomy Advisory Panel\\
ACES -- ALMA Central Molecular Zone Exploration Survey\\
ACA -- Atacama Compact Array (also known as the Morita Array)\\
ACES -- ALMA CMZ Exploration Survey\\
ACSIS -- Auto Correlation Spectral Imaging System\\
ACT -- Atacama Cosmology Telescope\\
ADMX -- Axion Dark Matter Experiment\\
ADS -- Astrophysics Data System\\
AGN -- Active Galactic Nucleus/Nuclei\\
AIG -- Astronomy and Earth Observation Instrumentation Group\\ 
ALMA -- Atacama Large Millimeter/submillimeter Array\\
ALMAGAL -- ALMA Evolutionary study of High Mass Protocluster Formation in the Galaxy\\
AMKID - Antenna-coupled MKID camera\\
AMT -- Africa Millimeter Telescope\\
APEX -- Atacama Pathfinder Experiment\\
ARC -- ALMA regional centre\\
ARIG -- Advanced Radio Instrumentation Group\\
ARKS -- ALMA survey to Resolve exoKuiper belt Substructures\\
ASIAA -- Academia Sinica Institute of Astronomy and Astrophysics\\
ASKAP -- Australian Square Kilometre Array Pathfinder\\
ASTHROS -- Astrophysics Stratospheric Telescope for High Spectral Resolution Observations at Submillimeter wavelengths\\
ASTRONET -- A planning and advisory network for European astronomy\\
ATAC -- Advanced Technology ALMA Correlator\\
ATHENA -- Advanced Telescope for High Energy Astrophysics\\
AtLAST -- Atacama Large Aperture Submillimeter Telescope\\

BHEX -- Black Hole Explorer\\
BISTRO -- B-fields In STar-forming Region Observations\\
BLAST -- Balloon-borne Large Aperture Submillimeter Telescope\\
BLAST-TNG -- Balloon-borne Large Aperture Submillimeter Telescope - The Next Generation\\
BOOMERanG -- Balloon Observations Of Millimetric Extragalactic Radiation And Geophysics\\

CAMELS -- Cambridge Emission Line Surveyor\\
CANFAR -- Canadian Advanced Network for Astronomy Research\\
CARUSO -- Cryogenic Array Receiver for Users of the Sardinia Observatory\\
CASA -- Common Astronomy Software Applications\\
CfA -- Harvard-Smithsonian Center for Astrophysics\\
CHIMPS -- CO Heterodyne Inner Milky Way Plane Survey\\
CLASS -- Cosmology Large Angular Scale Surveyor\\
CLOGS -- CO Large Outer-Galaxy Survey\\
CMB -- Cosmic Microwave Background\\
CMB-S4 -- Cosmic Microwave Background Stage 4 experiments\\
CMF -- Core Mass Function\\
CMZ -- Central Molecular Zone\\
COHRS -- CO High-Resolution Survey\\
CTT -- Celtic Terahertz Technology Ltd.\\

DDSI -- Direct Detection Spectrometer Instrument\\
DECO -- The ALMA Disk-Exoplanet COnnection\\
DESHIMA -- DEep Spectroscopic HIgh-redshift MApper\\
DSA -- Deep Synoptic Array\\

e-MERLIN -- enhanced Multi Element Remotely Linked Interferometer Network\\
EAO -- East Asian Observatory\\
EBEX -- E and B Experiment\\
EHT -- Event Horizon Telescope\\
EHTC -- Event Horizon Telescope Collaboration\\
ELT -- Extremely Large Telescope\\
EPSRC -- Engineering and Physical Sciences Research Council\\
ERC -- European Research Council\\
ESA -- European Space Agency\\
ESO -- European Southern Observatory\\
EU -- European Union\\ 

FIR -- Far Infrared\\
FIRESS -- Far Infrared Enhanced Survey Spectrometer\\
FIRSST -- Far-IR Spectroscopy Space Telescope\\
FPA -- focal plane array\\
FRB -- Fast Radio Burst\\ 
FYST -- Fred Young Submillimeter Telescope\\

GAIA -- Graphical Astronomy and Image Analysis\\
GLT -- Greenland Telescope\\
GPU -- Graphics Processing Unit\\
GUSTO -- Galactic / Extragalactic Ultra-long duration balloon Spectroscopic Terahertz Observatory\\

HARP -- Heterodyne Array Receiver Program\\
HASHTAG -- HARP And Scuba-2 High-resolution Terahertz Andromeda Galaxy survey\\
HAYSTAC -- Haloscope At Yale Sensitive To Axion CDM\\
HEMT -- High Electron Mobility Transistor\\
HFI -- High Frequency Instrument (Planck)\\
HFSs -- Hub-Filament Systems\\
HiRX -- High-Resolution Receiver\\
HSI -- Heterodyne Spectroscopy Instrument\\

IF -- Intermediate Frequency\\
IGO -- Inter-Governmental Organisation\\
IMF -- Initial Mass Function\\
IRAM -- Institut de radioastronomie millim\'{e}trique\\
IRIS -- eInfrastructure for Research and Inovation at STFC\\
IRAS -- Infrared Astronomical Satellite\\

JADES -- JWST Advanced Deep Extragalactic Survey\\
JAXA -- Japan Aerospace Exploration Agency\\
JCMT -- James Clerk Maxwell Telescope\\
JINGLE -- JCMT dust and gas In Nearby Galaxies Legacy Exploration\\
JWST -- James Webb Space Telescope\\

KID -- Kinetic Inductance Detector\\
KVN - Korean VLBI Network\\

LAT -- Large Aperture Telescope\\
LEKID -- Lumped-Element KID\\
LJMU -- Liverpool John Moores University\\
LMT -- Large Millimeter Telescope\\
LNA -- low noise amplifier\\
LO -- local oscillator\\
LST -- Large Submillimeter Telescope\\

MAJORS -- Massive, Active, JCMT-Observed Regions of Star formation\\
MCMC -- Markov chain Monte Carlo\\
MetOp(SG) -- ESA’s Second Generation series of operational meteorology satellites\\
MHD -- Magnetohydrocynamics\\
MIRI - Mid-Infrared Instrument (JWST)\\
MKID -- Microwave Kinetic Inductance Detector\\
MMT -- Millimetre Wave Technology\\
MSSL - Mullard Space Science Laboratory\\
MUSCAT -- Mexico-UK Sub-millimetre Camera for Astronomy\\
MW -- Milky Way\\

NARIT -- National Astronomical Research Institute of Thailand\\
NASA -- National Aeronautics and Space Administration\\
ngEHT -- next generation Event Horizon Telescope\\
ngVLA -- next generation Very Large Array\\
NIKA -- New IRAM KID Array\\
NIRSpec -- Near-Infrared Spectrometer (JWST)\\
NOEMA -- Northern Extended Millimetre Array\\
NOVA -- Nederlandse Onderzoekschool Voor Astronomie\\

OMT -- Orthomode Transducer\\
OSAS-B --  oxygen spectrometer for atmospheric science on a balloon\\
OSF -- Operations Support Facility\\
OVRO -- Owens Valley Radio Observatory\\\

PHANGS -- Physics at High Angular resolution in Nearby GalaxieS\\
PI -- Principal Investigator\\
PILOT -- Polarized Instrument for the Long-wavelength Observations of the Tenuous ISM\\
POL-2 -- a half-wave plate and grid analyser allowing dual-band polarimetry at the JCMT\\
PPAN -- Particle Physics, Astronomy and Nuclear Physics\\
PPD -- proto-planetary disc\\
PPRP -- Projects Peer Review Panel\\
PRIMA -- PRobe for-Infrared Mission for Astrophysics\\
PRIMER -- Public Release IMaging for Extragalactic Research\\

QMC -- Queen Mary College (now Queen Mary University of London)\\

R\&D -- Research and Development\\
RAGERS -- RAdio-Galaxy Environment Reference Survey\\
RAL -- Rutherford Appleton Laboratory\\
RF -- radio frequency\\
ROE -- Royal Observatory Edinburgh\\
ROVER -- Roving Polarimeter\\

SAFARI -- SpicA FAR-infrared Instrument\\
SALTUS -- Single aperture large telescope for universe studies\\
SCUBA -- Submillimetre Common User Bolometer Array\\
SCUBA-2 -- Submillimetre Common User Bolometer Array 2\\
SEDIGISM -- Structure, Excitation, and Dynamics of the Inner Galactic InterStellar Medium\\
SFE -- Star Formation Efficiency\\
SFR -- Star Formation Rate\\
SIS -- superconductor-insulator-superconductor\\
SKA -- Square Kilometre Array\\
SMA -- Sub-millimeter Array\\
SMBH -- Super-Massive Black Hole\\
SMG -- Sub-millimetre galaxy\\
SO -- Simons Observatory\\
SOFTS -- Superconducting On-chip Fourier Transform Spectrometer\\
SOI -- Statement of Interest\\
SPHEREx -- Spectro-Photometer for the History of the Universe, Epoch of Reionization, and Ices Explorer\\
SPICA -- Space Infrared telescope for Cosmology and Astrophysics\\
SPIRE -- Spectral and Photometric Imaging Receiver\\
SPT -- South Pole Telescope\\
SRON -- Netherlands Institute for Space Research\\
SST -- Swedish 1-m Solar Telescope\\
STFC -- Science and Technology Facilities Council\\

TAC -- Time Allocation Committee\\
TolTEC -- three-band imaging polarimeter at the LMT\\
TPGS -- Total Power GPU Spectrometer\\
TRAO -- Taeduk Radio Astronomy Observatory\\
TRL -- Technology Readiness Level\\
TWPA -- Traveling Wave Parametric Amplifier\\

UCL -- University College London\\
UCLan -- University of Central Lancashire\\
UK -- United Kingdom\\
UKATC - UK Astronomy Technology Centre\\
UKIRT -- UK Infrared Telescope\\
UKRI -- UK Research and Innovation\\
UKSA -- UK Space Agency\\
UKT14 -- UKIRT instrument 14 (originally on UKIRT, later on JCMT)\\ 

VLA -- Very Large Array\\
VLBA -- Very Long Baseline Array\\
VLBI -- Very Long Baseline Interferometry\\ 

WSU -- Wideband Sensitivity Upgrade\\

\clearpage
\bibliographystyle{mnras}
\begin{multicols}{2}
\phantomsection 
\addcontentsline{toc}{section}{References}
%\bibliography{bibliography}

\begin{thebibliography}{}
\makeatletter
\relax
\def\mn@urlcharsother{\let\do\@makeother \do\$\do\&\do\#\do\^\do\_\do\%\do\~}
\def\mn@doi{\begingroup\mn@urlcharsother \@ifnextchar [ {\mn@doi@}
  {\mn@doi@[]}}
\def\mn@doi@[#1]#2{\def\@tempa{#1}\ifx\@tempa\@empty \href
  {http://dx.doi.org/#2} {doi:#2}\else \href {http://dx.doi.org/#2} {#1}\fi
  \endgroup}
\def\mn@eprint#1#2{\mn@eprint@#1:#2::\@nil}
\def\mn@eprint@arXiv#1{\href {http://arxiv.org/abs/#1} {{\tt arXiv:#1}}}
\def\mn@eprint@dblp#1{\href {http://dblp.uni-trier.de/rec/bibtex/#1.xml}
  {dblp:#1}}
\def\mn@eprint@#1:#2:#3:#4\@nil{\def\@tempa {#1}\def\@tempb {#2}\def\@tempc
  {#3}\ifx \@tempc \@empty \let \@tempc \@tempb \let \@tempb \@tempa \fi \ifx
  \@tempb \@empty \def\@tempb {arXiv}\fi \@ifundefined
  {mn@eprint@\@tempb}{\@tempb:\@tempc}{\expandafter \expandafter \csname
  mn@eprint@\@tempb\endcsname \expandafter{\@tempc}}}

\bibitem[\protect\citeauthoryear{{ALMA Partnership} et~al.,}{{ALMA Partnership}
  et~al.}{2015a}]{alma2015}
{ALMA Partnership} et~al., 2015a, \mn@doi [\apjl] {10.1088/2041-8205/808/1/L3},
  \href {https://ui.adsabs.harvard.edu/abs/2015ApJ...808L...3A} {808, L3}

\bibitem[\protect\citeauthoryear{{ALMA Partnership} et~al.,}{{ALMA Partnership}
  et~al.}{2015b}]{vlahakis+15}
{ALMA Partnership} et~al., 2015b, \mn@doi [\apjl] {10.1088/2041-8205/808/1/L4},
  \href {https://ui.adsabs.harvard.edu/abs/2015ApJ...808L...4A} {808, L4}

\bibitem[\protect\citeauthoryear{{Ade}, {Beckman}  \& {Clark}}{{Ade}
  et~al.}{1971}]{ade1971}
{Ade} P.~A.~R.,  {Beckman} J.~E.,   {Clark} C.~D.,  1971, \mn@doi [Nature
  Physical Science] {10.1038/physci231055a0}, \href
  {https://ui.adsabs.harvard.edu/abs/1971NPhS..231...55A} {231, 55}

\bibitem[\protect\citeauthoryear{{Ade} et~al.,}{{Ade}
  et~al.}{2019}]{2019JCAP...02..056A}
{Ade} P.,  et~al., 2019, \mn@doi [\jcap] {10.1088/1475-7516/2019/02/056}, \href
  {https://ui.adsabs.harvard.edu/abs/2019JCAP...02..056A} {2019, 056}

\bibitem[\protect\citeauthoryear{{Akiyama}, {Kauffmann}, {Matthews},
  {Moriyama}, {Koyama}  \& {Hada}}{{Akiyama} et~al.}{2023}]{Akiyama2023}
{Akiyama} K.,  {Kauffmann} J.,  {Matthews} L.~D.,  {Moriyama} K.,  {Koyama} S.,
    {Hada} K.,  2023, \mn@doi [Galaxies] {10.3390/galaxies11010001}, \href
  {https://ui.adsabs.harvard.edu/abs/2023Galax..11....1A} {11, 1}

\bibitem[\protect\citeauthoryear{{Amblard} et~al.,}{{Amblard}
  et~al.}{2011}]{amblard+11}
{Amblard} A.,  et~al., 2011, \mn@doi [\nat] {10.1038/nature09771}, \href
  {https://ui.adsabs.harvard.edu/abs/2011Natur.470..510A} {470, 510}

\bibitem[\protect\citeauthoryear{{Anderson} et~al.,}{{Anderson}
  et~al.}{2021}]{Anderson+21}
{Anderson} M.,  et~al., 2021, \mn@doi [\mnras] {10.1093/mnras/stab2674}, \href
  {https://ui.adsabs.harvard.edu/abs/2021MNRAS.508.2964A} {508, 2964}

\bibitem[\protect\citeauthoryear{{Andersson}, {Lazarian}  \&
  {Vaillancourt}}{{Andersson} et~al.}{2015}]{andersson2015}
{Andersson} B.~G.,  {Lazarian} A.,   {Vaillancourt} J.~E.,  2015, \mn@doi
  [\araa] {10.1146/annurev-astro-082214-122414}, \href
  {https://ui.adsabs.harvard.edu/abs/2015ARA&A..53..501A} {53, 501}

\bibitem[\protect\citeauthoryear{{Andr{\'e}}, {Di Francesco}, {Ward-Thompson},
  {Inutsuka}, {Pudritz}  \& {Pineda}}{{Andr{\'e}} et~al.}{2014}]{Andre+14}
{Andr{\'e}} P.,  {Di Francesco} J.,  {Ward-Thompson} D.,  {Inutsuka} S.~I.,
  {Pudritz} R.~E.,   {Pineda} J.~E.,  2014, in {Beuther} H.,  {Klessen} R.~S.,
  {Dullemond} C.~P.,   {Henning} T.,  eds, Protostars and Planets VI. pp 27--51
  (\mn@eprint {arXiv} {1312.6232}),
  \mn@doi{10.2458/azu_uapress_9780816531240-ch002}

\bibitem[\protect\citeauthoryear{{Andrews} et~al.,}{{Andrews}
  et~al.}{2018}]{andrews2018}
{Andrews} S.~M.,  et~al., 2018, \mn@doi [\apjl] {10.3847/2041-8213/aaf741},
  \href {https://ui.adsabs.harvard.edu/abs/2018ApJ...869L..41A} {869, L41}

\bibitem[\protect\citeauthoryear{{Ansdell} et~al.,}{{Ansdell}
  et~al.}{2016}]{ansdell2016}
{Ansdell} M.,  et~al., 2016, \mn@doi [\apj] {10.3847/0004-637X/828/1/46}, \href
  {https://ui.adsabs.harvard.edu/abs/2016ApJ...828...46A} {828, 46}

\bibitem[\protect\citeauthoryear{{Ansdell} et~al.,}{{Ansdell}
  et~al.}{2018}]{ansdell2018}
{Ansdell} M.,  et~al., 2018, \mn@doi [\apj] {10.3847/1538-4357/aab890}, \href
  {https://ui.adsabs.harvard.edu/abs/2018ApJ...859...21A} {859, 21}

\bibitem[\protect\citeauthoryear{{Arribas} et~al.,}{{Arribas}
  et~al.}{2023}]{Arribas+23_z6.9_protocluster_SPT0311-58}
{Arribas} S.,  et~al., 2023, \mn@doi [arXiv e-prints]
  {10.48550/arXiv.2312.00899}, \href
  {https://ui.adsabs.harvard.edu/abs/2023arXiv231200899A} {p. arXiv:2312.00899}

\bibitem[\protect\citeauthoryear{{Arzoumanian} et~al.,}{{Arzoumanian}
  et~al.}{2021}]{Arzoumanian+21}
{Arzoumanian} D.,  et~al., 2021, \mn@doi [\aap] {10.1051/0004-6361/202038624},
  \href {https://ui.adsabs.harvard.edu/abs/2021A&A...647A..78A} {647, A78}

\bibitem[\protect\citeauthoryear{{Asboth} et~al.,}{{Asboth}
  et~al.}{2016}]{asboth+16_hermes_ultrared}
{Asboth} V.,  et~al., 2016, \mn@doi [\mnras] {10.1093/mnras/stw1769}, \href
  {https://ui.adsabs.harvard.edu/abs/2016MNRAS.462.1989A} {462, 1989}

\bibitem[\protect\citeauthoryear{{Backes} et~al.,}{{Backes}
  et~al.}{2016}]{backes2016}
{Backes} M.,  et~al., 2016, in The 4th Annual Conference on High Energy
  Astrophysics in Southern Africa (HEASA 2016). p.~29,
  \mn@doi{10.22323/1.275.0029}

\bibitem[\protect\citeauthoryear{{Bakx} et~al.,}{{Bakx} et~al.}{2018}]{bakx+18}
{Bakx} T. J.~L.~C.,  et~al., 2018, \mn@doi [\mnras] {10.1093/mnras/stx2267},
  \href {https://ui.adsabs.harvard.edu/abs/2018MNRAS.473.1751B} {473, 1751}

\bibitem[\protect\citeauthoryear{{Bakx} et~al.,}{{Bakx} et~al.}{2023}]{Bakx+23}
{Bakx} T. J.~L.~C.,  et~al., 2023, \mn@doi [\mnras] {10.1093/mnras/stac3723},
  \href {https://ui.adsabs.harvard.edu/abs/2023MNRAS.519.5076B} {519, 5076}

\bibitem[\protect\citeauthoryear{{Bakx}, {Gray}, {Gonz{\'a}lez-Nuevo},
  {Bonavera}, {Amvrosiadis}, {Eales}, {Hagimoto}  \& {Serjeant}}{{Bakx}
  et~al.}{2024a}]{bakx+24_flash}
{Bakx} T. J.~L.~C.,  {Gray} B.~S.,  {Gonz{\'a}lez-Nuevo} J.,  {Bonavera} L.,
  {Amvrosiadis} A.,  {Eales} S.,  {Hagimoto} M.,   {Serjeant} S.,  2024a,
  \mn@doi [\mnras] {10.1093/mnras/stad3759}, \href
  {https://ui.adsabs.harvard.edu/abs/2024MNRAS.527.8865B} {527, 8865}

\bibitem[\protect\citeauthoryear{{Bakx} et~al.,}{{Bakx}
  et~al.}{2024b}]{bakx+24_herbs70_protocluster_scuba2_noema}
{Bakx} T. J.~L.~C.,  et~al., 2024b, \mn@doi [\mnras] {10.1093/mnras/stae1155},
  \href {https://ui.adsabs.harvard.edu/abs/2024MNRAS.530.4578B} {530, 4578}

\bibitem[\protect\citeauthoryear{Baron \& Kloppenborg}{Baron \&
  Kloppenborg}{2010}]{Baron2010}
Baron F.,  Kloppenborg B.,  2010, \mn@doi [Proceedings of SPIE - The
  International Society for Optical Engineering] {10.1117/12.856713}, pp 139--

\bibitem[\protect\citeauthoryear{{Barry} et~al.,}{{Barry}
  et~al.}{2022}]{Barry2022}
{Barry} P.~S.,  et~al., 2022, \mn@doi [Journal of Low Temperature Physics]
  {10.1007/s10909-022-02843-4}, \href
  {https://ui.adsabs.harvard.edu/abs/2022JLTP..209..879B} {209, 879}

\bibitem[\protect\citeauthoryear{Basu~Thakur, Klimovich, Day, Shirokoff,
  Mauskopf, Faramarzi  \& Barry}{Basu~Thakur et~al.}{2020}]{SOFTS}
Basu~Thakur R.,  Klimovich N.,  Day P.~K.,  Shirokoff E.,  Mauskopf P.~D.,
  Faramarzi F.,   Barry P.~S.,  2020, \mn@doi [Journal of Low Temperature
  Physics] {10.1007/s10909-020-02490-7}, 200, 342–352

\bibitem[\protect\citeauthoryear{{Beeston} et~al.,}{{Beeston}
  et~al.}{2018}]{Beeston2018}
{Beeston} R.~A.,  et~al., 2018, \mn@doi [\mnras] {10.1093/mnras/sty1460}, \href
  {https://ui.adsabs.harvard.edu/abs/2018MNRAS.479.1077B} {479, 1077}

\bibitem[\protect\citeauthoryear{{Bendo} et~al.,}{{Bendo}
  et~al.}{2023}]{bears_2_dust}
{Bendo} G.~J.,  et~al., 2023, \mn@doi [\mnras] {10.1093/mnras/stac3771}, \href
  {https://ui.adsabs.harvard.edu/abs/2023MNRAS.522.2995B} {522, 2995}

\bibitem[\protect\citeauthoryear{{Benisty} et~al.,}{{Benisty}
  et~al.}{2021}]{benisty2021}
{Benisty} M.,  et~al., 2021, \mn@doi [\apjl] {10.3847/2041-8213/ac0f83}, \href
  {https://ui.adsabs.harvard.edu/abs/2021ApJ...916L...2B} {916, L2}

\bibitem[\protect\citeauthoryear{{Beno{\^\i}t} \& {ARCHEOPS
  Collaboration}}{{Beno{\^\i}t} \& {ARCHEOPS Collaboration}}{2004}]{Benoit2004}
{Beno{\^\i}t} A.,  {ARCHEOPS Collaboration} 2004, \mn@doi [Advances in Space
  Research] {10.1016/j.asr.2003.05.021}, \href
  {https://ui.adsabs.harvard.edu/abs/2004AdSpR..33.1790B} {33, 1790}

\bibitem[\protect\citeauthoryear{{Berta} et~al.,}{{Berta}
  et~al.}{2023}]{zgal_3_properties}
{Berta} S.,  et~al., 2023, \mn@doi [\aap] {10.1051/0004-6361/202346803}, \href
  {https://ui.adsabs.harvard.edu/abs/2023A&A...678A..28B} {678, A28}

\bibitem[\protect\citeauthoryear{{Biver} et~al.,}{{Biver}
  et~al.}{1999}]{biver1999}
{Biver} N.,  et~al., 1999, \mn@doi [\aj] {10.1086/301033}, \href
  {https://ui.adsabs.harvard.edu/abs/1999AJ....118.1850B} {118, 1850}

\bibitem[\protect\citeauthoryear{{Bogd{\'a}n} et~al.,}{{Bogd{\'a}n}
  et~al.}{2024}]{Bogdan+24_z10_xray}
{Bogd{\'a}n} {\'A}.,  et~al., 2024, \mn@doi [Nature Astronomy]
  {10.1038/s41550-023-02111-9}, \href
  {https://ui.adsabs.harvard.edu/abs/2024NatAs...8..126B} {8, 126}

\bibitem[\protect\citeauthoryear{{B{\o}gelund} \& {Hogerheijde}}{{B{\o}gelund}
  \& {Hogerheijde}}{2017}]{bogelund2017}
{B{\o}gelund} E.~G.,  {Hogerheijde} M.~R.,  2017, \mn@doi [\aap]
  {10.1051/0004-6361/201629197}, \href
  {https://ui.adsabs.harvard.edu/abs/2017A&A...604A.131B} {604, A131}

\bibitem[\protect\citeauthoryear{{Booth} et~al.,}{{Booth}
  et~al.}{2024}]{booth2024}
{Booth} M.,  et~al., 2024, \mn@doi [arXiv e-prints]
  {10.48550/arXiv.2407.01413}, \href
  {https://ui.adsabs.harvard.edu/abs/2024arXiv240701413B} {p. arXiv:2407.01413}

\bibitem[\protect\citeauthoryear{{Bouwens} et~al.,}{{Bouwens}
  et~al.}{2022}]{Bouwens+22_REBELS}
{Bouwens} R.~J.,  et~al., 2022, \mn@doi [\apj] {10.3847/1538-4357/ac5a4a},
  \href {https://ui.adsabs.harvard.edu/abs/2022ApJ...931..160B} {931, 160}

\bibitem[\protect\citeauthoryear{{Bouwens} et~al.,}{{Bouwens}
  et~al.}{2023}]{Bouwens+23}
{Bouwens} R.~J.,  et~al., 2023, \mn@doi [\mnras] {10.1093/mnras/stad1145},
  \href {https://ui.adsabs.harvard.edu/abs/2023MNRAS.523.1036B} {523, 1036}

\bibitem[\protect\citeauthoryear{{Boyden} \& {Eisner}}{{Boyden} \&
  {Eisner}}{2020}]{boyden2020}
{Boyden} R.~D.,  {Eisner} J.~A.,  2020, \mn@doi [\apj]
  {10.3847/1538-4357/ab86b7}, \href
  {https://ui.adsabs.harvard.edu/abs/2020ApJ...894...74B} {894, 74}

\bibitem[\protect\citeauthoryear{{Buckle} et~al.,}{{Buckle}
  et~al.}{2009}]{buckle2009}
{Buckle} J.~V.,  et~al., 2009, \mn@doi [\mnras]
  {10.1111/j.1365-2966.2009.15347.x}, \href
  {https://ui.adsabs.harvard.edu/abs/2009MNRAS.399.1026B} {399, 1026}

\bibitem[\protect\citeauthoryear{{Burgarella} et~al.,}{{Burgarella}
  et~al.}{2013}]{burgarella+13_hermes}
{Burgarella} D.,  et~al., 2013, \mn@doi [\aap] {10.1051/0004-6361/201321651},
  \href {https://ui.adsabs.harvard.edu/abs/2013A&A...554A..70B} {554, A70}

\bibitem[\protect\citeauthoryear{{Ca{\~n}ameras} et~al.,}{{Ca{\~n}ameras}
  et~al.}{2015}]{Planck_GEMS_1}
{Ca{\~n}ameras} R.,  et~al., 2015, \mn@doi [\aap]
  {10.1051/0004-6361/201425128}, \href
  {https://ui.adsabs.harvard.edu/abs/2015A&A...581A.105C} {581, A105}

\bibitem[\protect\citeauthoryear{{Ca{\~n}ameras} et~al.,}{{Ca{\~n}ameras}
  et~al.}{2017a}]{Planck_GEMS_3}
{Ca{\~n}ameras} R.,  et~al., 2017a, \mn@doi [\aap]
  {10.1051/0004-6361/201630359}, \href
  {https://ui.adsabs.harvard.edu/abs/2017A&A...600L...3C} {600, L3}

\bibitem[\protect\citeauthoryear{{Ca{\~n}ameras} et~al.,}{{Ca{\~n}ameras}
  et~al.}{2017b}]{Planck_GEMS_4}
{Ca{\~n}ameras} R.,  et~al., 2017b, \mn@doi [\aap]
  {10.1051/0004-6361/201630186}, \href
  {https://ui.adsabs.harvard.edu/abs/2017A&A...604A.117C} {604, A117}

\bibitem[\protect\citeauthoryear{{Ca{\~n}ameras} et~al.,}{{Ca{\~n}ameras}
  et~al.}{2018a}]{Planck_GEMS_5}
{Ca{\~n}ameras} R.,  et~al., 2018a, \mn@doi [\aap]
  {10.1051/0004-6361/201833679}, \href
  {https://ui.adsabs.harvard.edu/abs/2018A&A...620A..60C} {620, A60}

\bibitem[\protect\citeauthoryear{{Ca{\~n}ameras} et~al.,}{{Ca{\~n}ameras}
  et~al.}{2018b}]{Planck_GEMS_6}
{Ca{\~n}ameras} R.,  et~al., 2018b, \mn@doi [\aap]
  {10.1051/0004-6361/201833625}, \href
  {https://ui.adsabs.harvard.edu/abs/2018A&A...620A..61C} {620, A61}

\bibitem[\protect\citeauthoryear{{Ca{\~n}ameras} et~al.,}{{Ca{\~n}ameras}
  et~al.}{2021}]{Planck_GEMS_8}
{Ca{\~n}ameras} R.,  et~al., 2021, \mn@doi [\aap]
  {10.1051/0004-6361/202038979}, \href
  {https://ui.adsabs.harvard.edu/abs/2021A&A...645A..45C} {645, A45}

\bibitem[\protect\citeauthoryear{{Calahan} et~al.,}{{Calahan}
  et~al.}{2021}]{calahan2021}
{Calahan} J.~K.,  et~al., 2021, \mn@doi [\apjs] {10.3847/1538-4365/ac143f},
  \href {https://ui.adsabs.harvard.edu/abs/2021ApJS..257...17C} {257, 17}

\bibitem[\protect\citeauthoryear{{Calvi}, {Castignani}  \&
  {Dannerbauer}}{{Calvi}
  et~al.}{2023}]{calvi+23_bright_smgs_signpost_protoclusters_goodsn}
{Calvi} R.,  {Castignani} G.,   {Dannerbauer} H.,  2023, \mn@doi [\aap]
  {10.1051/0004-6361/202346200}, \href
  {https://ui.adsabs.harvard.edu/abs/2023A&A...678A..15C} {678, A15}

\bibitem[\protect\citeauthoryear{{Calvo} et~al.,}{{Calvo}
  et~al.}{2016}]{Calvo2016}
{Calvo} M.,  et~al., 2016, \mn@doi [Journal of Low Temperature Physics]
  {10.1007/s10909-016-1582-0}, \href
  {https://ui.adsabs.harvard.edu/abs/2016JLTP..184..816C} {184, 816}

\bibitem[\protect\citeauthoryear{{Carlstrom} et~al.,}{{Carlstrom}
  et~al.}{2011}]{carlstrom2011}
{Carlstrom} J.~E.,  et~al., 2011, \mn@doi [\pasp] {10.1086/659879}, \href
  {https://ui.adsabs.harvard.edu/abs/2011PASP..123..568C} {123, 568}

\bibitem[\protect\citeauthoryear{{Carpenter}, {Iono}, {Kemper}  \&
  {Wootten}}{{Carpenter} et~al.}{2020}]{2020arXiv200111076C}
{Carpenter} J.,  {Iono} D.,  {Kemper} F.,   {Wootten} A.,  2020, \mn@doi [arXiv
  e-prints] {10.48550/arXiv.2001.11076}, \href
  {https://ui.adsabs.harvard.edu/abs/2020arXiv200111076C} {p. arXiv:2001.11076}

\bibitem[\protect\citeauthoryear{{Carpenter}, {Brogan}, {Iono}  \&
  {Mroczkowski}}{{Carpenter} et~al.}{2023}]{2023pcsf.conf..304C}
{Carpenter} J.,  {Brogan} C.,  {Iono} D.,   {Mroczkowski} T.,  2023, in
  {Ossenkopf-Okada} V.,  {Schaaf} R.,  {Breloy} I.,   {Stutzki} J.,  eds,
  Physics and Chemistry of Star Formation: The Dynamical ISM Across Time and
  Spatial Scales. p.~304 (\mn@eprint {arXiv} {2211.00195}),
  \mn@doi{10.48550/arXiv.2211.00195}

\bibitem[\protect\citeauthoryear{{Casey}, {Narayanan}  \& {Cooray}}{{Casey}
  et~al.}{2014}]{casey+14}
{Casey} C.~M.,  {Narayanan} D.,   {Cooray} A.,  2014, \mn@doi [\physrep]
  {10.1016/j.physrep.2014.02.009}, \href
  {https://ui.adsabs.harvard.edu/abs/2014PhR...541...45C} {541, 45}

\bibitem[\protect\citeauthoryear{{Castillo-Dominguez}
  et~al.,}{{Castillo-Dominguez} et~al.}{2018}]{Muscat2018}
{Castillo-Dominguez} E.,  et~al., 2018, \mn@doi [Journal of Low Temperature
  Physics] {10.1007/s10909-018-2018-9}, \href
  {https://ui.adsabs.harvard.edu/abs/2018JLTP..193.1010C} {193, 1010}

\bibitem[\protect\citeauthoryear{{Chael} et~al.,}{{Chael}
  et~al.}{2019}]{Chael2019}
{Chael} A.,  et~al., 2019, {eht-imaging: v1.1.0: Imaging interferometric data
  with regularized maximum likelihood}, \mn@doi{10.5281/zenodo.2614016}

\bibitem[\protect\citeauthoryear{{Chawner} et~al.,}{{Chawner}
  et~al.}{2019}]{Chawner2019}
{Chawner} H.,  et~al., 2019, \mn@doi [\mnras] {10.1093/mnras/sty2942}, \href
  {https://ui.adsabs.harvard.edu/abs/2019MNRAS.483...70C} {483, 70}

\bibitem[\protect\citeauthoryear{{Chen} et~al.,}{{Chen} et~al.}{2015}]{chen+15}
{Chen} C.-C.,  et~al., 2015, \mn@doi [\apj] {10.1088/0004-637X/799/2/194},
  \href {https://ui.adsabs.harvard.edu/abs/2015ApJ...799..194C} {799, 194}

\bibitem[\protect\citeauthoryear{{Cheng} et~al.,}{{Cheng}
  et~al.}{2019}]{Cheng+19_Planck_SCUBA2_images}
{Cheng} T.,  et~al., 2019, \mn@doi [\mnras] {10.1093/mnras/stz2640}, \href
  {https://ui.adsabs.harvard.edu/abs/2019MNRAS.490.3840C} {490, 3840}

\bibitem[\protect\citeauthoryear{{Cheng} et~al.,}{{Cheng}
  et~al.}{2020}]{Cheng+20_Planck_SCUBA2_overdensities}
{Cheng} T.,  et~al., 2020, \mn@doi [\mnras] {10.1093/mnras/staa1096}, \href
  {https://ui.adsabs.harvard.edu/abs/2020MNRAS.494.5985C} {494, 5985}

\bibitem[\protect\citeauthoryear{{Chiang}, {Overzier}, {Gebhardt}  \&
  {Henriques}}{{Chiang} et~al.}{2017}]{Chiang+17_protoclusters}
{Chiang} Y.-K.,  {Overzier} R.~A.,  {Gebhardt} K.,   {Henriques} B.,  2017,
  \mn@doi [\apjl] {10.3847/2041-8213/aa7e7b}, \href
  {https://ui.adsabs.harvard.edu/abs/2017ApJ...844L..23C} {844, L23}

\bibitem[\protect\citeauthoryear{{Chin} et~al.,}{{Chin}
  et~al.}{2024}]{chen2024}
{Chin} G.,  et~al., 2024, \mn@doi [arXiv e-prints] {10.48550/arXiv.2405.12829},
  \href {https://ui.adsabs.harvard.edu/abs/2024arXiv240512829C} {p.
  arXiv:2405.12829}

\bibitem[\protect\citeauthoryear{{Chyba} \& {Sagan}}{{Chyba} \&
  {Sagan}}{1992}]{chyba1992}
{Chyba} C.,  {Sagan} C.,  1992, \mn@doi [\nat] {10.1038/355125a0}, \href
  {https://ui.adsabs.harvard.edu/abs/1992Natur.355..125C} {355, 125}

\bibitem[\protect\citeauthoryear{{Cockell}, {McMahon}  \& {Biddle}}{{Cockell}
  et~al.}{2021}]{cockell2021}
{Cockell} C.~S.,  {McMahon} S.,   {Biddle} J.~F.,  2021, \mn@doi [Astrobiology]
  {10.1089/ast.2020.2390}, \href
  {https://ui.adsabs.harvard.edu/abs/2021AsBio..21..261C} {21, 261}

\bibitem[\protect\citeauthoryear{{Colombo} et~al.,}{{Colombo}
  et~al.}{2019}]{Colombo19}
{Colombo} D.,  et~al., 2019, \mn@doi [\mnras] {10.1093/mnras/sty3283}, \href
  {https://ui.adsabs.harvard.edu/abs/2019MNRAS.483.4291C} {483, 4291}

\bibitem[\protect\citeauthoryear{{Colombo} et~al.,}{{Colombo}
  et~al.}{2022}]{Colombo22}
{Colombo} D.,  et~al., 2022, \mn@doi [\aap] {10.1051/0004-6361/202141287},
  \href {https://ui.adsabs.harvard.edu/abs/2022A&A...658A..54C} {658, A54}

\bibitem[\protect\citeauthoryear{{Coppi} et~al.,}{{Coppi}
  et~al.}{2020}]{Coppi2020}
{Coppi} G.,  et~al., 2020, in {Marshall} H.~K.,  {Spyromilio} J.,   {Usuda} T.,
   eds,  Society of Photo-Optical Instrumentation Engineers (SPIE) Conference
  Series Vol. 11445, Ground-based and Airborne Telescopes VIII. p. 1144526
  (\mn@eprint {arXiv} {2012.01039}), \mn@doi{10.1117/12.2560849}

\bibitem[\protect\citeauthoryear{{Cordiner} et~al.,}{{Cordiner}
  et~al.}{2014}]{cordiner2014}
{Cordiner} M.~A.,  et~al., 2014, \mn@doi [\apjl] {10.1088/2041-8205/792/1/L2},
  \href {https://ui.adsabs.harvard.edu/abs/2014ApJ...792L...2C} {792, L2}

\bibitem[\protect\citeauthoryear{{Cordiner} et~al.,}{{Cordiner}
  et~al.}{2015}]{cordiner2015}
{Cordiner} M.~A.,  et~al., 2015, \mn@doi [\apjl] {10.1088/2041-8205/800/1/L14},
  \href {https://ui.adsabs.harvard.edu/abs/2015ApJ...800L..14C} {800, L14}

\bibitem[\protect\citeauthoryear{{Cordiner}, {Teanby}, {Nixon}, {Vuitton},
  {Thelen}  \& {Charnley}}{{Cordiner} et~al.}{2019}]{cordiner2019}
{Cordiner} M.~A.,  {Teanby} N.~A.,  {Nixon} C.~A.,  {Vuitton} V.,  {Thelen}
  A.~E.,   {Charnley} S.~B.,  2019, \mn@doi [\aj] {10.3847/1538-3881/ab2d20},
  \href {https://ui.adsabs.harvard.edu/abs/2019AJ....158...76C} {158, 76}

\bibitem[\protect\citeauthoryear{{Cordiner} et~al.,}{{Cordiner}
  et~al.}{2022}]{cordiner2022}
{Cordiner} M.~A.,  et~al., 2022, \mn@doi [\grl] {10.1029/2022GL101055}, \href
  {https://ui.adsabs.harvard.edu/abs/2022GeoRL..4901055C} {49, e2022GL101055}

\bibitem[\protect\citeauthoryear{{Cordiner} et~al.,}{{Cordiner}
  et~al.}{2024}]{cordiner2024}
{Cordiner} M.~A.,  et~al., 2024, \mn@doi [Open Res Europe]
  {10.12688/openreseurope.17473.1}, \href
  {https://ui.adsabs.harvard.edu/abs/2024arXiv240302258C} {4, 78}

\bibitem[\protect\citeauthoryear{{Cornish} et~al.,}{{Cornish}
  et~al.}{2024}]{RAGERS_Cornish}
{Cornish} T.~M.,  et~al., 2024, \mn@doi [arXiv e-prints]
  {10.48550/arXiv.2407.21099}, \href
  {https://ui.adsabs.harvard.edu/abs/2024arXiv240721099C} {p. arXiv:2407.21099}

\bibitem[\protect\citeauthoryear{{Cortese} et~al.,}{{Cortese}
  et~al.}{2012}]{Cortese2012}
{Cortese} L.,  et~al., 2012, \mn@doi [\aap] {10.1051/0004-6361/201118499},
  \href {https://ui.adsabs.harvard.edu/abs/2012A&A...540A..52C} {540, A52}

\bibitem[\protect\citeauthoryear{{Coulson} et~al.,}{{Coulson}
  et~al.}{2020}]{coulson2020}
{Coulson} I.~M.,  et~al., 2020, \mn@doi [\aj] {10.3847/1538-3881/abafc0}, \href
  {https://ui.adsabs.harvard.edu/abs/2020AJ....160..182C} {160, 182}

\bibitem[\protect\citeauthoryear{{Cowie}, {Gonz{\'a}lez-L{\'o}pez}, {Barger},
  {Bauer}, {Hsu}  \& {Wang}}{{Cowie} et~al.}{2018}]{cowie+18}
{Cowie} L.~L.,  {Gonz{\'a}lez-L{\'o}pez} J.,  {Barger} A.~J.,  {Bauer} F.~E.,
  {Hsu} L.~Y.,   {Wang} W.~H.,  2018, \mn@doi [\apj]
  {10.3847/1538-4357/aadc63}, \href
  {https://ui.adsabs.harvard.edu/abs/2018ApJ...865..106C} {865, 106}

\bibitem[\protect\citeauthoryear{{Cox} et~al.,}{{Cox} et~al.}{2023}]{Cox+23}
{Cox} P.,  et~al., 2023, \mn@doi [\aap] {10.1051/0004-6361/202346801}, \href
  {https://ui.adsabs.harvard.edu/abs/2023A&A...678A..26C} {678, A26}

\bibitem[\protect\citeauthoryear{{Cronin-Coltsmann} et~al.,}{{Cronin-Coltsmann}
  et~al.}{2021}]{2021CroninColtsmann}
{Cronin-Coltsmann} P.~F.,  et~al., 2021, \mn@doi [\mnras]
  {10.1093/mnras/stab1237}, \href
  {https://ui.adsabs.harvard.edu/abs/2021MNRAS.504.4497C} {504, 4497}

\bibitem[\protect\citeauthoryear{{Cuadrado-Calle} et~al.,}{{Cuadrado-Calle}
  et~al.}{2017}]{2017ITMTT..65.1589C}
{Cuadrado-Calle} D.,  et~al., 2017, \mn@doi [IEEE Transactions on Microwave
  Theory Techniques] {10.1109/TMTT.2016.2639018}, \href
  {https://ui.adsabs.harvard.edu/abs/2017ITMTT..65.1589C} {65, 1589}

\bibitem[\protect\citeauthoryear{{Cunningham}, {Gear}, {Duncan}, {Hastings}  \&
  {Holland}}{{Cunningham} et~al.}{1994}]{cunningham1994}
{Cunningham} C.~R.,  {Gear} W.~K.,  {Duncan} W.~D.,  {Hastings} P.~R.,
  {Holland} W.~S.,  1994, in {Crawford} D.~L.,  {Craine} E.~R.,  eds,  Society
  of Photo-Optical Instrumentation Engineers (SPIE) Conference Series Vol.
  2198, Instrumentation in Astronomy VIII. pp 638--649,
  \mn@doi{10.1117/12.176772}

\bibitem[\protect\citeauthoryear{{Dannerbauer} et~al.,}{{Dannerbauer}
  et~al.}{2014}]{Dannerbauer+14_spiderweb}
{Dannerbauer} H.,  et~al., 2014, \mn@doi [\aap] {10.1051/0004-6361/201423771},
  \href {https://ui.adsabs.harvard.edu/abs/2014A&A...570A..55D} {570, A55}

\bibitem[\protect\citeauthoryear{{Davis} et~al.,}{{Davis}
  et~al.}{2022}]{2022Davis}
{Davis} T.~A.,  et~al., 2022, \mn@doi [\mnras] {10.1093/mnras/stac600}, \href
  {https://ui.adsabs.harvard.edu/abs/2022MNRAS.512.1522D} {512, 1522}

\bibitem[\protect\citeauthoryear{{Dayal}}{{Dayal}}{2024}]{dayal_2024_smbh_jwst_highz_models}
{Dayal} P.,  2024, arXiv e-prints, \href
  {https://ui.adsabs.harvard.edu/abs/2024arXiv240707162D} {p. arXiv:2407.07162}

\bibitem[\protect\citeauthoryear{{De Looze} et~al.,}{{De Looze}
  et~al.}{2019}]{DeLooze2019}
{De Looze} I.,  et~al., 2019, \mn@doi [\mnras] {10.1093/mnras/stz1533}, \href
  {https://ui.adsabs.harvard.edu/abs/2019MNRAS.488..164D} {488, 164}

\bibitem[\protect\citeauthoryear{{Decin}, {Danilovich}, {Gobrecht}, {Plane},
  {Richards}, {Gottlieb}  \& {Lee}}{{Decin} et~al.}{2018}]{Decin2018}
{Decin} L.,  {Danilovich} T.,  {Gobrecht} D.,  {Plane} J.~M.~C.,  {Richards}
  A.~M.~S.,  {Gottlieb} C.~A.,   {Lee} K.~L.~K.,  2018, \mn@doi [\apj]
  {10.3847/1538-4357/aaab6a}, \href
  {https://ui.adsabs.harvard.edu/abs/2018ApJ...855..113D} {855, 113}

\bibitem[\protect\citeauthoryear{{Decin} et~al.,}{{Decin}
  et~al.}{2020}]{Decin2020}
{Decin} L.,  et~al., 2020, \mn@doi [Science] {10.1126/science.abb1229}, \href
  {https://ui.adsabs.harvard.edu/abs/2020Sci...369.1497D} {369, 1497}

\bibitem[\protect\citeauthoryear{{Despali}, {Vegetti}, {White}, {Giocoli}  \&
  {van den Bosch}}{{Despali} et~al.}{2018}]{Despali+18}
{Despali} G.,  {Vegetti} S.,  {White} S. D.~M.,  {Giocoli} C.,   {van den
  Bosch} F.~C.,  2018, \mn@doi [\mnras] {10.1093/mnras/sty159}, \href
  {https://ui.adsabs.harvard.edu/abs/2018MNRAS.475.5424D} {475, 5424}

\bibitem[\protect\citeauthoryear{{Dharmawardena} et~al.,}{{Dharmawardena}
  et~al.}{2018}]{Dharmawardena2018}
{Dharmawardena} T.~E.,  et~al., 2018, \mn@doi [\mnras] {10.1093/mnras/sty1422},
  \href {https://ui.adsabs.harvard.edu/abs/2018MNRAS.479..536D} {479, 536}

\bibitem[\protect\citeauthoryear{{Dharmawardena} et~al.,}{{Dharmawardena}
  et~al.}{2019}]{Dharmawardena2019}
{Dharmawardena} T.~E.,  et~al., 2019, \mn@doi [\mnras] {10.1093/mnras/stz2334},
  \href {https://ui.adsabs.harvard.edu/abs/2019MNRAS.489.3218D} {489, 3218}

\bibitem[\protect\citeauthoryear{{Di Francesco}, {Evans}, {Caselli}, {Myers},
  {Shirley}, {Aikawa}  \& {Tafalla}}{{Di Francesco}
  et~al.}{2007}]{difrancesco2007}
{Di Francesco} J.,  {Evans} N.~J. I.,  {Caselli} P.,  {Myers} P.~C.,  {Shirley}
  Y.,  {Aikawa} Y.,   {Tafalla} M.,  2007, in {Reipurth} B.,  {Jewitt} D.,
  {Keil} K.,  eds, Protostars and Planets V. p.~17 (\mn@eprint {arXiv}
  {astro-ph/0602379}), \mn@doi{10.48550/arXiv.astro-ph/0602379}

\bibitem[\protect\citeauthoryear{{Di Mascolo} et~al.,}{{Di Mascolo}
  et~al.}{2023}]{DiMascolo+23_spiderweb_SZ}
{Di Mascolo} L.,  et~al., 2023, \mn@doi [\nat] {10.1038/s41586-023-05761-x},
  \href {https://ui.adsabs.harvard.edu/abs/2023Natur.615..809D} {615, 809}

\bibitem[\protect\citeauthoryear{{Disney} \& {Wallace}}{{Disney} \&
  {Wallace}}{1982}]{starlink}
{Disney} M.~J.,  {Wallace} P.~T.,  1982, \qjras, \href
  {https://ui.adsabs.harvard.edu/abs/1982QJRAS..23..485D} {23, 485}

\bibitem[\protect\citeauthoryear{{Donnan} et~al.,}{{Donnan}
  et~al.}{2024}]{donnan+24_primer}
{Donnan} C.~T.,  et~al., 2024, \mn@doi [arXiv e-prints]
  {10.48550/arXiv.2403.03171}, \href
  {https://ui.adsabs.harvard.edu/abs/2024arXiv240303171D} {p. arXiv:2403.03171}

\bibitem[\protect\citeauthoryear{{Doyle}, {Mauskopf}, {Naylon}, {Porch}  \&
  {Duncombe}}{{Doyle} et~al.}{2008}]{LEKID}
{Doyle} S.,  {Mauskopf} P.,  {Naylon} J.,  {Porch} A.,   {Duncombe} C.,  2008,
  \mn@doi [Journal of Low Temperature Physics] {10.1007/s10909-007-9685-2},
  \href {https://ui.adsabs.harvard.edu/abs/2008JLTP..151..530D} {151, 530}

\bibitem[\protect\citeauthoryear{{Drabek-Maunder}, {Greaves}, {Fraser},
  {Clements}  \& {Alconcel}}{{Drabek-Maunder} et~al.}{2019}]{drabekmaunder2019}
{Drabek-Maunder} E.,  {Greaves} J.,  {Fraser} H.~J.,  {Clements} D.~L.,
  {Alconcel} L.~N.,  2019, \mn@doi [International Journal of Astrobiology]
  {10.1017/S1473550417000428}, \href
  {https://ui.adsabs.harvard.edu/abs/2019IJAsB..18...25D} {18, 25}

\bibitem[\protect\citeauthoryear{{Duarte-Cabral} et~al.,}{{Duarte-Cabral}
  et~al.}{2021}]{Duarte-Cabral+21}
{Duarte-Cabral} A.,  et~al., 2021, \mn@doi [MNRAS] {10.1093/mnras/staa2480},
  500, 3027

\bibitem[\protect\citeauthoryear{{Dulk}}{{Dulk}}{1985}]{1985ARA&A..23..169D}
{Dulk} G.~A.,  1985, \mn@doi [\araa] {10.1146/annurev.aa.23.090185.001125},
  \href {https://ui.adsabs.harvard.edu/abs/1985ARA&A..23..169D} {23, 169}

\bibitem[\protect\citeauthoryear{{Dullemond}, {Isella}, {Andrews}, {Skobleva}
  \& {Dzyurkevich}}{{Dullemond} et~al.}{2020}]{dullemond2020}
{Dullemond} C.~P.,  {Isella} A.,  {Andrews} S.~M.,  {Skobleva} I.,
  {Dzyurkevich} N.,  2020, \mn@doi [\aap] {10.1051/0004-6361/201936438}, \href
  {https://ui.adsabs.harvard.edu/abs/2020A&A...633A.137D} {633, A137}

\bibitem[\protect\citeauthoryear{{Duncan}, {Robson}, {Ade}, {Sandell}  \&
  {Griffin}}{{Duncan} et~al.}{1990}]{duncan1990}
{Duncan} W.~D.,  {Robson} E.~I.,  {Ade} P.~A.~R.,  {Sandell} G.,   {Griffin}
  M.~J.,  1990, \mn@doi [\mnras] {10.1093/mnras/243.1.126}, \href
  {https://ui.adsabs.harvard.edu/abs/1990MNRAS.243..126D} {243, 126}

\bibitem[\protect\citeauthoryear{{Dunlop} et~al.,}{{Dunlop}
  et~al.}{2021}]{dunlop+21_primer}
{Dunlop} J.~S.,  et~al., 2021, {PRIMER: Public Release IMaging for
  Extragalactic Research}, JWST Proposal. Cycle 1, ID. \#1837

\bibitem[\protect\citeauthoryear{{Dunne}, {Maddox}, {Vlahakis}  \&
  {Gomez}}{{Dunne} et~al.}{2021}]{dunne+21_co_dark}
{Dunne} L.,  {Maddox} S.~J.,  {Vlahakis} C.,   {Gomez} H.~L.,  2021, \mn@doi
  [\mnras] {10.1093/mnras/staa3526}, \href
  {https://ui.adsabs.harvard.edu/abs/2021MNRAS.501.2573D} {501, 2573}

\bibitem[\protect\citeauthoryear{{Dye} et~al.,}{{Dye} et~al.}{2022}]{dye+22}
{Dye} S.,  et~al., 2022, \mn@doi [\mnras] {10.1093/mnras/stab3569}, \href
  {https://ui.adsabs.harvard.edu/abs/2022MNRAS.510.3734D} {510, 3734}

\bibitem[\protect\citeauthoryear{{Eales} \& {Ward}}{{Eales} \&
  {Ward}}{2024}]{Eales+24}
{Eales} S.,  {Ward} B.,  2024, \mn@doi [\mnras] {10.1093/mnras/stae403}, \href
  {https://ui.adsabs.harvard.edu/abs/2024MNRAS.529.1130E} {529, 1130}

\bibitem[\protect\citeauthoryear{{Eales} et~al.,}{{Eales}
  et~al.}{2010}]{eales+10}
{Eales} S.,  et~al., 2010, \mn@doi [\pasp] {10.1086/653086}, \href
  {https://ui.adsabs.harvard.edu/abs/2010PASP..122..499E} {122, 499}

\bibitem[\protect\citeauthoryear{{Eales} et~al.,}{{Eales}
  et~al.}{2018a}]{Eales+18_paradigm}
{Eales} S.,  et~al., 2018a, \mn@doi [\mnras] {10.1093/mnras/stx2548}, \href
  {https://ui.adsabs.harvard.edu/abs/2018MNRAS.473.3507E} {473, 3507}

\bibitem[\protect\citeauthoryear{{Eales} et~al.,}{{Eales}
  et~al.}{2018b}]{Eales+18_green_mountain}
{Eales} S.~A.,  et~al., 2018b, \mn@doi [\mnras] {10.1093/mnras/sty2220}, \href
  {https://ui.adsabs.harvard.edu/abs/2018MNRAS.481.1183E} {481, 1183}

\bibitem[\protect\citeauthoryear{{Eden} et~al.,}{{Eden} et~al.}{2020}]{Eden20}
{Eden} D.~J.,  et~al., 2020, \mn@doi [\mnras] {10.1093/mnras/staa2734}, \href
  {https://ui.adsabs.harvard.edu/abs/2020MNRAS.498.5936E} {498, 5936}

\bibitem[\protect\citeauthoryear{{Eden} et~al.,}{{Eden} et~al.}{2021}]{Eden21}
{Eden} D.~J.,  et~al., 2021, \mn@doi [\mnras] {10.1093/mnras/staa3188}, \href
  {https://ui.adsabs.harvard.edu/abs/2021MNRAS.500..191E} {500, 191}

\bibitem[\protect\citeauthoryear{{Edmunds}}{{Edmunds}}{2024}]{Edmunds2024}
{Edmunds} M.,  2024, Astronomy and Geophysics, 65, in press

\bibitem[\protect\citeauthoryear{{Eftekhari} et~al.,}{{Eftekhari}
  et~al.}{2022}]{2022ApJ...935...16E}
{Eftekhari} T.,  et~al., 2022, \mn@doi [\apj] {10.3847/1538-4357/ac7ce8}, \href
  {https://ui.adsabs.harvard.edu/abs/2022ApJ...935...16E} {935, 16}

\bibitem[\protect\citeauthoryear{{Eisenstein} et~al.,}{{Eisenstein}
  et~al.}{2023}]{JADES_overview_Eisenstein+23}
{Eisenstein} D.~J.,  et~al., 2023, \mn@doi [arXiv e-prints]
  {10.48550/arXiv.2306.02465}, \href
  {https://ui.adsabs.harvard.edu/abs/2023arXiv230602465E} {p. arXiv:2306.02465}

\bibitem[\protect\citeauthoryear{{Elbaz} et~al.,}{{Elbaz}
  et~al.}{2018}]{Elbaz+18}
{Elbaz} D.,  et~al., 2018, \mn@doi [\aap] {10.1051/0004-6361/201732370}, \href
  {https://ui.adsabs.harvard.edu/abs/2018A&A...616A.110E} {616, A110}

\bibitem[\protect\citeauthoryear{{Emonts} et~al.,}{{Emonts}
  et~al.}{2023}]{Emonts+21_CI_4c41.17}
{Emonts} B. H.~C.,  et~al., 2023, \mn@doi [Science] {10.1126/science.abh2150},
  \href {https://ui.adsabs.harvard.edu/abs/2023Sci...379.1323E} {379, 1323}

\bibitem[\protect\citeauthoryear{{Enoch}, {Evans}, {Sargent}  \&
  {Glenn}}{{Enoch} et~al.}{2009}]{enoch2009}
{Enoch} M.~L.,  {Evans} Neal~J. I.,  {Sargent} A.~I.,   {Glenn} J.,  2009,
  \mn@doi [\apj] {10.1088/0004-637X/692/2/973}, \href
  {https://ui.adsabs.harvard.edu/abs/2009ApJ...692..973E} {692, 973}

\bibitem[\protect\citeauthoryear{{Essinger-Hileman} et~al.,}{{Essinger-Hileman}
  et~al.}{2014}]{essinger2014}
{Essinger-Hileman} T.,  et~al., 2014, in {Holland} W.~S.,  {Zmuidzinas} J.,
  eds,  Society of Photo-Optical Instrumentation Engineers (SPIE) Conference
  Series Vol. 9153, Millimeter, Submillimeter, and Far-Infrared Detectors and
  Instrumentation for Astronomy VII. p. 91531I (\mn@eprint {arXiv}
  {1408.4788}), \mn@doi{10.1117/12.2056701}

\bibitem[\protect\citeauthoryear{{Event Horizon Telescope Collaboration}
  et~al.,}{{Event Horizon Telescope Collaboration} et~al.}{2019}]{eht2019}
{Event Horizon Telescope Collaboration} et~al., 2019, \mn@doi [\apjl]
  {10.3847/2041-8213/ab0ec7}, \href
  {https://ui.adsabs.harvard.edu/abs/2019ApJ...875L...1E} {875, L1}

\bibitem[\protect\citeauthoryear{{Event Horizon Telescope Collaboration}
  et~al.,}{{Event Horizon Telescope Collaboration} et~al.}{2022}]{eht2022}
{Event Horizon Telescope Collaboration} et~al., 2022, \mn@doi [\apjl]
  {10.3847/2041-8213/ac6674}, \href
  {https://ui.adsabs.harvard.edu/abs/2022ApJ...930L..12E} {930, L12}

\bibitem[\protect\citeauthoryear{{Fairhurst}, {Burdin}, {Daw}, {Kormos},
  {Lapington}, {McCabe}  \& {Sherwin}}{{Fairhurst}
  et~al.}{2022}]{STFC_PAAP_Roadmap_2022}
{Fairhurst} S.,  {Burdin} S.,  {Daw} E.,  {Kormos} L.,  {Lapington} J.,
  {McCabe} C.,   {Sherwin} B.,  2022, {Roadmap for UK Particle Astrophysics
  2022}, \url
  {https://www.ukri.org/publications/particle-astrophysics-advisory-board-roadmap-2022/}

\bibitem[\protect\citeauthoryear{{Fern{\'a}ndez-L{\'o}pez}
  et~al.,}{{Fern{\'a}ndez-L{\'o}pez} et~al.}{2021}]{fernandezlopez2021}
{Fern{\'a}ndez-L{\'o}pez} M.,  et~al., 2021, \mn@doi [\apj]
  {10.3847/1538-4357/abf2b6}, \href
  {https://ui.adsabs.harvard.edu/abs/2021ApJ...913...29F} {913, 29}

\bibitem[\protect\citeauthoryear{{Fischer}, {Hillenbrand}, {Herczeg},
  {Johnstone}, {Kospal}  \& {Dunham}}{{Fischer} et~al.}{2023}]{fischer2023}
{Fischer} W.~J.,  {Hillenbrand} L.~A.,  {Herczeg} G.~J.,  {Johnstone} D.,
  {Kospal} A.,   {Dunham} M.~M.,  2023, in {Inutsuka} S.,  {Aikawa} Y.,  {Muto}
  T.,  {Tomida} K.,   {Tamura} M.,  eds,  Astronomical Society of the Pacific
  Conference Series Vol. 534, Protostars and Planets VII. p.~355 (\mn@eprint
  {arXiv} {2203.11257}), \mn@doi{10.48550/arXiv.2203.11257}

\bibitem[\protect\citeauthoryear{{Fleishman} \& {Kontar}}{{Fleishman} \&
  {Kontar}}{2010}]{2010ApJ...709L.127F}
{Fleishman} G.~D.,  {Kontar} E.~P.,  2010, \mn@doi [\apjl]
  {10.1088/2041-8205/709/2/L127}, \href
  {https://ui.adsabs.harvard.edu/abs/2010ApJ...709L.127F} {709, L127}

\bibitem[\protect\citeauthoryear{{Fleishman}, {Martinez Oliveros}, {Landi}  \&
  {Glesener}}{{Fleishman} et~al.}{2022}]{fleishman2022}
{Fleishman} G.~D.,  {Martinez Oliveros} J.~C.,  {Landi} E.,   {Glesener} L.,
  2022, \mn@doi [Frontiers in Astronomy and Space Sciences]
  {10.3389/fspas.2022.966444}, \href
  {https://ui.adsabs.harvard.edu/abs/2022FrASS...9.6444F} {9, 966444}

\bibitem[\protect\citeauthoryear{{Flett} \& {Murray}}{{Flett} \&
  {Murray}}{1991}]{flett1991}
{Flett} A.~M.,  {Murray} A.~G.,  1991, \mn@doi [\mnras]
  {10.1093/mnras/249.1.4P}, \href
  {https://ui.adsabs.harvard.edu/abs/1991MNRAS.249P...4F} {249, 4P}

\bibitem[\protect\citeauthoryear{{Foreman-Mackey}, {Hogg}, {Lang}  \&
  {Goodman}}{{Foreman-Mackey} et~al.}{2013}]{emcee}
{Foreman-Mackey} D.,  {Hogg} D.~W.,  {Lang} D.,   {Goodman} J.,  2013, \mn@doi
  [\pasp] {10.1086/670067}, \href
  {https://ui.adsabs.harvard.edu/abs/2013PASP..125..306F} {125, 306}

\bibitem[\protect\citeauthoryear{{Friberg}, {Bastien}, {Berry}, {Savini},
  {Graves}  \& {Pattle}}{{Friberg} et~al.}{2016}]{friberg2016}
{Friberg} P.,  {Bastien} P.,  {Berry} D.,  {Savini} G.,  {Graves} S.~F.,
  {Pattle} K.,  2016, {POL-2: a polarimeter for the James-Clerk-Maxwell
  telescope}.
p. 991403, \mn@doi{10.1117/12.2231943}

\bibitem[\protect\citeauthoryear{{Fujimoto} et~al.,}{{Fujimoto}
  et~al.}{2023}]{Fujimoto+23}
{Fujimoto} S.,  et~al., 2023, \mn@doi [\apj] {10.3847/1538-4357/aceb67}, \href
  {https://ui.adsabs.harvard.edu/abs/2023ApJ...955..130F} {955, 130}

\bibitem[\protect\citeauthoryear{{Furuya} et~al.,}{{Furuya}
  et~al.}{2020}]{furuya2020}
{Furuya} R.~S.,  et~al., 2020, \mn@doi [arXiv e-prints]
  {10.48550/arXiv.2001.05753}, \href
  {https://ui.adsabs.harvard.edu/abs/2020arXiv200105753F} {p. arXiv:2001.05753}

\bibitem[\protect\citeauthoryear{{Gao}, {Carilli}, {Solomon}  \& {Vanden
  Bout}}{{Gao} et~al.}{2007}]{gao+07}
{Gao} Y.,  {Carilli} C.~L.,  {Solomon} P.~M.,   {Vanden Bout} P.~A.,  2007,
  \mn@doi [\apjl] {10.1086/518244}, \href
  {https://ui.adsabs.harvard.edu/abs/2007ApJ...660L..93G} {660, L93}

\bibitem[\protect\citeauthoryear{{Geach} et~al.,}{{Geach}
  et~al.}{2017}]{geach+17}
{Geach} J.~E.,  et~al., 2017, \mn@doi [\mnras] {10.1093/mnras/stw2721}, \href
  {https://ui.adsabs.harvard.edu/abs/2017MNRAS.465.1789G} {465, 1789}

\bibitem[\protect\citeauthoryear{{Gouin}, {Aghanim}, {Dole}, {Polletta}  \&
  {Park}}{{Gouin} et~al.}{2022}]{Gouin+22_protocluster_theory}
{Gouin} C.,  {Aghanim} N.,  {Dole} H.,  {Polletta} M.,   {Park} C.,  2022,
  \mn@doi [\aap] {10.1051/0004-6361/202243677}, \href
  {https://ui.adsabs.harvard.edu/abs/2022A&A...664A.155G} {664, A155}

\bibitem[\protect\citeauthoryear{{Greaves} et~al.,}{{Greaves}
  et~al.}{1998}]{1998Greaves}
{Greaves} J.~S.,  et~al., 1998, \mn@doi [\apjl] {10.1086/311652}, \href
  {https://ui.adsabs.harvard.edu/abs/1998ApJ...506L.133G} {506, L133}

\bibitem[\protect\citeauthoryear{{Greaves} et~al.,}{{Greaves}
  et~al.}{2003}]{greaves2003}
{Greaves} J.~S.,  et~al., 2003, \mn@doi [\mnras]
  {10.1046/j.1365-8711.2003.06230.x}, \href
  {https://ui.adsabs.harvard.edu/abs/2003MNRAS.340..353G} {340, 353}

\bibitem[\protect\citeauthoryear{{Greaves} et~al.,}{{Greaves}
  et~al.}{2021}]{greaves2021}
{Greaves} J.~S.,  et~al., 2021, \mn@doi [Nature Astronomy]
  {10.1038/s41550-020-1174-4}, \href
  {https://ui.adsabs.harvard.edu/abs/2021NatAs...5..655G} {5, 655}

\bibitem[\protect\citeauthoryear{Greve}{Greve}{2024}]{greve_ragers}
Greve T.,  in\ prep. 2024, Monthly Notices of the Royal Astronomical Society

\bibitem[\protect\citeauthoryear{{Griffin} et~al.,}{{Griffin}
  et~al.}{2010}]{Griffin_SPIRE}
{Griffin} M.~J.,  et~al., 2010, \mn@doi [\aap] {10.1051/0004-6361/201014519},
  \href {https://ui.adsabs.harvard.edu/abs/2010A&A...518L...3G} {518, L3}

\bibitem[\protect\citeauthoryear{{Grimes} et~al.,}{{Grimes}
  et~al.}{2024}]{2024arXiv240617192G}
{Grimes} P.~K.,  et~al., 2024, \mn@doi [arXiv e-prints]
  {10.48550/arXiv.2406.17192}, \href
  {https://ui.adsabs.harvard.edu/abs/2024arXiv240617192G} {p. arXiv:2406.17192}

\bibitem[\protect\citeauthoryear{{Gruchola}, {Galli}, {Vorburger}  \&
  {Wurz}}{{Gruchola} et~al.}{2021}]{gruchola2021}
{Gruchola} S.,  {Galli} A.,  {Vorburger} A.,   {Wurz} P.,  2021, \mn@doi
  [Advances in Space Research] {10.1016/j.asr.2021.07.024}, \href
  {https://ui.adsabs.harvard.edu/abs/2021AdSpR..68.3205G} {68, 3205}

\bibitem[\protect\citeauthoryear{{Gun{\'a}r}, {Heinzel}, {Mackay}  \&
  {Anzer}}{{Gun{\'a}r} et~al.}{2016}]{gunar2016}
{Gun{\'a}r} S.,  {Heinzel} P.,  {Mackay} D.~H.,   {Anzer} U.,  2016, \mn@doi
  [\apj] {10.3847/1538-4357/833/2/141}, \href
  {https://ui.adsabs.harvard.edu/abs/2016ApJ...833..141G} {833, 141}

\bibitem[\protect\citeauthoryear{{Gun{\'a}r}, {Heinzel}, {Anzer}  \&
  {Mackay}}{{Gun{\'a}r} et~al.}{2018}]{gunar2018}
{Gun{\'a}r} S.,  {Heinzel} P.,  {Anzer} U.,   {Mackay} D.~H.,  2018, \mn@doi
  [\apj] {10.3847/1538-4357/aaa001}, \href
  {https://ui.adsabs.harvard.edu/abs/2018ApJ...853...21G} {853, 21}

\bibitem[\protect\citeauthoryear{{Hacar}, {Clark}, {Heitsch}, {Kainulainen},
  {Panopoulou}, {Seifried}  \& {Smith}}{{Hacar} et~al.}{2023}]{Hacar+23}
{Hacar} A.,  {Clark} S.~E.,  {Heitsch} F.,  {Kainulainen} J.,  {Panopoulou}
  G.~V.,  {Seifried} D.,   {Smith} R.,  2023, in {Inutsuka} S.,  {Aikawa} Y.,
  {Muto} T.,  {Tomida} K.,   {Tamura} M.,  eds,  Astronomical Society of the
  Pacific Conference Series Vol. 534, Protostars and Planets VII. p.~153
  (\mn@eprint {arXiv} {2203.09562}), \mn@doi{10.48550/arXiv.2203.09562}

\bibitem[\protect\citeauthoryear{{Hagimoto} et~al.,}{{Hagimoto}
  et~al.}{2023}]{bears_3_properties}
{Hagimoto} M.,  et~al., 2023, \mn@doi [\mnras] {10.1093/mnras/stad784}, \href
  {https://ui.adsabs.harvard.edu/abs/2023MNRAS.521.5508H} {521, 5508}

\bibitem[\protect\citeauthoryear{{Han}}{{Han}}{2017}]{han2017}
{Han} J.~L.,  2017, \mn@doi [\araa] {10.1146/annurev-astro-091916-055221},
  \href {https://ui.adsabs.harvard.edu/abs/2017ARA&A..55..111H} {55, 111}

\bibitem[\protect\citeauthoryear{{Harikane} et~al.,}{{Harikane}
  et~al.}{2024a}]{Harikane+24_ALMA_Keck}
{Harikane} Y.,  et~al., 2024a, \mn@doi [arXiv e-prints]
  {10.48550/arXiv.2406.18352}, \href
  {https://ui.adsabs.harvard.edu/abs/2024arXiv240618352H} {p. arXiv:2406.18352}

\bibitem[\protect\citeauthoryear{{Harikane}, {Nakajima}, {Ouchi}, {Umeda},
  {Isobe}, {Ono}, {Xu}  \& {Zhang}}{{Harikane} et~al.}{2024b}]{Harikane+24}
{Harikane} Y.,  {Nakajima} K.,  {Ouchi} M.,  {Umeda} H.,  {Isobe} Y.,  {Ono}
  Y.,  {Xu} Y.,   {Zhang} Y.,  2024b, \mn@doi [\apj]
  {10.3847/1538-4357/ad0b7e}, \href
  {https://ui.adsabs.harvard.edu/abs/2024ApJ...960...56H} {960, 56}

\bibitem[\protect\citeauthoryear{{Harrington} et~al.,}{{Harrington}
  et~al.}{2016}]{harrington2016}
{Harrington} K.,  et~al., 2016, in {Holland} W.~S.,  {Zmuidzinas} J.,  eds,
  Society of Photo-Optical Instrumentation Engineers (SPIE) Conference Series
  Vol. 9914, Millimeter, Submillimeter, and Far-Infrared Detectors and
  Instrumentation for Astronomy VIII. p. 99141K (\mn@eprint {arXiv}
  {1608.08234}), \mn@doi{10.1117/12.2233125}

\bibitem[\protect\citeauthoryear{{Hartmann} \& {Kenyon}}{{Hartmann} \&
  {Kenyon}}{1985}]{hartmann1985}
{Hartmann} L.,  {Kenyon} S.~J.,  1985, \mn@doi [\apj] {10.1086/163713}, \href
  {https://ui.adsabs.harvard.edu/abs/1985ApJ...299..462H} {299, 462}

\bibitem[\protect\citeauthoryear{{Hensley} \& {Draine}}{{Hensley} \&
  {Draine}}{2023}]{Hensley2023}
{Hensley} B.~S.,  {Draine} B.~T.,  2023, \mn@doi [\apj]
  {10.3847/1538-4357/acc4c2}, \href
  {https://ui.adsabs.harvard.edu/abs/2023ApJ...948...55H} {948, 55}

\bibitem[\protect\citeauthoryear{{Hensley} et~al.,}{{Hensley}
  et~al.}{2022}]{2022ApJ...929..166H}
{Hensley} B.~S.,  et~al., 2022, \mn@doi [\apj] {10.3847/1538-4357/ac5e36},
  \href {https://ui.adsabs.harvard.edu/abs/2022ApJ...929..166H} {929, 166}

\bibitem[\protect\citeauthoryear{{Herczeg} et~al.,}{{Herczeg}
  et~al.}{2017}]{herczeg2017}
{Herczeg} G.~J.,  et~al., 2017, \mn@doi [\apj] {10.3847/1538-4357/aa8b62},
  \href {https://ui.adsabs.harvard.edu/abs/2017ApJ...849...43H} {849, 43}

\bibitem[\protect\citeauthoryear{{Hezaveh} et~al.,}{{Hezaveh}
  et~al.}{2016}]{hezaveh+16_sdp81}
{Hezaveh} Y.~D.,  et~al., 2016, \mn@doi [\apj] {10.3847/0004-637X/823/1/37},
  \href {https://ui.adsabs.harvard.edu/abs/2016ApJ...823...37H} {823, 37}

\bibitem[\protect\citeauthoryear{{Hidayat}, {Marten}, {B{\'e}zard}, {Gautier},
  {Owen}, {Matthews}  \& {Paubert}}{{Hidayat}
  et~al.}{1997}]{1997Icar..126..170H}
{Hidayat} T.,  {Marten} A.,  {B{\'e}zard} B.,  {Gautier} D.,  {Owen} T.,
  {Matthews} H.~E.,   {Paubert} G.,  1997, \mn@doi [\icarus]
  {10.1006/icar.1996.5640}, \href
  {https://ui.adsabs.harvard.edu/abs/1997Icar..126..170H} {126, 170}

\bibitem[\protect\citeauthoryear{{Hodge} \& {da Cunha}}{{Hodge} \& {da
  Cunha}}{2020}]{Hodge_daCunha_2020}
{Hodge} J.~A.,  {da Cunha} E.,  2020, \mn@doi [Royal Society Open Science]
  {10.1098/rsos.200556}, \href
  {https://ui.adsabs.harvard.edu/abs/2020RSOS....700556H} {7, 200556}

\bibitem[\protect\citeauthoryear{{Holland} et~al.,}{{Holland}
  et~al.}{1998}]{Holland1998}
{Holland} W.~S.,  et~al., 1998, \mn@doi [\nat] {10.1038/33874}, \href
  {https://ui.adsabs.harvard.edu/abs/1998Natur.392..788H} {392, 788}

\bibitem[\protect\citeauthoryear{{Holland} et~al.,}{{Holland}
  et~al.}{1999}]{scuba-1}
{Holland} W.~S.,  et~al., 1999, \mn@doi [\mnras]
  {10.1046/j.1365-8711.1999.02111.x}, \href
  {https://ui.adsabs.harvard.edu/abs/1999MNRAS.303..659H} {303, 659}

\bibitem[\protect\citeauthoryear{Holland et~al.,}{Holland
  et~al.}{2013}]{scuba-2}
Holland W.~S.,  et~al., 2013, \mn@doi [Monthly Notices of the Royal
  Astronomical Society] {10.1093/mnras/sts612}, 430, 2513

\bibitem[\protect\citeauthoryear{{Holland} et~al.,}{{Holland}
  et~al.}{2017}]{2017Holland}
{Holland} W.~S.,  et~al., 2017, \mn@doi [\mnras] {10.1093/mnras/stx1378}, \href
  {https://ui.adsabs.harvard.edu/abs/2017MNRAS.470.3606H} {470, 3606}

\bibitem[\protect\citeauthoryear{{Holland}, {Booth}, {Dent}, {Duchene},
  {Klaassen}, {Lestrate}, {Marshall}  \& {Matthews}}{{Holland}
  et~al.}{2019}]{2019Holland}
{Holland} W.,  {Booth} M.,  {Dent} W.,  {Duchene} G.,  {Klaassen} P.,
  {Lestrate} J.-F.,  {Marshall} J.,   {Matthews} B.,  2019, Astro2020: Decadal
  Survey on Astronomy and Astrophysics, \href
  {https://ui.adsabs.harvard.edu/abs/2019astro2020T..80H} {2020, 80}

\bibitem[\protect\citeauthoryear{{Holman} et~al.,}{{Holman}
  et~al.}{2011}]{2011SSRv..159..107H}
{Holman} G.~D.,  et~al., 2011, \mn@doi [\ssr] {10.1007/s11214-010-9680-9},
  \href {https://ui.adsabs.harvard.edu/abs/2011SSRv..159..107H} {159, 107}

\bibitem[\protect\citeauthoryear{{Hughes} et~al.,}{{Hughes}
  et~al.}{1998}]{Hughes+98}
{Hughes} D.~H.,  et~al., 1998, \mn@doi [\nat] {10.1038/28328}, \href
  {https://ui.adsabs.harvard.edu/abs/1998Natur.394..241H} {394, 241}

\bibitem[\protect\citeauthoryear{{Hull} et~al.,}{{Hull}
  et~al.}{2017}]{hull2017}
{Hull} C. L.~H.,  et~al., 2017, \mn@doi [\apj] {10.3847/1538-4357/aa7fe9},
  \href {https://ui.adsabs.harvard.edu/abs/2017ApJ...847...92H} {847, 92}

\bibitem[\protect\citeauthoryear{{Ikarashi} et~al.,}{{Ikarashi}
  et~al.}{2015}]{Ikarishi+15}
{Ikarashi} S.,  et~al., 2015, \mn@doi [\apj] {10.1088/0004-637X/810/2/133},
  \href {https://ui.adsabs.harvard.edu/abs/2015ApJ...810..133I} {810, 133}

\bibitem[\protect\citeauthoryear{{Ikarashi} et~al.,}{{Ikarashi}
  et~al.}{2017a}]{ikarishi+17}
{Ikarashi} S.,  et~al., 2017a, \mn@doi [\apj] {10.3847/1538-4357/835/2/286},
  \href {https://ui.adsabs.harvard.edu/abs/2017ApJ...835..286I} {835, 286}

\bibitem[\protect\citeauthoryear{{Ikarashi} et~al.,}{{Ikarashi}
  et~al.}{2017b}]{Ikarashi+17}
{Ikarashi} S.,  et~al., 2017b, \mn@doi [\apjl] {10.3847/2041-8213/aa9572},
  \href {https://ui.adsabs.harvard.edu/abs/2017ApJ...849L..36I} {849, L36}

\bibitem[\protect\citeauthoryear{{Imaz Blanco} et~al.,}{{Imaz Blanco}
  et~al.}{2023}]{2023ImazBlanco}
{Imaz Blanco} A.,  et~al., 2023, \mn@doi [\mnras] {10.1093/mnras/stad1221},
  \href {https://ui.adsabs.harvard.edu/abs/2023MNRAS.522.6150I} {522, 6150}

\bibitem[\protect\citeauthoryear{{Indebetouw} et~al.,}{{Indebetouw}
  et~al.}{2014}]{Indebetouw2014}
{Indebetouw} R.,  et~al., 2014, \mn@doi [\apjl] {10.1088/2041-8205/782/1/L2},
  \href {https://ui.adsabs.harvard.edu/abs/2014ApJ...782L...2I} {782, L2}

\bibitem[\protect\citeauthoryear{{Inoue} et~al.,}{{Inoue}
  et~al.}{2016}]{Inoue+16_88um_z=7.2120}
{Inoue} A.~K.,  et~al., 2016, \mn@doi [Science] {10.1126/science.aaf0714},
  \href {https://ui.adsabs.harvard.edu/abs/2016Sci...352.1559I} {352, 1559}

\bibitem[\protect\citeauthoryear{{Ismail} et~al.,}{{Ismail}
  et~al.}{2023}]{zgal_2_dust}
{Ismail} D.,  et~al., 2023, \mn@doi [\aap] {10.1051/0004-6361/202346804}, \href
  {https://ui.adsabs.harvard.edu/abs/2023A&A...678A..27I} {678, A27}

\bibitem[\protect\citeauthoryear{{Janssen} et~al.,}{{Janssen}
  et~al.}{2021}]{janssen2021}
{Janssen} M.,  et~al., 2021, \mn@doi [Nature Astronomy]
  {10.1038/s41550-021-01417-w}, \href
  {https://ui.adsabs.harvard.edu/abs/2021NatAs...5.1017J} {5, 1017}

\bibitem[\protect\citeauthoryear{{Johnson} et~al.,}{{Johnson}
  et~al.}{2020}]{Johnson2020}
{Johnson} M.~D.,  et~al., 2020, \mn@doi [Science Advances]
  {10.1126/sciadv.aaz1310}, \href
  {https://ui.adsabs.harvard.edu/abs/2020SciA....6.1310J} {6, eaaz1310}

\bibitem[\protect\citeauthoryear{{Jones}}{{Jones}}{2023}]{2023atyp.confE..12J}
{Jones} B.,  2023, in ALMA at 10 years: Past, Present, and Future. p.~12,
  \mn@doi{10.5281/zenodo.10230411}

\bibitem[\protect\citeauthoryear{{J{\o}rgensen}, {Belloche}  \&
  {Garrod}}{{J{\o}rgensen} et~al.}{2020}]{jorgensen2020}
{J{\o}rgensen} J.~K.,  {Belloche} A.,   {Garrod} R.~T.,  2020, \mn@doi [\araa]
  {10.1146/annurev-astro-032620-021927}, \href
  {https://ui.adsabs.harvard.edu/abs/2020ARA&A..58..727J} {58, 727}

\bibitem[\protect\citeauthoryear{{Jorstad} et~al.,}{{Jorstad}
  et~al.}{2023}]{jorstad2023}
{Jorstad} S.,  et~al., 2023, \mn@doi [\apj] {10.3847/1538-4357/acaea8}, \href
  {https://ui.adsabs.harvard.edu/abs/2023ApJ...943..170J} {943, 170}

\bibitem[\protect\citeauthoryear{Jovanovic et~al.,}{Jovanovic
  et~al.}{2023}]{Jovanovic2023}
Jovanovic N.,  et~al., 2023, \mn@doi [Journal of Physics: Photonics]
  {10.1088/2515-7647/ace869}, 5, 042501

\bibitem[\protect\citeauthoryear{{Kama} et~al.,}{{Kama}
  et~al.}{2016}]{kama2016}
{Kama} M.,  et~al., 2016, \mn@doi [\aap] {10.1051/0004-6361/201526991}, \href
  {https://ui.adsabs.harvard.edu/abs/2016A&A...592A..83K} {592, A83}

\bibitem[\protect\citeauthoryear{{Karoly} et~al.,}{{Karoly}
  et~al.}{2023}]{karoly2023}
{Karoly} J.,  et~al., 2023, \mn@doi [\apj] {10.3847/1538-4357/acd6f2}, \href
  {https://ui.adsabs.harvard.edu/abs/2023ApJ...952...29K} {952, 29}

\bibitem[\protect\citeauthoryear{{Katz} et~al.,}{{Katz} et~al.}{2022}]{Katz+22}
{Katz} H.,  et~al., 2022, \mn@doi [\mnras] {10.1093/mnras/stac028}, \href
  {https://ui.adsabs.harvard.edu/abs/2022MNRAS.510.5603K} {510, 5603}

\bibitem[\protect\citeauthoryear{{Kaufmann} et~al.,}{{Kaufmann}
  et~al.}{2008}]{kaufmann2008}
{Kaufmann} P.,  et~al., 2008, in {Stepp} L.~M.,  {Gilmozzi} R.,  eds,  Society
  of Photo-Optical Instrumentation Engineers (SPIE) Conference Series Vol.
  7012, Ground-based and Airborne Telescopes II. p. 70120L,
  \mn@doi{10.1117/12.788889}

\bibitem[\protect\citeauthoryear{{Kennedy}, {Marino}, {Matr{\`a}}, {Pani{\'c}},
  {Wilner}, {Wyatt}  \& {Yelverton}}{{Kennedy} et~al.}{2018}]{2018Kennedy}
{Kennedy} G.~M.,  {Marino} S.,  {Matr{\`a}} L.,  {Pani{\'c}} O.,  {Wilner} D.,
  {Wyatt} M.~C.,   {Yelverton} B.,  2018, \mn@doi [\mnras]
  {10.1093/mnras/sty135}, \href
  {https://ui.adsabs.harvard.edu/abs/2018MNRAS.475.4924K} {475, 4924}

\bibitem[\protect\citeauthoryear{{Kenyon}, {Hartmann}, {Strom}  \&
  {Strom}}{{Kenyon} et~al.}{1990}]{kenyon1990}
{Kenyon} S.~J.,  {Hartmann} L.~W.,  {Strom} K.~M.,   {Strom} S.~E.,  1990,
  \mn@doi [\aj] {10.1086/115380}, \href
  {https://ui.adsabs.harvard.edu/abs/1990AJ.....99..869K} {99, 869}

\bibitem[\protect\citeauthoryear{{Kirchschlager}, {Barlow}  \&
  {Schmidt}}{{Kirchschlager} et~al.}{2020}]{Kirchschlager2020}
{Kirchschlager} F.,  {Barlow} M.~J.,   {Schmidt} F.~D.,  2020, \mn@doi [\apj]
  {10.3847/1538-4357/ab7db8}, \href
  {https://ui.adsabs.harvard.edu/abs/2020ApJ...893...70K} {893, 70}

\bibitem[\protect\citeauthoryear{{Klaassen} et~al.,}{{Klaassen}
  et~al.}{2020}]{klaassen2020}
{Klaassen} P.~D.,  et~al., 2020, in {Marshall} H.~K.,  {Spyromilio} J.,
  {Usuda} T.,  eds,  Society of Photo-Optical Instrumentation Engineers (SPIE)
  Conference Series Vol. 11445, Ground-based and Airborne Telescopes VIII. p.
  114452F (\mn@eprint {arXiv} {2011.07974}), \mn@doi{10.1117/12.2561315}

\bibitem[\protect\citeauthoryear{{Klaassen} et~al.,}{{Klaassen}
  et~al.}{2024}]{Klaassen2024}
{Klaassen} P.,  et~al., 2024, \mn@doi [\ore] {10.12688/openreseurope.17450.1},
  \href {https://ui.adsabs.harvard.edu/abs/2024arXiv240300917K} {4, 112}

\bibitem[\protect\citeauthoryear{{Kocevski} et~al.,}{{Kocevski}
  et~al.}{2024}]{kocevski+24_lrd}
{Kocevski} D.~D.,  et~al., 2024, \mn@doi [arXiv e-prints]
  {10.48550/arXiv.2404.03576}, \href
  {https://ui.adsabs.harvard.edu/abs/2024arXiv240403576K} {p. arXiv:2404.03576}

\bibitem[\protect\citeauthoryear{Kogut et~al.,}{Kogut
  et~al.}{2014}]{kogut2014primordial}
Kogut A.,  et~al., 2014, in Space Telescopes and Instrumentation 2014: Optical,
  Infrared, and Millimeter Wave. pp 378--394

\bibitem[\protect\citeauthoryear{{Kokubo} \& {Harikane}}{{Kokubo} \&
  {Harikane}}{2024}]{Kokubo+24_lrd}
{Kokubo} M.,  {Harikane} Y.,  2024, \mn@doi [arXiv e-prints]
  {10.48550/arXiv.2407.04777}, \href
  {https://ui.adsabs.harvard.edu/abs/2024arXiv240704777K} {p. arXiv:2407.04777}

\bibitem[\protect\citeauthoryear{{Kontar}, {Motorina}, {Jeffrey}, {Tsap},
  {Fleishman}  \& {Stepanov}}{{Kontar} et~al.}{2018}]{2018A&A...620A..95K}
{Kontar} E.~P.,  {Motorina} G.~G.,  {Jeffrey} N.~L.~S.,  {Tsap} Y.~T.,
  {Fleishman} G.~D.,   {Stepanov} A.~V.,  2018, \mn@doi [\aap]
  {10.1051/0004-6361/201834124}, \href
  {https://ui.adsabs.harvard.edu/abs/2018A&A...620A..95K} {620, A95}

\bibitem[\protect\citeauthoryear{{K{\"o}nyves} et~al.,}{{K{\"o}nyves}
  et~al.}{2015}]{Konyves+15}
{K{\"o}nyves} V.,  et~al., 2015, \mn@doi [\aap] {10.1051/0004-6361/201525861},
  \href {https://ui.adsabs.harvard.edu/abs/2015A&A...584A..91K} {584, A91}

\bibitem[\protect\citeauthoryear{{K{\"o}nyves} et~al.,}{{K{\"o}nyves}
  et~al.}{2020}]{Konyves+20}
{K{\"o}nyves} V.,  et~al., 2020, \mn@doi [\aap] {10.1051/0004-6361/201834753},
  \href {https://ui.adsabs.harvard.edu/abs/2020A&A...635A..34K} {635, A34}

\bibitem[\protect\citeauthoryear{{Kovalenko}, {Doressoundiram}, {Lellouch},
  {Vilenius}, {M{\"u}ller}  \& {Stansberry}}{{Kovalenko}
  et~al.}{2017}]{kovalenko2017}
{Kovalenko} I.~D.,  {Doressoundiram} A.,  {Lellouch} E.,  {Vilenius} E.,
  {M{\"u}ller} T.,   {Stansberry} J.,  2017, \mn@doi [\aap]
  {10.1051/0004-6361/201730588}, \href
  {https://ui.adsabs.harvard.edu/abs/2017A&A...608A..19K} {608, A19}

\bibitem[\protect\citeauthoryear{{Kral}}{{Kral}}{2016}]{2016Kral}
{Kral} Q.,  2016, in {Reyl{\'e}} C.,  {Richard} J.,  {Cambr{\'e}sy} L.,
  {Deleuil} M.,  {P{\'e}contal} E.,  {Tresse} L.,   {Vauglin} I.,  eds,
  SF2A-2016: Proceedings of the Annual meeting of the French Society of
  Astronomy and Astrophysics. pp 463--472 (\mn@eprint {arXiv} {1611.06751}),
  \mn@doi{10.48550/arXiv.1611.06751}

\bibitem[\protect\citeauthoryear{{Kral}, {Matr{\`a}}, {Wyatt}  \&
  {Kennedy}}{{Kral} et~al.}{2017}]{2017Kral}
{Kral} Q.,  {Matr{\`a}} L.,  {Wyatt} M.~C.,   {Kennedy} G.~M.,  2017, \mn@doi
  [\mnras] {10.1093/mnras/stx730}, \href
  {https://ui.adsabs.harvard.edu/abs/2017MNRAS.469..521K} {469, 521}

\bibitem[\protect\citeauthoryear{{Kral}, {Clarke}  \& {Wyatt}}{{Kral}
  et~al.}{2018}]{2018kral}
{Kral} Q.,  {Clarke} C.,   {Wyatt} M.~C.,  2018, in {Deeg} H.~J.,  {Belmonte}
  J.~A.,  eds, , Handbook of Exoplanets.
p.~165, \mn@doi{10.1007/978-3-319-55333-7_165}

\bibitem[\protect\citeauthoryear{{Kruijssen} \& {Longmore}}{{Kruijssen} \&
  {Longmore}}{2014}]{2014KruijssenLongmore}
{Kruijssen} J.~M.~D.,  {Longmore} S.~N.,  2014, \mn@doi [\mnras]
  {10.1093/mnras/stu098}, \href
  {https://ui.adsabs.harvard.edu/abs/2014MNRAS.439.3239K} {439, 3239}

\bibitem[\protect\citeauthoryear{{Kubo} et~al.,}{{Kubo}
  et~al.}{2019}]{Kuko+19_planck_stacks_of_hsc_protoclusters}
{Kubo} M.,  et~al., 2019, \mn@doi [\apj] {10.3847/1538-4357/ab5a80}, \href
  {https://ui.adsabs.harvard.edu/abs/2019ApJ...887..214K} {887, 214}

\bibitem[\protect\citeauthoryear{{Kudriashov} et~al.,}{{Kudriashov}
  et~al.}{2021}]{kudriashov2021}
{Kudriashov} V.,  et~al., 2021, \mn@doi [Chinese Journal of Space Science]
  {10.3724/SP.J.0254-6124.2021.0202}, \href
  {https://ui.adsabs.harvard.edu/abs/2021ChJSS..41..211K} {41, 211}

\bibitem[\protect\citeauthoryear{{Ladjelate} et~al.,}{{Ladjelate}
  et~al.}{2020}]{Ladjelate+20}
{Ladjelate} B.,  et~al., 2020, \mn@doi [\aap] {10.1051/0004-6361/201936442},
  \href {https://ui.adsabs.harvard.edu/abs/2020A&A...638A..74L} {638, A74}

\bibitem[\protect\citeauthoryear{{Lagache}, {B{\'e}thermin}, {Montier}, {Serra}
   \& {Tucci}}{{Lagache} et~al.}{2020}]{2020A&A...642A.232L}
{Lagache} G.,  {B{\'e}thermin} M.,  {Montier} L.,  {Serra} P.,   {Tucci} M.,
  2020, \mn@doi [\aap] {10.1051/0004-6361/201937147}, \href
  {https://ui.adsabs.harvard.edu/abs/2020A&A...642A.232L} {642, A232}

\bibitem[\protect\citeauthoryear{{Lammers}, {Hill}, {Lim}, {Scott},
  {Ca{\~n}ameras}  \& {Dole}}{{Lammers}
  et~al.}{2022}]{Lammers+22_Planck_SPIRE_protocluster_candidates}
{Lammers} C.,  {Hill} R.,  {Lim} S.,  {Scott} D.,  {Ca{\~n}ameras} R.,   {Dole}
  H.,  2022, \mn@doi [\mnras] {10.1093/mnras/stac1555}, \href
  {https://ui.adsabs.harvard.edu/abs/2022MNRAS.514.5004L} {514, 5004}

\bibitem[\protect\citeauthoryear{{Lamperti} et~al.,}{{Lamperti}
  et~al.}{2019}]{Lamperti2019}
{Lamperti} I.,  et~al., 2019, \mn@doi [\mnras] {10.1093/mnras/stz2311}, \href
  {https://ui.adsabs.harvard.edu/abs/2019MNRAS.489.4389L} {489, 4389}

\bibitem[\protect\citeauthoryear{{Laporte} et~al.,}{{Laporte}
  et~al.}{2019}]{Laporte+19}
{Laporte} N.,  et~al., 2019, \mn@doi [\mnras] {10.1093/mnrasl/slz094}, \href
  {https://ui.adsabs.harvard.edu/abs/2019MNRAS.487L..81L} {487, L81}

\bibitem[\protect\citeauthoryear{{Laporte}, {Meyer}, {Ellis}, {Robertson},
  {Chisholm}  \& {Roberts-Borsani}}{{Laporte} et~al.}{2021}]{Laporte+21}
{Laporte} N.,  {Meyer} R.~A.,  {Ellis} R.~S.,  {Robertson} B.~E.,  {Chisholm}
  J.,   {Roberts-Borsani} G.~W.,  2021, \mn@doi [\mnras]
  {10.1093/mnras/stab1239}, \href
  {https://ui.adsabs.harvard.edu/abs/2021MNRAS.505.3336L} {505, 3336}

\bibitem[\protect\citeauthoryear{{Lee} et~al.,}{{Lee} et~al.}{2014}]{lee2014}
{Lee} S.-S.,  et~al., 2014, \mn@doi [\aj] {10.1088/0004-6256/147/4/77}, \href
  {https://ui.adsabs.harvard.edu/abs/2014AJ....147...77L} {147, 77}

\bibitem[\protect\citeauthoryear{{Lee} et~al.,}{{Lee} et~al.}{2020}]{lee2020}
{Lee} Y.-H.,  et~al., 2020, \mn@doi [\apj] {10.3847/1538-4357/abb6fe}, \href
  {https://ui.adsabs.harvard.edu/abs/2020ApJ...903....5L} {903, 5}

\bibitem[\protect\citeauthoryear{{Lee} et~al.,}{{Lee} et~al.}{2021}]{lee2021}
{Lee} Y.-H.,  et~al., 2021, \mn@doi [\apj] {10.3847/1538-4357/ac1679}, \href
  {https://ui.adsabs.harvard.edu/abs/2021ApJ...920..119L} {920, 119}

\bibitem[\protect\citeauthoryear{{Leech}, {Dewitt}, {Jenness}, {Greaves}  \&
  {Lightfoot}}{{Leech} et~al.}{2005}]{leech2005}
{Leech} J.,  {Dewitt} S.,  {Jenness} T.,  {Greaves} J.,   {Lightfoot} J.~F.,
  2005, in {Adamson} A.,  {Aspin} C.,  {Davis} C.,   {Fujiyoshi} T.,  eds,
  Astronomical Society of the Pacific Conference Series Vol. 343, Astronomical
  Polarimetry: Current Status and Future Directions. p.~83

\bibitem[\protect\citeauthoryear{{Leisawitz} et~al.,}{{Leisawitz}
  et~al.}{2019}]{origins_space_telescope_2019}
{Leisawitz} D.,  et~al., 2019, in {Barto} A.~A.,  {Breckinridge} J.~B.,
  {Stahl} H.~P.,  eds,  Society of Photo-Optical Instrumentation Engineers
  (SPIE) Conference Series Vol. 11115, UV/Optical/IR Space Telescopes and
  Instruments: Innovative Technologies and Concepts IX. p. 111150Q,
  \mn@doi{10.1117/12.2530514}

\bibitem[\protect\citeauthoryear{{Lellouch} et~al.,}{{Lellouch}
  et~al.}{2017}]{lellouch2017}
{Lellouch} E.,  et~al., 2017, \mn@doi [\icarus] {10.1016/j.icarus.2016.10.013},
  \href {https://ui.adsabs.harvard.edu/abs/2017Icar..286..289L} {286, 289}

\bibitem[\protect\citeauthoryear{{Leroy} et~al.,}{{Leroy}
  et~al.}{2021}]{2021Leroy}
{Leroy} A.~K.,  et~al., 2021, \mn@doi [\apjs] {10.3847/1538-4365/ac17f3}, \href
  {https://ui.adsabs.harvard.edu/abs/2021ApJS..257...43L} {257, 43}

\bibitem[\protect\citeauthoryear{{Lewis} et~al.,}{{Lewis}
  et~al.}{2018}]{Lewis+18_ultrareds_signpost_protoclusters}
{Lewis} A.~J.~R.,  et~al., 2018, \mn@doi [\apj] {10.3847/1538-4357/aacc25},
  \href {https://ui.adsabs.harvard.edu/abs/2018ApJ...862...96L} {862, 96}

\bibitem[\protect\citeauthoryear{{Li} \& {Henning}}{{Li} \&
  {Henning}}{2011}]{2011LiHenning}
{Li} H.-B.,  {Henning} T.,  2011, \mn@doi [\nat] {10.1038/nature10551}, \href
  {https://ui.adsabs.harvard.edu/abs/2011Natur.479..499L} {479, 499}

\bibitem[\protect\citeauthoryear{{Li}, {Frenk}, {Cole}, {Gao}, {Bose}  \&
  {Hellwing}}{{Li} et~al.}{2016}]{Li+16}
{Li} R.,  {Frenk} C.~S.,  {Cole} S.,  {Gao} L.,  {Bose} S.,   {Hellwing} W.~A.,
   2016, \mn@doi [\mnras] {10.1093/mnras/stw939}, \href
  {https://ui.adsabs.harvard.edu/abs/2016MNRAS.460..363L} {460, 363}

\bibitem[\protect\citeauthoryear{{Li}, {Frenk}, {Cole}, {Wang}  \& {Gao}}{{Li}
  et~al.}{2017}]{Li+17}
{Li} R.,  {Frenk} C.~S.,  {Cole} S.,  {Wang} Q.,   {Gao} L.,  2017, \mn@doi
  [\mnras] {10.1093/mnras/stx554}, \href
  {https://ui.adsabs.harvard.edu/abs/2017MNRAS.468.1426L} {468, 1426}

\bibitem[\protect\citeauthoryear{{Li} et~al.,}{{Li}
  et~al.}{2023a}]{Li+23_SCUBA2_quasars_z6}
{Li} Q.,  et~al., 2023a, \mn@doi [\apj] {10.3847/1538-4357/acd7f3}, \href
  {https://ui.adsabs.harvard.edu/abs/2023ApJ...954..174L} {954, 174}

\bibitem[\protect\citeauthoryear{{Li} et~al.,}{{Li}
  et~al.}{2023b}]{2023ApJ...956...36L}
{Li} Y.,  et~al., 2023b, \mn@doi [\apj] {10.3847/1538-4357/ace599}, \href
  {https://ui.adsabs.harvard.edu/abs/2023ApJ...956...36L} {956, 36}

\bibitem[\protect\citeauthoryear{{Li} et~al.,}{{Li} et~al.}{2024}]{li2024}
{Li} S.,  et~al., 2024, in Bryant J.~J.,  Motohara K.,   Vernet J. R.~D.,  eds,
   Proc. SPIE Vol. 13096, Ground-based and Airborne Instrumentation for
  Astronomy X. SPIE, p. 130967K, \mn@doi{10.1117/12.3018648}, \url
  {https://doi.org/10.1117/12.3018648}

\bibitem[\protect\citeauthoryear{{Lincowski} et~al.,}{{Lincowski}
  et~al.}{2021}]{lincowski2021}
{Lincowski} A.~P.,  et~al., 2021, \mn@doi [\apjl] {10.3847/2041-8213/abde47},
  \href {https://ui.adsabs.harvard.edu/abs/2021ApJ...908L..44L} {908, L44}

\bibitem[\protect\citeauthoryear{{Lindsey}, {Kopp}, {Clark}  \&
  {Watt}}{{Lindsey} et~al.}{1995}]{lindsey1995}
{Lindsey} C.,  {Kopp} G.,  {Clark} T.~A.,   {Watt} G.,  1995, \mn@doi [\apj]
  {10.1086/176412}, \href
  {https://ui.adsabs.harvard.edu/abs/1995ApJ...453..511L} {453, 511}

\bibitem[\protect\citeauthoryear{{Liu} et~al.,}{{Liu} et~al.}{2020}]{Liu20}
{Liu} T.,  et~al., 2020, \mn@doi [\mnras] {10.1093/mnras/staa1577}, \href
  {https://ui.adsabs.harvard.edu/abs/2020MNRAS.496.2790L} {496, 2790}

\bibitem[\protect\citeauthoryear{{Lodders} \& {Fegley}}{{Lodders} \&
  {Fegley}}{1994}]{lodders1994}
{Lodders} K.,  {Fegley} B. J.,  1994, \mn@doi [\icarus]
  {10.1006/icar.1994.1190}, \href
  {https://ui.adsabs.harvard.edu/abs/1994Icar..112..368L} {112, 368}

\bibitem[\protect\citeauthoryear{Lourie et~al.,}{Lourie
  et~al.}{2018}]{Blast_TNG}
Lourie N.~P.,  et~al., 2018, in Zmuidzinas J.,  Gao J.-R.,  eds, Millimeter,
  Submillimeter, and Far-Infrared Detectors and Instrumentation for Astronomy
  IX. SPIE, \mn@doi{10.1117/12.2314396}, \url
  {http://dx.doi.org/10.1117/12.2314396}

\bibitem[\protect\citeauthoryear{{MacTavish} et~al.,}{{MacTavish}
  et~al.}{2006}]{McTavish2006}
{MacTavish} C.~J.,  et~al., 2006, \mn@doi [\apj] {10.1086/505558}, \href
  {https://ui.adsabs.harvard.edu/abs/2006ApJ...647..799M} {647, 799}

\bibitem[\protect\citeauthoryear{{Madden} et~al.,}{{Madden}
  et~al.}{2020}]{2020Madden}
{Madden} S.~C.,  et~al., 2020, \mn@doi [\aap] {10.1051/0004-6361/202038860},
  \href {https://ui.adsabs.harvard.edu/abs/2020A&A...643A.141M} {643, A141}

\bibitem[\protect\citeauthoryear{{Madhusudhan}}{{Madhusudhan}}{2019}]{madhu2019}
{Madhusudhan} N.,  2019, \mn@doi [\araa] {10.1146/annurev-astro-081817-051846},
  \href {https://ui.adsabs.harvard.edu/abs/2019ARA&A..57..617M} {57, 617}

\bibitem[\protect\citeauthoryear{{Maercker}, {Khouri}, {De Beck}, {Brunner},
  {Mecina}  \& {Jaldehag}}{{Maercker} et~al.}{2018}]{Maercker2018}
{Maercker} M.,  {Khouri} T.,  {De Beck} E.,  {Brunner} M.,  {Mecina} M.,
  {Jaldehag} O.,  2018, \mn@doi [\aap] {10.1051/0004-6361/201833665}, \href
  {https://ui.adsabs.harvard.edu/abs/2018A&A...620A.106M} {620, A106}

\bibitem[\protect\citeauthoryear{{Maercker}, {Khouri}, {Mecina}  \& {De
  Beck}}{{Maercker} et~al.}{2022}]{Maercker2022}
{Maercker} M.,  {Khouri} T.,  {Mecina} M.,   {De Beck} E.,  2022, \mn@doi
  [\aap] {10.1051/0004-6361/202142117}, \href
  {https://ui.adsabs.harvard.edu/abs/2022A&A...663A..64M} {663, A64}

\bibitem[\protect\citeauthoryear{{Maffei} et~al.,}{{Maffei}
  et~al.}{2022}]{Maffei2022}
{Maffei} B.,  et~al., 2022, in {Zmuidzinas} J.,  {Gao} J.-R.,  eds,  Society of
  Photo-Optical Instrumentation Engineers (SPIE) Conference Series Vol. 12190,
  Millimeter, Submillimeter, and Far-Infrared Detectors and Instrumentation for
  Astronomy XI. p. 121900A, \mn@doi{10.1117/12.2630136}

\bibitem[\protect\citeauthoryear{{Maiolino} \& {Mannucci}}{{Maiolino} \&
  {Mannucci}}{2019}]{Maiolino_Mannucci_2019}
{Maiolino} R.,  {Mannucci} F.,  2019, \mn@doi [\aapr]
  {10.1007/s00159-018-0112-2}, \href
  {https://ui.adsabs.harvard.edu/abs/2019A&ARv..27....3M} {27, 3}

\bibitem[\protect\citeauthoryear{{Maiolino} et~al.,}{{Maiolino}
  et~al.}{2024}]{maiolino+24_jwst_smbh_gnz11}
{Maiolino} R.,  et~al., 2024, \mn@doi [\nat] {10.1038/s41586-024-07052-5},
  \href {https://ui.adsabs.harvard.edu/abs/2024Natur.627...59M} {627, 59}

\bibitem[\protect\citeauthoryear{{Mairs} et~al.,}{{Mairs}
  et~al.}{2019}]{mairs2019}
{Mairs} S.,  et~al., 2019, \mn@doi [\apj] {10.3847/1538-4357/aaf3b1}, \href
  {https://ui.adsabs.harvard.edu/abs/2019ApJ...871...72M} {871, 72}

\bibitem[\protect\citeauthoryear{{Manara}, {Morbidelli}  \& {Guillot}}{{Manara}
  et~al.}{2018}]{manara2018}
{Manara} C.~F.,  {Morbidelli} A.,   {Guillot} T.,  2018, \mn@doi [\aap]
  {10.1051/0004-6361/201834076}, \href
  {https://ui.adsabs.harvard.edu/abs/2018A&A...618L...3M} {618, L3}

\bibitem[\protect\citeauthoryear{{Mangilli} et~al.,}{{Mangilli}
  et~al.}{2019}]{Bernard2019}
{Mangilli} A.,  et~al., 2019, \mn@doi [Experimental Astronomy]
  {10.1007/s10686-019-09648-6}, \href
  {https://ui.adsabs.harvard.edu/abs/2019ExA....48..265M} {48, 265}

\bibitem[\protect\citeauthoryear{{Marino}}{{Marino}}{2022}]{2022Marino}
{Marino} S.,  2022, \mn@doi [arXiv e-prints] {10.48550/arXiv.2202.03053}, \href
  {https://ui.adsabs.harvard.edu/abs/2022arXiv220203053M} {p. arXiv:2202.03053}

\bibitem[\protect\citeauthoryear{{Marino} et~al.,}{{Marino}
  et~al.}{2018}]{2018Marino}
{Marino} S.,  et~al., 2018, \mn@doi [\mnras] {10.1093/mnras/sty1790}, \href
  {https://ui.adsabs.harvard.edu/abs/2018MNRAS.479.5423M} {479, 5423}

\bibitem[\protect\citeauthoryear{{Marrone} et~al.,}{{Marrone}
  et~al.}{2022}]{Marrone2022}
{Marrone} D.~P.,  et~al., 2022, in {Zmuidzinas} J.,  {Gao} J.-R.,  eds,
  Society of Photo-Optical Instrumentation Engineers (SPIE) Conference Series
  Vol. 12190, Millimeter, Submillimeter, and Far-Infrared Detectors and
  Instrumentation for Astronomy XI. p. 1219008, \mn@doi{10.1117/12.2630644}

\bibitem[\protect\citeauthoryear{{Marten}, {Gautier}, {Owen}, {Sanders},
  {Matthews}, {Atreya}, {Tilanus}  \& {Deane}}{{Marten}
  et~al.}{1993}]{marten1993}
{Marten} A.,  {Gautier} D.,  {Owen} T.,  {Sanders} D.~B.,  {Matthews} H.~E.,
  {Atreya} S.~K.,  {Tilanus} R.~P.~J.,   {Deane} J.~R.,  1993, \mn@doi [\apj]
  {10.1086/172440}, \href
  {https://ui.adsabs.harvard.edu/abs/1993ApJ...406..285M} {406, 285}

\bibitem[\protect\citeauthoryear{{Matr{\`a}} et~al.,}{{Matr{\`a}}
  et~al.}{2017}]{2017Matra}
{Matr{\`a}} L.,  et~al., 2017, \mn@doi [\mnras] {10.1093/mnras/stw2415}, \href
  {https://ui.adsabs.harvard.edu/abs/2017MNRAS.464.1415M} {464, 1415}

\bibitem[\protect\citeauthoryear{{Matsuura} et~al.,}{{Matsuura}
  et~al.}{2011}]{Matsuura2011}
{Matsuura} M.,  et~al., 2011, \mn@doi [Science] {10.1126/science.1205983},
  \href {https://ui.adsabs.harvard.edu/abs/2011Sci...333.1258M} {333, 1258}

\bibitem[\protect\citeauthoryear{{Menezes}, {Selhorst}, {Gim{\'e}nez de Castro}
   \& {Valio}}{{Menezes} et~al.}{2021}]{menezes2021}
{Menezes} F.,  {Selhorst} C.~L.,  {Gim{\'e}nez de Castro} C.~G.,   {Valio} A.,
  2021, \mn@doi [\apj] {10.3847/1538-4357/abe41c}, \href
  {https://ui.adsabs.harvard.edu/abs/2021ApJ...910...77M} {910, 77}

\bibitem[\protect\citeauthoryear{{Miotello} et~al.,}{{Miotello}
  et~al.}{2017}]{miotello2017}
{Miotello} A.,  et~al., 2017, \mn@doi [\aap] {10.1051/0004-6361/201629556},
  \href {https://ui.adsabs.harvard.edu/abs/2017A&A...599A.113M} {599, A113}

\bibitem[\protect\citeauthoryear{{Miotello}, {Kamp}, {Birnstiel}, {Cleeves}  \&
  {Kataoka}}{{Miotello} et~al.}{2023}]{miotello2023}
{Miotello} A.,  {Kamp} I.,  {Birnstiel} T.,  {Cleeves} L.~C.,   {Kataoka} A.,
  2023, in {Inutsuka} S.,  {Aikawa} Y.,  {Muto} T.,  {Tomida} K.,   {Tamura}
  M.,  eds,  Astronomical Society of the Pacific Conference Series Vol. 534,
  Protostars and Planets VII. p.~501 (\mn@eprint {arXiv} {2203.09818}),
  \mn@doi{10.48550/arXiv.2203.09818}

\bibitem[\protect\citeauthoryear{{Mizuno} et~al.,}{{Mizuno}
  et~al.}{2020}]{mizuno2020}
{Mizuno} I.,  et~al., 2020, in {Zmuidzinas} J.,  {Gao} J.-R.,  eds,  Society of
  Photo-Optical Instrumentation Engineers (SPIE) Conference Series Vol. 11453,
  Millimeter, Submillimeter, and Far-Infrared Detectors and Instrumentation for
  Astronomy X. p. 114533T (\mn@eprint {arXiv} {2012.07349}),
  \mn@doi{10.1117/12.2561742}

\bibitem[\protect\citeauthoryear{{Molinari} et~al.,}{{Molinari}
  et~al.}{2016}]{2016molinari}
{Molinari} S.,  et~al., 2016, \mn@doi [\aap] {10.1051/0004-6361/201526380},
  \href {https://ui.adsabs.harvard.edu/abs/2016A&A...591A.149M} {591, A149}

\bibitem[\protect\citeauthoryear{{Moncelsi} et~al.,}{{Moncelsi}
  et~al.}{2014}]{moncelsi2014}
{Moncelsi} L.,  et~al., 2014, \mn@doi [\mnras] {10.1093/mnras/stt2090}, \href
  {https://ui.adsabs.harvard.edu/abs/2014MNRAS.437.2772M} {437, 2772}

\bibitem[\protect\citeauthoryear{{Moore} et~al.,}{{Moore}
  et~al.}{2015}]{Moore15}
{Moore} T.~J.~T.,  et~al., 2015, \mn@doi [\mnras] {10.1093/mnras/stv1833},
  \href {https://ui.adsabs.harvard.edu/abs/2015MNRAS.453.4264M} {453, 4264}

\bibitem[\protect\citeauthoryear{{Motte} et~al.,}{{Motte}
  et~al.}{2022}]{Motte22}
{Motte} F.,  et~al., 2022, \mn@doi [\aap] {10.1051/0004-6361/202141677}, \href
  {https://ui.adsabs.harvard.edu/abs/2022A&A...662A...8M} {662, A8}

\bibitem[\protect\citeauthoryear{{Moullet} et~al.,}{{Moullet}
  et~al.}{2023}]{moullet2023}
{Moullet} A.,  et~al., 2023, \mn@doi [arXiv e-prints]
  {10.48550/arXiv.2310.20572}, \href
  {https://ui.adsabs.harvard.edu/abs/2023arXiv231020572M} {p. arXiv:2310.20572}

\bibitem[\protect\citeauthoryear{{Mr{\'a}zikov{\'a}}
  et~al.,}{{Mr{\'a}zikov{\'a}} et~al.}{2024}]{mrazikova2024}
{Mr{\'a}zikov{\'a}} K.,  et~al., 2024, \mn@doi [Astrobiology]
  {10.1089/ast.2023.0046}, \href
  {https://ui.adsabs.harvard.edu/abs/2024AsBio..24..407M} {24, 407}

\bibitem[\protect\citeauthoryear{{Mroczkowski} et~al.,}{{Mroczkowski}
  et~al.}{2024}]{mroczkowski2024}
{Mroczkowski} T.,  et~al., 2024, \mn@doi [arXiv e-prints]
  {10.48550/arXiv.2402.18645}, \href
  {https://ui.adsabs.harvard.edu/abs/2024arXiv240218645M} {p. arXiv:2402.18645}

\bibitem[\protect\citeauthoryear{{Mumma} \& {Charnley}}{{Mumma} \&
  {Charnley}}{2011}]{mumma2011}
{Mumma} M.~J.,  {Charnley} S.~B.,  2011, \mn@doi [\araa]
  {10.1146/annurev-astro-081309-130811}, \href
  {https://ui.adsabs.harvard.edu/abs/2011ARA&A..49..471M} {49, 471}

\bibitem[\protect\citeauthoryear{{Murray}, {Nartallo}, {Haynes}, {Gannaway}  \&
  {Ade}}{{Murray} et~al.}{1997}]{murray1997}
{Murray} A.~G.,  {Nartallo} R.,  {Haynes} C.~V.,  {Gannaway} F.,   {Ade}
  P.~A.~R.,  1997, in {Wilson} A.,  ed.,  ESA Special Publication Vol. 401, The
  Far Infrared and Submillimetre Universe.. p.~405

\bibitem[\protect\citeauthoryear{{Nagai} et~al.,}{{Nagai}
  et~al.}{2016}]{nagai2016}
{Nagai} H.,  et~al., 2016, \mn@doi [\apj] {10.3847/0004-637X/824/2/132}, \href
  {https://ui.adsabs.harvard.edu/abs/2016ApJ...824..132N} {824, 132}

\bibitem[\protect\citeauthoryear{{Negrello} et~al.,}{{Negrello}
  et~al.}{2010}]{negrello+10}
{Negrello} M.,  et~al., 2010, \mn@doi [Science] {10.1126/science.1193420},
  \href {https://ui.adsabs.harvard.edu/abs/2010Sci...330..800N} {330, 800}

\bibitem[\protect\citeauthoryear{{Negrello} et~al.,}{{Negrello}
  et~al.}{2017}]{negrello+17}
{Negrello} M.,  et~al., 2017, \mn@doi [\mnras] {10.1093/mnras/stw2911}, \href
  {https://ui.adsabs.harvard.edu/abs/2017MNRAS.465.3558N} {465, 3558}

\bibitem[\protect\citeauthoryear{{Neri} et~al.,}{{Neri} et~al.}{2020}]{Neri+20}
{Neri} R.,  et~al., 2020, \mn@doi [\aap] {10.1051/0004-6361/201936988}, \href
  {https://ui.adsabs.harvard.edu/abs/2020A&A...635A...7N} {635, A7}

\bibitem[\protect\citeauthoryear{{Nesvadba} et~al.,}{{Nesvadba}
  et~al.}{2016}]{Planck_GEMS_2}
{Nesvadba} N.,  et~al., 2016, \mn@doi [\aap] {10.1051/0004-6361/201629037},
  \href {https://ui.adsabs.harvard.edu/abs/2016A&A...593L...2N} {593, L2}

\bibitem[\protect\citeauthoryear{{Nesvadba}, {Ca{\~n}ameras}, {Kneissl},
  {Koenig}, {Yang}, {Le Floc'h}, {Omont}  \& {Scott}}{{Nesvadba}
  et~al.}{2019}]{Planck_GEMS_7}
{Nesvadba} N.~P.~H.,  {Ca{\~n}ameras} R.,  {Kneissl} R.,  {Koenig} S.,  {Yang}
  C.,  {Le Floc'h} E.,  {Omont} A.,   {Scott} D.,  2019, \mn@doi [\aap]
  {10.1051/0004-6361/201833777}, \href
  {https://ui.adsabs.harvard.edu/abs/2019A&A...624A..23N} {624, A23}

\bibitem[\protect\citeauthoryear{{Nindos}, {Kontar}  \& {Oberoi}}{{Nindos}
  et~al.}{2019}]{2019AdSpR..63.1404N}
{Nindos} A.,  {Kontar} E.~P.,   {Oberoi} D.,  2019, \mn@doi [Advances in Space
  Research] {10.1016/j.asr.2018.10.023}, \href
  {https://ui.adsabs.harvard.edu/abs/2019AdSpR..63.1404N} {63, 1404}

\bibitem[\protect\citeauthoryear{{Nixon} et~al.,}{{Nixon}
  et~al.}{2020}]{nixon2020}
{Nixon} C.~A.,  et~al., 2020, \mn@doi [\aj] {10.3847/1538-3881/abb679}, \href
  {https://ui.adsabs.harvard.edu/abs/2020AJ....160..205N} {160, 205}

\bibitem[\protect\citeauthoryear{{Noboriguchi} et~al.,}{{Noboriguchi}
  et~al.}{2022}]{Noboriguchi+22}
{Noboriguchi} A.,  et~al., 2022, \mn@doi [\apj] {10.3847/1538-4357/aca403},
  \href {https://ui.adsabs.harvard.edu/abs/2022ApJ...941..195N} {941, 195}

\bibitem[\protect\citeauthoryear{{{\"O}berg} et~al.,}{{{\"O}berg}
  et~al.}{2021}]{oberg2021}
{{\"O}berg} K.~I.,  et~al., 2021, \mn@doi [\apjs] {10.3847/1538-4365/ac1432},
  \href {https://ui.adsabs.harvard.edu/abs/2021ApJS..257....1O} {257, 1}

\bibitem[\protect\citeauthoryear{{Ohashi} et~al.,}{{Ohashi}
  et~al.}{2023}]{ohashi2023}
{Ohashi} N.,  et~al., 2023, \mn@doi [\apj] {10.3847/1538-4357/acd384}, \href
  {https://ui.adsabs.harvard.edu/abs/2023ApJ...951....8O} {951, 8}

\bibitem[\protect\citeauthoryear{{Oliver} et~al.,}{{Oliver}
  et~al.}{2010}]{Oliver+10}
{Oliver} S.~J.,  et~al., 2010, \mn@doi [\aap] {10.1051/0004-6361/201014697},
  \href {https://ui.adsabs.harvard.edu/abs/2010A&A...518L..21O} {518, L21}

\bibitem[\protect\citeauthoryear{{Oteo} et~al.,}{{Oteo} et~al.}{2017}]{Oteo+17}
{Oteo} I.,  et~al., 2017, \mn@doi [arXiv e-prints] {10.48550/arXiv.1709.04191},
  \href {https://ui.adsabs.harvard.edu/abs/2017arXiv170904191O} {p.
  arXiv:1709.04191}

\bibitem[\protect\citeauthoryear{{Overzier}}{{Overzier}}{2016}]{Overzier+16_protoclusters}
{Overzier} R.~A.,  2016, \mn@doi [\aapr] {10.1007/s00159-016-0100-3}, \href
  {https://ui.adsabs.harvard.edu/abs/2016A&ARv..24...14O} {24, 14}

\bibitem[\protect\citeauthoryear{{Palmer}, {Cordiner}, {Nixon}, {Charnley},
  {Teanby}, {Kisiel}, {Irwin}  \& {Mumma}}{{Palmer} et~al.}{2017}]{palmer2017}
{Palmer} M.~Y.,  {Cordiner} M.~A.,  {Nixon} C.~A.,  {Charnley} S.~B.,  {Teanby}
  N.~A.,  {Kisiel} Z.,  {Irwin} P. G.~J.,   {Mumma} M.~J.,  2017, \mn@doi
  [Science Advances] {10.1126/sciadv.1700022}, \href
  {https://ui.adsabs.harvard.edu/abs/2017SciA....3E0022P} {3, e1700022}

\bibitem[\protect\citeauthoryear{{Parker}, {Ward-Thompson}  \& {Kirk}}{{Parker}
  et~al.}{2022}]{parker2022}
{Parker} R.,  {Ward-Thompson} D.,   {Kirk} J.,  2022, \mn@doi [\mnras]
  {10.1093/mnras/stac152}, \href
  {https://ui.adsabs.harvard.edu/abs/2022MNRAS.511.2453P} {511, 2453}

\bibitem[\protect\citeauthoryear{{Paspaliaris}, {Xilouris}, {Nersesian},
  {Bianchi}, {Georgantopoulos}, {Masoura}, {Magdis}  \&
  {Plionis}}{{Paspaliaris} et~al.}{2023}]{Paspaliaris+23}
{Paspaliaris} E.~D.,  {Xilouris} E.~M.,  {Nersesian} A.,  {Bianchi} S.,
  {Georgantopoulos} I.,  {Masoura} V.~A.,  {Magdis} G.~E.,   {Plionis} M.,
  2023, \mn@doi [\aap] {10.1051/0004-6361/202244796}, \href
  {https://ui.adsabs.harvard.edu/abs/2023A&A...669A..11P} {669, A11}

\bibitem[\protect\citeauthoryear{{Pattle} et~al.,}{{Pattle}
  et~al.}{2017}]{pattle2017}
{Pattle} K.,  et~al., 2017, \mn@doi [\apj] {10.3847/1538-4357/aa80e5}, \href
  {https://ui.adsabs.harvard.edu/abs/2017ApJ...846..122P} {846, 122}

\bibitem[\protect\citeauthoryear{{Pattle} et~al.,}{{Pattle}
  et~al.}{2021a}]{pattle2021a}
{Pattle} K.,  et~al., 2021a, \mn@doi [\mnras] {10.1093/mnras/stab608}, \href
  {https://ui.adsabs.harvard.edu/abs/2021MNRAS.503.3414P} {503, 3414}

\bibitem[\protect\citeauthoryear{{Pattle}, {Gear}, {Redman}, {Smith}  \&
  {Greaves}}{{Pattle} et~al.}{2021b}]{2021Pattle}
{Pattle} K.,  {Gear} W.,  {Redman} M.,  {Smith} M. W.~L.,   {Greaves} J.,
  2021b, \mn@doi [\mnras] {10.1093/mnras/stab1300}, \href
  {https://ui.adsabs.harvard.edu/abs/2021MNRAS.505..684P} {505, 684}

\bibitem[\protect\citeauthoryear{{Pattle}, {Fissel}, {Tahani}, {Liu}  \&
  {Ntormousi}}{{Pattle} et~al.}{2023}]{pattle2023}
{Pattle} K.,  {Fissel} L.,  {Tahani} M.,  {Liu} T.,   {Ntormousi} E.,  2023, in
  {Inutsuka} S.,  {Aikawa} Y.,  {Muto} T.,  {Tomida} K.,   {Tamura} M.,  eds,
  Astronomical Society of the Pacific Conference Series Vol. 534, Protostars
  and Planets VII. p.~193 (\mn@eprint {arXiv} {2203.11179}),
  \mn@doi{10.48550/arXiv.2203.11179}

\bibitem[\protect\citeauthoryear{{Pawlyk} et~al.,}{{Pawlyk}
  et~al.}{2018}]{Pawlyk2018}
{Pawlyk} S.,  et~al., 2018, in {Zmuidzinas} J.,  {Gao} J.-R.,  eds,  Society of
  Photo-Optical Instrumentation Engineers (SPIE) Conference Series Vol. 10708,
  Millimeter, Submillimeter, and Far-Infrared Detectors and Instrumentation for
  Astronomy IX. p. 1070806, \mn@doi{10.1117/12.2313874}

\bibitem[\protect\citeauthoryear{{Pearce}}{{Pearce}}{2024}]{2024Pearce}
{Pearce} T.~D.,  2024, \mn@doi [arXiv e-prints] {10.48550/arXiv.2403.11804},
  \href {https://ui.adsabs.harvard.edu/abs/2024arXiv240311804P} {p.
  arXiv:2403.11804}

\bibitem[\protect\citeauthoryear{{Pearson} et~al.,}{{Pearson}
  et~al.}{2024}]{Pearson+24}
{Pearson} J.,  et~al., 2024, \mn@doi [\mnras] {10.1093/mnras/stad3916}, \href
  {https://ui.adsabs.harvard.edu/abs/2024MNRAS.52712044P} {527, 12044}

\bibitem[\protect\citeauthoryear{{Peretto} et~al.,}{{Peretto}
  et~al.}{2013}]{Peretto+13}
{Peretto} N.,  et~al., 2013, \mn@doi [\aap] {10.1051/0004-6361/201321318},
  \href {https://ui.adsabs.harvard.edu/abs/2013A&A...555A.112P} {555, A112}

\bibitem[\protect\citeauthoryear{{P{\'e}roux} \& {Howk}}{{P{\'e}roux} \&
  {Howk}}{2020}]{Celine_Cristopher_2020_ARAA}
{P{\'e}roux} C.,  {Howk} J.~C.,  2020, \mn@doi [\araa]
  {10.1146/annurev-astro-021820-120014}, \href
  {https://ui.adsabs.harvard.edu/abs/2020ARA&A..58..363P} {58, 363}

\bibitem[\protect\citeauthoryear{{Pilbratt} et~al.,}{{Pilbratt}
  et~al.}{2010}]{Pilbratt+10_Herschel}
{Pilbratt} G.~L.,  et~al., 2010, \mn@doi [\aap] {10.1051/0004-6361/201014759},
  \href {https://ui.adsabs.harvard.edu/abs/2010A&A...518L...1P} {518, L1}

\bibitem[\protect\citeauthoryear{{Pineda}, {Siles}, {Groppi}, {Kawamura},
  {Bernasconi}  \& {Goldsmith}}{{Pineda} et~al.}{2022}]{Pineda2022}
{Pineda} J.,  {Siles} J.,  {Groppi} C.,  {Kawamura} J.,  {Bernasconi} P.,
  {Goldsmith} P.,  2022, in American Astronomical Society Meeting \#240. p.
  314.02

\bibitem[\protect\citeauthoryear{{Pineda} et~al.,}{{Pineda}
  et~al.}{2023}]{pineda2023}
{Pineda} J.~E.,  et~al., 2023, in {Inutsuka} S.,  {Aikawa} Y.,  {Muto} T.,
  {Tomida} K.,   {Tamura} M.,  eds,  Astronomical Society of the Pacific
  Conference Series Vol. 534, Protostars and Planets VII. p.~233 (\mn@eprint
  {arXiv} {2205.03935}), \mn@doi{10.48550/arXiv.2205.03935}

\bibitem[\protect\citeauthoryear{{Pisano}, {Savini}, {Ade}, {Haynes}  \&
  {Gear}}{{Pisano} et~al.}{2006}]{pisano2006}
{Pisano} G.,  {Savini} G.,  {Ade} P. A.~R.,  {Haynes} V.,   {Gear} W.~K.,
  2006, \mn@doi [\ao] {10.1364/AO.45.006982}, \href
  {https://ui.adsabs.harvard.edu/abs/2006ApOpt..45.6982P} {45, 6982}

\bibitem[\protect\citeauthoryear{{Planck Collaboration} et~al.,}{{Planck
  Collaboration} et~al.}{2015}]{planckintXIX}
{Planck Collaboration} et~al., 2015, \mn@doi [\aap]
  {10.1051/0004-6361/201424082}, \href
  {https://ui.adsabs.harvard.edu/abs/2015A&A...576A.104P} {576, A104}

\bibitem[\protect\citeauthoryear{{Planck Collaboration} et~al.,}{{Planck
  Collaboration} et~al.}{2016}]{Planck_PHz}
{Planck Collaboration} et~al., 2016, \mn@doi [\aap]
  {10.1051/0004-6361/201527206}, \href
  {https://ui.adsabs.harvard.edu/abs/2016A&A...596A.100P} {596, A100}

\bibitem[\protect\citeauthoryear{{Polletta} et~al.,}{{Polletta}
  et~al.}{2021}]{Polletta+21_Planck_protocluster_cosmos}
{Polletta} M.,  et~al., 2021, \mn@doi [\aap] {10.1051/0004-6361/202140612},
  \href {https://ui.adsabs.harvard.edu/abs/2021A&A...654A.121P} {654, A121}

\bibitem[\protect\citeauthoryear{{Polletta}, {Dole}, {Martinache}, {Lehnert},
  {Frye}  \& {Kneissl}}{{Polletta}
  et~al.}{2022}]{Polletta+22_protocluster_molecular_gas}
{Polletta} M.,  {Dole} H.,  {Martinache} C.,  {Lehnert} M.~D.,  {Frye} B.~L.,
  {Kneissl} R.,  2022, \mn@doi [\aap] {10.1051/0004-6361/202142255}, \href
  {https://ui.adsabs.harvard.edu/abs/2022A&A...662A..85P} {662, A85}

\bibitem[\protect\citeauthoryear{{Priestley}, {Barlow}  \& {De
  Looze}}{{Priestley} et~al.}{2019}]{Priestley2019}
{Priestley} F.~D.,  {Barlow} M.~J.,   {De Looze} I.,  2019, \mn@doi [\mnras]
  {10.1093/mnras/stz414}, \href
  {https://ui.adsabs.harvard.edu/abs/2019MNRAS.485..440P} {485, 440}

\bibitem[\protect\citeauthoryear{{Priestley}, {Barlow}, {De Looze}  \&
  {Chawner}}{{Priestley} et~al.}{2020}]{Priestley2020}
{Priestley} F.~D.,  {Barlow} M.~J.,  {De Looze} I.,   {Chawner} H.,  2020,
  \mn@doi [\mnras] {10.1093/mnras/stz3434}, \href
  {https://ui.adsabs.harvard.edu/abs/2020MNRAS.491.6020P} {491, 6020}

\bibitem[\protect\citeauthoryear{{Quir{\'o}s-Rojas}, {Monta{\~n}a}, {Zavala},
  {Aretxaga}  \& {Hughes}}{{Quir{\'o}s-Rojas} et~al.}{2024}]{3000_ultra_reds}
{Quir{\'o}s-Rojas} M.,  {Monta{\~n}a} A.,  {Zavala} J.~A.,  {Aretxaga} I.,
  {Hughes} D.~H.,  2024, \mn@doi [arXiv e-prints] {10.48550/arXiv.2406.15729},
  \href {https://ui.adsabs.harvard.edu/abs/2024arXiv240615729Q} {p.
  arXiv:2406.15729}

\bibitem[\protect\citeauthoryear{RAL}{RAL}{2023}]{jupiter_moon}
RAL 2023, {STFC spin-out supports mission to Jupiter’s moons},
  \url{https://www.ukri.org/blog/stfc-spin-out-supports-mission-to-jupiters-moons/}

\bibitem[\protect\citeauthoryear{{Ramasawmy} et~al.,}{{Ramasawmy}
  et~al.}{2022}]{ramasawmy2020}
{Ramasawmy} J.,  et~al., 2022, in {Zmuidzinas} J.,  {Gao} J.-R.,  eds,  Society
  of Photo-Optical Instrumentation Engineers (SPIE) Conference Series Vol.
  12190, Millimeter, Submillimeter, and Far-Infrared Detectors and
  Instrumentation for Astronomy XI. p. 1219007 (\mn@eprint {arXiv}
  {2207.03914}), \mn@doi{10.1117/12.2627505}

\bibitem[\protect\citeauthoryear{{Raymond} et~al.,}{{Raymond}
  et~al.}{2021}]{raymond2021}
{Raymond} A.~W.,  et~al., 2021, \mn@doi [\apjs] {10.3847/1538-3881/abc3c3},
  \href {https://ui.adsabs.harvard.edu/abs/2021ApJS..253....5R} {253, 5}

\bibitem[\protect\citeauthoryear{{Reichborn-Kjennerud}
  et~al.,}{{Reichborn-Kjennerud} et~al.}{2010}]{Reichborn-Kjennerud2010}
{Reichborn-Kjennerud} B.,  et~al., 2010, in {Holland} W.~S.,  {Zmuidzinas} J.,
  eds,  Society of Photo-Optical Instrumentation Engineers (SPIE) Conference
  Series Vol. 7741, Millimeter, Submillimeter, and Far-Infrared Detectors and
  Instrumentation for Astronomy V. p. 77411C (\mn@eprint {arXiv} {1007.3672}),
  \mn@doi{10.1117/12.857138}

\bibitem[\protect\citeauthoryear{{Reid}, {Haschick}, {Burke}, {Moran},
  {Johnston}  \& {Swenson}}{{Reid} et~al.}{1980}]{reid1980}
{Reid} M.~J.,  {Haschick} A.~D.,  {Burke} B.~F.,  {Moran} J.~M.,  {Johnston}
  K.~J.,   {Swenson} G.~W. J.,  1980, \mn@doi [\apj] {10.1086/158092}, \href
  {https://ui.adsabs.harvard.edu/abs/1980ApJ...239...89R} {239, 89}

\bibitem[\protect\citeauthoryear{{Reuter} et~al.,}{{Reuter}
  et~al.}{2020}]{Reuter+20}
{Reuter} C.,  et~al., 2020, \mn@doi [\apj] {10.3847/1538-4357/abb599}, \href
  {https://ui.adsabs.harvard.edu/abs/2020ApJ...902...78R} {902, 78}

\bibitem[\protect\citeauthoryear{{Reuter} et~al.,}{{Reuter}
  et~al.}{2023}]{reuter+23}
{Reuter} C.,  et~al., 2023, \mn@doi [\apj] {10.3847/1538-4357/acaf51}, \href
  {https://ui.adsabs.harvard.edu/abs/2023ApJ...948...44R} {948, 44}

\bibitem[\protect\citeauthoryear{{Rigby} et~al.,}{{Rigby}
  et~al.}{2016}]{2016rigby}
{Rigby} A.~J.,  et~al., 2016, \mn@doi [\mnras] {10.1093/mnras/stv2808}, \href
  {https://ui.adsabs.harvard.edu/abs/2016MNRAS.456.2885R} {456, 2885}

\bibitem[\protect\citeauthoryear{{Rigby} et~al.,}{{Rigby}
  et~al.}{2019}]{Rigby19}
{Rigby} A.~J.,  et~al., 2019, \mn@doi [\aap] {10.1051/0004-6361/201935236},
  \href {https://ui.adsabs.harvard.edu/abs/2019A&A...632A..58R} {632, A58}

\bibitem[\protect\citeauthoryear{{Rigby} et~al.,}{{Rigby}
  et~al.}{2024}]{Rigby+24}
{Rigby} A.~J.,  et~al., 2024, \mn@doi [\mnras] {10.1093/mnras/stae030}, \href
  {https://ui.adsabs.harvard.edu/abs/2024MNRAS.528.1172R} {528, 1172}

\bibitem[\protect\citeauthoryear{{Ritacco} et~al.,}{{Ritacco}
  et~al.}{2017}]{ritacco2017}
{Ritacco} A.,  et~al., 2017, \mn@doi [\aap] {10.1051/0004-6361/201629666},
  \href {https://ui.adsabs.harvard.edu/abs/2017A&A...599A..34R} {599, A34}

\bibitem[\protect\citeauthoryear{{Ritacco} et~al.,}{{Ritacco}
  et~al.}{2020}]{ritacco2020}
{Ritacco} A.,  et~al., 2020, in mm Universe @ NIKA2 - Observing the mm Universe
  with the NIKA2 Camera. p. 00022 (\mn@eprint {arXiv} {1912.07894}),
  \mn@doi{10.1051/epjconf/202022800022}

\bibitem[\protect\citeauthoryear{{Rodger}, {Labrosse}, {Wedemeyer},
  {Szydlarski}, {Sim{\~o}es}  \& {Fletcher}}{{Rodger}
  et~al.}{2019}]{2019ApJ...875..163R}
{Rodger} A.~S.,  {Labrosse} N.,  {Wedemeyer} S.,  {Szydlarski} M.,
  {Sim{\~o}es} P. J.~A.,   {Fletcher} L.,  2019, \mn@doi [\apj]
  {10.3847/1538-4357/aafdfb}, \href
  {https://ui.adsabs.harvard.edu/abs/2019ApJ...875..163R} {875, 163}

\bibitem[\protect\citeauthoryear{{Roth} et~al.,}{{Roth}
  et~al.}{2021}]{roth2021}
{Roth} N.~X.,  et~al., 2021, \mn@doi [\apj] {10.3847/1538-4357/ac0441}, \href
  {https://ui.adsabs.harvard.edu/abs/2021ApJ...921...14R} {921, 14}

\bibitem[\protect\citeauthoryear{{Saintonge} et~al.,}{{Saintonge}
  et~al.}{2018}]{2018saintonge}
{Saintonge} A.,  et~al., 2018, \mn@doi [\mnras] {10.1093/mnras/sty2499}, \href
  {https://ui.adsabs.harvard.edu/abs/2018MNRAS.481.3497S} {481, 3497}

\bibitem[\protect\citeauthoryear{{Savini}, {Pisano}  \& {Ade}}{{Savini}
  et~al.}{2006}]{savini2006}
{Savini} G.,  {Pisano} G.,   {Ade} P. A.~R.,  2006, \mn@doi [\ao]
  {10.1364/AO.45.008907}, \href
  {https://ui.adsabs.harvard.edu/abs/2006ApOpt..45.8907S} {45, 8907}

\bibitem[\protect\citeauthoryear{{Savini}, {Ade}, {House}, {Pisano}, {Haynes}
  \& {Bastien}}{{Savini} et~al.}{2009}]{savini2009}
{Savini} G.,  {Ade} P. A.~R.,  {House} J.,  {Pisano} G.,  {Haynes} V.,
  {Bastien} P.,  2009, \mn@doi [\ao] {10.1364/AO.48.002006}, \href
  {https://ui.adsabs.harvard.edu/abs/2009ApOpt..48.2006S} {48, 2006}

\bibitem[\protect\citeauthoryear{{Schruba}, {Leroy}, {Walter}, {Sandstrom}  \&
  {Rosolowsky}}{{Schruba} et~al.}{2010}]{2010Schruba}
{Schruba} A.,  {Leroy} A.~K.,  {Walter} F.,  {Sandstrom} K.,   {Rosolowsky} E.,
   2010, \mn@doi [\apj] {10.1088/0004-637X/722/2/1699}, \href
  {https://ui.adsabs.harvard.edu/abs/2010ApJ...722.1699S} {722, 1699}

\bibitem[\protect\citeauthoryear{{Schuh} \& {Behrend}}{{Schuh} \&
  {Behrend}}{2012}]{schuh2012}
{Schuh} H.,  {Behrend} D.,  2012, \mn@doi [Journal of Geodynamics]
  {10.1016/j.jog.2012.07.007}, \href
  {https://ui.adsabs.harvard.edu/abs/2012JGeo...61...68S} {61, 68}

\bibitem[\protect\citeauthoryear{{Schuller} et~al.,}{{Schuller}
  et~al.}{2021}]{schuller2021}
{Schuller} F.,  et~al., 2021, \mn@doi [\mnras] {10.1093/mnras/staa2369}, \href
  {https://ui.adsabs.harvard.edu/abs/2021MNRAS.500.3064S} {500, 3064}

\bibitem[\protect\citeauthoryear{{Schulze-Makuch}, {Irwin}  \&
  {Irwin}}{{Schulze-Makuch} et~al.}{2024}]{schulzemakuch2024}
{Schulze-Makuch} D.,  {Irwin} L.~N.,   {Irwin} T.,  2024, \mn@doi
  [Astrobiology] {10.1089/ast.2022.0134}, \href
  {https://ui.adsabs.harvard.edu/abs/2024AsBio..24..397S} {24, 397}

\bibitem[\protect\citeauthoryear{{Schwarz}, {Bergin}, {Cleeves}, {Blake},
  {Zhang}, {{\"O}berg}, {van Dishoeck}  \& {Qi}}{{Schwarz}
  et~al.}{2016}]{schwarz2016}
{Schwarz} K.~R.,  {Bergin} E.~A.,  {Cleeves} L.~I.,  {Blake} G.~A.,  {Zhang}
  K.,  {{\"O}berg} K.~I.,  {van Dishoeck} E.~F.,   {Qi} C.,  2016, \mn@doi
  [\apj] {10.3847/0004-637X/823/2/91}, \href
  {https://ui.adsabs.harvard.edu/abs/2016ApJ...823...91S} {823, 91}

\bibitem[\protect\citeauthoryear{{Scicluna} et~al.,}{{Scicluna}
  et~al.}{2022}]{Scicluna2022}
{Scicluna} P.,  et~al., 2022, \mn@doi [\mnras] {10.1093/mnras/stab2860}, \href
  {https://ui.adsabs.harvard.edu/abs/2022MNRAS.512.1091S} {512, 1091}

\bibitem[\protect\citeauthoryear{{Scott} et~al.,}{{Scott}
  et~al.}{2002}]{scott+02}
{Scott} S.~E.,  et~al., 2002, \mn@doi [\mnras]
  {10.1046/j.1365-8711.2002.05193.x}, \href
  {https://ui.adsabs.harvard.edu/abs/2002MNRAS.331..817S} {331, 817}

\bibitem[\protect\citeauthoryear{{Serjeant} \& {Bakx}}{{Serjeant} \&
  {Bakx}}{2023}]{SerjeantBakx23}
{Serjeant} S.,  {Bakx} T. J.~L.~C.,  2023, \mn@doi [Nature Astronomy]
  {10.1038/s41550-023-02093-8}, \href
  {https://ui.adsabs.harvard.edu/abs/2023NatAs...7.1143S} {7, 1143}

\bibitem[\protect\citeauthoryear{{Serjeant}, {Bolton}, {Gandhi}, {Stappers},
  {Mazzali}, {Verma}  \& {No{\"e}l}}{{Serjeant}
  et~al.}{2023}]{STFC_AAP_2022_Roadmap}
{Serjeant} S.,  {Bolton} J.,  {Gandhi} P.,  {Stappers} B.,  {Mazzali} P.,
  {Verma} A.,   {No{\"e}l} N. E.~D.,  2023, \mn@doi [arXiv e-prints]
  {10.48550/arXiv.2301.05457}, \href
  {https://ui.adsabs.harvard.edu/abs/2023arXiv230105457S} {p. arXiv:2301.05457}

\bibitem[\protect\citeauthoryear{{Serjeant}, {Pearson}, {Dickinson}  \&
  {Jarvis}}{{Serjeant} et~al.}{2024}]{Serjeant+24}
{Serjeant} S.,  {Pearson} J.,  {Dickinson} H.,   {Jarvis} J.,  2024, \mn@doi
  [European Physical Journal Plus] {10.1140/epjp/s13360-024-05223-x}, \href
  {https://ui.adsabs.harvard.edu/abs/2024EPJP..139..418S} {139, 418}

\bibitem[\protect\citeauthoryear{{Shimakawa} et~al.,}{{Shimakawa}
  et~al.}{2018}]{Shimakawa+18}
{Shimakawa} R.,  et~al., 2018, \mn@doi [\mnras] {10.1093/mnras/sty2618}, \href
  {https://ui.adsabs.harvard.edu/abs/2018MNRAS.481.5630S} {481, 5630}

\bibitem[\protect\citeauthoryear{{Shimojo} et~al.,}{{Shimojo}
  et~al.}{2017a}]{shimojo2017}
{Shimojo} M.,  et~al., 2017a, \mn@doi [\solphys] {10.1007/s11207-017-1095-2},
  \href {https://ui.adsabs.harvard.edu/abs/2017SoPh..292...87S} {292, 87}

\bibitem[\protect\citeauthoryear{{Shimojo}, {Hudson}, {White}, {Bastian}  \&
  {Iwai}}{{Shimojo} et~al.}{2017b}]{shimojo2017a}
{Shimojo} M.,  {Hudson} H.~S.,  {White} S.~M.,  {Bastian} T.~S.,   {Iwai} K.,
  2017b, \mn@doi [\apjl] {10.3847/2041-8213/aa70e3}, \href
  {https://ui.adsabs.harvard.edu/abs/2017ApJ...841L...5S} {841, L5}

\bibitem[\protect\citeauthoryear{{Simpson} et~al.,}{{Simpson}
  et~al.}{2014}]{simpson+14}
{Simpson} J.~M.,  et~al., 2014, \mn@doi [\apj] {10.1088/0004-637X/788/2/125},
  \href {https://ui.adsabs.harvard.edu/abs/2014ApJ...788..125S} {788, 125}

\bibitem[\protect\citeauthoryear{{Skoki{\'c}}, {Benz}, {Braj{\v{s}}a}, {Sudar},
  {Matkovi{\'c}}  \& {B{\'a}rta}}{{Skoki{\'c}} et~al.}{2023}]{skokic2023}
{Skoki{\'c}} I.,  {Benz} A.~O.,  {Braj{\v{s}}a} R.,  {Sudar} D.,
  {Matkovi{\'c}} F.,   {B{\'a}rta} M.,  2023, \mn@doi [\aap]
  {10.1051/0004-6361/202244532}, \href
  {https://ui.adsabs.harvard.edu/abs/2023A&A...669A.156S} {669, A156}

\bibitem[\protect\citeauthoryear{{Smail}, {Ivison}  \& {Blain}}{{Smail}
  et~al.}{1997}]{smail+97}
{Smail} I.,  {Ivison} R.~J.,   {Blain} A.~W.,  1997, \mn@doi [\apjl]
  {10.1086/311017}, \href
  {https://ui.adsabs.harvard.edu/abs/1997ApJ...490L...5S} {490, L5}

\bibitem[\protect\citeauthoryear{{Smail} et~al.,}{{Smail}
  et~al.}{2021}]{smail+21}
{Smail} I.,  et~al., 2021, \mn@doi [\mnras] {10.1093/mnras/stab283}, \href
  {https://ui.adsabs.harvard.edu/abs/2021MNRAS.502.3426S} {502, 3426}

\bibitem[\protect\citeauthoryear{{Smith} et~al.,}{{Smith}
  et~al.}{2012}]{Smith2012}
{Smith} M.~W.~L.,  et~al., 2012, \mn@doi [\apj] {10.1088/0004-637X/756/1/40},
  \href {https://ui.adsabs.harvard.edu/abs/2012ApJ...756...40S} {756, 40}

\bibitem[\protect\citeauthoryear{{Smith} et~al.,}{{Smith}
  et~al.}{2021}]{2021smith}
{Smith} M. W.~L.,  et~al., 2021, \mn@doi [\apjs] {10.3847/1538-4365/ac23d0},
  \href {https://ui.adsabs.harvard.edu/abs/2021ApJS..257...52S} {257, 52}

\bibitem[\protect\citeauthoryear{{Stach} et~al.,}{{Stach}
  et~al.}{2021}]{stach+21}
{Stach} S.~M.,  et~al., 2021, \mn@doi [\mnras] {10.1093/mnras/stab714}, \href
  {https://ui.adsabs.harvard.edu/abs/2021MNRAS.504..172S} {504, 172}

\bibitem[\protect\citeauthoryear{{Stangalini} et~al.,}{{Stangalini}
  et~al.}{2022}]{2022NatCo..13..479S}
{Stangalini} M.,  et~al., 2022, \mn@doi [Nature Communications]
  {10.1038/s41467-022-28136-8}, \href
  {https://ui.adsabs.harvard.edu/abs/2022NatCo..13..479S} {13, 479}

\bibitem[\protect\citeauthoryear{{Stevens} et~al.,}{{Stevens}
  et~al.}{2003}]{Stevens+03_signposts}
{Stevens} J.~A.,  et~al., 2003, \mn@doi [\nat]
  {10.48550/arXiv.astro-ph/0309495}, \href
  {https://ui.adsabs.harvard.edu/abs/2003Natur.425..264S} {425, 264}

\bibitem[\protect\citeauthoryear{{Sun} et~al.,}{{Sun} et~al.}{2020}]{2020Sun}
{Sun} J.,  et~al., 2020, \mn@doi [\apjl] {10.3847/2041-8213/abb3be}, \href
  {https://ui.adsabs.harvard.edu/abs/2020ApJ...901L...8S} {901, L8}

\bibitem[\protect\citeauthoryear{{Swinbank} et~al.,}{{Swinbank}
  et~al.}{2015}]{swinbank+15}
{Swinbank} A.~M.,  et~al., 2015, \mn@doi [\apjl] {10.1088/2041-8205/806/1/L17},
  \href {https://ui.adsabs.harvard.edu/abs/2015ApJ...806L..17S} {806, L17}

\bibitem[\protect\citeauthoryear{{Taniguchi}, {Tamura}, {Ikeda}, {Takekoshi}
  \& {Kawabe}}{{Taniguchi} et~al.}{2021}]{Taniguchi2021}
{Taniguchi} A.,  {Tamura} Y.,  {Ikeda} S.,  {Takekoshi} T.,   {Kawabe} R.,
  2021, \mn@doi [\aj] {10.3847/1538-3881/ac11f7}, \href
  {https://ui.adsabs.harvard.edu/abs/2021AJ....162..111T} {162, 111}

\bibitem[\protect\citeauthoryear{Tapia et~al.,}{Tapia
  et~al.}{2020}]{tapia2020muscat}
Tapia M.,  et~al., 2020, MUSCAT focal plane verification (\mn@eprint {arXiv}
  {2012.05126})

\bibitem[\protect\citeauthoryear{{Tazzari} et~al.,}{{Tazzari}
  et~al.}{2017}]{tazzari2017}
{Tazzari} M.,  et~al., 2017, \mn@doi [\aap] {10.1051/0004-6361/201730890},
  \href {https://ui.adsabs.harvard.edu/abs/2017A&A...606A..88T} {606, A88}

\bibitem[\protect\citeauthoryear{Terris, Dabbech, Tang  \& Wiaux}{Terris
  et~al.}{2022}]{Terris2022}
Terris M.,  Dabbech A.,  Tang C.,   Wiaux Y.,  2022, \mn@doi [Monthly Notices
  of the Royal Astronomical Society] {10.1093/mnras/stac2672}, 518, 604

\bibitem[\protect\citeauthoryear{{Testi} et~al.,}{{Testi}
  et~al.}{2014}]{Testi2014}
{Testi} L.,  et~al., 2014, in {Beuther} H.,  {Klessen} R.~S.,  {Dullemond}
  C.~P.,   {Henning} T.,  eds, Protostars and Planets VI. pp 339--361
  (\mn@eprint {arXiv} {1402.1354}),
  \mn@doi{10.2458/azu_uapress_9780816531240-ch015}

\bibitem[\protect\citeauthoryear{Thomas et~al.,}{Thomas et~al.}{2014}]{CAMELS}
Thomas C.~N.,  et~al., 2014, The CAMbridge Emission Line Surveyor (CAMELS)
  (\mn@eprint {arXiv} {1401.4395}), \url {https://arxiv.org/abs/1401.4395}

\bibitem[\protect\citeauthoryear{{Tiede}, {Johnson}, {Pesce}, {Palumbo},
  {Chang}  \& {Galison}}{{Tiede} et~al.}{2022}]{Tiede2022}
{Tiede} P.,  {Johnson} M.~D.,  {Pesce} D.~W.,  {Palumbo} D. C.~M.,  {Chang}
  D.~O.,   {Galison} P.,  2022, \mn@doi [Galaxies] {10.3390/galaxies10060111},
  \href {https://ui.adsabs.harvard.edu/abs/2022Galax..10..111T} {10, 111}

\bibitem[\protect\citeauthoryear{{Truch} et~al.,}{{Truch}
  et~al.}{2009}]{Truch2009}
{Truch} M. D.~P.,  et~al., 2009, \mn@doi [\apj] {10.1088/0004-637X/707/2/1723},
  \href {https://ui.adsabs.harvard.edu/abs/2009ApJ...707.1723T} {707, 1723}

\bibitem[\protect\citeauthoryear{{Urquhart} et~al.,}{{Urquhart}
  et~al.}{2018}]{urquhart18}
{Urquhart} J.~S.,  et~al., 2018, \mn@doi [\mnras] {10.1093/mnras/stx2258},
  \href {https://ui.adsabs.harvard.edu/abs/2018MNRAS.473.1059U} {473, 1059}

\bibitem[\protect\citeauthoryear{{Urquhart} et~al.,}{{Urquhart}
  et~al.}{2021}]{Urquhart21}
{Urquhart} J.~S.,  et~al., 2021, \mn@doi [\mnras] {10.1093/mnras/staa2512},
  \href {https://ui.adsabs.harvard.edu/abs/2021MNRAS.500.3050U} {500, 3050}

\bibitem[\protect\citeauthoryear{{Urquhart} et~al.,}{{Urquhart}
  et~al.}{2022}]{Urquhart+22}
{Urquhart} S.~A.,  et~al., 2022, \mn@doi [\mnras] {10.1093/mnras/stac150},
  \href {https://ui.adsabs.harvard.edu/abs/2022MNRAS.511.3017U} {511, 3017}

\bibitem[\protect\citeauthoryear{Valenzuela-Venegas, Lode, Viole, Felice,
  Martinez~Alonso, Ramirez~Camargo, Sartori  \& Zeyringer}{Valenzuela-Venegas
  et~al.}{2023}]{valenzuela2023}
Valenzuela-Venegas G.,  Lode M.~L.,  Viole I.,  Felice A.,  Martinez~Alonso A.,
   Ramirez~Camargo L.,  Sartori S.,   Zeyringer M.,  2023, Designing renewable
  and socially accepted energy systems for astronomical telescopes: A move
  towards energy justice, \mn@doi{10.21203/rs.3.rs-3181969/v1}

\bibitem[\protect\citeauthoryear{{Vegetti} \& {Vogelsberger}}{{Vegetti} \&
  {Vogelsberger}}{2014}]{Vegetti+14}
{Vegetti} S.,  {Vogelsberger} M.,  2014, \mn@doi [\mnras]
  {10.1093/mnras/stu1284}, \href
  {https://ui.adsabs.harvard.edu/abs/2014MNRAS.442.3598V} {442, 3598}

\bibitem[\protect\citeauthoryear{{Viole}, {Valenzuela-Venegas}, {Zeyringer}  \&
  {Sartori}}{{Viole} et~al.}{2023}]{viole2023}
{Viole} I.,  {Valenzuela-Venegas} G.,  {Zeyringer} M.,   {Sartori} S.,  2023,
  \mn@doi [Energy] {10.1016/j.energy.2023.128570}, \href
  {https://ui.adsabs.harvard.edu/abs/2023Ene...28228570V} {282, 128570}

\bibitem[\protect\citeauthoryear{{Walker} et~al.,}{{Walker}
  et~al.}{2022}]{Walker2022}
{Walker} C.,  et~al., 2022, in {Zmuidzinas} J.,  {Gao} J.-R.,  eds,  Society of
  Photo-Optical Instrumentation Engineers (SPIE) Conference Series Vol. 12190,
  Millimeter, Submillimeter, and Far-Infrared Detectors and Instrumentation for
  Astronomy XI. p. 121900E, \mn@doi{10.1117/12.2629051}

\bibitem[\protect\citeauthoryear{{Wang} et~al.,}{{Wang}
  et~al.}{2019}]{wang+19_dark_smgs}
{Wang} T.,  et~al., 2019, \mn@doi [\nat] {10.1038/s41586-019-1452-4}, \href
  {https://ui.adsabs.harvard.edu/abs/2019Natur.572..211W} {572, 211}

\bibitem[\protect\citeauthoryear{{Wang} et~al.,}{{Wang}
  et~al.}{2021a}]{wang+21_spt_unlensed_smgs_signpost_protoclusters}
{Wang} G. C.~P.,  et~al., 2021a, \mn@doi [\mnras] {10.1093/mnras/stab2800},
  \href {https://ui.adsabs.harvard.edu/abs/2021MNRAS.508.3754W} {508, 3754}

\bibitem[\protect\citeauthoryear{{Wang} et~al.,}{{Wang}
  et~al.}{2021b}]{wang+21_hermes_hlirgs}
{Wang} L.,  et~al., 2021b, \mn@doi [\aap] {10.1051/0004-6361/202038811}, \href
  {https://ui.adsabs.harvard.edu/abs/2021A&A...648A...8W} {648, A8}

\bibitem[\protect\citeauthoryear{{Ward-Thompson} et~al.,}{{Ward-Thompson}
  et~al.}{2017}]{wardthompson2017}
{Ward-Thompson} D.,  et~al., 2017, \mn@doi [\apj] {10.3847/1538-4357/aa70a0},
  \href {https://ui.adsabs.harvard.edu/abs/2017ApJ...842...66W} {842, 66}

\bibitem[\protect\citeauthoryear{{Ward}, {Eales}, {Ivison}  \&
  {Arumugam}}{{Ward} et~al.}{2024a}]{Ward2024}
{Ward} B.~A.,  {Eales} S.~A.,  {Ivison} R.~J.,   {Arumugam} V.,  2024a, \mn@doi
  [\mnras] {10.1093/mnras/stae405}, \href
  {https://ui.adsabs.harvard.edu/abs/2024MNRAS.530.4887W} {530, 4887}

\bibitem[\protect\citeauthoryear{{Ward}, {Eales}, {Ivison}  \&
  {Arumugam}}{{Ward} et~al.}{2024b}]{Ward+24}
{Ward} B.~A.,  {Eales} S.~A.,  {Ivison} R.~J.,   {Arumugam} V.,  2024b, \mn@doi
  [\mnras] {10.1093/mnras/stae405}, \href
  {https://ui.adsabs.harvard.edu/abs/2024MNRAS.530.4887W} {530, 4887}

\bibitem[\protect\citeauthoryear{{Wardlow} et~al.,}{{Wardlow}
  et~al.}{2013}]{wardlow+13}
{Wardlow} J.~L.,  et~al., 2013, \mn@doi [\apj] {10.1088/0004-637X/762/1/59},
  \href {https://ui.adsabs.harvard.edu/abs/2013ApJ...762...59W} {762, 59}

\bibitem[\protect\citeauthoryear{{Wedemeyer} et~al.,}{{Wedemeyer}
  et~al.}{2016}]{2016SSRv..200....1W}
{Wedemeyer} S.,  et~al., 2016, \mn@doi [\ssr] {10.1007/s11214-015-0229-9},
  \href {https://ui.adsabs.harvard.edu/abs/2016SSRv..200....1W} {200, 1}

\bibitem[\protect\citeauthoryear{Wenninger, Boussaha, Chaumont, Tan  \&
  Yassin}{Wenninger et~al.}{2023}]{wenninger2023design}
Wenninger J.,  Boussaha F.,  Chaumont C.,  Tan B.-K.,   Yassin G.,  2023,
  Superconductor Science and Technology, 36, 055012

\bibitem[\protect\citeauthoryear{{Williams}, {Gear}  \& {Smith}}{{Williams}
  et~al.}{2018}]{2018Williams}
{Williams} T.~G.,  {Gear} W.~K.,   {Smith} M. W.~L.,  2018, \mn@doi [\mnras]
  {10.1093/mnras/sty1476}, \href
  {https://ui.adsabs.harvard.edu/abs/2018MNRAS.479..297W} {479, 297}

\bibitem[\protect\citeauthoryear{{Williams} et~al.,}{{Williams}
  et~al.}{2022}]{2022Williams}
{Williams} T.~G.,  et~al., 2022, \mn@doi [\apjl] {10.3847/2041-8213/aca674},
  \href {https://ui.adsabs.harvard.edu/abs/2022ApJ...941L..27W} {941, L27}

\bibitem[\protect\citeauthoryear{{Williams} et~al.,}{{Williams}
  et~al.}{2023}]{2023Williams}
{Williams} T.~G.,  et~al., 2023, \mn@doi [\mnras] {10.1093/mnras/stad2455},
  \href {https://ui.adsabs.harvard.edu/abs/2023MNRAS.525.4270W} {525, 4270}

\bibitem[\protect\citeauthoryear{{Wyatt}}{{Wyatt}}{2006}]{2006Wyatt}
{Wyatt} M.~C.,  2006, \mn@doi [\apj] {10.1086/499487}, \href
  {https://ui.adsabs.harvard.edu/abs/2006ApJ...639.1153W} {639, 1153}

\bibitem[\protect\citeauthoryear{{Wyatt}}{{Wyatt}}{2021}]{2021Wyatt}
{Wyatt} M.~C.,  2021, in {Madhusudhan} N.,  ed., , ExoFrontiers; Big Questions
  in Exoplanetary Science.
pp 15--1, \mn@doi{10.1088/2514-3433/abfa8fch15}

\bibitem[\protect\citeauthoryear{{Yagoubov} et~al.,}{{Yagoubov}
  et~al.}{2020}]{2020A&A...634A..46Y}
{Yagoubov} P.,  et~al., 2020, \mn@doi [\aap] {10.1051/0004-6361/201936777},
  \href {https://ui.adsabs.harvard.edu/abs/2020A&A...634A..46Y} {634, A46}

\bibitem[\protect\citeauthoryear{{Yang}, {Jewitt}, {Zhao}, {Jiang}, {Ye}  \&
  {Chen}}{{Yang} et~al.}{2021}]{yang2021}
{Yang} B.,  {Jewitt} D.,  {Zhao} Y.,  {Jiang} X.,  {Ye} Q.,   {Chen} Y.-T.,
  2021, \mn@doi [\apjl] {10.3847/2041-8213/ac03b7}, \href
  {https://ui.adsabs.harvard.edu/abs/2021ApJ...914L..17Y} {914, L17}

\bibitem[\protect\citeauthoryear{{Yoo} et~al.,}{{Yoo} et~al.}{2017}]{yoo2017}
{Yoo} H.,  et~al., 2017, \mn@doi [\apj] {10.3847/1538-4357/aa8c0a}, \href
  {https://ui.adsabs.harvard.edu/abs/2017ApJ...849...69Y} {849, 69}

\bibitem[\protect\citeauthoryear{{Yoon} et~al.,}{{Yoon}
  et~al.}{2022}]{yoon2022}
{Yoon} S.-Y.,  et~al., 2022, \mn@doi [\apj] {10.3847/1538-4357/ac5632}, \href
  {https://ui.adsabs.harvard.edu/abs/2022ApJ...929...60Y} {929, 60}

\bibitem[\protect\citeauthoryear{{Zavala} et~al.,}{{Zavala}
  et~al.}{2023}]{Zavala+23}
{Zavala} J.~A.,  et~al., 2023, \mn@doi [\apjl] {10.3847/2041-8213/acacfe},
  \href {https://ui.adsabs.harvard.edu/abs/2023ApJ...943L...9Z} {943, L9}

\bibitem[\protect\citeauthoryear{{Zhang} et~al.,}{{Zhang}
  et~al.}{2021}]{Zhang2021}
{Zhang} K.,  et~al., 2021, \mn@doi [\apjs] {10.3847/1538-4365/ac1580}, \href
  {https://ui.adsabs.harvard.edu/abs/2021ApJS..257....5Z} {257, 5}

\bibitem[\protect\citeauthoryear{{Zhou} et~al.,}{{Zhou}
  et~al.}{2024a}]{RAGERS_Zhou}
{Zhou} D.,  et~al., 2024a, \mn@doi [arXiv e-prints]
  {10.48550/arXiv.2408.02177}, \href
  {https://ui.adsabs.harvard.edu/abs/2024arXiv240802177Z} {p. arXiv:2408.02177}

\bibitem[\protect\citeauthoryear{{Zhou} et~al.,}{{Zhou}
  et~al.}{2024b}]{Zhou+24_herschel_irac_overdensity_noema_protocluster}
{Zhou} L.,  et~al., 2024b, \mn@doi [\aap] {10.1051/0004-6361/202348351}, \href
  {https://ui.adsabs.harvard.edu/abs/2024A&A...684A.196Z} {684, A196}

\bibitem[\protect\citeauthoryear{{Zouganelis} et~al.,}{{Zouganelis}
  et~al.}{2020}]{2020A&A...642A...3Z}
{Zouganelis} I.,  et~al., 2020, \mn@doi [\aap] {10.1051/0004-6361/202038445},
  \href {https://ui.adsabs.harvard.edu/abs/2020A&A...642A...3Z} {642, A3}

\makeatother
\end{thebibliography}

%%%% BELOW IS THE CONTENTS OF THE main.bbl FILE
%%%% THAT HAS BEEN COMPILED OFFLINE. THIS IS NEEDED
%%%% HERE BECAUSE ARXIV WON'T COMPILE THE BBL. 

 \newcommand{\noop}[1]{}

\end{multicols}

\end{document}